\documentclass[12pt,a4paper]{report}

\usepackage[T2A]{fontenc}
\usepackage[utf8]{inputenc}
\usepackage[english]{babel}

\usepackage{lmodern}
\usepackage{graphicx}
\usepackage{amsmath,amssymb,bm}
\usepackage{siunitx}
\usepackage{array,tabularx}
\usepackage{enumitem}
\usepackage{placeins}
\usepackage{braket}
\usepackage{listings}
\usepackage{fancyvrb}

\usepackage{caption} 

\usepackage{hyperref} 
\newcommand{\centerfloat}{\centering}

\usepackage{nomencl}
\makenomenclature

\usepackage{mathtools}
\usepackage[backend=biber,style=gost-numeric,sorting=none]{biblatex} 
\begin{document}

\begin{titlepage}
	\centering
	
	\vspace*{3cm}
	
	{\LARGE \textbf{Dispersive Microwave Sensing for Quantum Computing with Floating Electrons}\par}
	\vspace{2cm}
	
	{\Large Tian\,Yiran\par}
	\vspace{2cm}
	
	\parbox{0.85\textwidth}{
		The work presented in this thesis was performed in the Floating Electron Based Quantum Information team.
	}
	
	\vfill

	{\large \today\par}
	
\end{titlepage}
\clearpage

\tableofcontents

\chapter{Introduction}\label{ch:ch1}

 This chapter introduces the concept of floating electrons on cryogenic substrates as a platform for quantum information processing. It presents the physical model of surface states, the quantization of vertical motion, and the structure of Rydberg energy levels. This chapter also describes the mechanisms by which these electrons interact with external environments and measurement circuits through their charge and spin degrees of freedom. These foundational results are used in subsequent chapters to guide device design, simulation, and experimental implementation.
 
 \section{Qubit} 
 
\begin{figure}[htbp]
	\centerfloat{
		\includegraphics[width=0.4\linewidth]{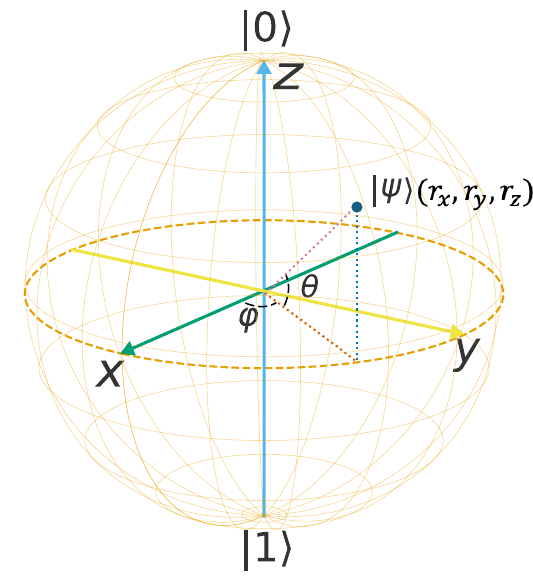}
	}
	\caption[Bloch sphere representation of a qubit]
	{Bloch sphere representation of a qubit. A pure state $\ket{\psi}$ is represented as a point on the Bloch sphere with coordinate $(r_x,r_y,r_z)$.}
	\label{fig:bloch}
\end{figure}

 A qubit, the fundamental unit of quantum computation, is a quantum system whose state is represented in a two-dimensional Hilbert space~\cite{nielsen2010quantum}. We define two orthonormal basis states, $|0\rangle$ and $|1\rangle$, which represent the logical states 0 and 1, respectively. An arbitrary pure state of a qubit is a normalized superposition of $|0\rangle$ and $|1\rangle$:
 
 \begin{equation}
 	|\psi\rangle = \alpha|0\rangle + \beta|1\rangle,
 \end{equation} 
 where $\alpha$ and $\beta$ are complex numbers, and $|\alpha|^2 + |\beta|^2 =1$. To represent a qubit geometrically, this equation can be rewritten as: $|\psi\rangle= e^{i\gamma}(\cos\!\frac{\theta}{2}\,|0\rangle+ e^{i\phi}\sin\!\frac{\theta}{2}\,|1\rangle)$, where $\theta$, $\phi$ and $\gamma$ are real numbers, we can drop the global phase $e^{i\gamma}$ as it has no observable effects, and get: 
 \begin{equation}
 	|\psi\rangle
 	= \cos\!\frac{\theta}{2}\,|0\rangle
 	+ e^{i\phi}\sin\!\frac{\theta}{2}\,|1\rangle,
 \end{equation} 
 where the $\theta$ and $\phi$ define a point on a unit 3D sphere whose north and south poles are $\lvert 0\rangle$ and $\lvert 1\rangle$, which is known as the Bloch sphere, see Fig.~\ref{fig:bloch} . This picture makes it easy to interpret single‑qubit gates as rotations of the Bloch vector~\cite{nielsen2010quantum}.

 Coherent superpositions and entanglements form the basis of quantum algorithms that outperform classical computers. However, when qubits couple to the environment, they undergo decoherence, causing a pure state to evolve into a mixed state, which is a major challenge for quantum computing. 
 To describe this process, a density matrix can be introduced. For a pure state $\lvert\psi\rangle$:
 
 \begin{equation}
 	\rho \equiv \lvert\psi\rangle\!\langle\psi\rvert
 	=\begin{pmatrix}
 		\lvert\alpha\rvert^2 & \alpha\,\beta^*\\
 		\alpha^*\,\beta & \lvert\beta\rvert^2.
 	\end{pmatrix},
 \end{equation} 
 where the diagonal elements are the populations and the off-diagonal terms represent quantum coherence. A mixed state can be represented as:
 \begin{equation}
 	\rho=\sum_i p_i\,\lvert\psi_i\rangle\!\langle\psi_i\rvert,\qquad
 	p_i\ge 0,\ \sum_i p_i=1,
 \end{equation} 
 where $i$ is an index. In Bloch representation, an arbitrary density matrix for a qubit can be written as:
 \begin{equation}
 	\rho=\tfrac{1}{2}\big(I+\bm{r}\cdot\boldsymbol\sigma\big),\quad
 	\bm{r}=(r_x,r_y,r_z),\ \ r_i=\mathrm{Tr}(\rho\,\sigma_i),
 \end{equation} 
 where $\bm{r}$ represents the Bloch vector with coordinates $(r_x, r_y, r_z)$ in three-dimensional space, see Fig.~\ref{fig:bloch}, and $\sigma_i\,(i = x,y,z)$ are Pauli matrices. The magnitude $|\bm{r}|$ indicates the distance from the origin: pure states have $|\bm{r}|=1$ (point on the Bloch sphere); mixed states have $|\bm{r}|<1 $(inside the sphere)~\cite{nielsen2010quantum}.
 
 Two timescales quantify the system decay: the longitudinal relaxation time \(T_1\), describing energy relaxation from the excited state $\ket{1}$ to the ground state $\ket{0}$ as $\Delta \rho_{11}(t) = \Delta \rho_{11}(0)\,e^{-t/T_1}$, and the transverse relaxation time $T_2$, which describes the loss of phase coherence in superpositions through decay of the off-diagonal density matrix elements: $\rho_{01}(t) = \rho_{01}(0)\,e^{-t/T_2}$.
 
 At typical experimental temperatures (millikelvin range), for floating-electron-based (FEB) qubits on solid neon, dominant decoherence mechanisms are (i) fluctuating diamagnetic susceptibility of the neon due to thermal phonons, (ii) fluctuating thermal current in nearby normal-metal electrodes (i.e., charge noise), and (iii) quasi‑statically fluctuating nuclear spins of the $^{21}\text{Ne}$ isotopes in the substrate~\cite{chen2022electron}; for FEB qubits on liquid helium, dominant decoherence mechanisms are (i) two-ripplon emission, (ii) scattering by bulk-helium phonons, and (iii) quasielastic ripplon-induced dephasing~\cite{monarkha2010decay,dykman2003qubits,jennings2024quantum}.
 
 Theoretical analyses predict spin-coherence times \(T_2\) up to $100~\text{s}$ for qubits on liquid helium~\cite{lyon2006spin} and  $0.1\,\text{ms}$ on solid neon~\cite{jennings2024quantum}, experiments have achieved charge-coherence times of about $100~\mu\text{s}$ on solid neon~\cite{zhou2022single,zhou2024electron}.

 \subsection{Physical realizations}
 Qubits can be physically realized in a variety of systems, each offering different advantages and limitations in terms of coherence, control, and scalability. Common platforms include superconducting circuits~\cite{huang2020superconducting}, trapped ions~\cite{bruzewicz2019trapped}, and cold neutral atoms~\cite{henriet2020quantum}, etc. In this thesis, the primary focus is on FEB qubits~\cite{jennings2024quantum}, which offer distinct advantages in coherence time and scalability.
 
 \paragraph{Superconducting qubits.}
 Superconducting qubits, first experimentally realized in 1999~\cite{nakamura1999coherent}, are the leading platform in quantum computing, using the macroscopic quantum states of superconducting circuits to process quantum information. These qubits operate at cryogenic temperatures, typically in dilution refrigerators at millikelvin scales, allowing for long coherence times and low-energy dissipation. Superconducting qubits are realized by integrating Josephson junctions with microwave resonators. The Josephson junction provides a nonlinear, non-dissipative inductive element that introduces anharmonocity into the energy spectrum of an LC oscillator, producing anharmonic level spacings (\(\omega_{01}\neq \omega_{12}\)). This anharmonicity allows selective driving of the \(\ket{0}\!\leftrightarrow\!\ket{1}\) transition with microwaves, enabling coherent control and high-fidelity quantum gates~\cite{kjaergaard2020superconducting}. This platform has scaled to processors with hundreds of qubits~\cite{kelly2015state,arute2019quantum}. Despite these advantages, superconducting qubits require relatively large nanocircuit structures, making it challenging to fit a large number of qubits within the limited space of a dilution refrigerator.
 
 In March 2023, RIKEN, in collaboration with Fujitsu and other Japanese research institutions, launched Japan's first domestically developed superconducting quantum computer, featuring a 64-qubit two-dimensional integrated circuit. In December 2024, the University of Science and Technology of China (USTC) unveiled the Zuchongzhi 3.0 superconducting quantum computer, featuring a 105-qubit architecture with high operational fidelities~\cite{gao2025establishing}. 
 
 \paragraph{Trapped-ion qubits.}
 Trapped-ion qubits use individual ions confined in electromagnetic traps, where quantum operations are performed using laser or microwave fields~\cite{bruzewicz2019trapped}. They exhibit long coherence times, often measured in seconds to minutes~\cite{ruster2016long,wang2017single}. High gate fidelities have been achieved, with single-qubit errors below $10^{-6}$ and two-qubit fidelities around $99.9\%$~\cite{schafer2018fast}. Despite these advantages, scalability is limited by slow gate speeds and challenges in integrating large numbers of qubits~\cite{bruzewicz2019trapped}.

 \paragraph{Cold-atom qubits.}
 Cold-atom quantum computers utilize neutral atoms, such as rubidium or cesium, cooled to near absolute zero and trapped using optical lattices or optical tweezers. This platform offers advantages in scalability due to the ability to arrange large numbers of atoms in well-ordered arrays and benefits from relatively long coherence times. However, challenges remain in achieving high-fidelity quantum gates and maintaining uniform control across extensive qubit arrays~\cite{wintersperger2023neutral,weiss2017quantum,negretti2011quantum}. Recent advancements have demonstrated entanglement between cold atoms in optical traps, highlighting progress toward practical quantum computing applications~\cite{evered2023high}.
 
 In December 2024, researchers from Lomonosov Moscow State University and the Russian Quantum Center developed a 50-qubit quantum computer based on neutral rubidium atoms. 
 
 \paragraph{FEB qubits.}
 In this work, the feasibility of using FEB qubits for quantum computing has been investigated. Floating electrons (FEs) on cryogenic substrates, such as liquid helium or solid neon, offer a promising platform for quantum computing due to their ultra-clean, low-noise environment, as electrons residing in vacuum and thus free from impurities and defects. Qubits operating in such an environment, minimizing decoherence sources and allowing for remarkably long coherence times, particularly for spin states, with theoretical predictions exceeding 100 seconds for qubits on liquid helium~\cite{lyon2006spin}. Unlike superconducting qubits, which are limited by fabrication disorder, or trapped ions, which require complex laser-based operations, FEB qubits use simpler electrode-based control for trapping and manipulation. Their scalability is enhanced by the potential for high-density qubit packing within a cryogenic system, avoiding the space limitations faced by superconducting circuits~\cite{jennings2024quantum}. Additionally, electron charge states couple to microwave photons, facilitating fast, high-fidelity quantum gate operations and integration with superconducting resonators~\cite{kawakami2023blueprint}. These advantages make electrons on cryogenic substrates a promising candidate for scalable quantum computing while maintaining long coherence and high gate fidelity.
 
 In 2022, a single-electron charge qubit on solid neon was realized, two-qubit gates and spin-qubit operation remain under active development~\cite{zhou2022single,zhou2024electron,jennings2024quantum}.
 
 \subsection{Qubit gates and fidelity}
 Having defined the qubit, the following section introduces the operations used to manipulate it. Quantum gates are unitary transformations that map qubit states to other states. In this section a brief introduction of single‑qubit and two‑qubit gates and fidelity will be shown.
 
 \subsubsection{Single-qubit gates}
 
 A single-qubit gate is represented by a $2\times 2$ unitary matrix $U$. Unitarity ensures that the transformation preserves state normalization and the inverse operation is given by $U^\dagger$~\cite{nielsen2010quantum}.
 Typical examples include the Pauli $X$ gate, which swaps $\ket{0}$ and $\ket{1}$, the Pauli $Z$ gate, which applies a relative phase to $\ket{1}$, and the Hadamard gate
 \begin{equation}
 	H=\frac{1}{\sqrt{2}}
 	\begin{pmatrix}
 		1 & 1\\[2pt]
 		1 & -1
 	\end{pmatrix},
 \end{equation}
 which maps $\ket{0}\mapsto(\ket{0}+\ket{1})/\sqrt{2}$ and $\ket{1}\mapsto(\ket{0}-\ket{1})/\sqrt{2}$.
 Any single‑qubit unitary can be decomposed into rotations about different axes of the Bloch sphere plus an overall phase.
 
 \subsubsection{Multi-qubit gates}
 Two-qubit gates generate entanglement and are therefore essential for universal quantum computation. The typical example is the controlled-NOT (CNOT), which acts in the computational basis as
 
 \begin{equation}
 	\mathrm{CNOT}=
 	\begin{pmatrix}
 		1&0&0&0\\
 		0&1&0&0\\
 		0&0&0&1\\
 		0&0&1&0
 	\end{pmatrix},
 \end{equation}
 flipping the second qubit if and only if the first qubit is $\ket{1}$.
 
 Another example is the iSWAP gate, which is a two-qubit gate that swaps the amplitudes of $|01\rangle$ and $|10\rangle$ and multiplies each swapped term by $i$. The states $|00\rangle$ and $|11\rangle$ stay the same.
 A standard way is to consider the exchange Hamiltonian:
 
 \begin{equation}
 	H=\hbar g\big(\sigma_1^{+}\sigma_2^{-}+\sigma_1^{-}\sigma_2^{+}\big)
 	=\frac{\hbar g}{2}\big(\sigma_x\!\otimes\!\sigma_x+\sigma_y\!\otimes\!\sigma_y\big).
 \end{equation}
 This represents an energy exchange between two qubits ($g$ is the coupling constant). The time evolution under this Hamiltonian is given by:
 
 \begin{equation}
 	U(t) = e^{-i H t/\hbar} =
 	\begin{pmatrix}
 		1 & 0 & 0 & 0 \\
 		0 & \cos(gt) & -i\sin(gt) & 0 \\
 		0 & -i\sin(gt) & \cos(gt) & 0 \\
 		0 & 0 & 0 & 1
 	\end{pmatrix},
 \end{equation} 
 written in the computational basis $\{|00\rangle, |01\rangle, |10\rangle, |11\rangle\}$. At time $t = \pi/(2g)$, this becomes
 
 \begin{equation}
 	U\left(\frac{\pi}{2g}\right) =
 	\begin{pmatrix}
 		1 & 0 & 0 & 0 \\
 		0 & 0 & i & 0 \\
 		0 & i & 0 & 0 \\
 		0 & 0 & 0 & 1
 	\end{pmatrix}
 	= U_{\mathrm{iSWAP}},
 \end{equation}  
 which defines the iSWAP gate.
 By linearity, for example, for an arbitrary superposition on the first qubit with the second in $|\!\uparrow\rangle$:
 \begin{equation}
 	(\alpha|g\rangle+\beta|e\rangle)\!\otimes\!|\!\uparrow\rangle
 	\;\longrightarrow\;
 	|g\rangle\!\otimes\!\big(\alpha|\!\uparrow\rangle+i\beta|\!\downarrow\rangle\big),
 \end{equation} 
 so the excitation amplitude $\beta$ is transferred to the second qubit and picks up a phase $i$.
 
 
 Arbitrary single-qubit gates plus two-qubit gate (e.g., CNOT) form a universal set, that is, any multiple‑qubit logic gate can be constructed from one- and two-qubit gates~\cite{barenco1995elementary,divincenzo1995two}
 
 \subsubsection{Fidelity}
 To measure how close two states are, we use the fidelity between density matrices $\rho$ and $\sigma$, defined by
 
 \begin{equation}
 	F(\rho, \sigma) \equiv \mathrm{Tr}\!\left(\sqrt{\sqrt{\rho} \, \sigma \, \sqrt{\rho}}\right),
 \end{equation}
 which lies between 0 and 1. A fidelity of one means the states are identical, while zero indicates that they are orthogonal~\cite{nielsen2010quantum}. 
 
 When evaluating a quantum gate rather than a single state preparation, we want a metric that captures how closely the implemented operation $\mathcal{E}$ approximates the ideal unitary gate $U$. One approach is to prepare a pure input state $\ket{\psi}$, apply the ideal gate and the experimental gate separately, and compute the fidelity between the resulting states. Because the result generally depends on $\ket{\psi}$, averaging this fidelity over all pure inputs one can get average gate fidelity~\cite{nielsen2002simple}:
 
 \begin{equation}
 	F_{\mathrm{gate}}(\mathcal{E}, U) = \int d\psi \, 
 	\langle \psi | U^\dagger \, \mathcal{E} \!\left( |\psi\rangle \langle\psi| \right) U | \psi \rangle.
 \end{equation}
 In Sec.~\ref{sec:ne}, we theoretically estimated the fidelities of single- and two-qubit gates of the spin qubits of electrons on solid neon.

 \section{Properties of Floating Electrons} 
 It is important to understand the behavior of electrons in a cryogenic environment. This section therefore discusses the fundamental properties of FEs on liquid helium and solid neon.
 
 \subsection{Bound States and Wave Function}\label{sec:bound_states}
 One key advantage of using liquid helium and solid neon as a cryogenic substrate for a 2DES is their extremely low polarizability. The dielectric constant of liquid helium is close to 1, with values of approximately 1.056 for $^{4}\text{He}$ and 1.042 for $^{3}\text{He}$, making them an ideal medium for surface-bound electrons. The dielectric constant of solid neon is 1.244, which provides stronger trapping and better confinement of the electrons~\cite{cole1969image,cole1971electronic,jin2020quantum}. 
 
\begin{figure}[htbp]
	\centerfloat{
		\includegraphics[width=0.4\linewidth]{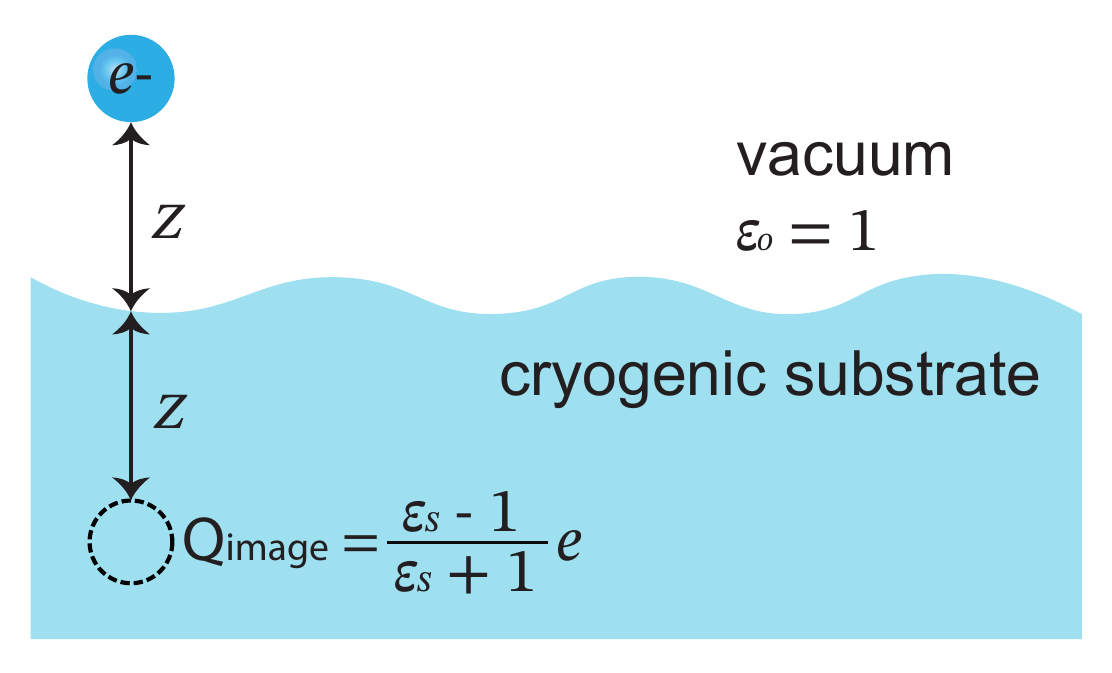}
	}
	\caption[Schematic representation of the image charge effect]
	{Schematic representation of the image charge effect for an electron hovering above a cryogenic substrate with dielectric constant $\varepsilon_s$.}
	\label{fig:image}
\end{figure}

 In FEB qubits system, image charge effects play a crucial role in trapping and manipulating electrons near the surface. When an electron is positioned close to a dielectric interface, it induces a polarization charge within the substrate, leading to an image potential that attracts the electron toward the surface, the scheme is demonstrated in Fig.~\ref{fig:image}. The effective image charge can be expressed as:
 \begin{equation}
 	q' = +e \frac{\varepsilon_{\text{s}} - 1}{\varepsilon_{\text{s}} + 1} =\;+e\,\Lambda,
 	\qquad \Lambda\equiv\frac{\varepsilon_s-1}{\varepsilon_s+1},
 \end{equation}
 where $\varepsilon_s$ is the relative permittivity of the cryogenic substrate. This produces an image potential:
 \begin{equation}
 	V_{\mathrm{im}}(z)\;=\;-\frac{e^{2}}{4\pi\varepsilon_0}\,\frac{\Lambda}{4z},
 \end{equation}
 where  $z$  is the electron’s distance from the cryogenic surface,  $e$  is the elementary charge, and  $\varepsilon_0$  is the permittivity of vacuum. This interaction leads to a hydrogen-like energy spectrum, the so-called Rydberg states which will be introduced later in this section.

 The potential energy of an electron at distance
 $z$ can be approximated as~\cite{jennings2024quantum}:
 \begin{equation} \label{eq:potential}
 	V(z) =
 	\begin{cases} 
 		V_0, & z \leq 0 \\[8pt]
 		-\frac{e^2}{4\pi\varepsilon_0} \frac{\Lambda}{4} \frac{1}{(z+z_0)} + e z E_{\perp}, & z > 0,
 	\end{cases}
 \end{equation}
 where $E_{\perp}$ is the magnitude of an external electrical field applied normal to the surface, \(z_0\) is a constant value introduced to be consistent with the spectroscopy measurement ($z_0=0.1\,\text{nm}$ liquid helium-4 and $=0.23~\text{nm}$ nm for solid neon)~\cite{jin2020quantum,monarkha2004two}. As a result, the electron experiences an attractive force toward the cryogenic substrate. However, due to the negative electron affinity of the substrate, the electron remains confined above the surface, experiencing a potential barrier \( V_0 \), which is approximately \(\sim 1\) eV for liquid \(^4\)He and \(\sim 0.7\) eV for solid neon~\cite{Monarkha2004chp1,jin2020quantum}.
 
 The 1D Schrödinger equation of a FE for motion normal to the surface is then~\cite{cole1997surface}:
 \begin{equation} \label{eq:schordinger}
 	\left( -\frac{d^2}{dz^2} + \frac{2m}{\hbar^2} V(z) \right) \psi = \frac{2m}{\hbar^2} E \psi.
 \end{equation}

\begin{figure}[htbp]
 	\centerfloat{
 		\includegraphics[width=0.6\linewidth]{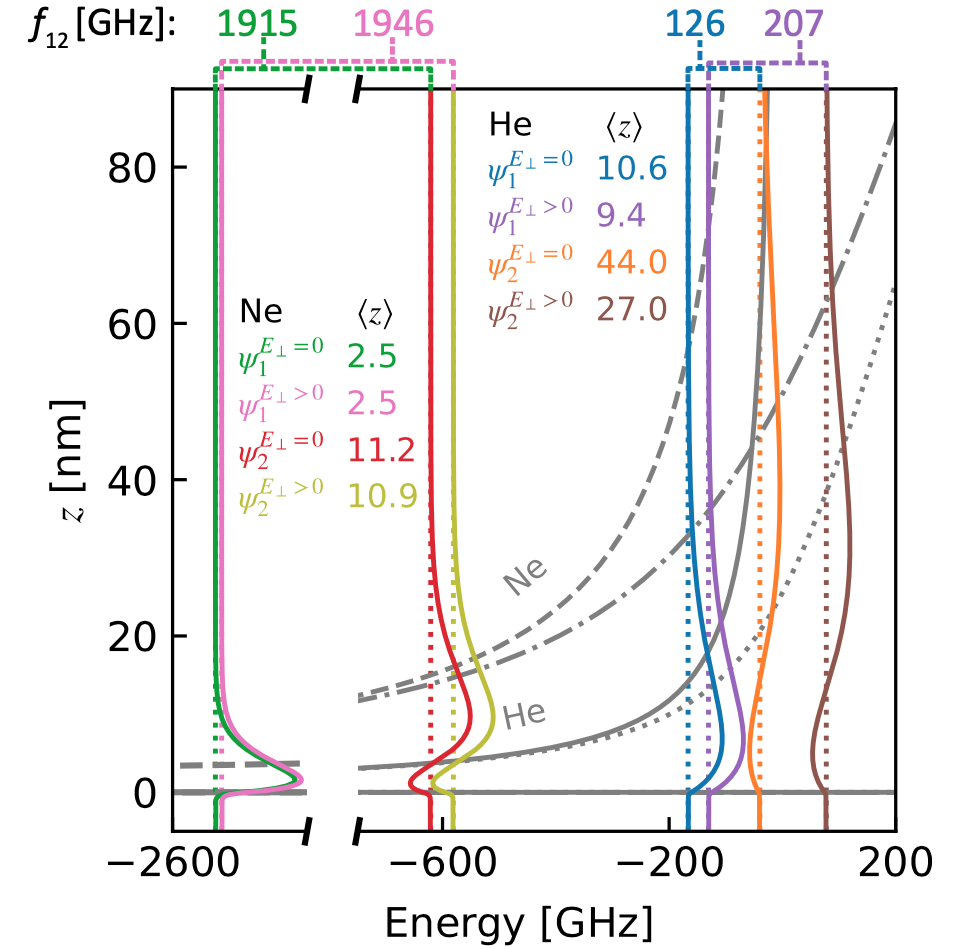}
 	}
 	\caption[Numerically solved eigenenergies and wavefunctions for electrons on helium and neon]{%
 		Numerically solved eigenenergies (colored dotted lines), wavefunctions \( \psi_n \) (colored solid lines), first-excited transition frequencies \( f_{12} \), and average electron position \( \langle z \rangle_n \) (values in the legend) for electrons confined above liquid helium and solid neon. The calculations correspond to the Rydberg ground state (\( n=1 \)) and the first-excited Rydberg state (\( n=2 \)). The solid, dashed, and dot-dashed gray curves represent the potential energy of an electron on helium and neon under perpendicular electric fields of \( E_\perp = 0 \) and \( E_\perp = 15 \) kV/m, respectively. For clarity, helium potentials are not shown below \(-800\) GHz. The figure is taken from~\cite{jennings2024quantum}.}
 	\label{fig:allstates}
 \end{figure}

 For electrons on liquid helium, the dielectric constant is close to 1, resulting in a relatively weak image potential. This leads to shallow bound states, with an electron average position $\langle z_1 \rangle \simeq 10.6\,\text{nm}$ above the helium surface at $E_\perp = 0$ (see Fig.~\ref{fig:allstates}). In contrast, solid neon, with a higher dielectric constant, produces a stronger image potential, binding electrons more tightly to the surface at a shorter distance, with $\langle z_1 \rangle \simeq 2.5\,\text{nm}$ at $E_\perp = 0$ (see Fig.~\ref{fig:allstates}) ~\cite{jennings2024quantum}.

 Since \( V_0 \) is much larger than the typical eigen-energies of the electron in the potential given by Eq.~\eqref{eq:potential}, it is common to adopt a simplified model by taking the limits \( V_0 \to \infty \) and \( z_0 \to 0 \). This approximation allows for an analytical solution to the Schrödinger equation, yielding the corresponding quantized energy levels alone $z$ direction:
 
 \begin{equation} \label{eq:energyz}
 	E_{n_z} = -R_{\infty} \left( \frac{\Lambda}{4} \right)^2 \frac{1}{n_z^2},
 \end{equation}
 where \( R_{\infty} = \frac{m_e e^4}{8 \varepsilon_0^2 h^2} \) is the Rydberg constant, \( m_e \) is the electron mass, \( h \) is the Planck constant, and \( n_z \) is the quantum number describing the direction perpendicular to the surface. These quantized eigenstates are called Rydberg states~\cite{Monarkha2004chp1}.
 
 The numerical solutions of the Schrödinger equation for electron on liquid helium and solid neon are presented in Figure~\ref{fig:allstates} which is taken from~\cite{jennings2024quantum}.

 \subsection{Mobility}


 To describe the in–plane transport of free electrons we use the Drude model.  
 In this model, electrons behave as classical particles. They move under the electric field and are scattered at random times, with an average time between collisions $\tau$.  
 The mobility $\mu$ characterizes how easily electrons move through a system and, in the Drude model, is defined as
 \begin{equation}
 	\mu = \frac{e\tau}{m_e},
 \end{equation}
 where $e$ is the elementary charge and $m_e$ is the electron mass.  For a
 two-dimensional electron layer with surface density $n_e$, the Drude model then
 gives the complex conductivity
 \begin{equation}
 	\sigma^{2\mathrm{D}}
 	= \frac{n_e e^2 \tau}{m_e}\,\frac{1}{1+i\omega\tau},
 	\label{eq:drude}
 \end{equation}
 which will be used later in this thesis.

\begin{figure}[htbp]
 	\centerfloat{
 		\includegraphics[width=0.6\linewidth]{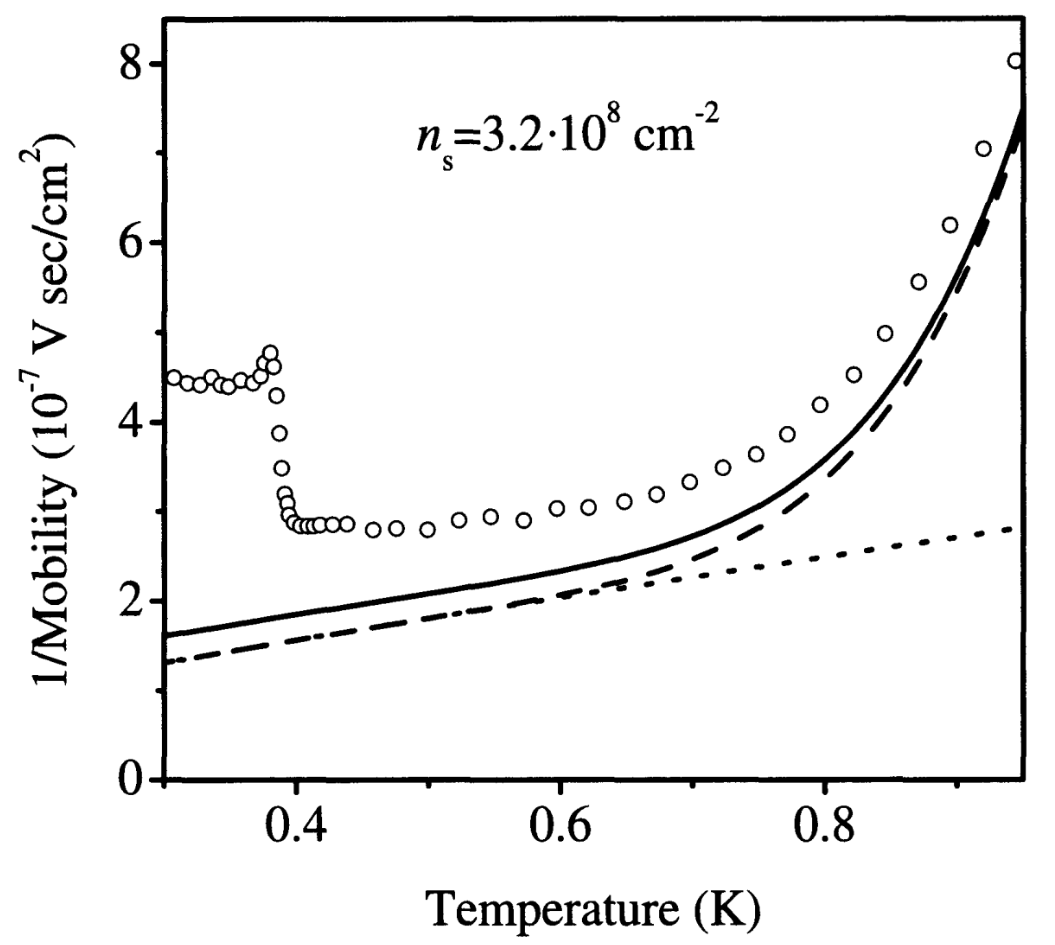}
 	}
 	\caption[Inverse mobility vs. temperature for FEs on liquid $^4$He]{%
 		Inverse mobility vs. temperature for FEs with electron density
 		$n_e = 3.2 \times 10^8\ \mathrm{cm}^{-2}$ on the surface of liquid $^4\mathrm{He}$.
 		The dashed curve represents the theoretical result of the single-electron approximation considering both ripplon and helium gas atom scattering. The dotted curve represents ripplon scattering only; the solid curve is the many-electron theory prediction, where the electron distribution function is shaped by electron-electron interactions; the open circles are the experimental data~\cite{mehrotra1984density}. The figure is taken from~\cite{monarkha2013two}.}
 	\label{fig:mobility}
 \end{figure}

 
 For electrons on liquid helium with density $n_e = 3.2 \times 10^8\,\text{cm}^{-2}$, the mobility shows a sudden decrease when the temperature is lowered to about $0.4\,\text{K}$ (see Fig.~\ref{fig:mobility}). This behavior indicates the formation of a Wigner solid (WS) -- a crystalline phase of electrons first predicted by Wigner in 1934~\cite{wigner1934interaction} and first experimentally observed by Grimes and Adams in 1980~\cite{grimes1979evidence,grimes1980crystallization}.

 In the WS regime, electrons form a hexagonal lattice when the plasma parameter $\Gamma = \frac{\pi^{1/2} n_s^{1/2} e^2}{k_B T}$ exceeds a critical value of around 135, depending on the electron density $n_s$ and temperature $T$~\cite{grimes1979evidence,kajita1985wigner,grimes1980crystallization}, where $k_B$ is the Boltzmann constant and $e$ is the elementary charge.This discovery became a key milestone in the study of strongly correlated 2D electron systems.

 At temperatures below $1\,\mathrm{K}$, electrons on liquid helium exhibit exceptionally high mobility, exceeding $10^6\,\mathrm{cm^2/V\,s}$ (Fig.~\ref{fig:mobility}). The mobility is mainly limited by scattering from ripplons and a small background of helium gas atoms~\cite{leiderer2025surface}. These values are substantially higher than those in typical semiconductors and GaAs/AlGaAs heterostructures, indicating an exceptionally clean two-dimensional transport environment.

 As the electron density increases, electron–electron interactions and surface effects become important and the mobility begins to drop.
 For an electron density of $n_s = 3.2\times10^8\,\mathrm{cm^{-2}}$, Fig.~\ref{fig:mobility} shows that the mobility drops sharply at $T \approx 0.4\,\mathrm{K}$, indicating the formation of a Wigner crystal.
 The maximum areal density of FEs on a liquid substrate is limited by a surface-charge instability that occurs at $q \sim \kappa$, where $q$ is the in-plane wave number and $\kappa$ is the inverse capillary length of the liquid~\cite{monarkha2004two}.
 This sets a critical density $n_s^{(c)}$, which for liquid $^4$He is $n_s^{(c)} \approx 2\times10^{9}\,\mathrm{cm^{-2}}$~\cite{monarkha2013two}. When the density approaches this value, the system becomes unstable and the mobility drops sharply.
 To push electron densities higher, we could replace the liquid dielectric with a solid dielectric whose permittivity is close to one.
 Since helium does not solidify at zero temperature, solid neon and solid hydrogen are considered suitable cryogenic solid substrates~\cite{monarkha2013two}.

\begin{figure}[htbp]
	\centerfloat{
		\includegraphics[width=0.6\linewidth]{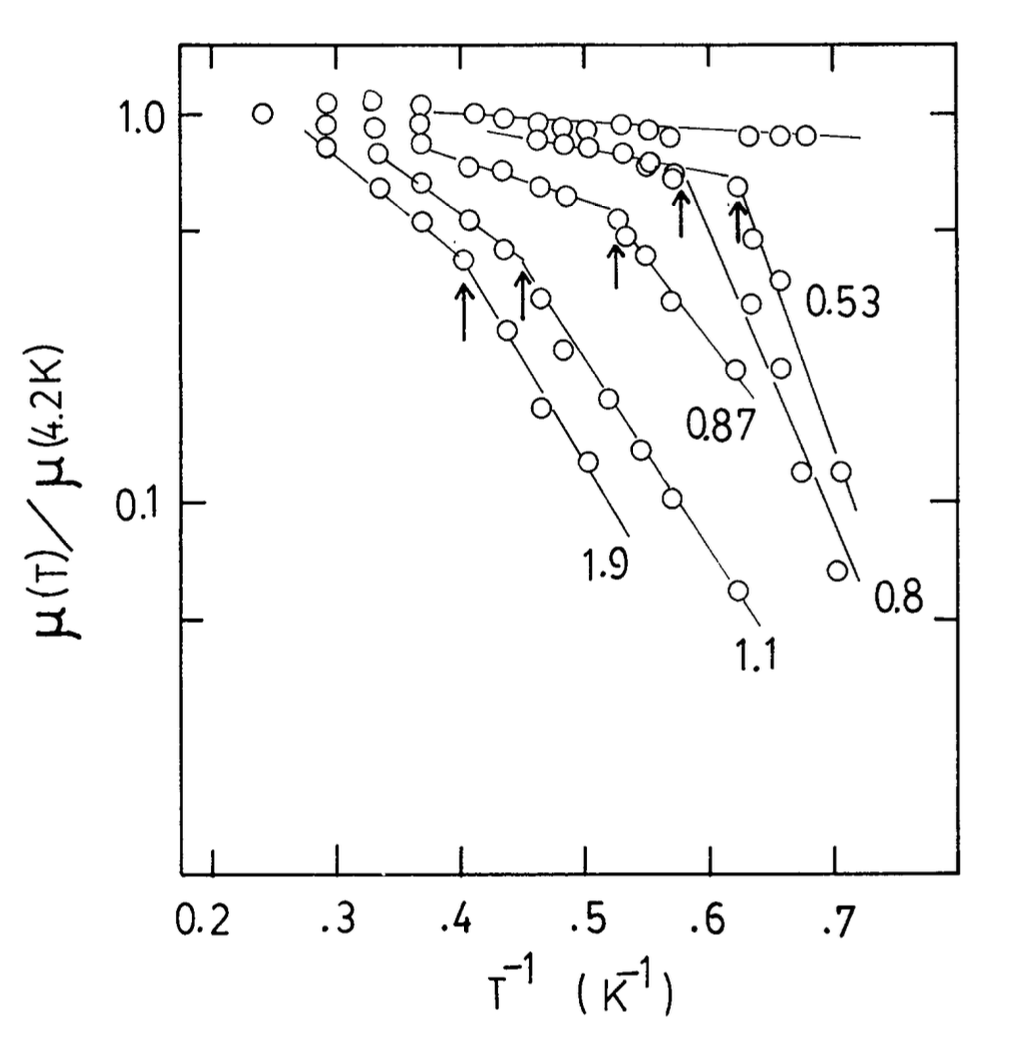}
	}
	\caption[Pinned Wigner crystal on solid neon]{%
		Pinned Wigner crystal on solid neon.
		\(\mu(T)/\mu(4.2\,\mathrm{K})\) versus \(T^{-1}\) for several datasets (arrows mark the crossover between high- and low-\(T\) regimes). The decrease of conductivity at low \(T\) is interpreted as pinning of the Wigner crystal by the random surface potential, so the pinned crystal contributes little to electronic conduction. The figure is taken from~\cite{kajita1985wigner}.}
	\label{fig:mobilityneon}
\end{figure}

 On solid neon, electrons are much more strongly bound in the vertical direction than on liquid helium, K. Kajita reported
 that a stable 2DES with areal density up to $3\times10^{10}\,\mathrm{cm^{-2}}$ can be formed on solid neon~\cite{kajita1984new}.
 The mobility of electrons on solid neon is strongly dependent on the electron density. For $n_s\gtrsim 10^{9}\,\mathrm{cm^{-2}}$, the conductivity of FEs increases nonlinearly with electron density and eventually saturates. Comparison with helium data shows that this nonlinearity arises mainly from many-electron (density) effects, rather than from changes in the plasma parameter $\Gamma$~\cite{monarkha2013two}.  At lower densities, the mobility is nearly independent of $n_s$ up to $\Gamma\approx 100$, while at higher densities ($\Gamma\approx 135$) the conductivity drops with cooling, which Kajita attributes to pinning of the Wigner crystal by the random surface potential, see Fig.~\ref{fig:mobilityneon}. The highest mobility measured on solid neon is $\mu \approx 8\times10^{3}\,\mathrm{cm^{2}/V\,s}$, which is significantly lower than the $\sim 10^{6}\,\mathrm{cm^{2}/V\,s}$ measured for electrons on the liquid helium surface below $1\,\mathrm{K}$, primarily due to surface roughness of the solid neon substrate~\cite{kajita1985wigner}.

 \subsection{Rydberg transitions of FEs on liquid helium}
 As noted in Sec.~\ref{sec:bound_states}, FEB qubits on liquid helium exploit Rydberg transitions: electrons above liquid helium occupy discrete energy states, and their transition frequencies lie in the microwave (GHz) range.
 
 In this thesis,  the Rydberg transition primarily refers to the excitation from the ground state \({n=1}\) to the first excited state \({n=2}\). The transition frequency, given by $f_{12} = \frac{E_2 - E_1}{h}$, depends on the confining potential $V(z)$.
 
 \subsubsection{Stark Shift}
 In most experimental cases, an electric field perpendicular to the surface of the substrates is applied to control the 2D electron density in xy plane. Solving Eq.~\ref{eq:schordinger} results in a linear Stark shift:
 
 \begin{equation} \label{eq:stark}
 	\Delta E_n \approx e E_\perp \langle n | z | n \rangle.
 \end{equation}
 
\begin{figure}[htbp]
	\centerfloat{
		\includegraphics[width=0.6\linewidth]{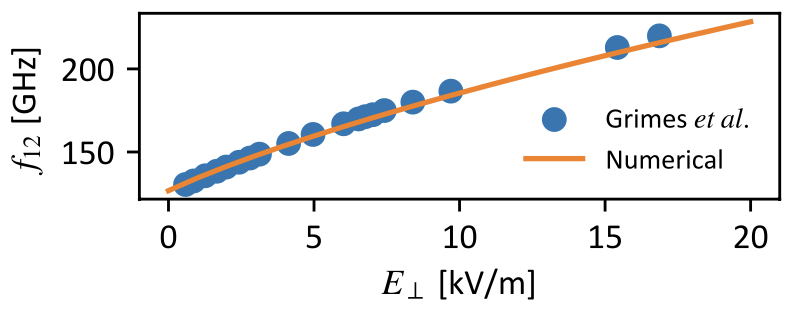}
	}
	\caption[Comparison of calculated transition energy and experimental data]{%
		Comparison of the numerically calculated transition energy and experimental data from Ref.~\cite{grimes1976spectroscopy} for the Rydberg ground to first-excited state transition. The figure is taken from~\cite{jennings2024quantum}.}
	\label{fig:stark}
\end{figure}

 It is important to note that the Stark shift is positive, indicating that the transition frequencies increase with the applied perpendicular electric field $E_\perp$. Experimental data supporting this behavior can be found in~\cite{grimes1976spectroscopy}. A comparison between the numerical calculations and the experimentally measured transition energy between the Rydberg ground state and the first-excited Rydberg state is presented in Fig.~\ref{fig:stark}.
 
 The Stark effect therefore provides a way to tune the Rydberg transition
 frequency $f_{12}$ with $E_\perp$.

 \section{Coupling Mechanisms}
 In this section, we examine the mechanisms by which FEs on cryogenic substrates interact with environments through their charge and spin degrees of freedom, enabling the control and manipulation of FEB qubits.

 \subsection{Charge-Photon Coupling}
\begin{figure}[htbp]
	\centerfloat{
		\includegraphics[width=0.6\linewidth]{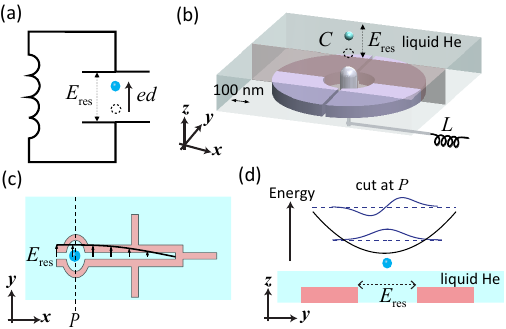}
	}
	\caption[Charge-photon coupling mechanisms for electrons on helium/neon]{%
		(a) Schematic representation of the charge-photon coupling mechanism for a floating electron (light blue circle) interacting with an LC resonator. The electron, possessing an electric dipole \(ed\), experiences an electric field \(E_{\text{res}}\) within the resonator capacitor. (b) Illustration of the Rydberg state transition coupling to a lumped-element LC resonator. The light blue sphere (dashed circle) represents an electron transitioning between the ground and first-excited Rydberg states. (c,d) Coupling of the orbital states of an electron to a superconducting quarter-wavelength resonator. The pink regions indicate the resonator structure, while the black line depicts the harmonic confinement potential along the y-axis. The blue curves represent the wavefunctions of the ground and first-excited orbital states. This setup is used to investigate charge and spin dynamics of electrons on helium/neon, facilitating their integration into hybrid quantum architectures. The figure is taken from~\cite{jennings2024quantum}.}
	\label{fig:dipole}
\end{figure}

 The interaction between a floating electron and an LC resonator is illustrated in Figure~\ref{fig:dipole}(a)~\cite{jennings2024quantum}. The electron experiences an electric field  $E_{\text{res}}$  in the capacitor and forms an electric dipole $ed$. This interaction is governed by the Hamiltonian:
 
 \begin{equation}
 	H_{\text{int}} = e E_{\text{res}} d = \hbar g_c (a + a^\dagger),
 \end{equation}
 where  $g_c$  represents the charge-photon coupling strength, given by  $g_c = e\alpha V_0 / \hbar$, here,  $\alpha$  is the differential lever arm~\cite{ibberson2021large}, and  $V_0$  denotes the vacuum fluctuation voltage.
 
 A lumped-element LC resonator, shown in Figure~\ref{fig:dipole}(b), has been proposed for detecting Rydberg state transitions of single electrons on helium~\cite{kawakami2023blueprint}. The transition between the Rydberg ground and first excited states is driven by a microwave field, and its occurrence is determined from changes in the resonator’s transmission or reflection. Due to the geometry of the resonator, the differential lever arm is relatively small ($\alpha = 0.01$)~\cite{kawakami2023blueprint}, resulting in a charge-photon coupling strength of  $g_c / 2\pi \approx 0.1 \,\text{MHz}$ with the help of a high-quality factor lumped LC resonator~\cite{ibberson2021large,vigneau2023probing,ahmed2018radio,apostolidis2020quantum}. Achieving strong coupling is difficult due to Rydberg state relaxation rates, but the system remains sensitive to electron transitions~\cite{kawakami2023blueprint}. Experiments with many electrons have validated the principle of this measurement method~\cite{jennings2024quantum}.
 
 An alternative approach employs a superconducting resonator, as depicted in Figure~\ref{fig:dipole}(c-d)~\cite{koolstra2019coupling,koolstra2019trapping}. The $y$-component of the electric field couples to the electron’s orbital motion along the $y$-axis. Comparing with the lumped-element LC circuit, this design increases  $\alpha$  to 0.03, allowing for a charge-photon coupling strength of  $g_c / 2\pi = 4.8 \text{MHz}$ at  $\omega_r / 2\pi \approx 6\,\text{GHz}$. However, strong coupling remains limited by a large charge linewidth ( $\gamma_c / 2\pi = 77 \text{MHz}$), likely caused by helium surface fluctuations~\cite{schuster2010proposal}. 
 
 A significant breakthrough was achieved using electrons on solid neon, where improved charge coherence eliminated major decoherence sources. Strong coupling was demonstrated with  $g_c / 2\pi = 3.5  \text{MHz}$,  $\gamma_c / 2\pi = 1.7 \text{MHz}$, and  $\kappa / 2\pi = 0.4 \text{MHz}$~\cite{zhou2022single}. Charge coherence times reached $100~\mu\text{s}$, enabling high-fidelity quantum operations with a $99.97\%$ gate fidelity~\cite{zhou2024electron,zhou2022single}. However, surface roughness in solid neon introduces potential scalability issues, requiring further improvements in fabrication and neon flim growth processes.

 \subsection{Spin-Charge Coupling}
 To access the spin state of an electron, it is necessary to establish coupling between the spin and charge degrees of freedom. Since there is no intrinsic spin-charge coupling for FEs, an artificial mechanism must be introduced. Two primary approaches have been proposed: spin-Rydberg coupling, which exploits the magnetic field dependence of Rydberg states~\cite{kawakami2023blueprint}, and spin-orbit coupling, where a magnetic-field gradient links spin to the orbital motion~\cite{pioro2008electrically,nowack2007coherent,tokura2006coherent}. Both mechanisms convert spin information into charge dynamics and can be used for electric–dipole spin resonance (EDSR) and dispersive spin readout .

 \subsubsection{Spin-Rydberg Coupling}
 Spin-Rydberg coupling introduced in Ref.~\cite{kawakami2023blueprint} relies on the fact that the energy of Rydberg states shifts with an applied magnetic field, making the transition frequency spin-dependent. This allows for dispersive readout of the spin state using an LC resonator~\cite{kawakami2023blueprint}. The concept is illustrated in Fig.~\ref{fig:dipole}(b), where an electron in a Rydberg state couples to an LC resonator.
 This method offers the potential for quantum-non-demolition (QND) spin readout by monitoring the reflection or transmission of a microwave signal. However, experimental challenges remain, primarily related to the finite lifetime of Rydberg states and the precise control required over the local magnetic environment~\cite{kawakami2023blueprint}.

 \subsubsection{Spin-Orbit Coupling}
 Spin-orbit coupling is induced by a magnetic field gradient transverse to the static Zeeman field, allowing spin manipulation via EDSR. There are two primary ways to generate this gradient: (1) using a current-carrying superconducting wire, which allows dynamic control~\cite{schuster2010proposal,zhang2012spin,dykman2023spin}, or (2) placing local ferromagnets, which provide a stronger but static gradient~\cite{kawakami2023blueprint,jennings2024quantum}.
 In the first approach, a superconducting wire carrying a current creates a magnetic field gradient that can be turned on and off, making it useful for selective qubit operations~\cite{zhang2012spin}. However, the field amplitude is limited by the critical current of the superconductor~\cite{dykman2023spin,kawakami2013excitation}. In the second approach, lithography-defined ferromagnets are placed near the electron. They provide strong, static transverse magnetic field gradients, which enhance spin–orbit coupling and allow for higher EDSR Rabi frequencies~\cite{pioro2008electrically}.
 The drawback is that the static gradient cannot be switched off, which makes the spin energy splitting more sensitive to charge noise, and can limit coherence and tunability~\cite{kawakami2016gate,kha2015micromagnets}.

 \subsection{Spin-Photon coupling}
 
 Spin–photon coupling links the electron spin to electromagnetic field modes. 
 For FEB qubits, spin-photon coupling is mediated by the electron charge: the resonator electric field couples first to the electron charge, and the spin degree of freedom is then coupled with the charge via spin-orbit or spin-Rydberg
 mechanisms, as discussed above. The resulting spin–photon coupling strength $g_s$ is set by the charge-photon coupling strength $g_c$ and the degree of spin–charge hybridization.
 
 In the following,the spin–photon coupling for a single electron on helium and then on solid neon is discussed.
 
 \subsubsection{Spin-photon coupling for a single electron on helium}
 
\begin{figure}[htbp]
	\centerfloat{
		\includegraphics[width=0.5\linewidth]{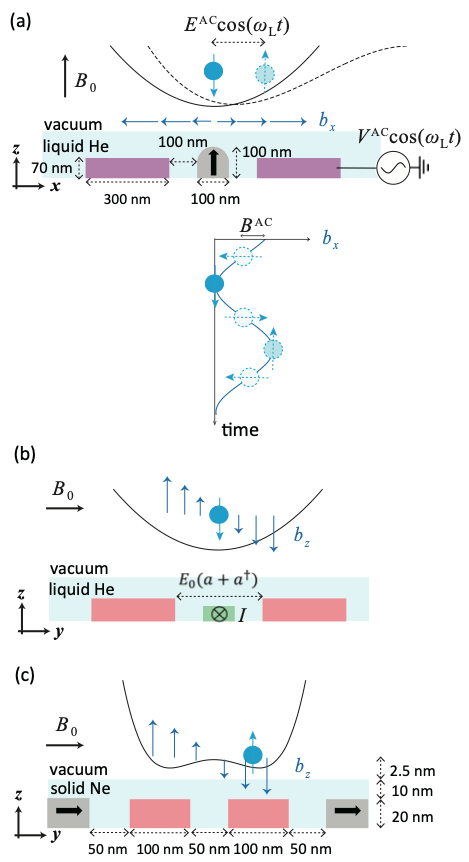}
	}
	\caption[Spin-orbit coupling schemes for electrons above helium or neon]{%
		Spin-orbit coupling schemes for electrons floating above helium or neon.
		(a) A central ferromagnetic pillar (dark gray) is surrounded by two outer electrodes (purple, see Ref.~\cite{kawakami2023blueprint}). An AC voltage \(V^{\mathrm{AC}}\cos(\omega_L t)\) applied to an electrode induces an AC electric field \(E^{\mathrm{AC}}\cos(\omega_L t)\), cyclically shifting the electron’s position. In the presence of an external field \(B_0\) along \(z\), the transverse field component \(b_x\) develops a gradient in \(x\), shown by the blue arrows. This converts the AC electric field into an effective AC magnetic field \(B^{\mathrm{AC}}\) driving EDSR.
		(b) A superconducting resonator (pink) generates a quantized electric field \(E_0 (a + a^\dagger)\) (indicated by arrows) for coupling to the electron~\cite{koolstra2019coupling,zhou2022single}. A current-carrying central pin (green) produces a stray field \(b_z\) with a gradient along \(y\), while an external field \(B_0\) is applied along \(y\)~\cite{schuster2010proposal}.
		(c) A similar configuration, but with two ferromagnets (dark gray) magnetized along the \(y\)-axis to generate \(b_z\). The field gradient \(\Delta b = \partial b_z / \partial y\) is measured as \(0.1\,\mathrm{mT/nm}\) for a cobalt structure with dimensions \(1.5\,\mu\mathrm{m} \times 1.5\,\mu\mathrm{m} \times 20\,\mu\mathrm{m}\)~\cite{pioro2008electrically}. The figure is taken from~\cite{jennings2024quantum}.}
	\label{fig:spinorbit}
\end{figure}

 In the presence of an external magnetic field  $B_0$  and a magnetic field gradient, spin-photon coupling can be achieved via EDSR. The coupling mechanism relies on an AC electric field  $E^{\mathrm{AC}} \cos(\omega_L t)$ , which modulates the electron’s lateral position. If the magnetic field gradient is perpendicular to the modulation direction, an effective AC magnetic field  $B^{\mathrm{AC}}$  is induced, driving spin transitions via EDSR~\cite{pioro2008electrically}. The magnitude of  $B^{\mathrm{AC}}$  is determined by the magnetic field gradient  $\Delta b$  and the zero-point electric field fluctuation, and it is given by~\cite{pioro2008selective,schuster2010proposal}:
 
 \begin{equation}\label{eq:Bac}
 	B^{AC} = \Delta b \frac{e E^{AC} l_0^2 \omega_0}{2 \hbar (\omega_0^2 - \omega_L^2)},
 \end{equation}
 where  $l_0$  is the characteristic confinement length and  $\omega_0$  is the orbital frequency. The schematic of the experimental setup is shown in Fig.~\ref{fig:spinorbit}(a). The ferromagnetic structure generates a field gradient parallel to the helium surface, enabling spin-orbit interaction. The Rabi frequency for this setup has been estimated to reach $100$ MHz, demonstrating its feasibility for fast spin control while maintaining long coherence times~\cite{kawakami2023blueprint}.

 Spin-photon coupling is further enhanced by tuning the spin transition frequency  $\omega_L$  to match the cavity resonance frequency. The strong coupling condition  $g_s > \gamma_s, \kappa$ , where  $\gamma_s$  and  $\kappa$  are the spin and cavity decay rates, respectively, can be satisfied by carefully engineering the system parameters~\cite{schuster2010proposal}. In the case of single-electron confinement on helium, spin coherence times are expected to be long due to the weak spin-orbit interaction~\cite{lyon2006spin}. Fig.~\ref{fig:spinorbit}(b) illustrates a superconducting coplanar resonator used to enhance spin-photon coupling~\cite{koolstra2019coupling,zhou2024electron}.

 \subsubsection{Spin-photon coupling for a single electron on neon}
 To achieve spin-photon coupling for electrons on neon, an artificial spin-orbit interaction must be introduced. Recent charge-qubit experiments show that single electrons on solid neon exhibit a spectrum resembling that of a semiconductor double quantum dot (DQD), likely due to surface roughness~\cite{zhou2022single,zhou2024electron}. Since charge-photon coupling—and consequently spin-photon coupling—is generally stronger in a DQD than in a single quantum dot~\cite{burkard2020superconductor,mi2018coherent}, leveraging this feature provides a promising pathway.
 
 For a DQD system, the spin-photon coupling strength is given by $
 g_s = \frac{g \mu_B B_0}{\hbar}$, where  $B_0$  is determined by:
 
 \begin{equation}\label{eq:B0}
 	B_0 = \Delta b \frac{e E_0 d^2}{4(2t - \hbar \omega_L)},
 \end{equation}
 and  $d$  is the interdot distance~\cite{benito2017input,samkharadze2018strong,reed2010fast}. Unlike semiconductor DQDs, quantum tunneling between potential wells is negligible for electrons on helium or neon~\cite{dykman2023spin,jeffrey2014fast,kanai2023single}, meaning  $t$  represents the orbital level spacing when the electron is located at the center between the two dots.
 
 Fig.~\ref{fig:spinorbit}(c) shows a magnetic field gradient is generated at the electron’s position,  $\Delta b = 0.1 \, \text{mT/nm}$  for cobalt sized $1.5\,\mu\text{m} \times 1.5\,\mu\text{m} \times 20\,\text{nm}
 $~\cite{pioro2008electrically}. Assuming  $d = 100 \text{ nm}$ and $(2t - \hbar \omega_L)/h = 1\,\text{ GHz}$, with a charge-photon coupling of  $g_c / 2\pi = 3.5  \text{ MHz}$~\cite{zhou2022single}, the spin-photon coupling is calculated as  $g_s / 2\pi = 0.2  \text{ MHz}$.
 
 Spin decoherence arises from nuclear spins in natural neon, with a linewidth broadening of  $\gamma_s / 2\pi = 10 \text{ kHz}$~\cite{bronn2015broadband}, and from charge-spin hybridization, given by:
 
 \begin{equation}
 	\left( \frac{g \mu_B \Delta b d}{2(2t - \hbar \omega_L)} \right)^2 \gamma_c,
 \end{equation}
 when  $g \mu_B \Delta b d / 2 \ll 2t - \hbar \omega_L$~\cite{samkharadze2018strong}. With  $\gamma_c / 2\pi = 0.36  \text{ MHz}$~\cite{zhou2024electron}, the spin decoherence due to hybridization is about $7 \text{ kHz}$, comparable to nuclear spin effects, meaning the spin coherence time remains largely unaffected. 
 
 For  $(2t - \hbar \omega_L)/h = 100 \text{ MHz}$,  $g_s / 2\pi$  increases to 2 MHz, but strong hybridization leads to a spin decoherence rate matching  $\gamma_c / 2\pi = 0.36 \text{ MHz}$, compromising spin coherence~\cite{jennings2024quantum}.
 
 Enhancing  $g_c$ can improve  $g_s$  while preserving spin coherence. This can be achieved by increasing the vacuum electric field  $E_0$ , where:
 
 \begin{equation}
 	E_0 d = \alpha V_0, \quad V_0 \propto \sqrt{L / C},
 \end{equation}
 with  $L$  and  $C$  being the resonator’s inductance and capacitance~\cite{blais2021circuit,hu2012strong}. Possible improvements, including optimizing $\alpha$, increasing $L$, and minimizing $C$, can enhance $V_0$.
 
 In this thesis, NbTiN is used due to its high kinetic inductance and robustness against magnetic fields~\cite{reed2010fast}, making it a more suitable choice for enhancing electron-photon coupling strength. The high kinetic inductance of NbTiN arises from the inertia of its Cooper pairs and is especially high due to its long London penetration depth $\lambda_L\approx 390\,\text{nm}$, similar to other high kinetic inductance materials such as TiN and NbN~\cite{lee2024penetration}. However, NbTiN yields more field-resilient resonators, maintaining higher quality factors in the in-plane magnetic fields required for spin qubits~\cite{muller2022magnetic,kim2025field}.

 \subsection{Superconducting resonators}
 Coplanar waveguide (CPW)-coupled superconducting resonators provide a
 high-coherence, on-chip platform that concentrates microwave fields in a
 well-defined geometry, enabling strong coupling and dispersive readout of
 quantum systems~\cite{simons2004coplanar,goppl2008coplanar,wallraff2004strong}.
 
 In this thesis, to access electrons on solid neon, superconducting resonators coupled to a  CPW feedline were used. The resonator is a high-impedance nanowire, electrically short compared with the guided wavelength $\lambda$; near resonance it is well described by a lumped LCR tank that is capacitively coupled to the CPW feedline. The effective inductance is dominated by the nanowire’s kinetic inductance, and the effective capacitance is set primarily by the small gap between the open end of the nanowire, see Fig. Neon paper 1a.

 A key parameter describing any resonator is its quality factor $Q$, which quantifies how sharp the resonance is. Physically, $Q$ is the ratio of energy stored in the resonator to energy lost per cycle. A large $Q$ means the resonator can store energy for many oscillation cycles before dissipating it, resulting in a sharp resonance. The internal decay rate is
 $\kappa_i = \omega_0 / Q_\text{int}$, where $\omega_0$ is the resonance
 frequency of the resonator and $Q_\text{int}$ is the internal quality factor.
 
 When the resonator is coupled to an external CPW, the total quality factor $Q_\text{tot}$
 is determined by the internal and external quality factors,
 \begin{equation}
 	\frac{1}{Q_\text{tot}} = \frac{1}{Q_\text{int}} + \frac{1}{Q_\text{ext}}.
 \end{equation}
 In practice, $Q_\text{int}$ is set by the resonator materials and geometry,
 while $Q_\text{ext}$ is set by the coupling strength to the feedline
 (e.g.\ via the coupling capacitor).
 By setting the ratio $Q_{\text{int}}/Q_{\text{ext}}$ (the coupling coefficient), the resonator can be tuned from the undercoupled regime to the overcoupled regime. To obtain maximum power transfer and the highest sensitivity, the resonator should be operated at critical coupling to the feedline, i.e., $Q_{\text{int}}/Q_{\text{ext}} = 1$~\cite{pozar2021microwave}.

 \section{Readout}
 Fast, non-destructive readout of a qubit’s state is essential for all quantum information platforms. Since the electron-escape readout scheme described in the initial qubit proposal~\cite{platzman1999quantum}, many approaches have been explored over the decades~\cite{chae2024elementary}. In this thesis, two methods are employed: radio-frequency (RF) reflectometry and microwave (MW) circuit-QED readout. Both methods convert qubit information into a measurable voltage signal.
 
 RF reflectometry uses a resonant circuit in the tens-hundreds-of-megahertz range to sense variations in impedance or capacitance; it was pioneered in 1998 with the RF single-electron transistor (rf-SET) that matched a high-resistance SET to a $50~\Omega$ environment using an LC tank circuit~\cite{schoelkopf1998radio}. MW readout, on the other hand, embeds the qubit in a superconducting resonator at a few gigahertz; the qubit imparts a dispersive shift to the cavity frequency, which is measured using homodyne or heterodyne detection~\cite{blais2021circuit}.
 
 In both schemes, the measurement relies on RF–MW waves that travel along on-chip conductors. At these frequencies, a piece of wire no longer behaves as a single lumped element. Rather, it acts as a transmission line characterized by a characteristic impedance $Z_0$, typically engineered to be $Z_0 \approx 50\,\Omega$~\cite{pozar2021microwave}. 
 
 When this line is connected to a load impedance  $Z_{\text{load}}$, the fraction of the incident signal that is reflected depends on how well  $Z_{\text{load}}$  matches the characteristic impedance  $Z_0$ . The reflection coefficient is given by~\cite{pozar2021microwave}:
 \begin{equation}
 	\Gamma(\omega) = \frac{Z_{\text{load}} - Z_0}{Z_{\text{load}} + Z_0}.
 \end{equation}
 If  $Z_{\text{load}} = Z_0$, there is no reflection, the power transfer is maximum.

 \subsection{Radio-Frequency Reflectometry}
 In RF reflectometry, the resonant circuit operates at tens to hundreds of megahertz and is typically implemented as an LC tank circuit connected directly to the device.
 
\begin{figure}[htbp]
	\centerfloat{
		\includegraphics[width=0.7\linewidth]{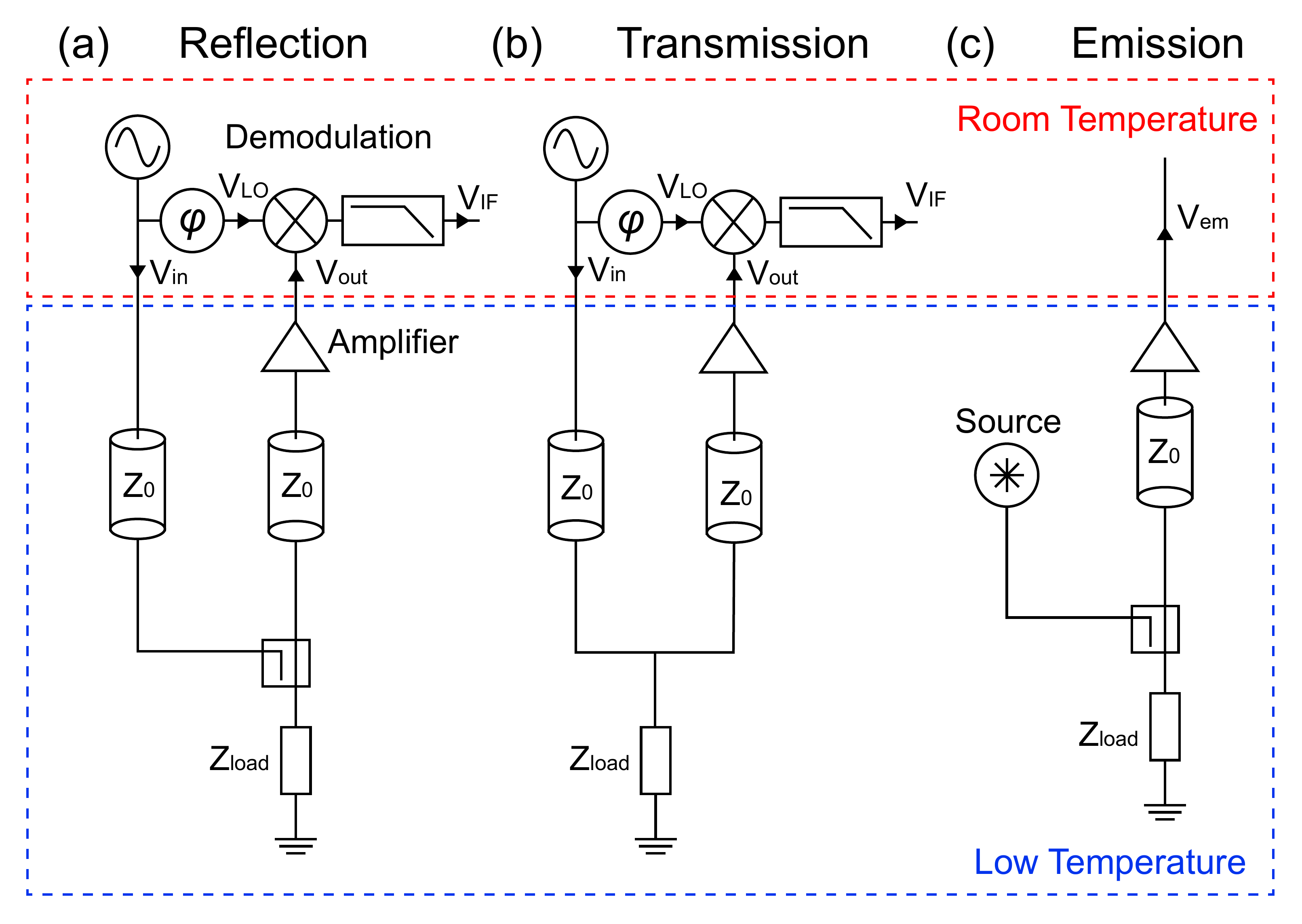}
	}
	\caption[Measurement setups for reflection, transmission, and emission]{%
		Measurement setups for reflection, transmission, and emission. In (a,b), the input signal \(V_{\text{in}}\) is generated by an RF source and propagates along transmission lines with a characteristic impedance \(Z_0\). After interacting with the load, the reflected signal is amplified to \(V_{\text{out}}\) and undergoes homodyne demodulation into \(V_{\text{IF}}\). In (c), the source emits a signal independently, which is demodulated using an external reference signal. These figures are adapted, with slight modifications, from~\cite{vigneau2023probing}.}
	\label{fig:3rf}
\end{figure}
 
 There are three main types of RF measurement: reflection, transmission and emission. In reflection and transmission measurements, see Fig.~\ref{fig:3rf} (a,b), impedance changes of the device under test are converted into voltage changes. The device, along with its tank circuit, forms a total impedance $Z_{\text{load}}$, which acts as a load on the transmission line. A carrier signal $V_{\text{in}}$ is injected, propagating toward the load, where it is either reflected (reflection mode) or transmitted (transmission mode). The outgoing signal, modified by $Z_{\text{load}}$, carries information about the device impedance. To ensure accurate analysis, the signal is amplified above the noise floor of subsequent electronics. These techniques have been applied to monitor single‑electron tunnelling~\cite{vandersypen2004real}, charge parity in superconducting circuits~\cite{schroer2012radio} and spin states in semiconductor qubits~\cite{connors2020rapid,crippa2019gate}, etc.  In our work~\cite{jennings2025probing}, we applied the reflection mode to detect the Rydberg transition of surface electrons on liquid helium (Ch. He) and transmission mode to measure the electrons on solid neon (Ch. Neon).
 
 The third type (Fig.~\ref{fig:3rf}(c)) uses a cryogenic source and does not require any external generator.In our implementation, the cryogenic source is realized by a tunnel-diode oscillator (see Ch.~TDO).

 \subsubsection{Tunnel Diode Oscillator (TDO)}
 \label{sec:1_TDO}
 
 A key challenge in qubit readout is sending RF/MW signals down to millikelvin temperatures. Conventionally, the source is placed at room temperature, in this case each qubit needs a delicated coaxial line with cold attenuators and amplifiers which becomes impractical as qubit number grows. Therefore, in this thesis, we propose to move the source to the cryogenic side, the approach is to embed a negative-resistance element, a tunnel-diode oscillator (TDO), into the experimental cell as a self-sustaining oscillator~\cite{chow1964principles,van1975tunnel}.
 
\begin{figure}[htbp]
	\centerfloat{
		\includegraphics[width=0.6\linewidth]{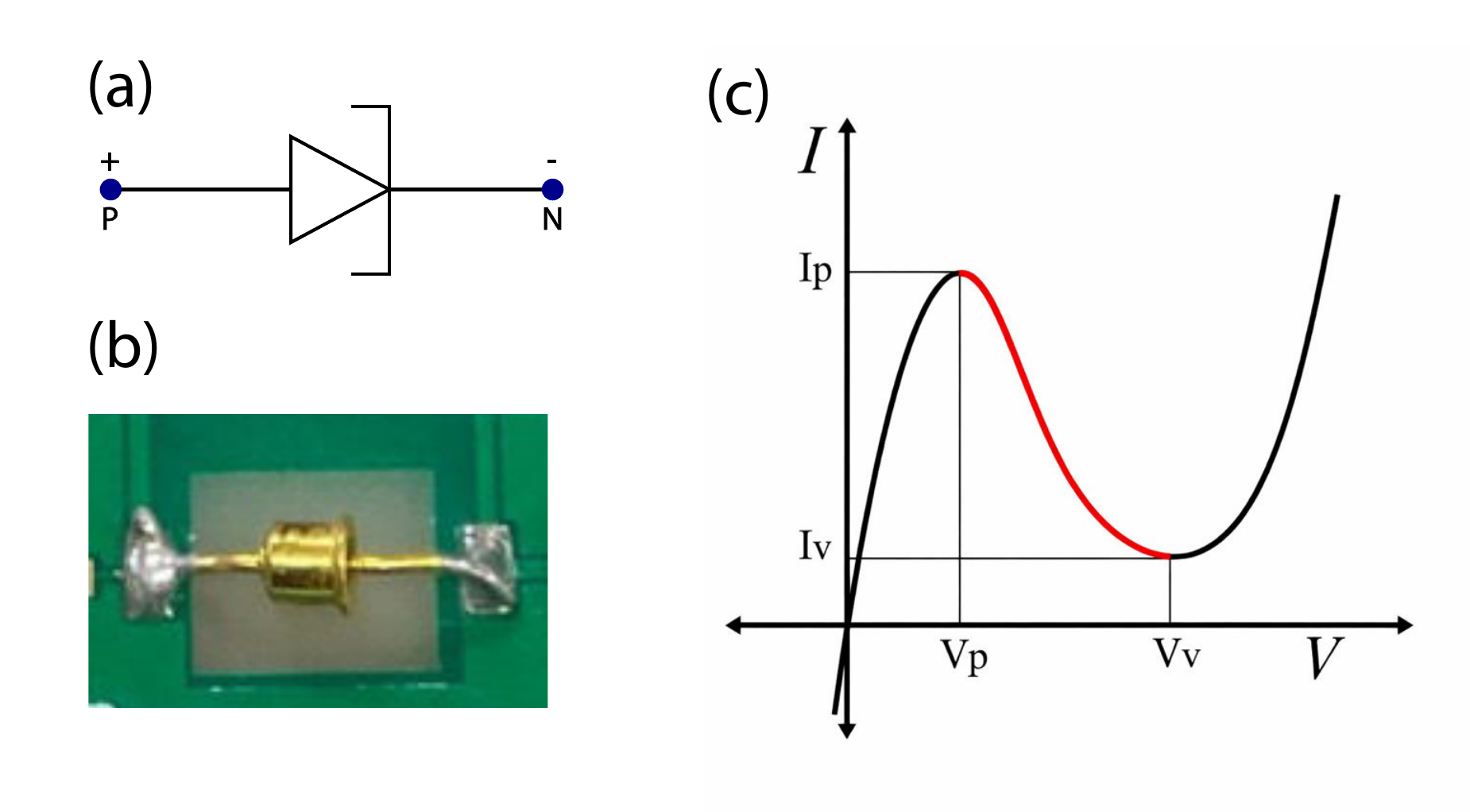}
	}
	\caption[Tunnel diode symbol, photo, and I--V curve]{%
		(a) Symbol of a tunnel diode with \(p\) and \(n\) terminals.
		(b) Photograph of a BD-6 Ge-backward tunnel diode used in this work.
		(c) Typical \(I\)--\(V\) curve of a tunnel diode showing a peak at \((V_p, I_p)\) and a valley at \((V_v, I_v)\); between them, the slope is the negative-resistance region.}
	\label{fig:tdoov}
\end{figure}
 
 A tunnel diode is a semiconductor diode in which the p–n junction is very heavily doped on both sides, operated in its negative-differential-resistance region, see Fig.~\ref{fig:tdoov}. In this region, over a small voltage range, the diode’s current decreases as the applied voltage increases. As a result, its differential conductance $g_d$ is negative. It can then deliver AC power by converting DC-bias voltage into oscillations~\cite{chen2004electrical}. A passive resonator stores oscillating electrical energy, but because it has internal resistance and other losses, its oscillations are damped and decay to zero. A negative resistance can cancel the resonator’s positive resistance, effectively creating a lossless resonator in which self-sustained oscillations occur at the resonant frequency~\cite{solymar2014electrical}.
 
 When a tunnel diode is connected to an LC circuit, it cancels the circuit’s resistive loss,  when the negative resistance is greater than the positive resistance of the circuit, we have a diode oscillator. The oscillation frequency is  determined by the inductance $L$ of the LC circuit, and total capacitance  $C_{\mathrm{tot}}=C_\text{LC}+C_\text{TD}$:
 \begin{equation}
 	f_0 = \frac{1}{2\pi\sqrt{L\,C_{\mathrm{tot}}}}.
 	\label{eq:ch1_tdo}
 \end{equation}
 The capacitance of the tunnel diode can be tuned by changing its bias voltage. In this thesis, a nanofabricated Nb spiral coil on sapphire is used (details is discussed in Capter. Fabrication), the inductance can be tuned by changing the number of turns in the spiral.

 \subsection{MW circuit‑QED readout}
 MW readout exploits the coupling between a qubit and a superconducting microwave resonator, operating in the dispersive regime. The qubit is embedded in a cavity with resonance frequency $\omega_0/2\pi$ of a few gigahertz. The system is described by the Jaynes–Cummings Hamiltonian~\cite{jaynes2005comparison}:
 \begin{equation}
 	H = \hbar\omega_0\bigl(a^{\dagger}a + \tfrac{1}{2}\bigr) + \tfrac{\hbar\omega_q}{2} \sigma_z + \hbar g \bigl(a^{\dagger}\sigma_- + a\,\sigma_+\bigr),
 \end{equation}
 where $\omega_0$ is the resonator frequency, $\omega_q$ is the qubit frequency, $a^\dagger$ and $a$ are the creation/annihilation operators for photons in the resonator, $\sigma_z, \sigma_\pm$ are Pauli operators, and $g$ is the coupling strength between qubit and resonator. The JC Hamiltonian describes the interaction between a qubit and a resonator. The first term describes the resonator as a harmonic oscillator that stores photons of energy $\hbar\omega_0$. The second term models the qubit as a two-level system with transition energy $\hbar\omega_q$. The third term describes the energy exchange between the qubit and resonator. 
 
 When $\omega_q \approx \omega_0$, coherent energy exchange occurs, leading to vacuum Rabi oscillations. In the dispersive regime, where $|\omega_q - \omega_0| \gg g$, the qubit and resonator do not exchange energy but the resonator frequency shifts by an amount $\pm \chi$ depending on the qubit state, with $\chi \approx g^2/(\omega_q - \omega_0)$. A microwave tone near $\omega_0$ picks up a phase shift that depends on the qubit state, which can then be measured~\cite{blais2021circuit}.
 
 In 2022, the first single-electron qubit on solid neon was read out using microwave dispersive techniques within a circuit-QED architecture coupled to the electron’s motional states.~\cite{zhou2022single}.
 
 Compared to RF reflectometry, MW readout often demands more complex microwave engineering, but can achieve higher signal-to-noise ratios. The choice between the two methods depends on the qubit platform and experimental requirements: in this thesis, RF reflectometry was employed for electrons on helium, while microwave readout was used for electrons on neon.

\nomenclature{%
    \( \begin{rcases}
        a_n \\
        b_n
    \end{rcases} \)%
}{коэффициенты разложения Ми в дальнем поле соответствующие электрическим и
    магнитным мультиполям}
\nomenclature[a\( e \)]{\( {\boldsymbol{\hat{\mathrm e}}} \)}{единичный вектор}
\nomenclature{\( E_0 \)}{амплитуда падающего поля}
\nomenclature{\( j \)}{тип функции Бесселя}
\nomenclature{\( k \)}{волновой вектор падающей волны}
\nomenclature{%
    \( \begin{rcases}
        a_n \\
        b_n
    \end{rcases} \)%
}{и снова коэффициенты разложения Ми в дальнем поле соответствующие
    электрическим и магнитным мультиполям. Добавлено много текста, так что
    описание группы условных обозначений значительно превысило высоту этой
    группы...}
\nomenclature{\( L \)}{общее число слоёв}
\nomenclature{\( l \)}{номер слоя внутри стратифицированной сферы}
\nomenclature{\( \lambda \)}{длина волны электромагнитного излучения в вакууме}
\nomenclature{\( n \)}{порядок мультиполя}
\nomenclature{%
    \( \begin{rcases}
        {\mathbf{N}}_{e1n}^{(j)} & {\mathbf{N}}_{o1n}^{(j)} \\
        {\mathbf{M}_{o1n}^{(j)}} & {\mathbf{M}_{e1n}^{(j)}}
    \end{rcases} \)%
}{сферические векторные гармоники}
\nomenclature{\( \mu \)}{магнитная проницаемость в вакууме}
\nomenclature{\( r, \theta, \phi \)}{полярные координаты}
\nomenclature{\( \omega \)}{частота падающей волны}

\nomenclature{FEM}{finite element method, метод конечных элементов}
\nomenclature{FIT}{finite integration technique, метод конечных интегралов}
\nomenclature{FMM}{fast multipole method, быстрый метод многополюсника}
\nomenclature{FVTD}{finite volume time-domain, метод конечных объёмов
    во~временной области}
\nomenclature{MLFMA}{multilevel fast multipole algorithm, многоуровневый
    быстрый алгоритм многополюсника}
\nomenclature{BEM}{boundary element method, метод граничных элементов}
\nomenclature{CST MWS}{Computer Simulation Technology Microwave Studio
    программа для компьютерного моделирования уравнен Максвелла}
\nomenclature{DDA}{discrete dipole approximation, приближение дискретиных
    диполей}
\nomenclature{FDFD}{finite difference frequency domain, метод конечных
    разностей в~частотной области}
\nomenclature{FDTD}{finite difference time domain, метод конечных разностей
    во~временной области}
\nomenclature{MoM}{method of moments, метод моментов}
\nomenclature{MSTM}{multiple sphere T-Matrix, метод Т-матриц для множества
    сфер}
\nomenclature{PSTD}{pseudospectral time domain method, псевдоспектральный метод
    во~временной области}
\nomenclature{TLM}{transmission line matrix method, метод матриц линий передач}

\FloatBarrier

\chapter{Experimental Setup} 

This chapter describes the experimental implementation of the work presented in this thesis, with a focus on the development and characterization of the cryogenic platforms for trapping and manipulating FEs above liquid helium and solid neon.

\section{Experimental Sample Cells} 

We use hermetically sealed sample cells that allow helium or neon to be condensed onto the device under cryogenic conditions. In this section, we describe two sample cells: one for electron-on-helium experiments, and the other for electron-on-neon experiments.

\begin{figure}[htbp]
	\centerfloat{
		\includegraphics[width=0.7\linewidth]{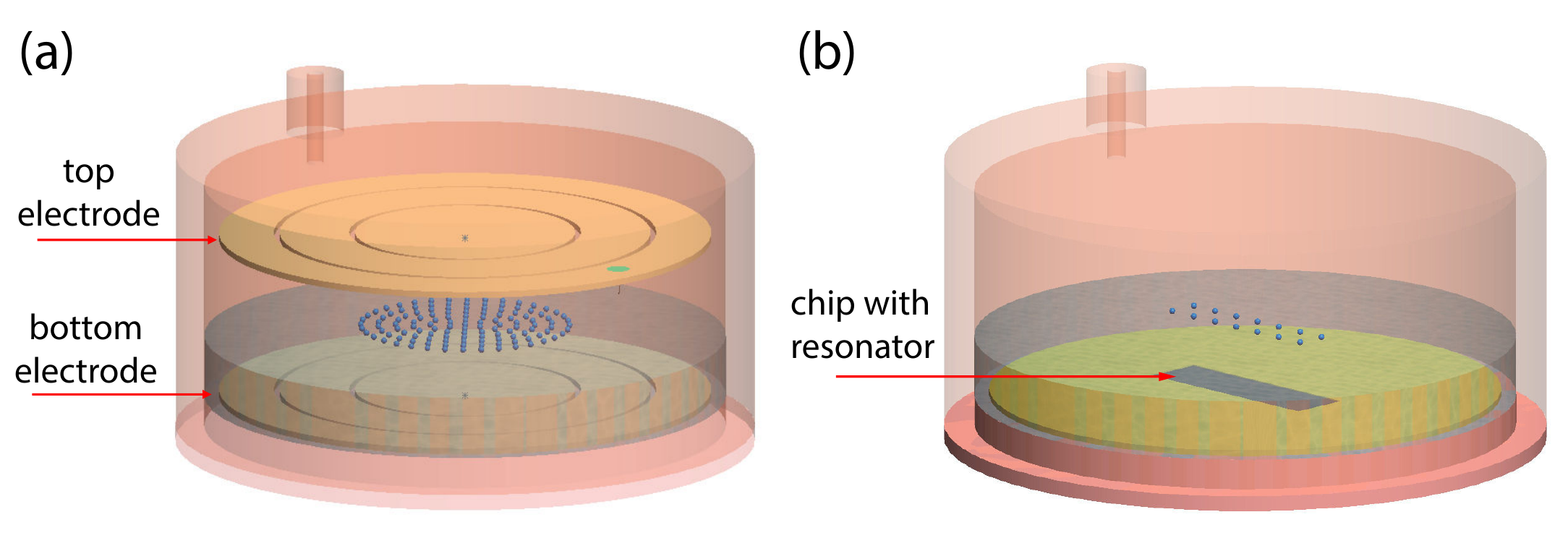}
	}
	\caption[Conceptual illustration of the cryogenic sample cells]{%
		Conceptual illustration of the cryogenic sample cells used in the experiments.
		(a) The helium cell incorporates a Corbino geometry formed by two parallel plates.
		(b) The neon cell features a nanofabricated chip mounted on the bottom plate.}
	\label{fig:mockcell}
\end{figure}

Both cells are made of oxygen-free copper and consist of two parts: a top flange and a bottom flange. Schematics of the sample cells are shown in Fig.~\ref{fig:mockcell}. The overall design is the same for both cells, but the internal structures differ. The helium cell incorporates a Corbino electrode geometry on the top flange, while the neon cell is configured such that a nanofabricated chip for electron-on-neon experiments can be mounted on the bottom flange. These design differences reflect the distinct experimental requirements of the helium and neon platforms: the helium experiments employ bulky Corbino electrodes to study many-electron systems, whereas the neon experiments rely on nanofabricated structures designed for single-electron experiments.

\subsection{Helium Experiment}

\subsubsection{Cell for Helium Experiment}

Below 2.17 K, liquid helium becomes superfluid with zero viscosity, thin-film creep, and quantized vortices~\cite{london1951superfluids}. These properties allow it to flow through nanometer-scale pores without resistance. A common method for hermetically sealing the sample cell is to use an indium wire between two flanges. Indium is a soft metal that ensures a leak-tight seal under cryogenic conditions when compressed between two mating surfaces~\cite{lim1986indium}. After closing the cell, we use a leak detector to ensure that there are no leaks in the experimental line at room temperature. To check for leaks that appear only at low temperature (cold leaks), i.e., helium passing through microscopic gaps that are sealed at room temperature, we performed an additional leak test after cooling to 4\,K.

\begin{figure}[htbp]
	\centerfloat{
		\includegraphics[width=0.6\linewidth]{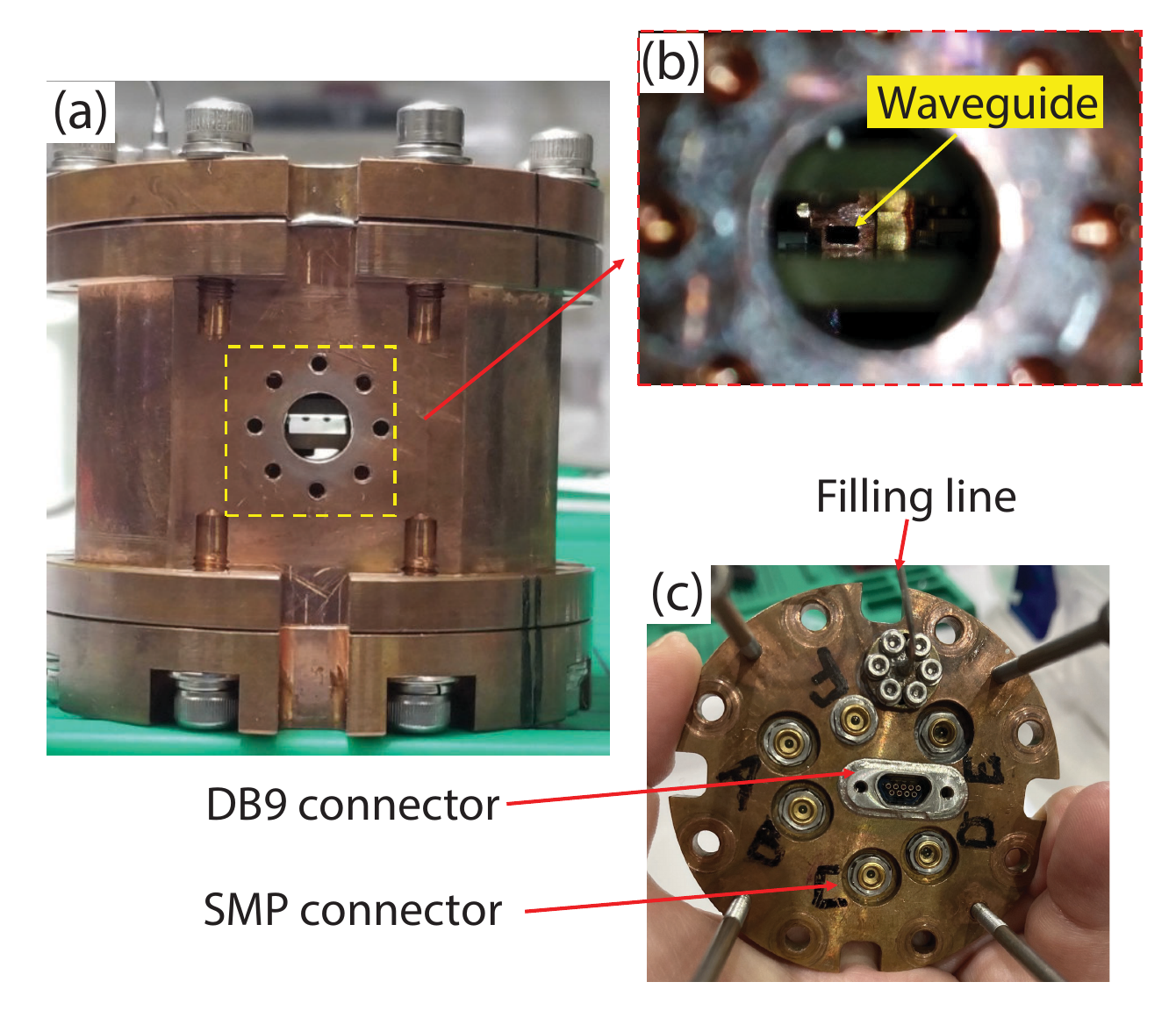}
	}
	\caption[Photographs of the helium sample cell]{%
		Photographs of the helium sample cell.
		(a) Assembled helium cell with a millimeter waveguide feedthrough.
		(b) WR6 waveguide from the opposite side of the MW feedthrough.
		(c) Top flange of the helium cell with hermetic SMP, 9-pin DC feedthrough (Micro-DB9), and a helium filling line.}
	\label{fig:hecell}
\end{figure}

Owing to the low thermal conductivity of liquid helium, a sintered Ag heat exchanger is integrated into the bottom flange of the helium cell to enhance heat exchange between the cell and the helium and to ensure efficient thermalization at cryogenic temperatures. As shown in Fig.~\ref{fig:hecell}(c), the top flange of the helium cell contains six hermetic SMP feedthroughs with indium seals and a 9-pin DC electrical feedthrough (Micro-DB9). A stainless-steel helium filling line, sealed with an indium wire, is used to condense helium into the cell. A millimeter waveguides extends into the cell, is positioned between the parallel-plate.

\subsubsection{PCB for Helium Experiment}

\begin{figure}[htbp]
	\centerfloat{
		\includegraphics[width=0.7\linewidth]{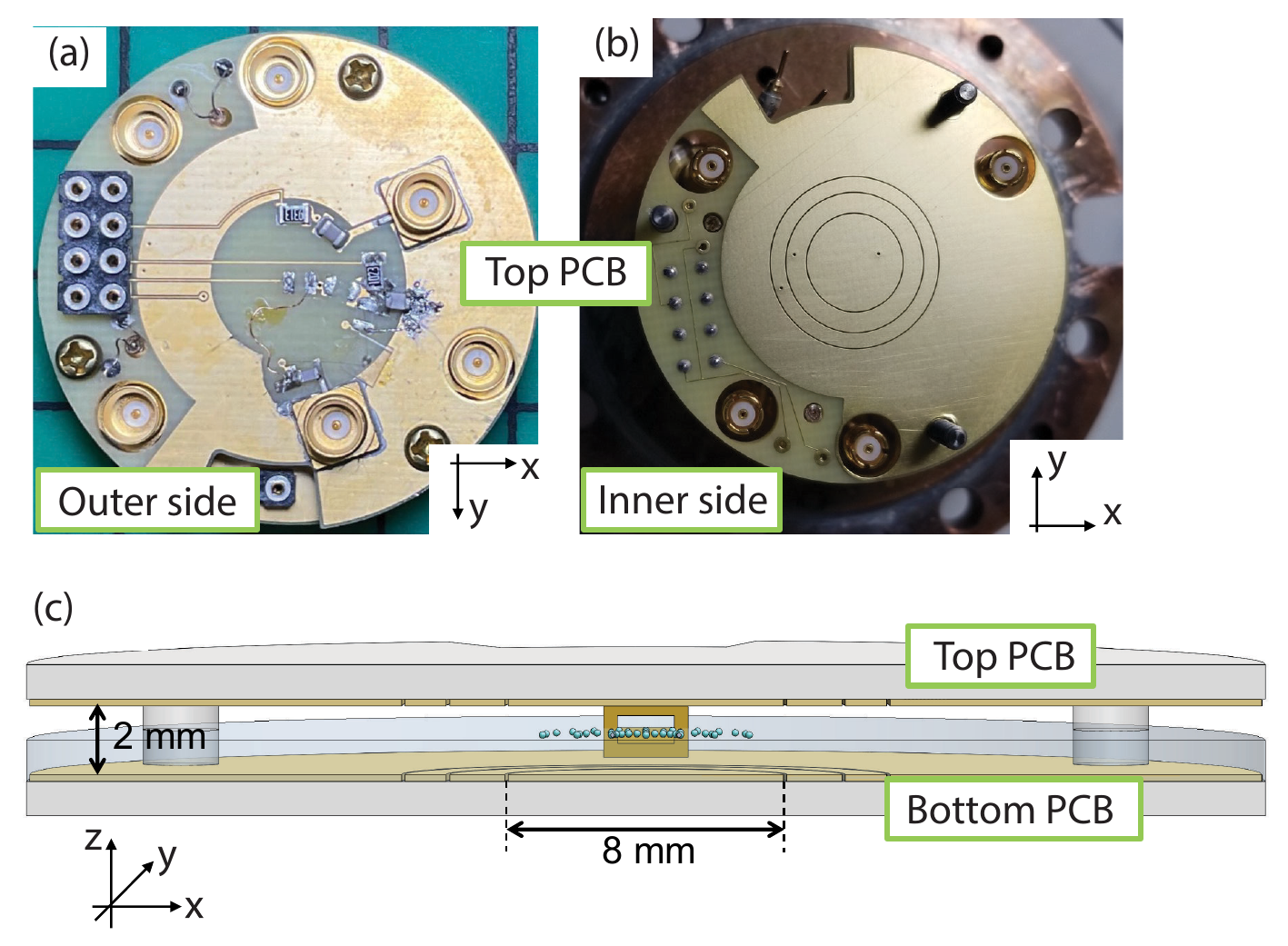}
	}
	\caption[PCB design for the helium experiment]{%
		\textbf{PCB design for the helium experiment.} (a) The outer side of the top PCB. (b) Bottom view of the top PCB with concentric electrodes. (c) The rectangular MW waveguide (brown color) located between the parallel plate electrodes. Liquid helium (transparent blue color) with a height of approximately 1\,mm. Electrons (blue spheres) floating above the helium surface.}
	\label{fig:hepcb}
\end{figure}

Inside the sample cell, the Corbino electrodes and other passive components are implemented on two printed circuit boards (PCBs). The PCBs are mounted parallel to each other with a $2\,\text{mm}$ sapphire spacer between them, with the Corbino electrode sides facing each other (see Fig.~\ref{fig:hepcb}(c)).

The inner sides of the two PCBs have three concentric ring electrodes (inner, middle, and outer), which define an in-plane Corbino geometry through a radial electrode configuration~\cite{iye1980mobility,mehrotra1987analysis,wilen1988impedance}. Opposing electrodes on the top and bottom PCBs form parallel-plate capacitors. The central opposing pair provides the capacitance of the LC resonator, whereas the middle and outer electrodes are biased with DC voltages to confine electrons toward the central electrode~\cite{ahmed2018radio,aassime2001radio,gonzalez2015probing,oakes2023fast,ibberson2021large,apostolidis2024quantum}.
On the outer side of the top PCB, the remaining LC elements are soldered (Fig.~\ref{fig:hepcb}(a)) and connected to the central Corbino electrode through the PCB. The Corbino electrodes are located on the inner sides of the top PCB (Fig.~\ref{fig:hepcb}(b)) and the bottom PCB. The concentric electrodes on the top and bottom PCBs form the capacitor of the resonant LC circuit, with a total capacitance $C = 2.131\,\text{pF}$. Details of the circuits are explained in Sec.~\ref{sec:he}.

Eight DC lines provide the bias voltages for the other Corbino electrodes and the filament current. The connections between the top cell flange and the cell PCBs are made through SMP connectors and SMP bullets.
The LC resonator is designed with a microfabricated Nb spiral inductor (40\,turns, 708\,nH at 4\,K, fabrication details are given in Chapter.~\ref{ch:fab}) in combination with surface-mounted capacitors, forming an LC tank circuit with a resonance frequency around 121.9\,MHz and a total quality factor $Q \sim 311$. The components are soldered directly onto the PCB surface, while the Nb spiral inductor is glued on the PCB with varnish and connected to the LC circuit by aluminum wire bonds made with a bonding machine (Westbond Model 7476D).

\subsection{Neon Experiment}
\subsubsection{Cell for Neon Experiment}

\begin{figure}[htbp]
	\centerfloat{
		\includegraphics[width=0.6\linewidth]{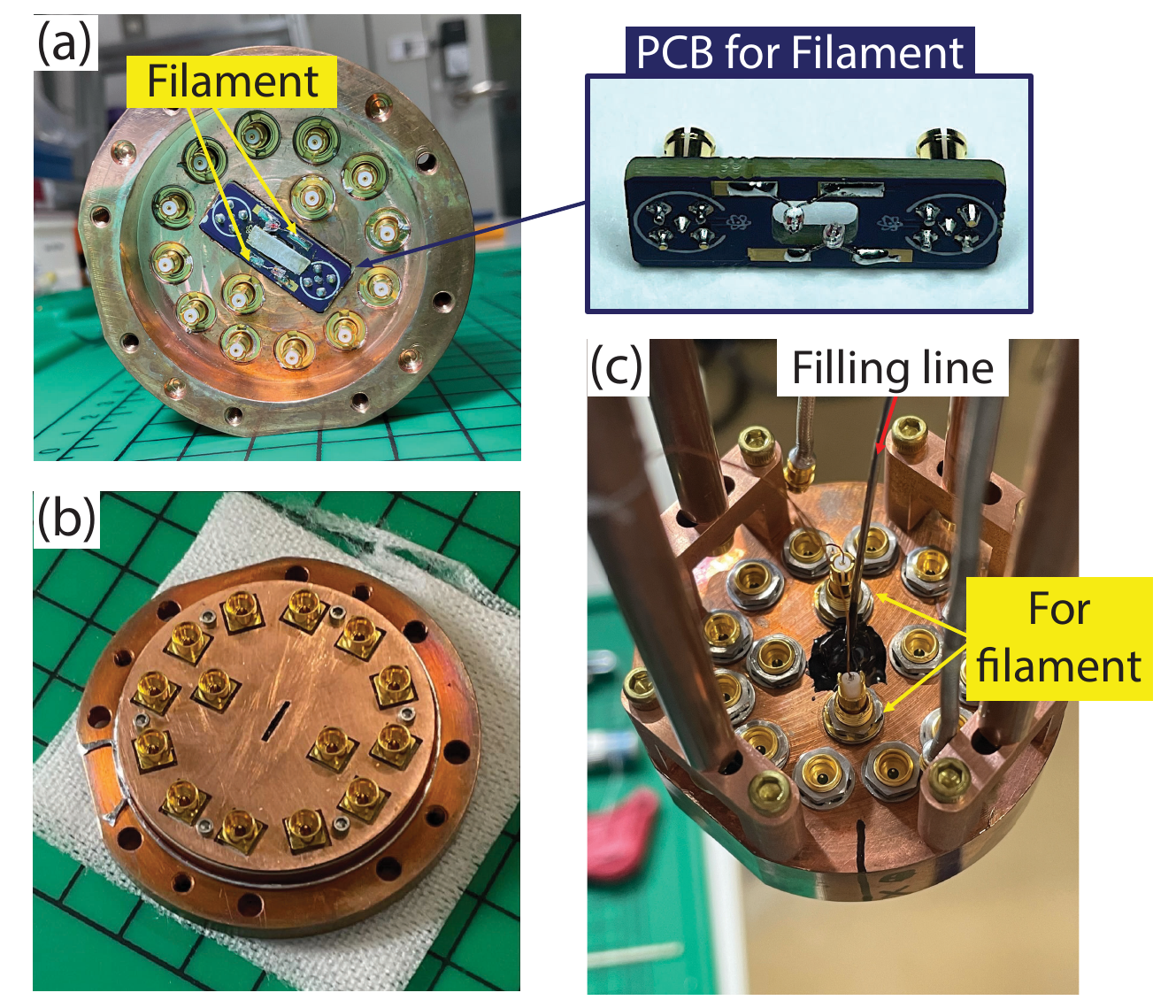}
	}
	\caption[Photographs of the neon sample cell]{%
		\textbf{Photographs of the neon sample cell.} (a) Inside of the top flange of the cell showing 14 hermetic SMP connectors and two filaments connected in parallel on a small rectangular PCB. The callout image shows the PCB for filaments. (b) Bottom flange of the cell with fixed PCB, nanofabricated chip, copper shield and 14 PCB surface mount jack SMP connectors arranged in a circular layout. (c) Assembled neon cell mounted on the mixing chamber plate cold finger. A capillary line in the center is used for neon filling. Two SMP connectors adjacent to the filling line are used for the electron-emitting filaments that are located inside the cell.}
	\label{fig:necell}
\end{figure}

The neon cell top flange contains 14 vacuum-tight hermetic SMP feedthroughs arranged in a circular configuration (see Fig.~\ref{fig:necell}). Neon gas is condensed into the cell through a capillary filling line. The line enters the cell through a vented screw that is hermetically sealed at the center of the top flange.

The neon sample cell is designed to suppress spurious microwave modes and enhance the quality factor of the on-chip resonator by minimizing coupling to unwanted electromagnetic cavity modes. The interior geometry, including the compact copper spacer structure, is expected to discretize the microwave density of states and shift the lowest eigenmode to higher frequencies. While no direct calculation or measurement of the cell’s eigenmodes has been performed, the structural design closely follows the configuration described in Ref.\cite{koolstra2019trapping}, where the lowest mode is expected to exceed 10 GHz.

\subsubsection{PCB for Neon Experiment}

Two electron-emission filaments are soldered onto a dedicated small rectangular PCB that is separate from the sample PCB (Fig.~\ref{fig:necell}(a)). This filament PCB is connected to a pair of vacuum-tight hermetic SMP connectors and serves exclusively as the filament base (Fig.~\ref{fig:necell}(c)). If the filaments burn out, the PCB with the filaments can be replaced easily.
The sample chip is mounted on the sample PCB installed on the bottom flange, and DC and high-frequency signals are routed into the cell through the electrical connectors on the top flange and wire-bonded to the sample PCB.

\begin{figure}[htbp]
	\centerfloat{
		\includegraphics[width=0.5\linewidth]{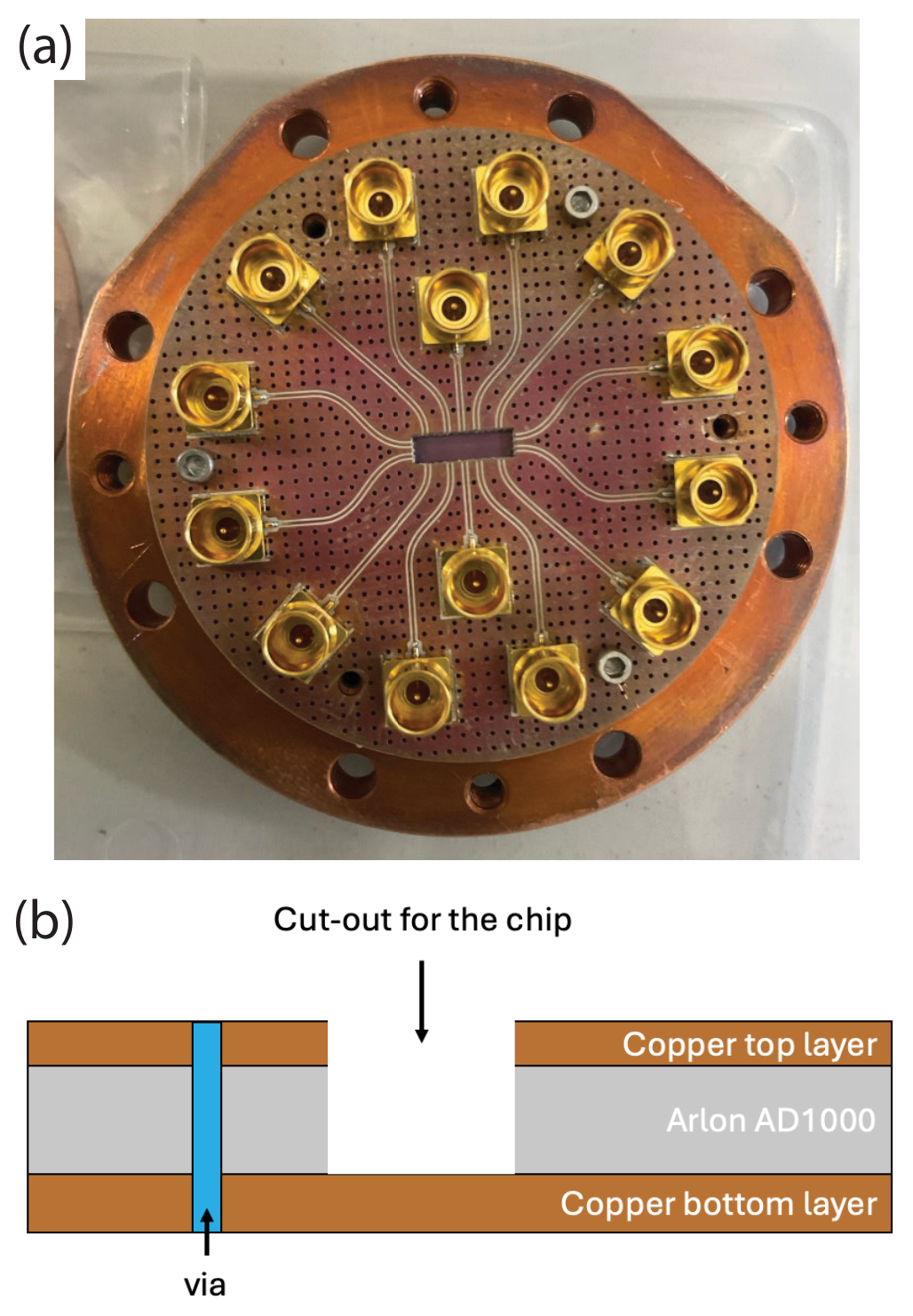}
	}
	\caption[PCB design for the neon experiment]{%
		\textbf{PCB design for the neon experiment.} (a) Top view of the sample PCB mounted on the bottom flange, showing the soldered SMP surface mount jacks and the designed high-frequency lines (white stripes). (b) Schematic view of the PCB, illustrating the multilayer structure composition of copper layers separated by an Arlon AD1000 dielectric. A cut-out cavity for mounting the nanofabricated chip, and plated vias connect the top and bottom copper planes~\cite{koolstra2019trapping}.}
	\label{fig:nepcb}
\end{figure}

The sample PCB is used to route high-frequency signals up to 10\,GHz and DC signals to the on-chip microwave resonator and bias electrodes. For the resonator used in this thesis (described in Sec.~\ref{sec:ne}) no electrodes for DC bias are included. However, for future implementations, DC voltage control will be necessary to tune the qubit frequency and define the electron trapping potential. As shown in Fig.~\ref{fig:nepcb}(a), the PCB includes soldering pads for 14 SMP connectors, plated through-hole vias to connect the front and back copper layers, and a central cutout designed to fix a 2 × 7 mm chip, allowing its top surface to be on the same level as the top copper layer of the PCB for efficient bonding of the chip.

The PCB is made from Arlon AD1000 substrate, which has a high dielectric constant ($\varepsilon = 10.2$) that remains stable across the 1--10~GHz frequency range, closely matching that of typical chip materials such as sapphire or silicon ($\varepsilon \approx 11$)~\cite{pozar2021microwave}, see Fig.~\ref{fig:nepcb}(b). The microstrip traces on the PCB are designed with a center conductor width of 5.9~mils and a gap width of 3.5~mils, chosen to maintain a characteristic impedance of 50~$\Omega$ to match the measurement line.

\section{Electron Source}

\begin{figure}[htbp]
	\centerfloat{
		\includegraphics[width=0.6\linewidth]{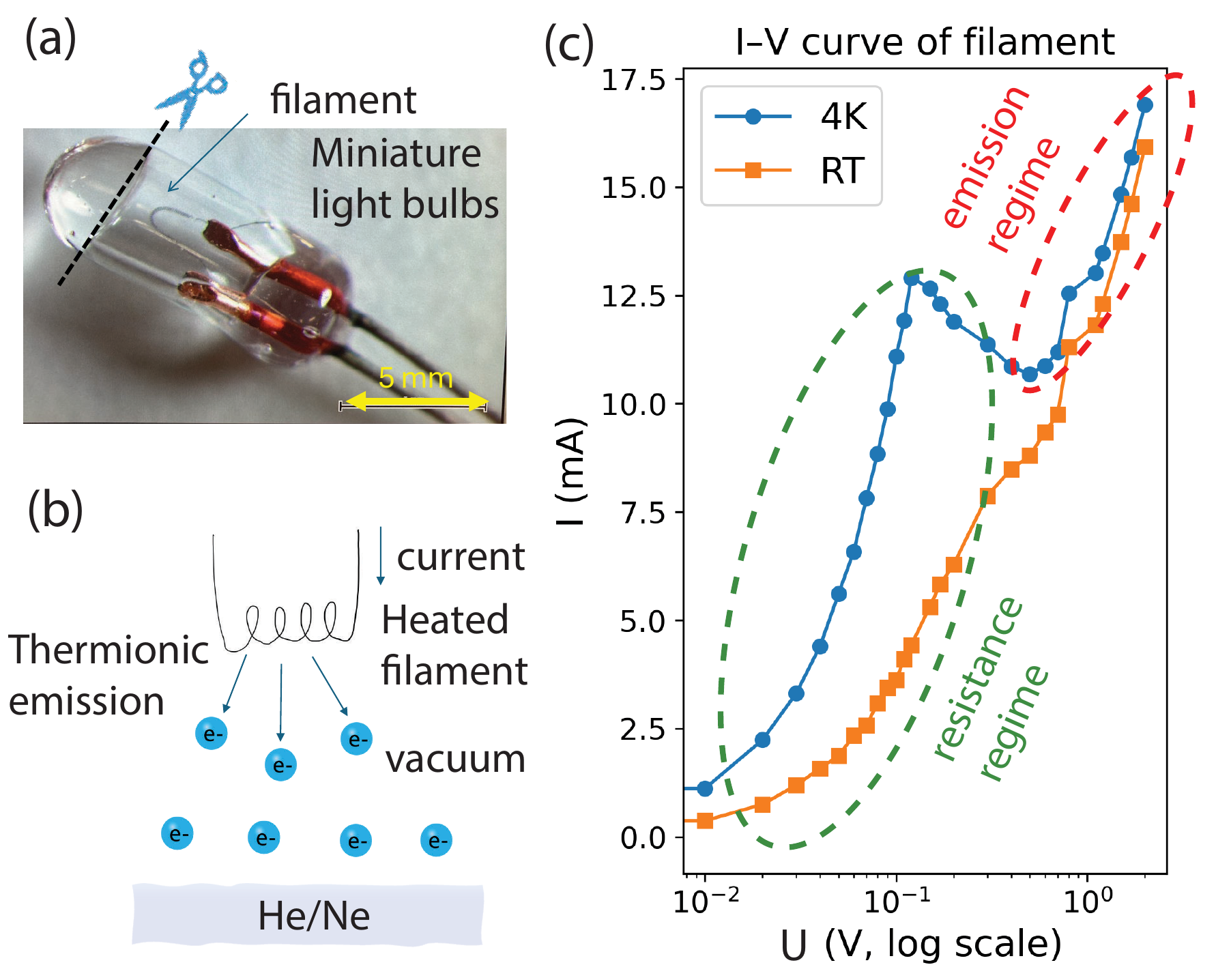}
	}
	\caption[Electron source used in the experiment]{%
		\textbf{Electron source used in the experiment.}
		(a) General view of tungsten bulb, which is used as the electron source in the experiments.
		(b) Illustration of spreading of the electrons above cryogenic substrates.
		(c) Typical I-V curve of tungsten filament at room temperature and low temperature in logarithm scale.}
	\label{fig:filament}
\end{figure}

To generate electrons into the cryogenic sample cell, we use a tungsten filament extracted from a miniature incandescent bulb, as shown in Fig.~\ref{fig:filament}(a). The filament is exposed by carefully breaking the glass envelope using pliers, after which the two metal leads are soldered to the DC lines on the PCB, and the PCB is mounted on the top flange inside the sample cell .

Electron emission is achieved through thermionic emission, as illustrated in Fig.~\ref{fig:filament}(b). 
Figure~\ref{fig:filament}(c) shows the I–V characteristics of the filament measured at 4\,K and room temperature (RT). In the low-voltage region, the current increases almost linearly with voltage, which corresponds to the resistance regime, where the filament behaves as an ohmic conductor. As the voltage increases further, the current begins to decrease, indicating a negative resistance region. This behavior is due to the space-charge effects around the filament. When the bias voltage is increased beyond a certain point, the system enters the emission regime, where electrons are emitted from the filament into the surrounding vacuum, and finally land on the surface of the cryogenic substrate.
Simultaneously, the heating generates a small amount of helium vapour inside the cell, which scatters and slows down the emitted electrons. These cooled electrons then accumulate above the surface of the helium or neon substrate, forming the electron layer.

\section{Cryogenic Setups}

The experiments in this thesis were carried out over a temperature range
from a few tenths of a kelvin to several millikelvin. To achieve this, we used three kinds of cryogenic refrigerators~\cite{pobell2007matter}:

\textbf{Bluefors~LD400 "dry" dilution refrigerator} – a cryogen‑free system that reaches several millikelvin temperatures with high cooling power.  It provides a fast cooling platform for the neon and TDO measurements.

\textbf{Oxford Instruments KelvinOx~400HA "wet" dilution refrigerator} – a liquid-helium-cooled system with a several millikelvin base temperature. It is mainly used for helium experiments because it introduces minimal mechanical vibrations to the free surface of bulk superfluid helium.

\textbf{Home‑built $^3$He refrigerator} – a continuous-circulation $^3$He refrigerator system, which is used for quick tests.  It can cool to 600\,mK without the complexity of a dilution unit in a short period of time.

\subsection{"Dry" Dilution Refrigerator}\label{ch:bluefors}
The Bluefors LD400 is a dry dilution refrigerator used for the neon and TDO experiments. Before each experiment, the vacuum can is evacuated, and the pulse-tube cooler cools down the inner parts below 4\,K. The $^3$He/$^4$He mixture is then condensed and starts circulating continuously, which cools the mixing chamber to around 7 millikelvin. This provides a stable base temperature for long measurements.

\begin{figure}[p]
	\centerfloat{
		\includegraphics[width=0.7\linewidth]{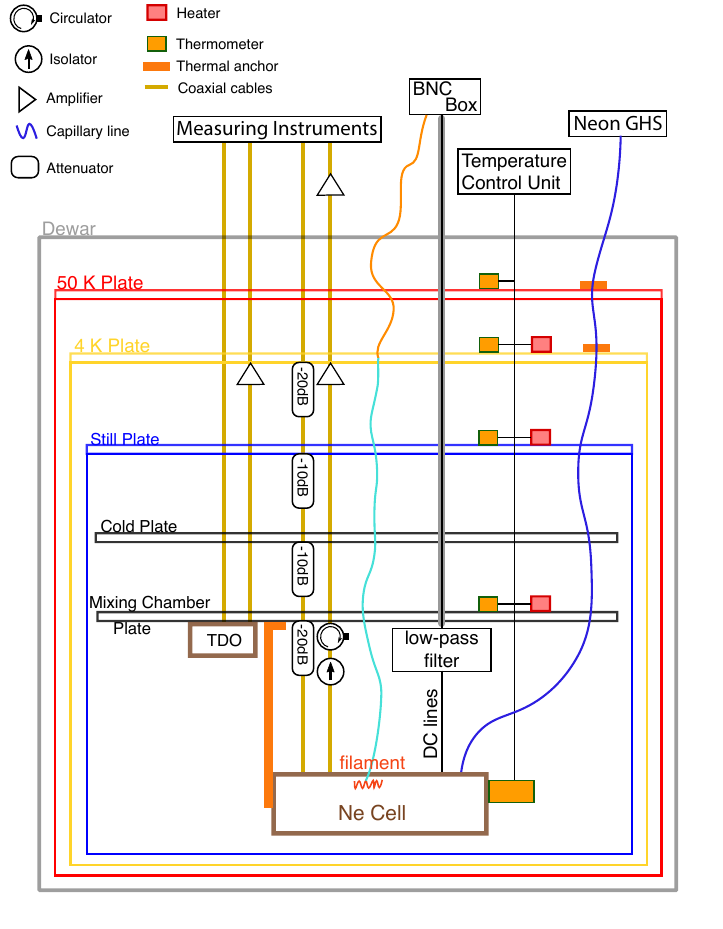}
	}
	\caption[Schematic of the Bluefors LD400 setup]{%
		\textbf{Schematic of the Bluefors LD400 setup.}
		Eight semi-rigid high-frequency lines are installed: several for the superconducting resonator and one pair for the TDO circuit. The high-frequency input is step-attenuated at each dedicated temperature stage (total attenuation about -60 dB), and the output passes through a circulator$\rightarrow$isolator pair, a 4\,K HEMT amplifier, and then a room temperature amplifier. DC control signals from the BNC box go to the cell through a 24-channel low-pass filter circuit made of $\pi$-filter and RC-filter in series. An additional DC source powers the temperature-control unit and the tungsten filament, the wire changes to superconducting wire (light blue) from the 4\,K plate. The stainless-steel capillary filling line (blue) connects the cell to a homemade neon gas-handling system (see Fig.~\ref{fig:negas}) and, together with all DC/high-frequency lines, the neon cell, and the TDO board, are thermally anchored at  each dedicated temperature stage (see Sec.~\ref{sec:tdo} namely~\cite{grytsenko2025characterization} for TDO circuit details).}
	\label{fig:BF400}
\end{figure}

A simplified wiring diagram is shown in Fig.~\ref{fig:BF400}. The fridge is equipped with coaxial cables to send microwave signals to the experimental cell.
To reduce high-frequency noise, DC control signals are sent from the BNC box to the cell through a low-pass filter circuit. 
This filter circuit provides 24 channels, each with a dedicated cutoff frequency, and each channel consists of a $\pi$-filter and an RC filter connected in series.
An additional DC source supplies voltage to the filament to generate electrons. To reduce heat from this DC line, the part of the filament lead from the 4\,K plate to the cell is made from a piece of superconducting wire, and no DC filters are used.

\begin{figure}[htbp]
	\centerfloat{
		\includegraphics[width=0.6\linewidth]{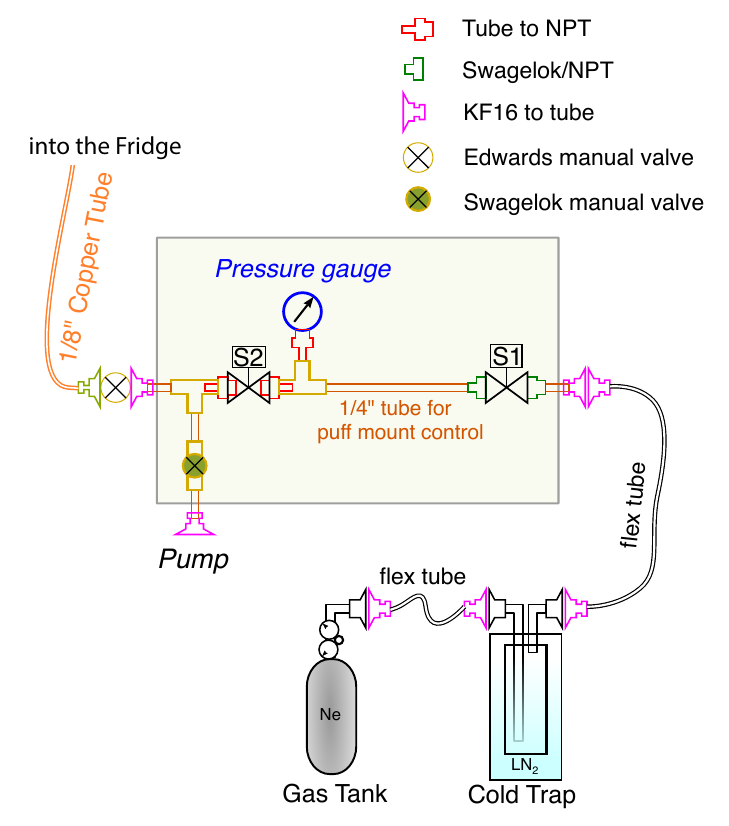}
	}
	\caption[Homemade neon gas handling system]{%
		\textbf{Homemade neon gas handling system.}
		Neon from the gas tank passes through an LN$_2$ cold trap and a control board equipped with two valves, and then enters the refrigerator via a 1/8'' copper tube (orange).A pump port is connected via a tee between the manual valve and S2, allowing an external vacuum pump to be attached for evacuating the line and retrieving neon gas after the experiments.See the main text for details of the control board operation.}
	\label{fig:negas}
\end{figure}

Eight semi-rigid high-frequency coaxial lines are installed for different purposes. Several of these lines are dedicated to the superconducting resonator used in the electron-on-neon experiment, and one pair is reserved for TDO measurements.
In the electron-on-neon experiment, the input high-frequency line (CuNi-CuNi coaxial cable, with both inner and outer conductors made of CuNi) passes through attenuators thermally anchored at each temperature stage, giving a total attenuation of about -60 dB. The resonator output is routed into the output line (SSS-SS, i.e., silver-plated stainless-steel/stainless-steel, coaxial from RT to 4\,K; NbTi-NbTi coaxial from 4\,K to base), passes through a circulator (LNF-CIC4\_8A) and an isolator (LNF-ISISC4\_8A) mounted on the mixing-chamber plate, is then amplified by a cryogenic HEMT amplifier (LNF-LNC4\_8C) thermally anchored to the 4\,K plate, and finally returns to room temperature for analysis using different measurement setups. This low-temperature amplifier operates in the 4–8 GHz range with a power consumption of about 10 mW, and its output is further amplified by a room-temperature amplifier (LNF-LNR4\_8ART).

For the TDO experiment, both the input and output lines use SSS–SS coaxial cable from RT to 4\,K and NbTi–NbTi coaxial cable from 4\,K to mixing-chamber plate temperature. The output signal from the TDO circuit is amplified by a low-temperature amplifier (Cosmic Microwave Technology CMT-BA1), which is thermally anchored to the 4\,K plate. 
All the DC and high-frequency lines are thermally anchored by feedthroughs on each stage. The electron-on-neon experimental cell and TDO circuit board are thermally anchored to the mixing-chamber plate.

Fig.~\ref{fig:negas} illustrates the homemade gas-handling system used to introduce neon into the fridge. Neon gas from the tank first passes through a liquid-nitrogen cold trap to remove possible impurities and prevent the capillary filling line from clogging, and then to the control board with two solenoid valves S1 and S2. By opening S1 while S2 is closed, and then closing S1 and opening S2, a controlled “puff” of neon is delivered to the fridge. By repeating this sequence, a controlled amount of neon gas can be introduced to the fridge. The neon gas enters the fridge through a stainless-steel capillary line to the cell via a vented screw. The filling line is thermally anchored at 50\,K and 4\,K (Fig.~\ref{fig:BF400}). A pressure gauge monitors the gas pressure, and a vacuum pump can be connected to the pump port to evacuate the line before the neon deposition and pump out/retrieve neon after experiments. The fridge is warmed up to 24\,K to condense neon into the cell near the neon triple point.

\subsection{KelvinOx~400HA Setup}

\begin{figure}[htbp]
	\centerfloat{
		\includegraphics[width=0.9\linewidth]{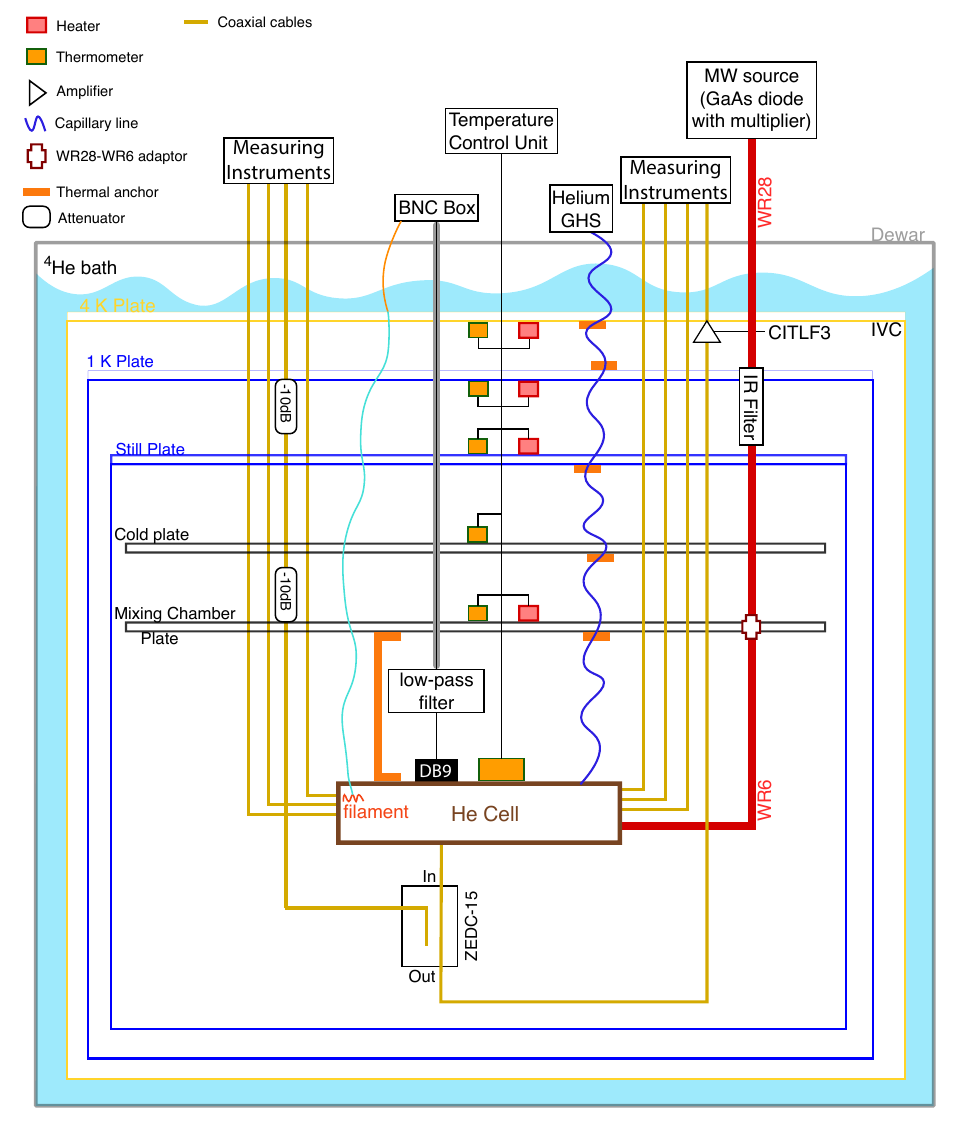}
	}
	\caption[Schematic of the KelvinOx~400HA setup]{%
		\textbf{Schematic of the KelvinOx~400HA setup.} Eight semi-rigid high-frequency lines (gold) are installed: several were used for the electron-on-helium experiments. The readout line includes a directional coupler (ZEDC-15) mounted on the cell. The input signal, attenuated by -20 dB at 1\,K plate and mixing chamber plate, is sent through the coupler to the cell, while the reflected signal is coupled out and amplified by the cryogenic amplifier (CITLF3). The millimeter waveguides are shown in red. DC lines pass a low-pass filter circuit and enter the cell through a Micro-DB9 connector, and additional DC source powers the temperature-control unit and the tungsten filament, the wire becomes superconducting (light blue) from 4\,K plate. The blue line refers to the CuNi capillary filling line.}
	\label{fig:kelvinox}
\end{figure}

\begin{figure}[htbp]
	\centerfloat{
		\includegraphics[width=0.8\linewidth]{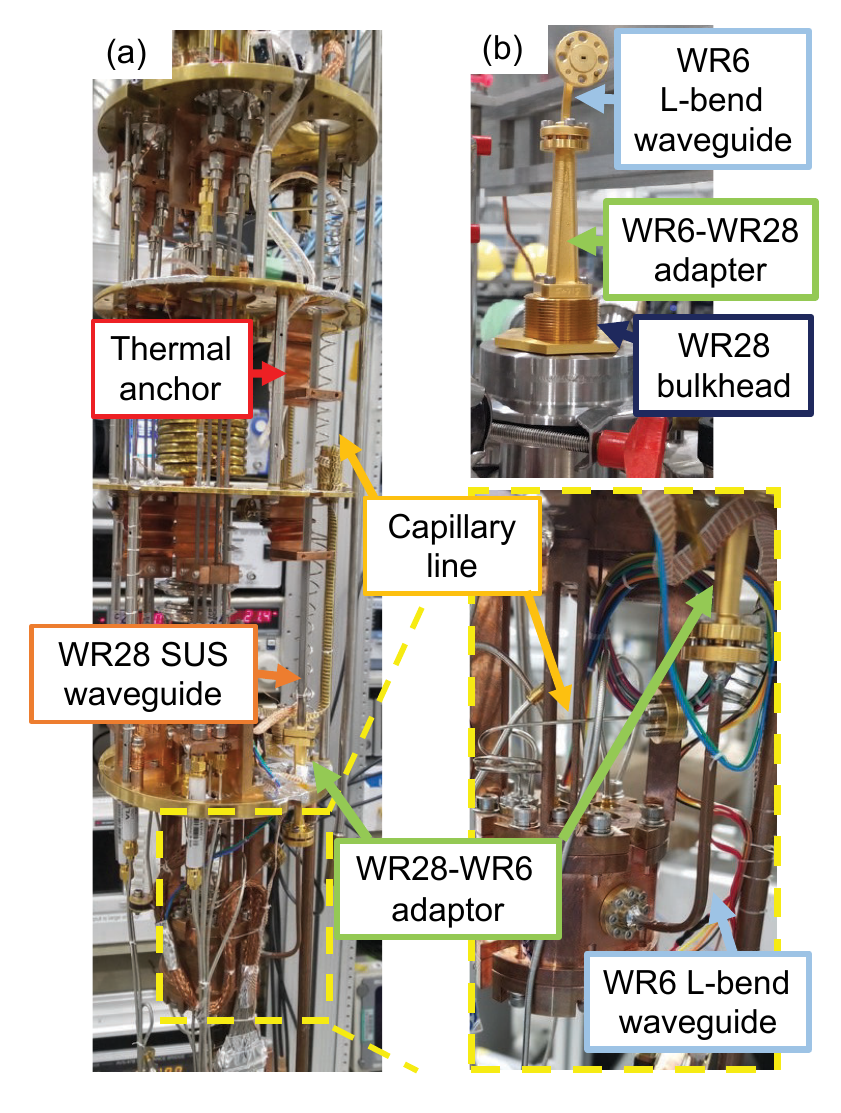}
	}
	\caption[Waveguides setup]{%
		\textbf{Waveguides setup.} (a) Interior of KelvinOx 400HA. 110-170 GHz millimeter waves are transmitted through WR28 SUS waveguides with $1\,\mu\mathrm{m}$ thick internal gold plating (Oshima Prototype Engineering). At the mixing-chamber plate, a WR28–WR6 adapter is used, and a WR6 copper waveguide is inserted into the experimental cell. Helium gas is transmitted through the capillary line (a CuNi capillary C7150, with an outer diameter of 1.0 mm and thickness of 0.2 mm). (b) At the top of the dilution refrigerator, the microwave from the room-temperature passes through a WR6 L-shape bend waveguide and a WR6 to WR28 adapter before entering the refrigerator.}
	\label{fig:waveguide}
\end{figure}

The KelvinOx 400HA is a "wet" dilution refrigerator manufactured by Oxford Instruments. Unlike dry fridges, it requires a liquid-helium bath to operate the 1\,K pot and reach a base temperature near 10\,mK.
For such a “wet” system, the inner vacuum chamber is first cooled to $4.2\, \text{K}$ with liquid $^4$He. Then, by operating the 1 K pot, the inner parts are cooled further to about $1.8\, \text{K}$. Under these conditions, circulation of the $^3$He/$^4$He mixture can be started easier than in a "dry" refrigerator. In addition, the absence of a pulse tube reduces vibrations of the fridge and of the bulk superfluid $^4$He in the cell.
To further improve stability, the whole fridge is placed on an air-suspension table.

A simplified fridge setup is shown in Fig.~\ref{fig:kelvinox}. The input high-frequency coaxial lines are installed from room temperature to the cell via two attenuators, -10 dB each, at the 1\,K plate and the mixing-chamber plate. The output coax returns the reflected signal through a directional coupler (ZEDC-15) outside the cell and then a cryogenic low noise amplifier (Cosmic Microwave Technology CITLF3). DC signal lines are connected via a low-pass filter (as described in Chap.~\ref{ch:bluefors}) and a Micro-DB9 connector. All the DC and high-frequency lines are thermally anchored by feedthrough on each stage.

Millimeter waves in the 110-170 GHz range are generated by GaAs Schottky Diode paired with a multiplier whose multiplication factor is 12 (Virginia Diodes Inc, WR6.5AMC-I) and are transmitted into the dilution refrigerator via a waveguide. The distance from the room temperature port at the top of the dilution refrigerator to the mixing chamber plate, where the sample cell is located, is approximately 1.5 meters. To minimize attenuation in the straight sections, a WR28 waveguide is utilized, while a WR6 waveguide is employed in the curved sections (Fig.~\ref{fig:waveguide}(a,b)). The waveguide is hermetically isolated from the cell which is filled with superfluid $^4$He by a glued Kapton window.

Helium inside the sample cell is condensed through a CuNi filling capillary line (outer diameter 1.0 mm, wall 0.2 mm). Gas enters from room temperature, cools as it passes down the stages, and condenses along the filling capillary line.
The amount of condensed liquid helium is controlled by monitoring the capacitance between the top and bottom electrodes (Ch.~Helium). The desired $^4$He level inside the sample cell is controlled by adding small portions of gaseous $^4$He from room temperature side from $^4$He gas handling storage tank. The gap between electrodes is 2 mm. Condensation is usually stopped when the liquid helium height above the bottom electrode reaches 1 mm. This helium level was used in all measurements with FEs described in this thesis.

\subsection{Home built $^3$He Refrigerator}

A home-built $^3$He refrigerator is used for rapid testing of TDO circuits and superconducting nanofabricated devices before installing them in the KelvinOx or Bluefors refrigerators. The required time for this home-built fridge to reach its base temperature (600\,mK) from room-temperature is about 12 to 16 hours.

Owing to the relatively high vapour pressure of $^3$He at sub-kelvin temperatures and its large latent heat of vaporization, efficient evaporative cooling can be achieved by pumping on liquid $^3$He, allowing refrigerator base temperatures well below 1 K with a simple design.

The $^3$He refrigerator insert is installed in a Dewar filled with liquid $^4$He. The inner part of the refrigerator cools down to a temperature of $\sim 1.9\,\text{K}$ by operation the 1 K pot, which is filled with liquid $^4$He from the outer bath.
After all inner parts cool down to around 1.9\,K, we start the circulation of $^3$He slowly.

An LN$_2$ trap installed in the condensing line cleans the circulating $^3$He. After passing the LN$_2$ trap, $^3$He gas is precooled in the condensing line to about 4\,K and then condensed in a capillary near the 1 K pot ($T \sim 1.9\,\text{K}$), since the condensation temperature of $^3$He is 3.0 K.

After condensation, liquid $^3$He flows to the $^3$He pot through an impedance capillary and is then pumped by two pumps located at room temperature. To reach the lowest possible temperature and maximize the cooling power, the $^3$He in the pot must be pumped to a low vapour pressure. This is achieved with two pumps connected in series: a turbomolecular pump and a scroll pump. After the scroll pump, the gaseous $^3$He passes again through the LN$_2$ trap. This closed cycle allows continuous operation of the $^3$He fridge and allows it to reach about $0.6\ \text{K}$ for cryogenic elements installed inside the fridge.

\begin{figure}[htbp]
	\centerfloat{
		\includegraphics[width=0.9\linewidth]{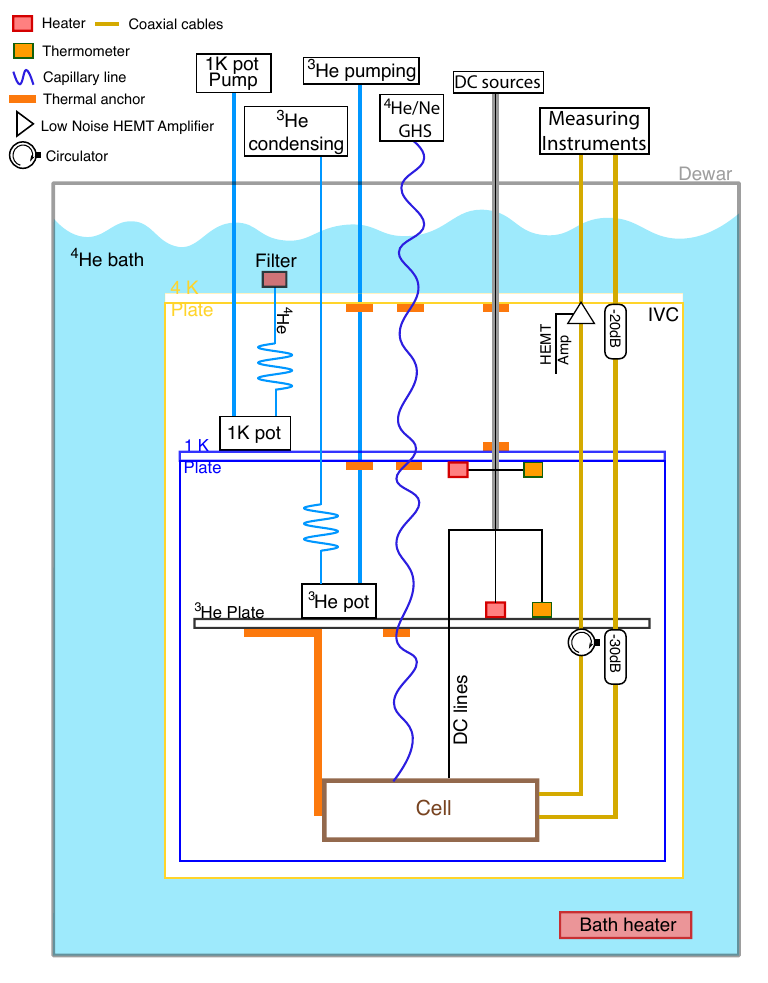}
	}
	\caption[$^3$He Fridge Setup]{%
		\textbf{$^3$He Fridge Setup.} Schematic of $^3$He evaporation refrigerator. The cell is thermally anchored to $^3$He plate. $^3$He circulation part at room temperature is not shown.}
	\label{fig:3He}
\end{figure}

Fig.~\ref{fig:3He} shows the fridge layout. The manganin DC wires are used for thermometers and heaters at the 1\,K plate and the $^3$He evaporation plate. The DC wires provide bias voltages to the sample cell. Both the DC wires and the Cu-Ni filling line are thermally anchored by flexible copper braid at each temperature stage.

The input high-frequency line has a total attenuation of -50 dB to reduce thermal noise from room temperature. The output high-frequency line contains an isolator and a low-temperature HEMT amplifier made by Low Noise Factory. This amplifier (LNF-LNC 4\_8F) has a nominal power consumption of approximately 8\,mW at 4\,K.
The cell is located at the bottom of the insert and is thermally anchored to the $^3$He plate. A bath heater is used to warm up the $^3$He refrigerator from 4\,K to room temperature in a relatively short time.

\FloatBarrier

\chapter{Sample Fabrication}\label{ch:fab}
Nanodevices form the basis for experiments for probing the behaviour of FEs on cryogenic substrates. This chapter introduces the major processes for creating nano-scale devices, provides details of the fabrication procedures and device designs for each sample.

All nanofabrication was carried out in a cleanroom environment. The RIKEN CEMS cleanroom operated by Semiconductor Science Research Support Team meets ISO 5 (Class 100) standards, which means there are fewer than 100,000 particles per cubic metre of air with diameters $\geq 0.1\,\mu\mathrm{m}$. The temperature and humidity inside the cleanroom are kept constant. For a more detailed description of nanofabrication techniques, see Ref.~\cite{chen2003vlsi}.

There are generally two ways of nanofabrication -- bottom up and top down. In the bottom-up (lift-off) process, the circuit pattern is first defined on the resist layer using lithography, followed by metal deposition. The resist is then removed, leaving the desired metal pattern on the substrate. In contrast, the top-down (etch) process begins with the deposition of the metal layer first, after which the circuit pattern is defined on the resist by lithography and transferred to the metal film by etching down the unwanted regions.

In this thesis, two types of devices were fabricated: NbTiN resonators on silicon (Si) serving as cavities for FEs on solid neon, and niobium (Nb) spiral inductors on sapphire functioning as the inductive element in the LC tank circuit for FEs on liquid helium. Both samples were fabricated using the top-down method. This chapter describes the step-by-step process of the sample fabrication in detail.

\section{Metal Deposition}

\subsection{Nb}\label{sec:nb_depo}

In the helium experiment, Nb on sapphire substrate is used for the spiral inductor for the LC circuit~\cite{colless2013dispersive}. Nb is a well-established superconducting material, in the superconducting state, Nb exhibits very low surface resistance and negligible ohmic loss~\cite{padamsee2014superconducting}, while sapphire exhibits low dielectric loss and high thermal conductivity at low temperatures~\cite{krupka2006frequency,berman1955thermal}. Together, Nb thin film on sapphire provides a good, low-loss, and thermally stable platform for a cryogenic spiral inductor.

The Nb thin film used in this work was deposited by electron‑beam evaporation, a physical vapour deposition method where a high‑energy electron beam melts and evaporates the target material in a crucible and the resulting vapour condenses on the substrate.
For the Nb metal layer deposition, a $20\times20~\mathrm{mm}$ sapphire substrate was used. A $100\,\mathrm{nm}$ Nb film was deposited by a custom measurement system built by Biemtron under a base pressure of $1.6\times10^{-7}\,\mathrm{Pa}$. The deposition was carried out with an electron-beam voltage of $8\,\mathrm{kV}$ and a beam current of approximately $450~\mathrm{mA}$. The deposition rate was maintained at $2\,\text{\AA}/\text{s}$ the total deposition time was approximately 9 min 45 s. The film parameters were set in the Inficon IC 6000 with a density of $2\,\mathrm{g/cm^3}$, a tooling factor of 137\%, and a $Z$-ratio of 0.493. This procedure produced clean, uniform films and supports good superconductivity.

\begin{figure}[htbp]
	\centerfloat{
		\includegraphics[width=0.8\linewidth]{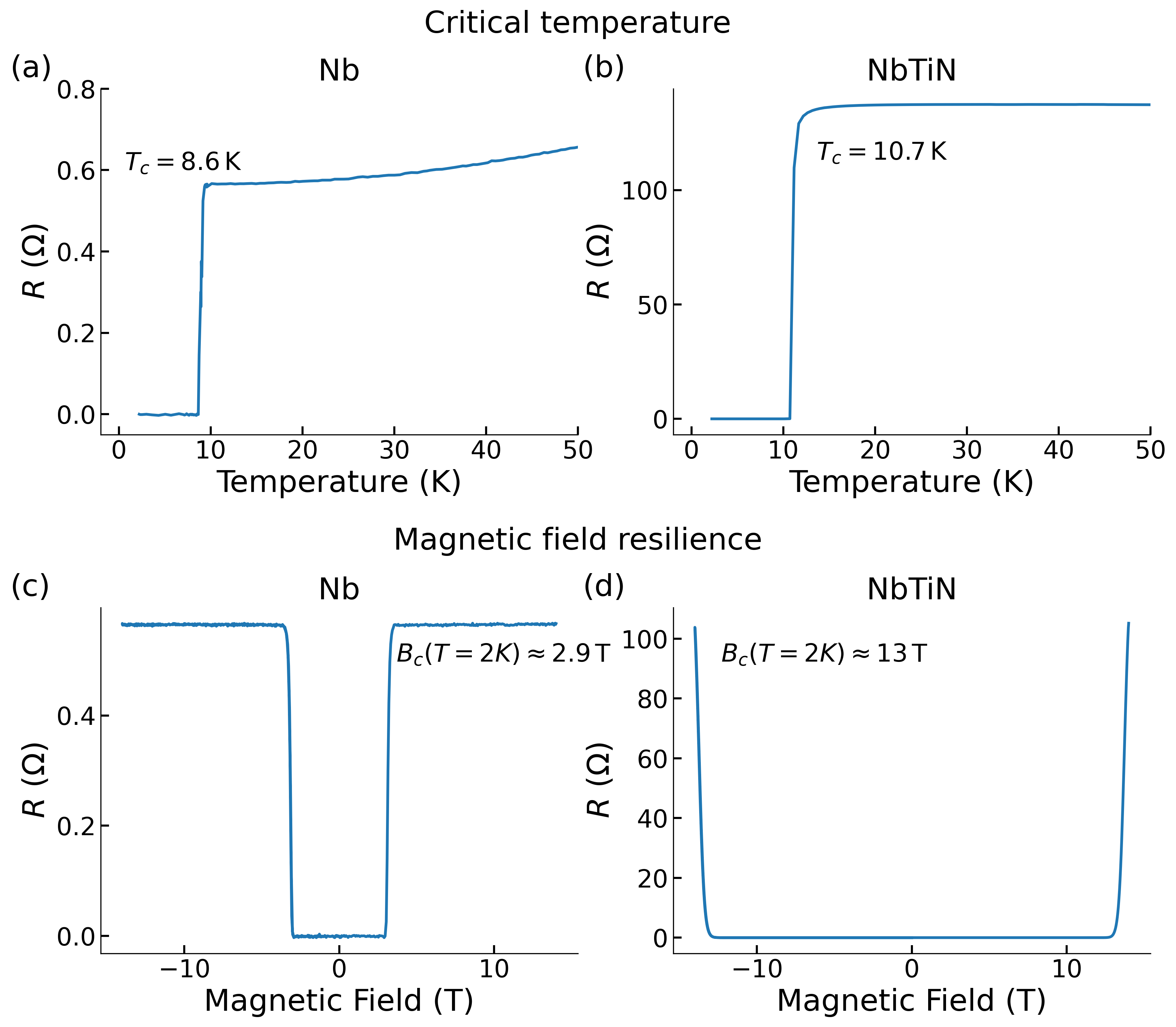}
	}
	\caption[Critical temperature and magnetic-field resilience of Nb and NbTiN films]{%
		\textbf{Critical temperature and magnetic-field resilience of Nb and NbTiN films.}
		(a,b) Resistance $R$ versus temperature $T$ for Nb and NbTiN films, measured in a PPMS.
		(c,d) Resistance $R$ versus magnetic field $B$ (applied perpendicular to the film plane) at $T=2\,\mathrm{K}$. The extracted critical temperatures are $T_c = 8.6\,\mathrm{K}$ for Nb and $T_c = 10.7\,\mathrm{K}$ for NbTiN, and the critical fields are
		$B_c \approx 2.9\,\mathrm{T}$ and $B_c \approx 13\,\mathrm{T}$, respectively.}
	\label{fig:ct_mr}
\end{figure}

The quality of the Nb film was characterised by four-terminal transport
measurements in a physical property measurement system (PPMS\textsuperscript{\textregistered}) from Quantum Design. Figure~\ref{fig:ct_mr}(a) shows the temperature
dependence of the resistance $R(T)$, from which we extract a critical
temperature $T_c = 8.6\,\mathrm{K}$, close to the bulk value of
$9.2\,\mathrm{K}$. The magnetic-field dependence $R(B)$ at $T = 2\,\mathrm{K}$ is plotted in Fig.~\ref{fig:ct_mr}(c), where the magnetic field is applied normal to the film plane, and from these data a critical field of
$B_c \approx 2.9\,\mathrm{T}$ is obtained. These measurements confirm that the evaporated Nb film has good superconducting properties and is suitable as the spiral inductor material in the LC circuit.

\subsection{NbTiN}\label{sec:nbtin_depo}

In the first electron-on-neon qubit experiments, the resonator was fabricated from superconducting Nb on high-resistivity Si, which is a standard choice for high-Q coplanar-waveguide (CPW) resonators~\cite{zhou2022single,zhou2024electron}.

NbTiN exhibits high magnetic-field tolerance, which is essential for spin-qubit operations~\cite{bruzewicz2019trapped,samkharadze2016high}, and high kinetic inductance which gives high impedance and thus stronger charge-photon coupling~\cite{bretz2022high}. Therefore, NbTiN is selected as the resonator metal for the electron-on-neon experiments in this thesis.

The NbTiN thin film used in this work was deposited by Dr. Hirotaka Terai of Advanced ICT Research Institute, National Institute of Information and Communications Technology (NICT).
For the NbTiN metal layer deposition, an intrinsic, (100)-oriented silicon wafer with high resistivity ($\rho \ge 10~\mathrm{k\Omega\!\cdot\!cm}$) was cleaned in a piranha solution for 15 min, rinsed with  deionzed (DI) water, and dipped in buffered HF (BHF) for 5 min to remove the oxide. A $20~\mathrm{nm}$ NbTiN film was deposited by reactive magnetron sputtering from an 8~inch Nb--Ti alloy target (Nb:Ti = 5:1). Deposition was performed at room temperature with a DC sputter current of $3~\mathrm{A}$ and a chamber pressure of $2~\mathrm{mTorr}$ using an Ar:$\mathrm{N_2}$ ratio of 100:37. 

The superconducting properties of the NbTiN film were evaluated in the same
PPMS\textsuperscript{\textregistered} setup. The temperature dependence $R(T)$ is shown in
Fig.~\ref{fig:ct_mr}(b), giving a critical temperature
$T_c = 10.7\,\mathrm{K}$. The magnetic-field dependence $R(B)$ at
$T = 2\,\mathrm{K}$ is plotted in Fig.~\ref{fig:ct_mr}(d), again with the
magnetic field applied normal to the film plane, and from which a critical field of $B_c \approx 13\,\mathrm{T}$ is obtained. From separate room-temperature measurements of the resistivity, we estimate the zero-temperature magnetic penetration depth of the sputtered $20~\mathrm{nm}$-thick NbTiN film to be $\lambda_0 \approx 390~\mathrm{nm}$.

\section{Electron-Beam Lithography}\label{sec:EBL}

\begin{figure}[htbp]
	\centerfloat{
		\includegraphics[width=0.6\linewidth]{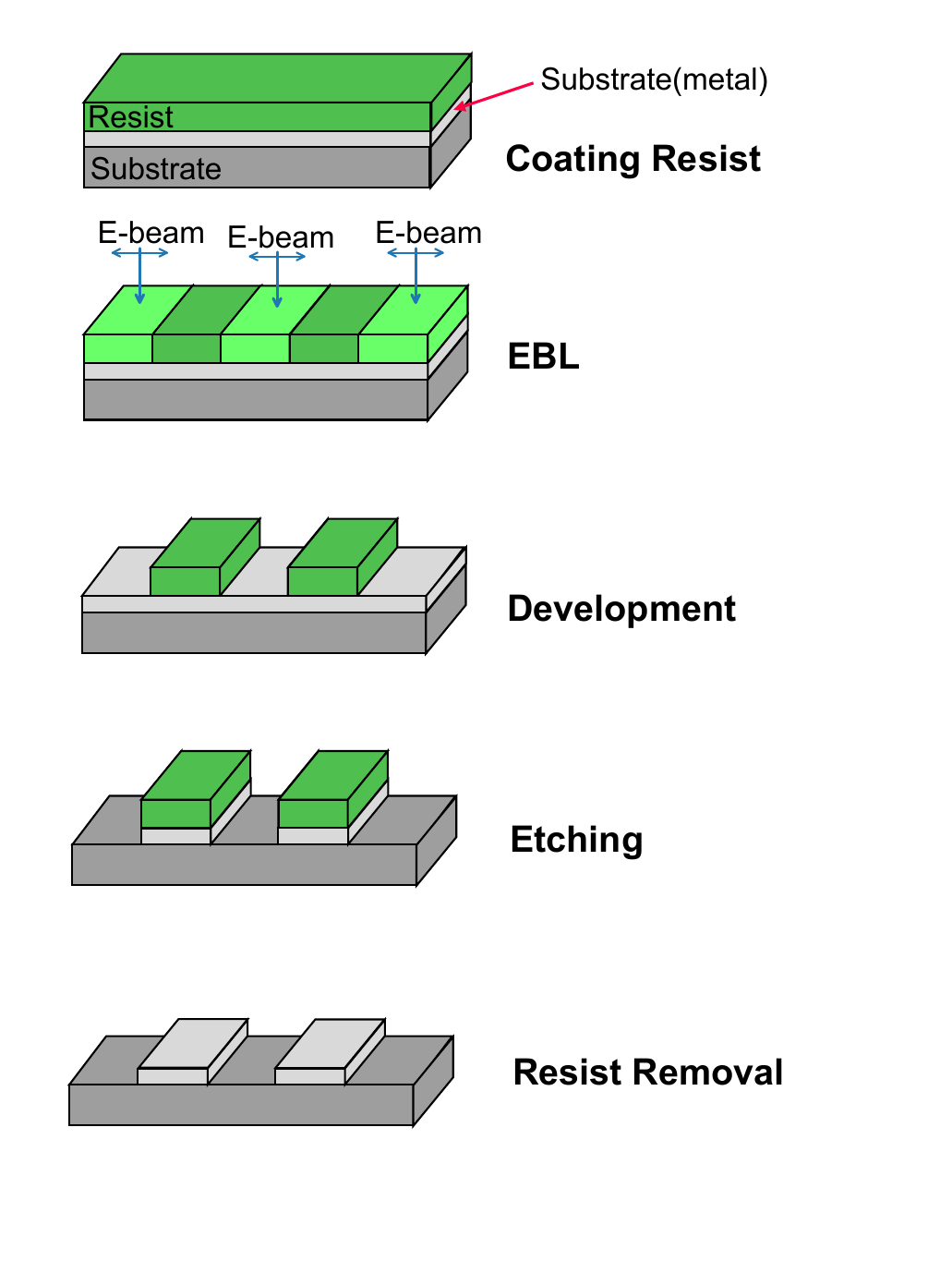}
	}
	\caption[Schematic workflow of the electron-beam lithography]{%
		\textbf{Schematic workflow of the electron-beam lithography in the case of positive resist.}	Steps are shown in processing order, from top to bottom.
	}
	\label{fig:ebl}
\end{figure}

Electron-beam lithography (EBL) is a maskless patterning method. A focused electron beam writes the pattern directly into an electron-sensitive resist. The beam changes the solubility of the exposed areas. For a positive resist, the exposed regions become soluble during development and are removed, leaving the desired pattern on the substrate (Fig.~\ref{fig:ebl}). In contrast, for a negative resist, the exposed regions remain after development. EBL is employed when sub-micron or nanometre-scale features are required that cannot be achieved using UV lithography (see Sec.~\ref{sec:UV}).

In this work, the NbTiN resonators were fabricated using an ELS-F125 e-beam lithography system (ELIONIX). The samples were coated with ZEP520A positive resist (diluted in anisole, weight ratio D.R.=2.4). During spin coating, the resist was spun at $\sim 6000$ rpm for $\sim 50$\,s, resulting in a resist thickness of $\sim 100$\,nm. The resist was prebaked at $180\,^{\circ}\mathrm{C}$ and after coating baked at $180\,^{\circ}\mathrm{C}$ again for 3 min.

During exposure, a focused high-energy electron beam is scanned across the sample to define the desired pattern. The conditions used for the for the electron-on-neon sample were a beam current of $\sim 1\,\mathrm{nA}$, an accelerating voltage of $125\,\mathrm{kV}$, and a dose of $\sim 230\,\mu\mathrm{C}\,\mathrm{cm}^{-2}$. These values were chosen to balance resolution and throughput.

After exposure, the samples were developed in ZED-N50 for 2 min, then rinsed in IPA for 30 s, and dried with nitrogen gas. Development removes the exposed resist and leaves clean openings with the designed geometry. The pattern was then transferred to the NbTiN film using dry etching, see section~\ref{sec:RIE}.

The final linewidth is determined by several factors, mainly the resist thickness, exposure dose, and development time. In addition, electron backscattering gives rise to the proximity effect, which can broaden narrow gaps~\cite{chang1975proximity}. To compensate for this, dose tests and simple proximity corrections were performed before writing the full devices patterns.

EBL offers high resolution, but its main limitation is the low writing speed compared with UV lithography (Sec.~\ref{sec:UV}).

\section{UV Lithography}\label{sec:UV}

UV photolithography (UVL) is a form of optical lithography that can be either mask-based or maskless. It follows the same process flow as EBL	: coat $\rightarrow$ expose $\rightarrow$ develop $\rightarrow$ transfer. Mask-based UV uses a fixed photomask to expose a whole field at once; maskless UV uses a digital micromirror device DMD to project tiles or a scanning UV laser to write directly, without a mask~\cite{owa2023review}. It is faster than EBL but offers lower resolution. It is well suited for our micrometre-scale Nb spiral inductors.

For fabrication of the spiral inductor, the UV lithography is performed on a sapphire substrate. The sapphire surface was deposited with 50 nm Nb thin flim (see Sec.~\ref{sec:nb_depo}) and was pre-coated with hexamethyldisilazane (HMDS) to improve resist adhesion and ensure accurate photolithography~\cite{Campbell2007-yx}. HMDS was spun with a two-step ramp (500~rpm for 5~s, then 4500~rpm for 45~s) and baked at $80\,^{\circ}\mathrm{C}$ for 10~min. A positive photoresist AZ~1500 was then spin-coated using the same ramp (500/5 $\rightarrow$ 4500/45) and soft-baked at $80\,^{\circ}\mathrm{C}$ for 10~min to stabilise the film.

Exposure was performed using a maskless UV lithography system DL-1000SG from PMT Corporation that uses a DMD, which provides a finest resolution of $1~\mu\mathrm{m}$. The exposure dose was 93~$\mathrm{mJ\,cm^{-2}}$. A sub-pixel exposure mode was applied and repeated four times to average edge errors and improve line uniformity across the field. After exposure, the wafer was developed for 45~s in AZ developer, rinsed in running de-ionised water for 2~min, and dried with nitrogen gas.

The pattern was then transferred to the Nb thin film by dry etching (RIE) as described in section~\ref{sec:RIE}. Finally, the remaining resist was stripped in NMP at $80\,^{\circ}\mathrm{C}$ overnight. The sample was rinsed, dried, and inspected by optical microscopy. This UV process is fast and highly reproducible for the Nb spiral geometry, while the EBL process (see Sec.~\ref{sec:EBL}) was reserved for the smaller nano-scale features.

\section{Reactive Ion Etching}\label{sec:RIE}
Reactive ion etching (RIE), a type of dry etching, is used to transfer the resist pattern into the functional layer. A low-pressure plasma of reactive gases is generated in the chamber, and ions are accelerated toward the substrate by an RF-induced bias. Material is removed through a combination of chemical reactions and physical sputtering. Because the ions are directional, RIE can produce nearly vertical sidewalls. 

\begin{figure}[htbp]
	\centerfloat{
		\includegraphics[width=0.5\linewidth]{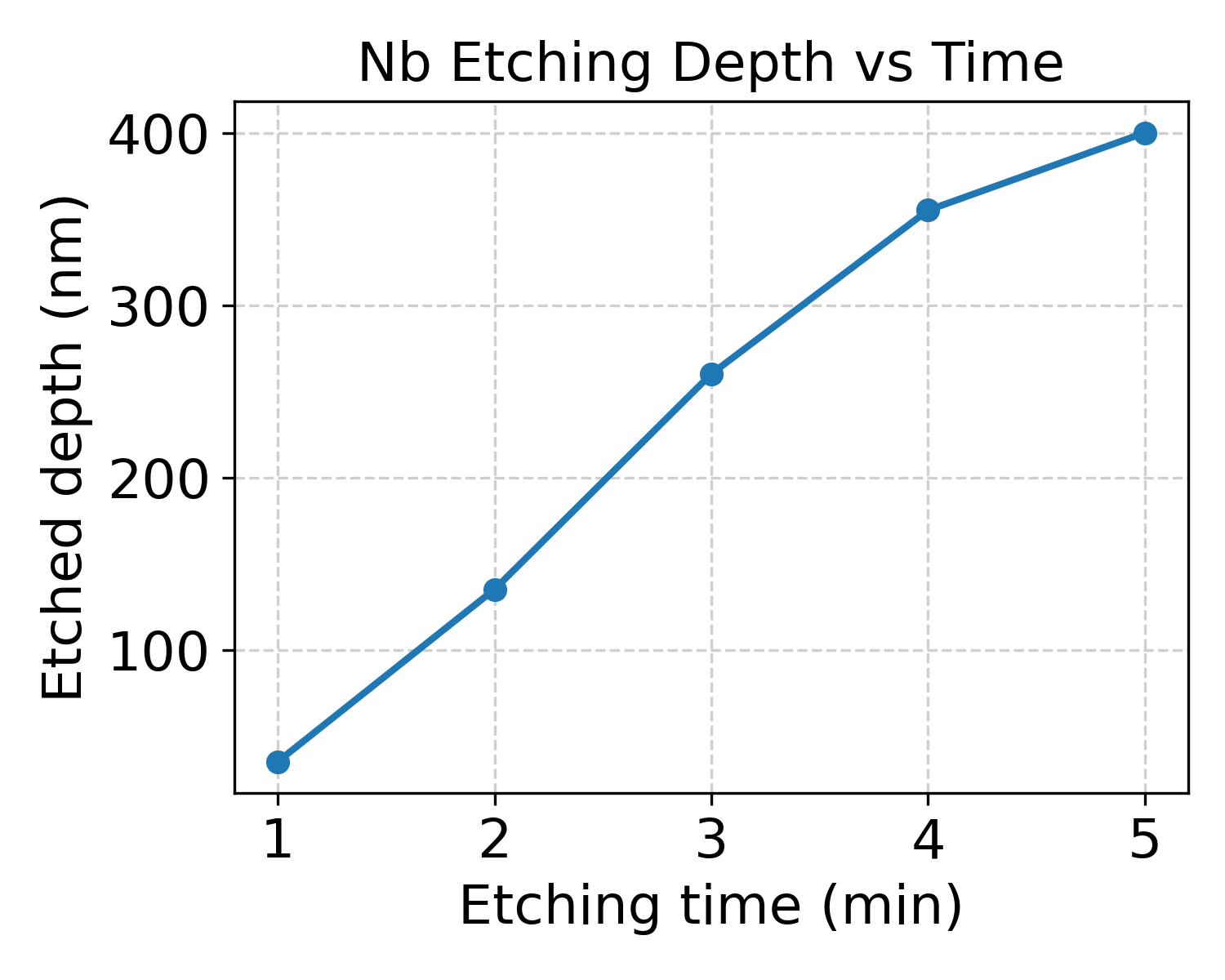}
	}
	\caption[Nb etching depth as a function of etching time]{%
		\textbf{ Nb etching depth as a function of etching time.} Etch rate increases nearly linearly with time, yielding an average rate of $80\,\text{nm}/\text{min}$.
	}
	\label{fig:nb_etch}
\end{figure}

In this work, we used a Samco RIE-10NR system. The resist served as the etch mask. We first performed a short $\mathrm{O_2}$ descum for 20~s to remove any resist residue in the openings. The main etching gas was $\mathrm{CF_4}$. The gas flow rate (sccm), chamber pressure (Pa), RF power (W), and etching time (s) were tuned to achieve optimal selectivity and profile control. For both NbTiN and Nb etching, the chamber pressure was set to 10 Pa, the gas flow to 50 sccm, and the RF power to 100 W. The only difference was the etching time, which set the final etch depth. For the 20 nm NbTiN film, the etching time was 65 seconds, while for the 100 nm Nb film, the etching time was 2 minutes. These settings gave good anisotropy and preserved the designed line width.

The Nb etching depth as a function of etching time is shown in Fig.~\ref{fig:nb_etch}. While for NbTiN on Si, 20 nm of NbTiN and 60 nm of Si were etched for 65 seconds etching time. In the case of Nb on sapphire, no sapphire was over-etched.

\begin{figure}[htbp]
	\centerfloat{
		\includegraphics[width=0.7\linewidth]{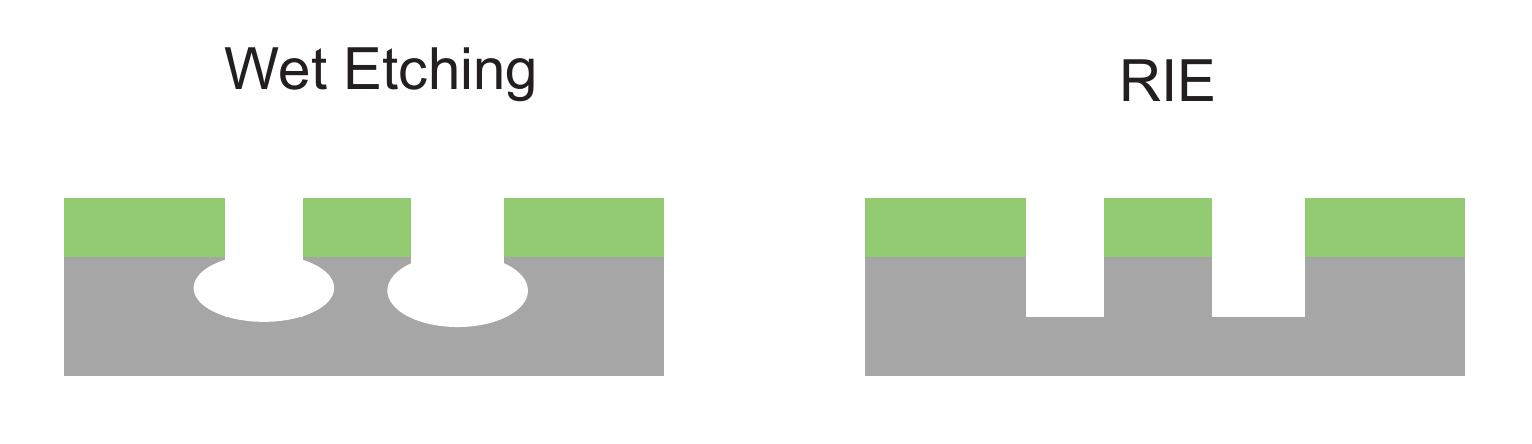}
	}
	\caption[Comparison of etch profiles]{%
		\textbf{Comparison of etch profiles.} Left (Wet etching): isotropic removal with lateral undercut. Right (RIE): anisotropic removal with near-vertical sidewalls and minimal undercut.
	}
	\label{fig:rie}
\end{figure}

After etching, the resist was removed in N‑methyl‑pyrrolidone
(NMP) at $80\,^{\circ}\mathrm{C}$ overnight (followed by an O$_2$ ashing or BHF cleaning if needed). The samples were rinsed, dried with nitrogen gas, and inspected by optical microscopy and scanning electron microscope (SEM). 

Compared with wet etching, which removes material in all directions and often causes undercut, RIE etches mostly downward. Therefore, it makes smaller features with straighter sidewalls and transfers the resist pattern as intended, see Fig.~\ref{fig:rie}.

\section{Fabrication Processes}
\label{sec:fab_pro}

As mentioned above, for electrons-on-neon experiments, NbTiN resonators are fabricated using EBL; and for electrons-on-helium experiments, Nb spiral inductors are fabricated using UV lithography. 

For the NbTiN resonators, after metal deposition (see Sec.~\ref{sec:nbtin_depo}), the 20 nm thin flim NbTiN wafer was then diced into $10\,\mathrm{mm}\times10~\mathrm{mm}$ chips, given a 1~min $\mathrm{O_2}$ plasma clean to remove organics, and dipped in BHF for 1~min to clear any remaining surface residues. Then the chip is ready for device fabrication, see Sec.~\ref{sec:EBL} and Sec.~\ref{sec:RIE}. A summary of the above described fabrication process is shown in Table.~\ref{tab:fabrication}.

\begin{figure}[htbp]
	\centerfloat{
		\includegraphics[width=0.5\linewidth]{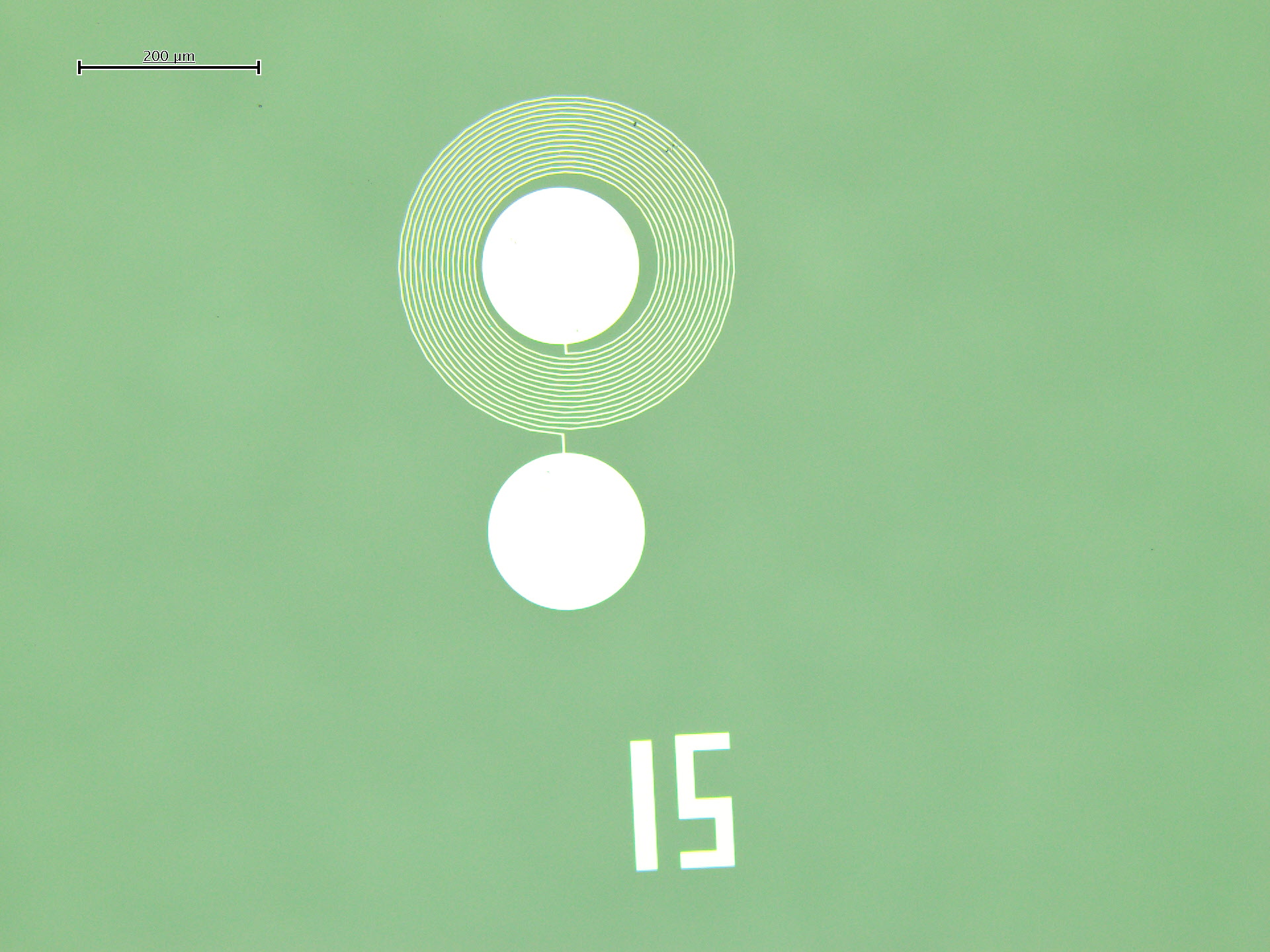}
	}
	\caption[Optical micrograph of a 15-turn spiral inductor]{%
		\textbf{Optical micrograph of a 15-turn spiral inductor}. The line width is  $2~\mu\mathrm{m}$, and the gap width is $4~\mu\mathrm{m}$. The scale bar shows $200\,\mu\text{m}$.
	}
	\label{fig:sp}
\end{figure}

For the Nb spiral inductor, a 50‑nm Nb layer is deposited on a $20~\mathrm{mm}\times20~\mathrm{mm}$ sapphire chip, metal deposition details see Sec.~\ref{sec:nb_depo}. After deposition, the Nb film is ready for UV lithography, see Sec.~\ref{sec:UV} and Sec.~\ref{sec:RIE}. A summary of the above described fabrication process is given in Table~\ref{tab:fabrication}. A photograph of 15-turns spiral inductor on sapphire taken with optical microscope is shown in Figure.~\ref{fig:sp}.

\begin{table}[h]
	\centering
	\setlength{\tabcolsep}{4pt}          
	\renewcommand{\arraystretch}{1.1}    
	\footnotesize                        
	\caption{Fabrication processes for the NbTiN nanowire resonator and Nb spiral inductor.}
	\begin{tabularx}{\textwidth}{l >{\raggedright\arraybackslash}X >{\raggedright\arraybackslash}X}
		\hline
		\textbf{Step} & \textbf{NbTiN nanowire resonator} & \textbf{Nb spiral inductor} \\
		\hline
		Substrate        & High‑resistivity Si,       $10\times10\,\mathrm{mm}$               & Sapphire, $20\times20\,\mathrm{mm}$ \\
		Film deposition  & Sputter; \SI{20}{\nano\meter} NbTiN                  & E-beam evaporator; \SI{50}{\nano\meter} Nb \\
		Spin-coating     & ZEP520A (D.R.=2.4); 500/5 $\rightarrow$ 6000/50; bake \SI{180}{\celsius} 3 min
		& HMDS 500/5 $\rightarrow$ 4500/45; bake \SI{80}{\celsius} 10 min; AZ1500 same ramp; bake \SI{80}{\celsius} 10 min \\
		Patterning       & EBL exposure (dose $\sim 230\,\mu\mathrm{C}\,\mathrm{cm}^{-2}$)                   & UV exposure  (dose \SI{93}{\milli\joule\per\centi\meter\squared}) sub-pixel $\times$4 \\
		Development      & ZED-N50 2 min; IPA rinse 30 s                        & AZ Developer 45 s; DI rinse 2 min \\
		Etching          & O$_2$ descum 25 s; CF$_4$ RIE 65 s (50 sccm, 100 W, 10 Pa)
		& O$_2$ descum 20 s; CF$_4$ RIE 2 min (50 sccm, 100 W, 10 Pa) \\
		Resist removal   & NMP \SI{80}{\celsius} overnight; rinse and dry          & NMP \SI{80}{\celsius} overnight; rinse and dry \\
		\hline
	\end{tabularx}
	\label{tab:fabrication}
\end{table}

\section{Problems and modification}

\begin{figure}[htbp]
	\centerfloat{
		\includegraphics[width=0.7\linewidth]{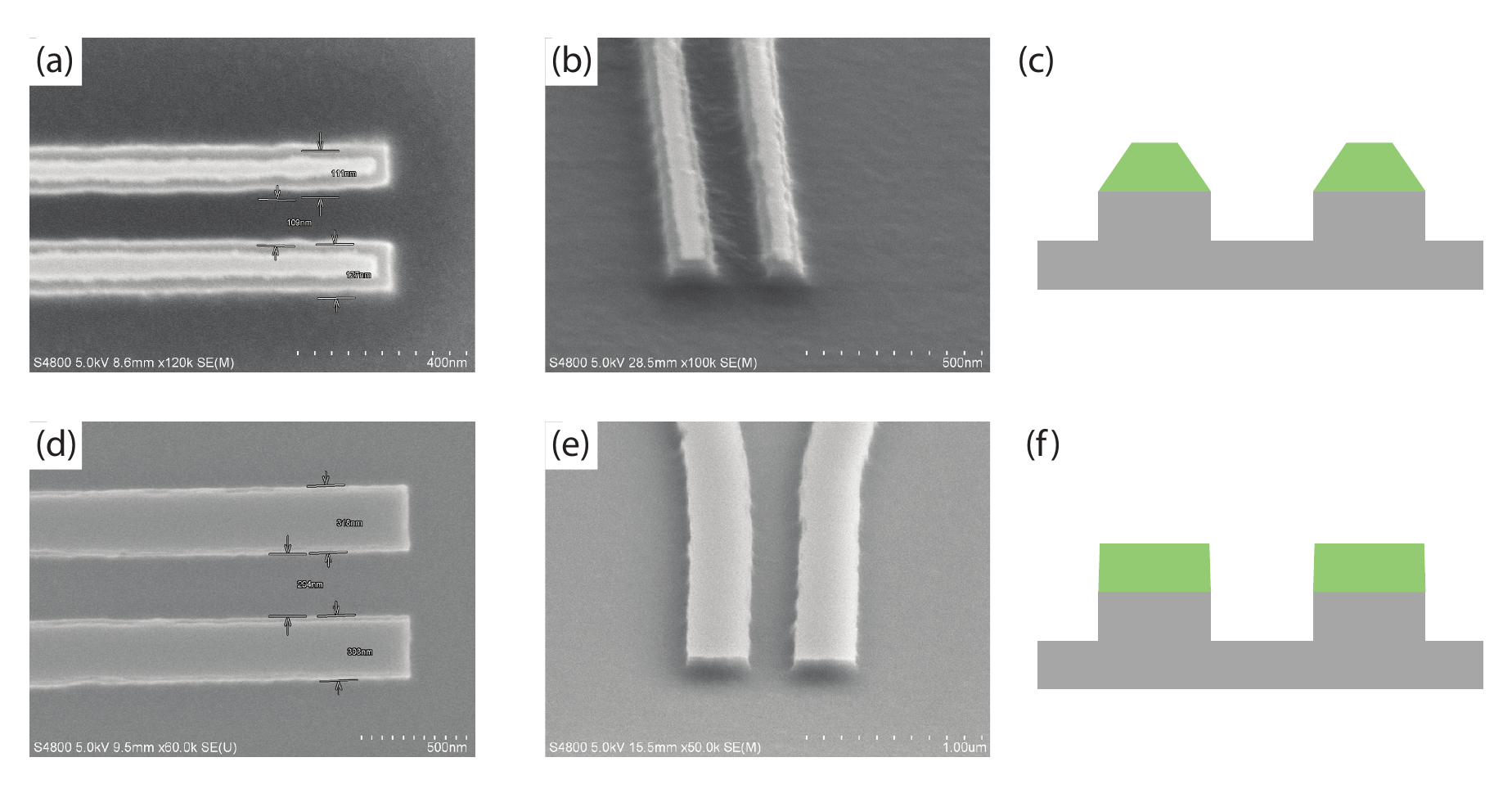}
	}
	\caption[Comparison of NbTiN nanowire patterns before and after optimization]{%
		\textbf{Comparison of NbTiN nanowire patterns before and after optimization.} (a,d) Plan-view SEM with measured gaps. (b,e) Tilted SEM. (c,f) Corresponding cross-section schematics to (a,b) and (d,e). SEM: Hitachi S-4800, 5.0\,kV.
	}
	\label{fig:nesem}
\end{figure}

Several issues were encountered during the fabrication process of the NbTiN resonators. Figure.~\ref{fig:nesem} compares SEM photographs of the resonator patterns before optimisation (a-c) and after optimisation (d-f). 

In the early attempts, the gaps between the nanowires were either collapsed and merged at low dose, or became wider than designed at higher dose. For example, a designed 100 nm gap grew to about 120–150 nm after development and etching.

To fix this critical-dimension bias, we added CAD pre-compensation in the layout. We drew the nominal 100:100 nm (line:gap) pattern as 200:60; the 200:200 nm pattern as 285:140; and the 300:300 nm pattern as 390:230. With this pre-compensation, the final line widths and gaps matched the targets, with deviations of about $\pm 10$ nm.

The second issue was the sidewall shape. After etching, the cross-section was trapezoidal, narrower at the top. This resulted from insufficient anisotropy in the etching process. We extended the O$_2$ descum time (from 10 seconds to 25 seconds) and tuned the CF$_4$ flow, chamber pressure, and RF power to increase directionality (from 20 sccm/5 Pa/50 W to 50 sccm/10 Pa/100 W). After optimisation, the gaps were uniform and the sidewalls were nearly vertical. Fig.~\ref{fig:nesem}(a-c) show the failed sample with tapered sidewalls, Fig.~\ref{fig:nesem}(d-f) show the optimised sample with design-matched gaps and straight sidewalls. The schematics on the right illustrate the change from trapezoidal to almost rectangular cross-sections. With these improvements, the NbTiN resonator met the experiment's requirements.

\clearpage

\chapter{Results and Discussions}

\section{Probing the Quantum Capacitance of Rydberg Transitions of Surface Electrons on Liquid Helium via Microwave Frequency Modulation}
\label{sec:he}

FEs floating on liquid helium form an exceptionally pure two-dimensional electron system, providing a promising platform for qubit implementation. Several theoretical proposals have been based on this system~\cite{lyon2006spin,platzman1999quantum,Lea2000,schuster2010proposal,dykman2023spin,zhang2012spin,jennings2024quantum}. The orbital states of the FEs perpendicular to the liquid helium surface are typically referred to as Rydberg states~\cite{Monarkha2004-un, Andrei1997Two-DimensionalSubstrates}, with the energy levels given by  \( E_{n_z} =  -  R_\infty \left(\frac{\Lambda}{4 }\right)^2  \frac{1}{ n_z^2} \), where \( R_{\infty}\) is the Rydberg constant, \( n_z \) is the quantum number of the Rydberg state, and \( \Lambda=0.0272 \) for liquid helium-4.

The initial proposal for quantum computing using FEs defined qubits using the two lowest Rydberg states~\cite{platzman1999quantum,dykman2003qubits}. Later proposals considered the use of electron spin as qubit states more advantageous, due to its predicted coherence time of several seconds~\cite{schuster2010proposal,Lyon2006spin}. However, direct spin readout is challenging because of its small magnetic moment. To address this, indirect detection schemes have been proposed,  coupling the spin state to either the orbital state associated with motion parallel to the helium surface (in-plane orbital state)~\cite{schuster2010proposal} or the Rydberg state~\cite{kawakami2023blueprint}, enabling spin detection through these coupled degrees of freedom.

For the former approach, experimental efforts have successfully demonstrated the coupling of the in-plane orbital state of a single electron to a microwave superconducting resonator~\cite{koolstra2019coupling}. Among related platforms, solid neon has shown more rapid progress, with recent experiments demonstrating high-fidelity orbital qubits~\cite{zhou2022single,zhou2024electron,Li2025-em}. In contrast, while qubits have not yet been realized on liquid helium, the system offers unique scalability advantages, particularly due to the absence of surface roughness—a key limitation in solid-state substrates.

We focus on the latter approach, which couples the spin state to the Rydberg state. In this scheme, the occurrence of a Rydberg transition reflects the spin state and enables spin readout~\cite{kawakami2023blueprint}. This readout scheme is analogous to spin-to-charge conversion in quantum dot-based systems, such as Pauli spin blockade in double quantum dots (DQDs)~\cite{Ono2002}, where the spin state determines whether a charge transition occurs. In this analogy, the ground and first excited Rydberg states of our system are mapped onto the DQD charge states. Instead of microwave superconducting resonators, radio-frequency (RF) LC tank circuits are used to detect the Rydberg transition, inspired by techniques in quantum dot systems where charge transitions in DQDs are detected via RF reflectometry~\cite{aassime2001radio,Brenning2006-wc,ahmed2018radio,gonzalez2015probing,oakes2023fast,ibberson2021large,apostolidis2024quantum}.  While this method does not allow coherent coupling to microwave photons, it offers better scalability due to the smaller footprint of LC circuits. As a proof-of-concept, we demonstrate the detection of Rydberg transitions from many electrons using an LC tank circuit, leveraging its high sensitivity as a step toward single-electron detection.

\subsection{Experimental setup}

\begin{figure}[htbp]
	\centering
	\centerfloat{%
		\includegraphics[width=0.6\linewidth]{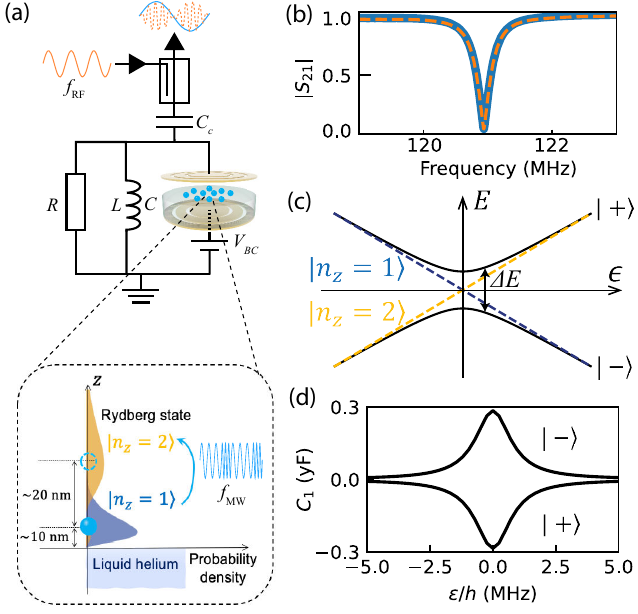}%
	}
	\caption[LC circuit and quantum capacitance measurement]{%
		(a) Schematic of the LC circuit incorporating the Corbino electrodes as a capacitor, with $10^7$ FEs (light blue circles) floating on liquid helium-4 (semi-transparent blue) half-way between. The capacitor is integrated into an LC circuit housed in a leak-tight cell on the mixing chamber plate of  a dilution refrigerator. See the main text for details on the components of the LC circuit. The inset shows the probability density of an electron in the Rydberg-ground state (blue) and in the Rydberg-1st-excited state (yellow), showing its average position from the liquid helium surface. The light blue curves show the FM-MW signal used to induce the Rydberg transition, with carrier frequency \( f_\mathrm{MW} \) and modulation frequency \( f_\mathrm{mf} \) as defined in Eq.~\ref{eq:FM-MW}. Returning to the main panel, the rectangle represents an RF coupler. The orange curves represent the incident RF signal with frequency \( f_\mathrm{RF} \), while the dashed orange curves depict the reflected RF signal. A light blue sinusoidal curve overlapping the orange dashed curves indicates that the reflected signal is amplitude-modulated with \( f_\mathrm{mf} \). (b) Normalized magnitude of the reflectance of the LC circuit measured with a VNA (blue dots) at 150~mK, along with a fit (orange line) without electrons and without MW irradiation. See the main text for details of the fitting results. (c) Schematic energy diagram as a function of the detuning from the Rydberg resonance \(\epsilon\), showing the energy levels of a single electron under microwave irradiation (solid black lines) and without irradiation (dashed lines). (d) Quantum capacitance of a single electron $C_1^{\pm}$ as a function of \( \epsilon/h \) for the eigenstates \( \ket{\pm} \), with the Rydberg transition rate \( 2t_c/h = 0.83~\mathrm{MHz} \).
	}
	\label{fig:he_fig1}
\end{figure}

We built a parallel LC circuit~\cite{ahmed2018radio,gonzalez2015probing,oakes2023fast,ibberson2021large,apostolidis2024quantum}, as shown in Fig.~\ref{fig:he_fig1}(a). In our experimental setup, two sets of parallel plate electrodes with a Corbino geometry~\cite{iye1980mobility,mehrotra1987analysis,wilen1988impedance} are used, with the plates separated by $D=2$~mm. Both the top and bottom plates consist of three concentric electrodes, each with an area of approximately $0.5$~cm$^2$, forming a capacitor. The top central electrode is connected to a niobium spiral inductor microfabricated on a sapphire substrate, which is used to suppress resonator losses thanks to its low dielectric loss and superconducting properties~\cite{colless2013dispersive}. In addition, low-pass filters are implemented on the DC voltage lines to minimize noise coupling to the resonator (Appendix.~\ref{sec:details_circuit}). The LC circuit resonance without electrons, measured using a vector network analyzer (VNA), is shown in Fig.~\ref{fig:he_fig1}(b). From separate measurements, the inductance was determined to be \( L = 708~\mathrm{nH} \). Fitting Eq.~B1 of the Appendix.~\cite{probst2015efficient,probst2024circlefit} gives a loaded quality factor \( Q_\mathrm{tot} = 311 \), an external quality factor \( Q_\mathrm{ext} = 648 \), a coupling capacitance \( C_\mathrm{c} = 0.315~\mathrm{pF} \), a total capacitance \( C = 2.131~\mathrm{pF} \), and an effective resistance \( R = 321~\mathrm{k}\Omega \). The resonance frequency is \( f_0 =\frac{1}{2\pi \sqrt{LC_\mathrm{t}}}= 120.946~\mathrm{MHz} \), where $C_\mathrm{t}=C+C_\mathrm{c}$. $R$ accounts for distributed circuit losses~\cite{ahmed2018radio,oakes2023fast,ibberson2021large} and \( C \) includes contributions from both parasitic and Corbino capacitance after the coupling capacitor.

Microwaves (MWs) are applied to the FEs via a TEM waveguide extending from room temperature into the leak-tight cell, positioned between the top and bottom electrodes. The Rydberg transition energy, \( h f_\mathrm{Ry} \) where \( h \) is Planck's constant, can be tuned by the electric field perpendicular to the helium surface via the Stark effect~\cite{grimes1976spectroscopy}. The MWs resonantly drive transitions between the Rydberg-ground state and the first excited state~\cite{Grimes1974-fg,Lambert1979ElectronsHelium,collin2002microwave,konstantinov2007investigation,kawakami2019image} when the MW frequency \( f_\mathrm{MW} \) matches the Rydberg transition frequency \( f_\mathrm{Ry} \). The bottom center electrode is labeled \(\mathrm{BC}\), and the combination of the bottom middle and outer electrodes is referred to as \(\mathrm{BG}\). DC voltages of \( V_\mathrm{BC} = 12~\mathrm{V} \) and \( V_\mathrm{BG} = -90~\mathrm{V} \) are applied, while all other electrodes are grounded. Note that this charge configuration is chosen to prevent interference from lateral modes of motion, known as plasmons~\cite{grimes1976spectroscopy}, with the Rydberg measurement. These settings produce a nominal perpendicular electric field \( E_z = V_\mathrm{BC}/D \) to the helium surface. This field sets the peak Rydberg transition frequency to approximately \( f_\mathrm{Ry}^0 \approx 165~\mathrm{GHz} \).

When an electron is excited from the Rydberg ground state to the Rydberg first-excited state, their average distance from the liquid surface increases by \( d = 20~\mathrm{nm} \) (Fig.~\ref{fig:he_fig1}(a)), altering the image charge induced on the top plate electrode by  \( \Delta q = \frac{d}{D}e = 10^{-5} e \), where \( e \) is the elementary charge~\cite{kawakami2019image}. Previously, such changes induced by many electrons were detected as a current~\cite{kawakami2019image} and as a voltage~\cite{kawakami2021relaxation}. In this work, we detect such changes as a quantum capacitance~\cite{smith1985direct,luryi1988quantum,duty2005observation,mizuta2017quantum,gonzalez2015probing,crippa2019gate,vigneau2023probing} with an LC circuit using RF reflectometry. This capacitance arises from the finite curvature of the Rydberg energy bands when the Rydberg transition is resonantly driven by microwaves. We assess the circuit’s performance by measuring its capacitance sensitivity in a controlled manner using a frequency-modulated microwave (FM-MW) technique.

\subsection{Quantum capacitance}

Here, we focus on the two lowest Rydberg states, and the Hamiltonian for an electron can be written as~\cite{kawakami2023blueprint}
\begin{equation}
	H = \frac{h f_{\mathrm{Ry}}}{2} \sigma_z + t_c \, \sigma_x \cos\left(2\pi f_{\mathrm{MW}} t\right),
\end{equation}
where $\sigma_z = \ket{n_z=2}\bra{n_z=2} - \ket{n_z=1}\bra{n_z=1}$ and 
$\sigma_x = \ket{x+}\bra{x+} - \ket{x-}\bra{x-}$. 
Here, $\ket{n_z=2}$ denotes the Rydberg first-excited state, 
$\ket{n_z=1}$ the Rydberg ground state, and 
$\ket{x\pm} = \frac{1}{\sqrt{2}} \left( \ket{n_z=2} \pm \ket{n_z=1} \right)$. The second term represents MW irradiation at frequency $f_{\mathrm{MW}}$, which induces transitions between $\ket{n_z=2}$ and $\ket{n_z=1}$. 
By moving to the rotating frame at frequency $f_{\mathrm{MW}}$, the Hamiltonian is transformed to
\begin{equation}
	H_\mathrm{r} = \frac{\epsilon}{2} \sigma_z + t_c \, \sigma_x,
\end{equation}
where $\epsilon = h (f_{\mathrm{Ry}} - f_{\mathrm{MW}})$ is the detuning from the Rydberg resonance, and $2t_c/h$ is the Rydberg transition rate. 
The eigenstates of $H_\mathrm{r}$ are denoted by $\ket{\pm}$ and their energies are given by $E_\pm= \pm \frac{1}{2}\sqrt{\epsilon^2+(2t_c)^2}$ . 
Figure~\ref{fig:he_fig1}(c) shows a schematic energy diagram of this two-level system. For RF reflectometry, an incident signal $ V_\mathrm{RF} \cos(2\pi f_\mathrm{RF} t)$ is injected into the LC circuit, and the reflected signal is measured using a spectrum analyzer. Throughout this work, we use a frequency of $f_\mathrm{RF} = 120.94~\mathrm{MHz}$, which is set to the LC resonance frequency $f_0$ in the presence of FEs.

\begin{figure}[htbp]
	\centering
	\centerfloat{%
		\includegraphics[width=0.6\linewidth]{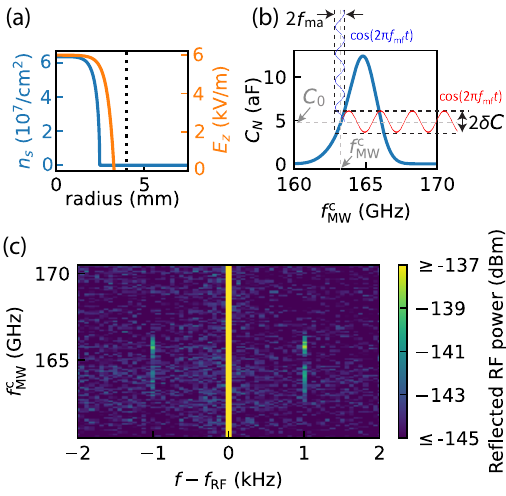}%
	}
	\caption[Electron density, quantum capacitance, and FM sidebands]{%
		(a) Calculated saturated electron density $n_s$ (blue, left axis) and vertical electric field $E_z$ (orange, right axis) as a function of radial distance from the center of the Corbino electrodes, for $V_\mathrm{BC} = 12$~V and $V_\mathrm{BG} = -90$~V. Given that the radius of the central electrode is 4~mm (vertical dotted line), all electrons are confined well within this region. Only the positive values of the electric field are shown for clarity.
		(b) Quantum capacitance of many electrons $C_N$ as a function of the microwave frequency $f_\mathrm{MW}$ with the Rydberg transition rate $2t_c/h=0.83$~MHz. The FM-MW signal in Eq.~\ref{eq:FM-MW} induces a modulation in the quantum capacitance at \( f_{\mathrm{mf}} \), as described in Eq.~\ref{eq:C_N}. (c) Reflected RF power measured with a spectrum analyzer as a function of the MW carrier frequency $f_\mathrm{MW}^c$. The FM modulation parameters are \( f_\mathrm{mf} = 1~\mathrm{kHz} \) and \( f_\mathrm{ma} = 768~\mathrm{MHz} \). In all experiments presented in the main text, the microwave power irradiated on the electrons is fixed (see Appendix.~\ref{sec:power_dependence} for the estimated value).
		Sideband signals appear at $f = f_\mathrm{RF} \pm f_\mathrm{mf}$ around $f_\mathrm{MW}^c = 165$~GHz, associated with the Rydberg transition. Only every 250\textsuperscript{th} data point is plotted along $f$ for clarity. The sharp signals at $f-f_\mathrm{RF}=\pm 1$~kHz lie on retained points, with no significant features missed due to downsampling.}
	\label{fig:he_fig2}
\end{figure}

When an electron is in the state \( \ket{\pm} \), the induced image charge on the top plate electrode is given by
\( Q_1^{\pm} = \Delta q \, |\braket{n_z=2|\pm}|^2 = \frac{\Delta q}{2} \left( 1 \mp \frac{\epsilon}{\Delta E} \right) \),
where \( \Delta E = E_+ - E_- \). The corresponding quantum capacitance (Appendix.~\ref{sec:quantum_C}) is proportional to the curvature of the energy bands:
\begin{equation}
	C_1^{\pm}(\epsilon) = \Delta q \, \frac{dQ_1^{\pm}}{d\epsilon} = \mp \Delta q^2 \frac{(2t_c)^2}{2\Delta E^3} = \mp \Delta q^2 \frac{d^2 E_\pm}{d \epsilon^2},
\end{equation}
and is shown in Fig.~\ref{fig:he_fig1}(d). Introducing the population difference between the \( \ket{+} \) and \( \ket{-} \) states as \( \chi \), the quantum capacitance of a single electron is expressed as
\begin{equation}
	C_1(\epsilon) = \chi \, \Delta q^2 \frac{(2t_c)^2}{2\Delta E^3},\label{eq:C_1}
\end{equation}
where $\chi\approx t_c/k_BT =1.2 \times 10^{-4}$  for $2t_c/h=0.83$~MHz and $T=160$~mK is the temperature (see Sec.~\ref{sec:LZ} for the derivation of $2t_c$).   To capture the quantum capacitance arising from a large number of FEs, we convolute the single-electron quantum capacitance \( C_1(\epsilon) \) with the distribution \( n(\epsilon) \), which represents the number of electrons per energy detuning (Appendix.~\ref{sec:Ez_fRy_distribution}):
\begin{equation}
	C_N(\epsilon^0) =  \int_{-\infty}^{+\infty} C_1(\epsilon^0 -\epsilon)\, n(\epsilon)\, d\epsilon , \label{eq:C_N}
\end{equation}

\noindent
where the detuning for the ensemble of FEs is defined as \( \epsilon^0 = h  (f^0_\mathrm{Ry} - f_\mathrm{MW}) \). Numerically calculated $C_N(f_\mathrm{MW})$ is plotted in Fig.~\ref{fig:he_fig2}(b). As shown in Fig.~\ref{fig:he_fig2}(a), both the electron density \( n_s \) and the perpendicular electric field \( E_z \) experienced by the electrons vary with radial distance from the center, leading to a corresponding variation in \( f_\mathrm{Ry} \). This spatial inhomogeneity results in an asymmetry in \( C_N(f_\mathrm{MW}) \) (Appendix.~\ref{sec:Ez_fRy_distribution}).

\subsection{Frequency modulation}
To probe the system, we irradiate the FEs with a FM-MW signal. The time-dependent frequency of the FM-MW is given by

\begin{equation}
	f_\mathrm{MW}(t) = f_\mathrm{MW}^\mathrm{c} + f_\mathrm{ma} \cos(2\pi f_\mathrm{mf} t),
	\label{eq:FM-MW}
\end{equation}

\noindent
where \( f_\mathrm{MW}^\mathrm{c} \) is the carrier frequency, \( f_\mathrm{ma} \) is the modulation amplitude, and \( f_\mathrm{mf} \) is the modulation frequency. Under FM modulation, the detuning \( \epsilon^0 \) is modulated at frequency \( f_\mathrm{mf} \) with amplitude \( h f_\mathrm{ma} \), inducing a time-dependent variation in the quantum capacitance \( C_N \). On either side of the peak of \( C_N \), and for sufficiently small \( f_\mathrm{ma} \), \( C_N \) is approximately linear in \( \epsilon^0 \) and can be expressed as
\begin{equation}
	C_N(t) \approx C_0 + \delta C \cos(2\pi f_\mathrm{mf} t),
\end{equation}
where \( C_0 \) is the average capacitance at a fixed detuning, and \( \delta C \) is the amplitude of capacitance modulation (Fig.~\ref{fig:he_fig2}(b), Appendix.~\ref{sec:quantum_C}). Expressing the reflection coefficient \( \Gamma_\mathrm{ref} \) as a function of capacitance (Appendix.~\ref{sec:reflectance_change}), the time-dependent reflection coefficient under FM-MW irradiation becomes
\begin{equation}
	\Gamma_\mathrm{ref}(C_\mathrm{t}+C_N(t))  
	\approx  \Gamma_\mathrm{ref}(C_\mathrm{t}) 
	- j\frac{2Q_\mathrm{tot}^2}{Q_\mathrm{ext}} \frac{C_0 + \delta C \cos(2\pi f_\mathrm{mf} t)}{C_\mathrm{t}}.
	\label{eq:GammaDiff}
\end{equation}

\noindent
Accordingly, the reflected signal is given by $ V_\mathrm{o}=\Gamma_\mathrm{ref}(C_\mathrm{t} + C_N(t))\, G\, V_\mathrm{RF} \cos (2\pi f_\mathrm{RF} t)$, where \( V_\mathrm{RF} \) is the voltage amplitude of the RF signal at the top-plate electrode, and \( G = 41 \) is the total gain from the output of the LC resonator to the input of the spectrum analyzer. A portion of the reflected signal is amplitude modulated at frequency \( f_\mathrm{mf} \), and thus contains sideband components at \( f_\mathrm{RF} \pm f_\mathrm{mf} \). The amplitude of the sideband signals is given by
\begin{equation}
	V_\mathrm{s}  = G  \frac{Q_\mathrm{tot}^2}{Q_\mathrm{ext}} \frac{  |\delta C|}{C_\mathrm{t}} V_\mathrm{RF}.
	\label{eq:Vs}
\end{equation}

\noindent Note that the Rydberg transition can also be detected under continuous-wave microwave irradiation without FM~\cite{jennings_inprep}, but this requires phase-sensitive detection (Appendix.~\ref{sec:quantum_C}). In contrast, FM shifts the quantum capacitance signal to distinct sideband frequencies, enabling detection via amplitude measurements. Furthermore, varying FM parameters allows systematic control over the signal strength, enabling more controlled and quantitative characterization.

Figure~\ref{fig:he_fig2}(c) shows the reflected RF power measured with the spectrum analyzer with \( f_\mathrm{mf} = 1~\mathrm{kHz} \). Sideband peaks are observed at \( f=f_\mathrm{RF} \pm 1~\mathrm{kHz} \) and around \( f_\mathrm{MW}^\mathrm{c} = 165~\mathrm{GHz} \). At exact Rydberg-resonance, \( f_\mathrm{MW}^\mathrm{c} = 165~\mathrm{GHz} \), the signal vanishes because \( \delta C= 0 \) at this point~\footnote{Sideband signals at \( f_\mathrm{RF} \pm 2~\mathrm{kHz} \) were expected to be observed at this point, but they were extremely small and difficult to distinguish from the noise.}. The signal exhibits peaks at \( \pm 1~\mathrm{GHz} \) offsets from the resonance, as shown in Fig.~\ref{fig:he_fig2}(c) and Fig.~\ref{fig:he_fig3}(a). The peak at \( 166~\mathrm{GHz} \) is larger than that at \( 164~\mathrm{GHz} \), which is attributed to the steeper slope in $C_N$ on the high-frequency side.

\subsection{Capacitance sensitivity}

\begin{figure}[htbp]
	\centering
	\centerfloat{%
		\includegraphics[width=0.6\linewidth]{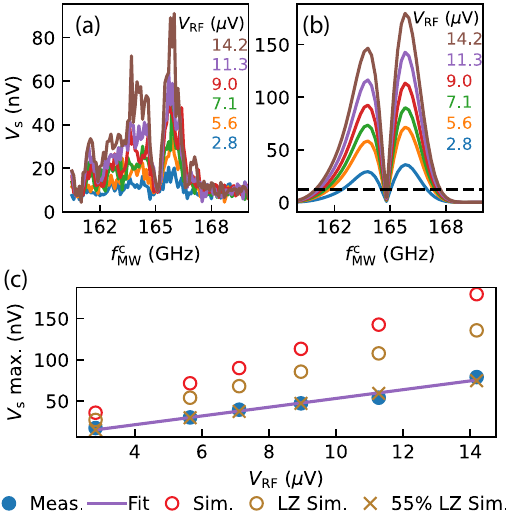}%
	}
	\caption[Sideband amplitude vs.\ microwave frequency and RF voltage]{%
		(a) Sideband amplitude $V_\mathrm{s}$ at $f = f_\mathrm{RF} + f_\mathrm{mf}$ as a function of the MW carrier frequency $f_\mathrm{MW}$ for different $V_\mathrm{RF}$, measured with $f_\mathrm{mf} = 1$~kHz, $f_\mathrm{ma} = 768$~MHz.  (b) Simulated counterpart of (a). At low $V_\mathrm{RF}$, the signal drops below the noise level $V_\mathrm{n}=12$~nV (dashed black line). (c) Blue points (red circles) represent the maximum sideband amplitude extracted from the experimental data in (a) (simulation in (b)) as a function of \( V_\mathrm{RF} \). Bronze circles (crosses) represent the simulation results including Landau--Zener (LZ) effects with the saturated electron density (55\% of the saturated electron density) (see Sec.~\ref{sec:LZ}). To determine the experimental maxima, a Gaussian rolling average with a width of 5 points was applied. The purple line is a fit of Eq.~(\ref{eq:Vs}) to the experimental data.}
	\label{fig:he_fig3}
\end{figure}

Figures~\ref{fig:he_fig3}(a) and (b) show the measured and simulated sideband amplitude $V_\mathrm{s}$ at $f=f_\mathrm{RF}+f_\mathrm{mf}$ as a function of $f_\mathrm{MW}^c$ for different \( V_\mathrm{RF} \), respectively. This simulation is performed by calculating the time-domain modulated $C_N(t)$ using Eq.~\ref{eq:C_N} and Eq.~\ref{eq:FM-MW} and taking the Fourier transform of the reflected signal $V_\mathrm{o} $ (Appendix. ~\ref{sec:Ez_fRy_distribution}). The simulation is performed without using any parameters extracted from the experimental data, except for the Rydberg transition rate, $2t_c/h = 0.83$~MHz, which is taken from Fig.~\ref{fig:he_fig4}(b). Figure~\ref{fig:he_fig3}(c) shows the maximum sideband amplitude measured and simulated around \( f_\mathrm{MW}^c = 166~\mathrm{GHz} \) as a function of \( V_\mathrm{RF} \). The experimental amplitude is smaller than the simulated one even if the effect of the Landau--Zener transition discussed in Sec.~\ref{sec:LZ} is also included (bronze circles in Fig.~\ref{fig:he_fig3}(c)), which could be attributed to a reduction in the electron density below the saturated value due to electron loss, or a higher electron temperature compared to the thermometer, which is located outside the experimental cell and reads approximately $T=$160~mK in the data shown in this manuscript. The bronze crosses in Fig.~\ref{fig:he_fig3}(c) indicate that the experimental value agrees with the simulation if the electron density is assumed to be 55\% of the saturated value.

The purple line in Fig.~\ref{fig:he_fig3}(c) shows fitting Eq.~\ref{eq:Vs} of the Appendix to the experimental data yields \(
\frac{|\delta C|}{C_\mathrm{t}} = 8.6 \times 10^{-7}
\). From this, we obtain  \( |\delta C| = 2.1~\mathrm{aF} \). The capacitance sensitivity is given by
\begin{equation}
	S_c = \frac{|\delta C| \, V_\mathrm{n}}{ \sqrt{B} \, V_\mathrm{s} } 
	= \frac{Q_\mathrm{ext} C_\mathrm{t} V_\mathrm{n}}{G Q_\mathrm{tot}^2 \sqrt{B} V_\mathrm{RF}} 
	= 0.34 ~\mathrm{aF/\sqrt{Hz}}
\end{equation}

\noindent for $V_\mathrm{RF} = 14~\mathrm{\mu V}$, where the voltage noise is $V_\mathrm{n} = 12$~nV and the measurement bandwidth is $B = 1$~Hz. These results demonstrate that RF reflectometry is well suited for detecting the Rydberg transition of a single SE, as the resulting capacitance change in nanoscale devices is expected to be 60~aF~\cite{kawakami2023blueprint}. With a measurement bandwidth of 10~Hz, detection with a signal-to-noise ratio of approximately 1 is feasible. Sensitivity could be further enhanced by using lossless variable capacitors based on ferroelectric materials such as STO~\cite{apostolidis2024quantum} to achieve critical coupling. At critical coupling, where \( Q_\mathrm{int} = Q_\mathrm{ext} \), the sensitivity coefficient scales as \( S_c \propto 1/Q_\mathrm{int} \). The internal quality factor \( Q_\mathrm{int} \) can be increased either by increasing the resonance frequency \( f_0 \) or by suppressing circuit losses, i.e., increasing \( R \) (Appendix.~\ref{sec:appendix_compare}).

\subsection{Landau--Zener transition}\label{sec:LZ}

\begin{figure}[htbp]
	\centering
	\centerfloat{%
		\includegraphics[width=0.6\linewidth]{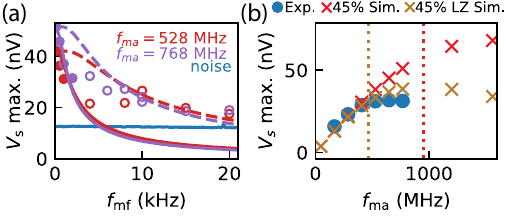}%
	}
	\caption[FM-dependence of sideband amplitude and extracted transition rates]{%
		(a) Maximum sideband amplitude as a function of FM frequency \( f_\mathrm{mf} \) for \( f_\mathrm{ma} = 528 \)~MHz (pink circles) and \( 768 \)~MHz (purple circles). Solid lines are a joint fit of both $f_\mathrm{ma}$ only to the data in the domain \( f_\mathrm{mf} \leq 3 \)~kHz (filled circles), while dashed lines are a joint fit for the domain up to \( f_\mathrm{mf} = 20 \)~kHz (solid and open circles). The extracted Rydberg transition rates are \( 2t_c/h = 0.83 \pm 0.29 \)~MHz and \( 1.74 \pm 0.22 \)~MHz, for the two fit domains respectively, with 99\% confidence intervals. The blue line indicates the measured noise level. Data for \( f_\mathrm{mf} > 20 \)~kHz fall below the noise floor and are omitted. (b) Maximum sideband amplitude as a function of FM amplitude \( f_\mathrm{ma} \), with \( f_\mathrm{mf} = 1 \)~kHz. Blue points (red crosses) represent the experimental data (simulation data). Bronze crosses represent the simulation results including Landau--Zener (LZ) effects. The dotted red and bronze lines show the values up to which there is linear relationship for $f_{ma}$ and $\delta C$, with and without the LZ transition, respectively.  The simulation results use 45\% of the saturated electron density. In both (a) and (b), \( V_\mathrm{RF} = 9~\mu\mathrm{V} \).}
	\label{fig:he_fig4}
\end{figure}

For a coupled two-level system with a modulated detuning, Landau--Zener (LZ) transitions need to be considered~\cite{shevchenko2010landau,Landau1932,Zener1932,Stuckelberg1932}. In this work, although the detuning is modulated continuously, it is sufficient to consider only a single passage through \(\epsilon = 0\), as \(f_\mathrm{mf}\) is much slower than the typical relaxation rate of the Rydberg state of electrons on liquid helium (1~MHz~\cite{Monarkha2007-el,monarkha2006decay,kawakami2021relaxation}), allowing the system to reach thermal equilibrium before the next passage. Starting from the state \( \ket{-} \) at $|\epsilon| \gg 0$, the system undergoes a LZ transition when passing through \( \epsilon = 0 \), switching to \( \ket{+} \) with probability \( P_{\mathrm{LZ}} = \exp(-2\pi\delta) \), where \( \delta = (2t_c)^2 / 4 f_{\mathrm{ma}} f_{\mathrm{mf}} \), or remaining in \( \ket{-} \) with probability \( 1 - P_{\mathrm{LZ}} \).  Only the population that stays in \( \ket{-} \) contributes to the change in quantum capacitance, and thus to the observed signal. An increase in \( f_{\mathrm{mf}} \) enhances the LZ transition probability \( P_{\mathrm{LZ}} \), which in turn suppresses the signal amplitude. Figure~\ref{fig:he_fig4}(a) shows the experimentally measured sideband peak amplitude as a function of \( f_{\mathrm{mf}} \). The data were fitted using the expression \( a(1 - P_{\mathrm{LZ}}) \), where \( a \) denotes the amplitude in the absence of LZ transitions and \( 2t_c \) was treated as a fitting parameter. From the fit, we extracted a Rydberg transition rate of \( 2t_c /h = 0.83~\mathrm{MHz} \) (\( 1.74~\mathrm{MHz} \)), using data points with \( f_{\mathrm{mf}} \leq3~\mathrm{kHz} \) (\( f_{\mathrm{mf}} \leq 20~\mathrm{kHz} \)). Slightly different Rydberg transition rates depending on \( f_{\mathrm{mf}}\) may arise from variation in the effective microwave power, possibly due to frequency-dependent transmission characteristics in the frequency multiplier or the waveguide. Except for Fig.~\ref{fig:he_fig4}(a), all experimental data were acquired at \( f_{\mathrm{mf}} = 1~\mathrm{kHz} \), and adopting \( 2t_c/h = 0.83~\mathrm{MHz} \) in the analysis yields better consistency with those data. Therefore, we adopt \( 2t_c/h = 0.83~\mathrm{MHz} \) throughout the main text. Additionally, as the FM modulation amplitude \( f_{\mathrm{ma}} \) increases, the signal begins to be suppressed, as seen in Fig.~\ref{fig:he_fig4}(b). Although this behavior is partly due to the system leaving the linear region of \( C_N \) at high \( f_\mathrm{ma} \), the earlier onset of the signal drop is consistent with the effect of LZ transitions. Note that the experiments shown in Fig.~\ref{fig:he_fig4}(b) (and Fig.~\ref{fig:power_dependence} of the Appendix) were conducted at a different time from those in Fig.~\ref{fig:he_fig3} and Fig.~\ref{fig:he_fig4}(a). Even under the same conditions, the signal was smaller, which we attribute to a loss of electrons. In the simulations in  Fig.~\ref{fig:he_fig4}(b) and Fig.~S3, this effect was included by assuming that the number of electrons had decreased to 45\% of the saturated density.

\subsection{Conclusion}

This section demonstrated the measurement of quantum capacitance induced by the Rydberg transition of surface electrons on liquid helium using FM-MW. The observed signal shape and amplitude are well explained by a model incorporating quantum capacitance and LZ transitions induced by FM. FM enables amplitude-based detection without phase-sensitive techniques and allows systematic tuning of the signal strength for quantitative characterization. Our LC circuit, which incorporates a microfabricated superconducting coil and employs filters to suppress external losses, achieves a high quality factor with minimal dissipation. Consequently, we attained a capacitance sensitivity of \( 0.34~\mathrm{aF}/\sqrt{\mathrm{Hz}} \), sufficient to resolve the Rydberg transition of a single electron and promising for future readout of qubit states.

\section{NbTiN Nanowire Resonators and Spin–Photon Coupling Prospects with Electrons on Solid Neon}\label{sec:ne}

Using cavity quantum electrodynamics (cQED), the quantum states of an electron can be coupled to a microwave photon in a superconducting resonator. This approach has been successfully implemented using electrons in semiconductors, for both charge~\cite{Petersson2012,Mi2017-sp,Stockklauser2017-ao} and spin~\cite{Landig2018-wt,samkharadze2016high,mi2018coherent} degrees of freedom. Recent experiments have demonstrated coupling between distant qubits~\cite{Borjans2020-do,Bottcher2022-dp,Harvey-Collard2022-jh} and, more recently, realized remote two-qubit gate operations~\cite{dijkema2025cavity}, contributing to the development of scalable electron spin qubit systems.

Recently, cQED experiments have been demonstrated using electrons on solid neon~\cite{zhou2022single}. The development of an electron charge qubit on solid neon is a significant breakthrough, reaching charge coherence times of about $100\,\mu\mathrm{s}$~\cite{zhou2022single,zhou2024electron}. Recent theoretical studies suggest that at temperatures around $\sim10\,\mathrm{mK}$, decoherence of the electron's charge degree of freedom is primarily limited by acoustic phonons in solid neon~\cite{Li2025-em}. However, the coherence time expected from this mechanism is longer than the measured value, which indicates that the observed decoherence may instead be dominated by stray electrons near the qubit~\cite{li2025electron}. Therefore, controlling the electron density on solid neon is a key step toward achieving qubits with longer coherence times.

Even longer coherence times, potentially reaching the order of $1\,\mathrm{s}$, are expected for the spin states of an electron on solid neon~\cite{chen2022electron}, making them attractive for spin-photon coupling schemes. Such coupling, mediated by charge-photon interaction, has been proposed~\cite{benito2017input,Benito2019-al} and experimentally demonstrated with electrons in semiconductors~\cite{mi2018coherent,samkharadze2018strong,dijkema2025cavity}. In this framework, maintaining long coherence times for both spin and charge states is crucial for achieving high spin qubit gate fidelities.

By combining cQED technology with these extended coherence times for both the charge and spin states of electrons on solid neon, electron qubit systems can be scaled into large networks, while simultaneously simplifying the complexity of control required for quantum error correction~\cite{fowler2012surface}. Along this line, the primary goal is to achieve strong coupling between the spin state of a single electron on solid neon and a microwave photon (spin-photon strong coupling). In order to realize spin-photon strong coupling, spin-charge interaction is required. The charge state can couple to a photon in a superconducting resonator, enabling spin-photon coupling via the spin-charge interaction. This spin-charge interaction can be artificially introduced using locally placed micro-ferromagnets~\cite{tokura2006coherent,jennings2024quantum,kawakami2023blueprint,schuster2010proposal}. An external magnetic field is applied to magnetize the ferromagnet and induce a finite Zeeman splitting.

As a first step toward realizing spin qubits using electrons on neon and a nanowire resonator, we fabricated NbTiN superconducting nanowire resonators without integrating micromagnets, and carried out foundational experiments. In these experiments, we deposited a thin layer of neon followed by electrons onto the resonator, gradually increased the electron density, and monitored the resulting resonance peak shift. Importantly, depositing neon and electrons did not degrade the internal quality factor of the resonators. Encouraged by these experimental results, we proceeded to a theoretical analysis of the next stage toward implementing spin qubits, using the experimentally obtained values as input. We investigated the optimized geometry for micromagnets required to induce local magnetic field gradients, and demonstrated that cooperativity between the resonator photons and the spin state can reach \(\gtrsim 10^7\). We theoretically identified optimal conditions that enable high-fidelity single- and two-qubit gate operations for spin qubits, and found that fidelities of 99.99\% and 99.9\%, respectively, are potentially achievable even when using natural neon.

\subsection{Nanowire resonator}
\label{sec:NWresonator_neon}

Since applying a magnetic field is necessary, we fabricated resonators using a $20\,\text{nm}$-thick NbTiN film sputtered onto high-resistivity silicon to enhance their resilience to magnetic fields. In addition, NbTiN has high kinetic inductance, which, in turn, strengthens the coupling between the resonator's photons and the charge state~\cite{samkharadze2016high}. The critical temperature of the NbTiN film was measured to be $10.7\,\text{K}$. Achieving strong coupling between the resonator's photons and the charge state is crucial for obtaining high spin-photon cooperativity, which is essential for realizing high spin qubit gate fidelities~\cite{jeffrey2014fast,dijkema2025cavity}. To further improve performance, a nanowire-shaped resonator was chosen to suppress the generation of vortices that could disrupt superconductivity~\cite{samkharadze2016high,kroll2019magnetic}.

\begin{figure}[htbp]
	\centerfloat{
		\includegraphics[width=0.9\linewidth]{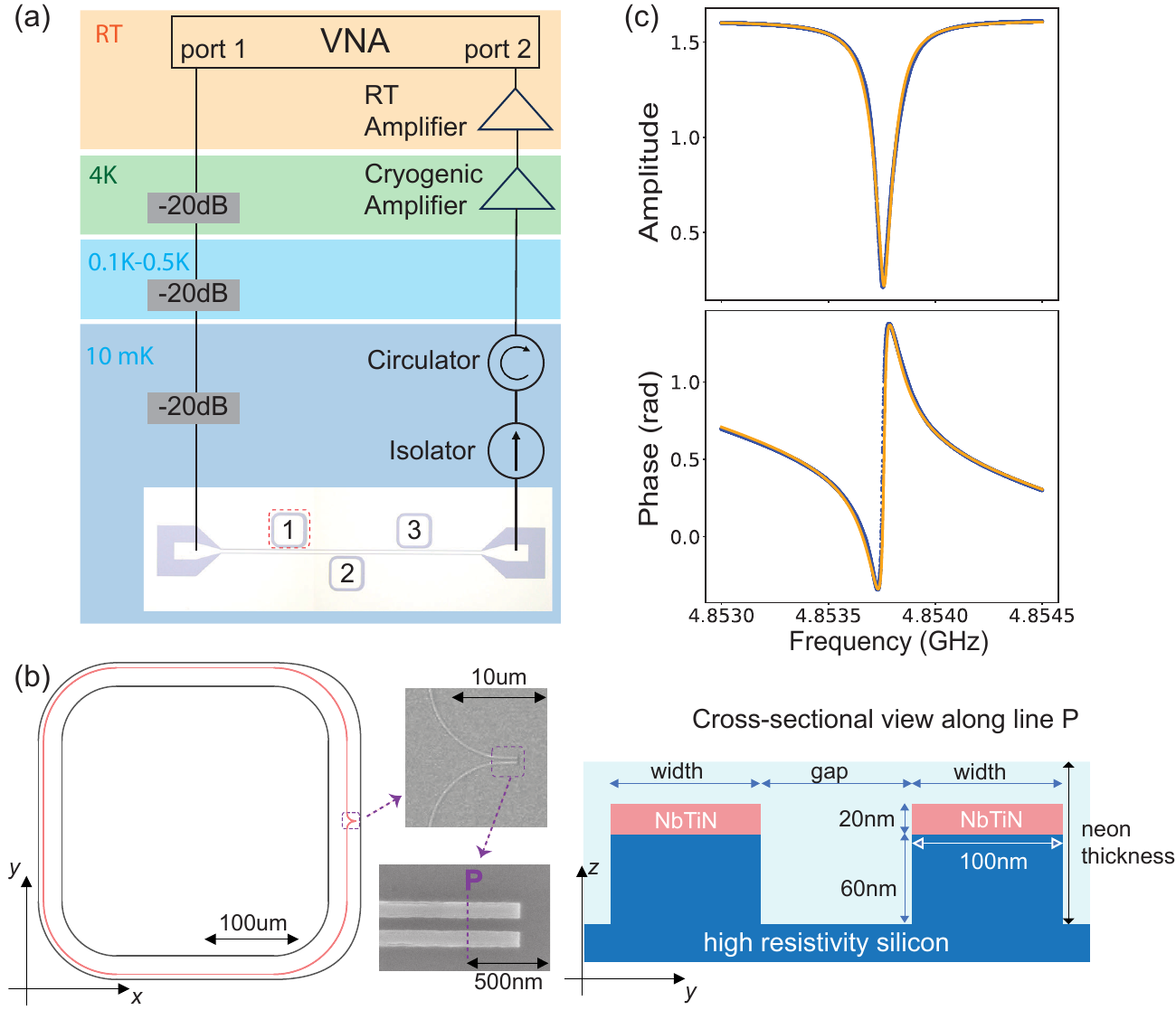}
	}
	\caption[Device overview and resonator characterization]{%
		(a) Electrical setup within the dilution refrigerator and at room temperature (RT), and an optical micrograph of the device on the $10\,\mathrm{mK}$ plate showing three resonators sharing a single microwave feedline. (b) Conceptual illustration of Resonator~1 and SEM zoom-ins near the resonator ends. (c) Transmission response of Resonator~1 measured at $10\,\mathrm{mK}$; data are fit following Refs.~\cite{probst2015efficient,probst2024circlefit}.%
	}
	\label{fig:sample_neon}
\end{figure}

We fabricated three resonators, all sharing a single feedline (Fig.~\ref{fig:sample_neon}). The two ends of each resonator meet with a small gap, where the electric field is maximized, enabling strong coupling between the electron's dipole moment and the electric field. Both the widths and the gaps were designed to be $100\,\mathrm{nm}$ for Resonator~1, $200\,\mathrm{nm}$ for Resonator~2, and $300\,\mathrm{nm}$ for Resonator~3. The resonance frequency of Resonator~1 was measured to be $f_\mathrm{r} = 4.81\,\mathrm{GHz}$ with $Q_\mathrm{ext} = 3.7 \times 10^4$, showing good agreement with simulations (Appendix~\ref{sec:TLC} and Table~\ref{tab:reso_table_neon}). $Q_\mathrm{int} = 2.3 \times 10^5$ and $Q_\mathrm{tot} = 3.2 \times 10^4$ were measured at $10\,\mathrm{mK}$. These values were obtained using an input power of $-21\, \text{dBm}$, corresponding to an estimated intracavity average photon number of $\langle n_{\text{ph}} \rangle \approx 10^5$. The resonance frequency of Resonator~2 was measured to be \(f_\mathrm{r} = 5.91\,\mathrm{GHz}\). Although both the width and gap were nominally designed to be \(200\,\mathrm{nm}\), over-etching appears to have reduced the resonator width, resulting in better agreement with simulation results for a width-gap configuration in the range of \(150\)–\(150\,\mathrm{nm}\) (Table~\ref{tab:reso_table_neon}). Other parameters for Resonators~2 and 3 are provided in Table~\ref{tab:reso_table_neon}. For all the resonators, the total loss rate $\kappa/2\pi= f_\mathrm{r} /Q_\mathrm{tot} \approx 0.1\,\mathrm{MHz}$.

\begin{table}[htbp]
	\centering
	\caption[Resonator parameters: simulation vs experiment]{COMSOL simulations and experimental data measured at $10\,\mathrm{mK}$ without neon or electrons. Intended width--gap dimensions for Resonators~1, 2, and 3 were 100\,nm/100\,nm, 200\,nm/200\,nm, and 300\,nm/300\,nm, respectively. For Resonator~2, simulation results for width--gap dimensions of 150\,nm/150\,nm are also included. See Appendix~\ref{sec:3D_RF_COMSOL} for more details of the simulation. For the experimental data, a power of $-21\,\text{dBm}$ was used, corresponding to an estimated photon number \(\braket{n}_\mathrm{ph} \approx 10^5\). For power dependence measurements, see Appendix~\ref{Power_Qint}.}
	\label{tab:reso_table_neon}
	\resizebox{\linewidth}{!}{%
		\begin{tabular}{ccc|cc|cccc}
			\hline
			\multicolumn{3}{c|}{Resonators} & \multicolumn{2}{c|}{COMSOL Simulation} & \multicolumn{4}{c}{Experimental Data} \\
			\hline
			Resonator & Width/Gap (nm)& Length (\(\mu\text{m}\)) & \( f_\text{r} \) (GHz) & \( Q_{\text{ext}} \) & \( f_\text{r} \) (GHz) & \( Q_{\text{int}} \)  & \( Q_{\text{ext}} \)  & \( Q_{\text{tot}} \) \\
			\hline
			Resonator 1 & 100/100 & 1450 & 4.72 & \( 4.0 \times 10^4 \) & 4.81 & \( 2.3 \times 10^5 \) & \( 3.7 \times 10^4 \) & \( 3.2 \times 10^4 \) \\
			Resonator 2 & 200/200 & 1450 & 6.42 & \( 2.7 \times 10^4 \) & 5.91 & \( 1.8 \times 10^5 \) & \( 2.5 \times 10^4 \) & \( 2.2 \times 10^4 \) \\
			& 150/150 & 1450 & 5.66 & \( 3.3 \times 10^4 \) &  &  &  &  \\
			Resonator 3 & 300/300 & 1450 & 7.66 & \( 2.0 \times 10^4 \) & 6.98 & \( 1.7 \times 10^5 \) & \( 3.8 \times 10^5 \) & \( 1.2 \times 10^5 \) \\
			\hline
		\end{tabular}%
	}
\end{table}

We note that this device design intentionally omitted DC bias lines to avoid degrading the $Q_\mathrm{int}$ of the NbTiN resonator, as the primary objective of this work was to establish a performance baseline of the resonators with solid neon and electrons.

\subsection{Neon and electron deposition}
\label{sec:NeAndEDeposition_neon}

We deposited neon at $25\,\mathrm{K}$. Based on the resonance frequency shifts of Resonator~1 and Resonator~2, measured at $10\,\mathrm{mK}$ as $-0.94\%$ and $-0.86\%$, respectively, we estimated that the neon thicknesses in the regions of Resonator~1 and Resonator~2 are $160\,\mathrm{nm}$ and $270\,\mathrm{nm}$ at most (Appendix~\ref{sec:neon_electron_sim}). From the volume of neon gas recovered after warming up, we estimated that the total amount of neon deposited in the experimental cell was $0.2\,\mathrm{mol}$. If we assume that the neon was deposited uniformly from the bottom of the experimental cell (surface area = $14\,\mathrm{cm}^2$), the expected neon thickness would be $1\,\mathrm{mm}$. This estimate is inconsistent with the resonance frequency shifts, suggesting that neon is primarily deposited on the walls of the experimental cell and capillary lines. This spread of neon to these surfaces may occur due to triple point wetting during the passage through the triple point when depositing solid neon~\cite{leiderer2025surface,migone1986triple}. Aside from the thickness estimation, we also found that compared to the bare resonator, \(Q_\mathrm{int}\) measured at $10\,\mathrm{mK}$ and at low photon number is increased with the presence of neon. This must be attributed to the fact that the higher dielectric constant of neon reducing the participation ratio in the region where the TLS exists.

\begin{figure}[htbp]
	\centerfloat{
		\includegraphics[width=0.6\linewidth]{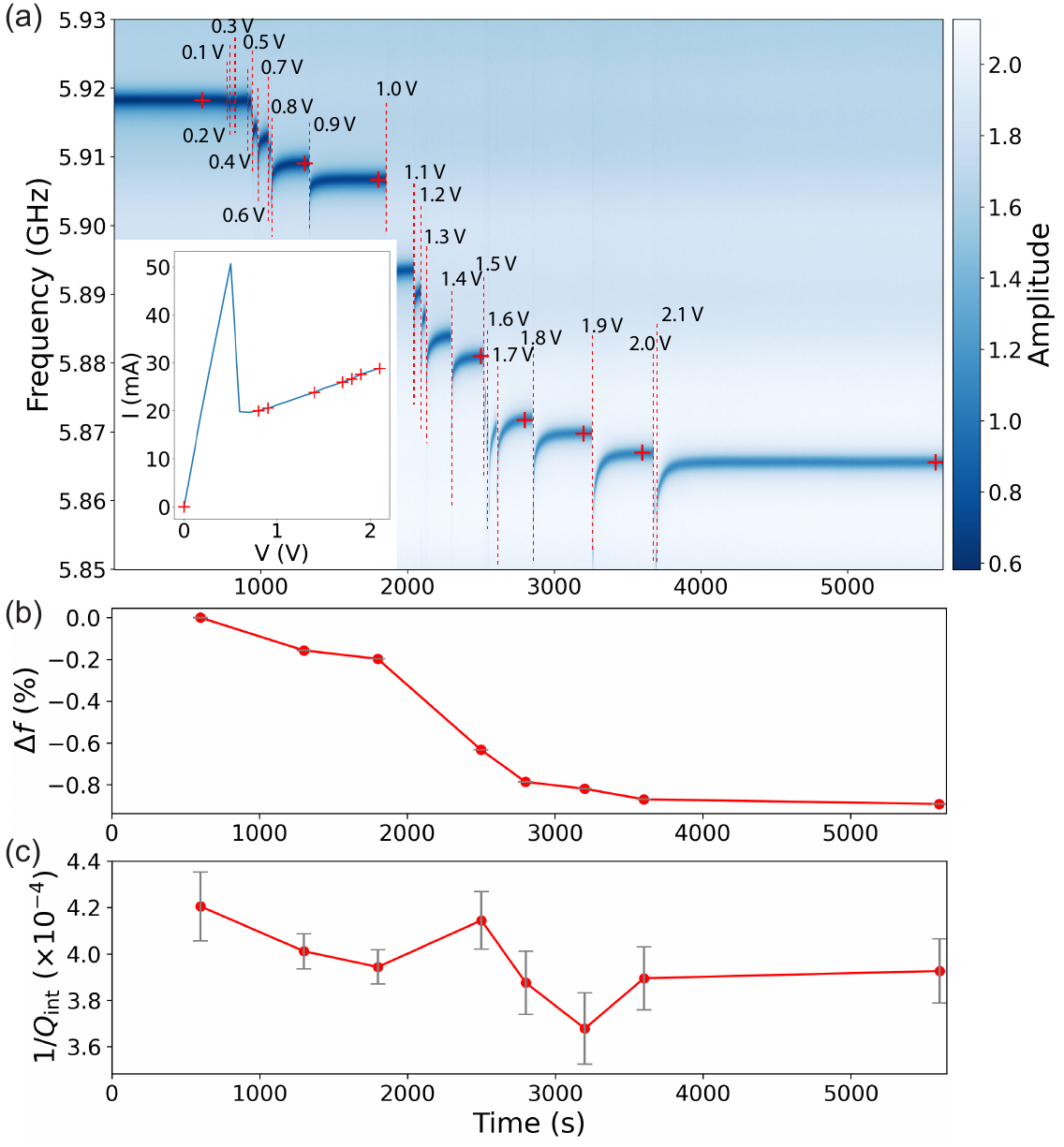}
	}
	\caption[Electron loading and resonator response]{%
		(a) Resonance peaks of Resonator~2 measured at $3.4\,\mathrm{K}$ over time during electron emission. (b) Resonance frequency shift \(\Delta f\). (c) Inverse internal quality factor \(1/Q_\mathrm{int}\) (95\% confidence).%
	}
	\label{fig:e_deposition_neon}
\end{figure}

After depositing neon, we proceeded with electron deposition by sending a voltage pulse to a filament placed in the vacuum cell. Figure~\ref{fig:e_deposition_neon}(a) shows the resonance frequency shifts measured at $3.4\,\mathrm{K}$ by increasing the height of the voltage pulse (duration \(\approx 3\,\text{s}\)) applied to the filament. As shown in Fig.~\ref{fig:e_deposition_neon}(b), the resonance frequency decreases with increasing electron number, while no change is observed in $Q_\mathrm{int}$, as shown in Fig.~\ref{fig:e_deposition_neon}(c). The maximum measured frequency shift was $-0.9\%$.

The decrease in resonance frequency with increasing electron number on the neon surface is attributed to the surface conductivity of the two-dimensional electron layer. To illustrate this, we first consider the Drude model:
\(
\sigma^{\mathrm{2D}} = \frac{e^2 n_e \tau}{m_e} \frac{1}{1 + i\omega \tau},
\)
where \(e\) is the elementary charge, \(n_e\) is the surface electron density, \(\tau\) is the scattering time, \(m_e\) is the electron mass, \(\omega\) is the angular frequency of the microwave signal, and \(i\) is the imaginary unit. The highest reported \(\tau\) for electrons on solid neon is \(4.7\,\mathrm{ps}\), based on low-frequency mobility measurements at \(4.2\,\mathrm{K}\) \cite{kajita1984new,kajita1985wigner}. A shorter value of \(1.9\,\mathrm{ps}\) has also been reported for rougher surfaces \cite{kajita1984new}. The simulations show that a shift of \(-0.9\%\) corresponds to \(n_e = 0.4\,(0.8) \times 10^9\,\mathrm{cm}^{-2}\) and \(1/Q_e = 4.0\,(3.9) \times 10^{-3}\) for \(\tau = 4.7\,(1.9)\,\mathrm{ps}\) (Appendix~\ref{sec:simu_with_e}). Since \(\tau\) is not directly measured and the Drude model does not accurately capture the measured \(Q_\mathrm{int}\), these values of \(n_e\) should be interpreted as indicative. The simulated \(1/Q_\mathrm{int}\) is nearly an order of magnitude larger than the experimentally measured value, which remains around \(4 \times 10^{-4}\) at \(3.4\,\mathrm{K}\) and shows no dependence on electron density. This discrepancy suggests that the Drude model fails to capture the measured $Q_\mathrm{int}$. We explore a model that incorporates surface disorder via harmonic confinement, which can localize electrons (Appendix~\ref{sec:simu_with_e_Lorentz}). The results show that in certain parameter regimes, this approach can account for the observed frequency shift without a significant reduction in $Q_\mathrm{e}$. These findings suggest that the limitations of the Drude model may arise from localization effects induced by surface roughness on a length scale smaller than the scattering length. In addition, the difference in the neon thickness observed between the two resonators can be attributed to large-scale nonuniformity of the neon deposition across the chip.

After depositing electrons at \(3.4\,\mathrm{K}\), the sample was cooled down to \(10\,\mathrm{mK}\), where two-tone spectroscopy was performed to confirm that the electrons remained trapped during the cooling process (Appendix~\ref{2_tone}).

\subsection{Magnetic field gradient}
\label{sec:Magnet_neon}

\begin{figure}[htbp]
	\centerfloat{
		\includegraphics[width=0.9\linewidth]{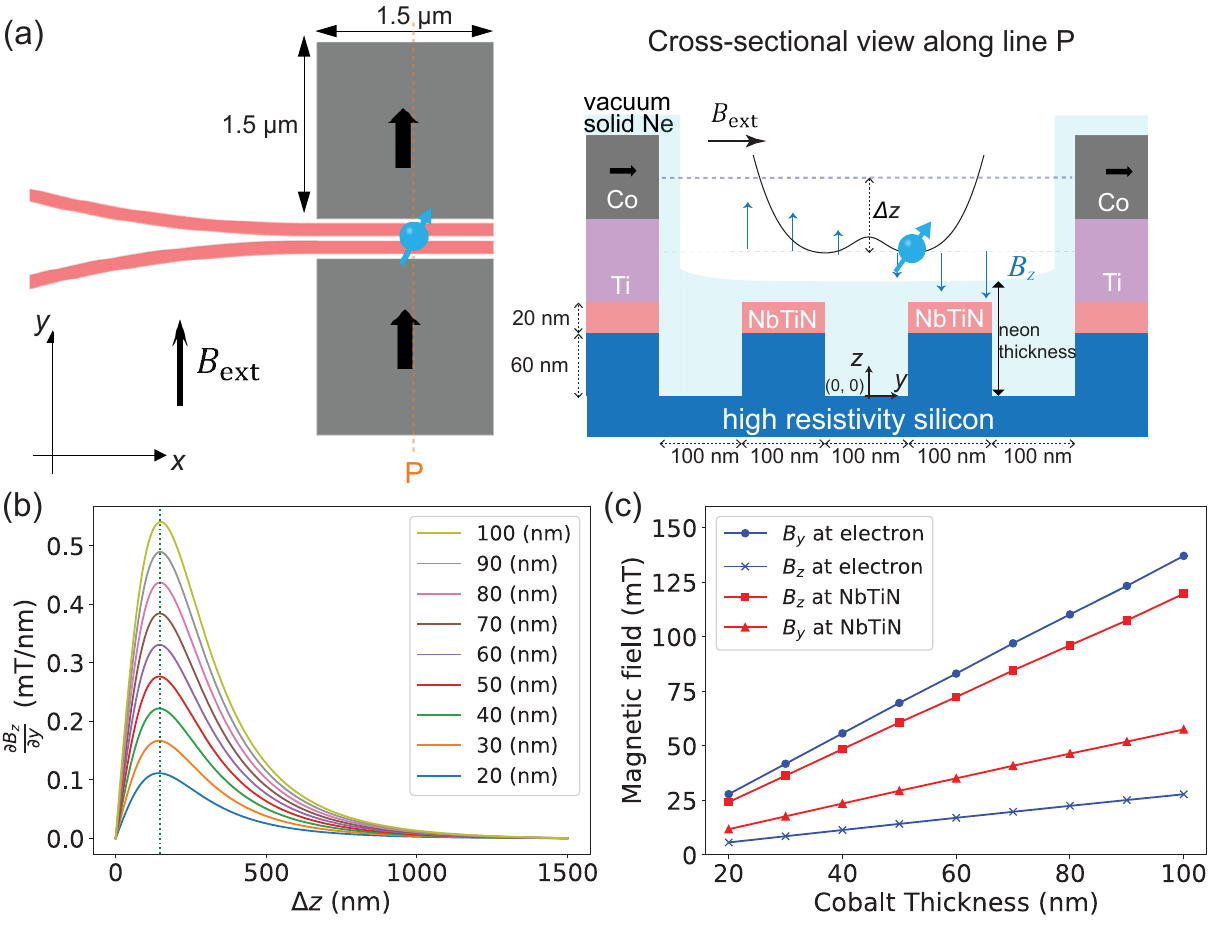}
	}
	\caption[Proposed micromagnet integration and field gradients]{%
		(a) Proposed addition of Co magnets near the gap between the two ends of Resonator~1. (b) Magnetic field gradient \(\partial B_z/\partial y\) versus vertical offset \(\Delta z\) for different Co thicknesses. (c) Offsets of the magnetic fields at the electron position and at the NbTiN resonator position as functions of Co thickness.%
	}
	\label{fig:micro_magnet_neon}
\end{figure}

The sample used in this work consists exclusively of resonators. As depicted in Fig.~\ref{fig:micro_magnet_neon}, we propose to add two Co/Ti/NbTiN electrodes at both ends of Resonator~1, on its upper and lower sides. Cobalt (Co) serves as a ferromagnetic layer, titanium (Ti) as an adhesion layer, and NbTiN as part of the resonator. These electrodes are designed to serve three roles: generating local magnetic field gradients, tuning the electron's confinement potential through the application of DC voltage to the electrodes, and performing Electric Dipole Spin Resonance (EDSR)~\cite{nowack2007coherent,tokura2006coherent,pioro2008electrically,dijkema2025cavity} by applying microwave (MW) signals to one of the electrodes.

In the charge-qubit experiments reported in Refs.~\cite{zhou2022single,zhou2024electron}, a single electron trapped on solid neon shows a spectrum that qualitatively equals that seen in semiconductor double quantum dots (i.e., the energy of the charge state follows $\sqrt{\epsilon^2 + 4t_c^2}$, where $\epsilon$ is the charge detuning and $t_c$ is the effective tunnel coupling~\cite{jennings2024quantum}), which may be due to the surface roughness of the solid neon~\cite{Kanai2024-bo}. Here, we assume that the two minima of the electrical potential are created at the edges of each end of the resonator. This assumption is plausible because these points are where the attraction towards the surface becomes maximum due to the image potential created by the resonators. This configuration is illustrated in Fig.~\ref{fig:micro_magnet_neon}, and the distance between the two minima, i.e., the inter-dot distance $d = 100\,\text{nm}$.

By applying an external magnetic field $B_\mathrm{ext}$ along the $y$-axis, the Co parts are magnetized along the same axis and induce local magnetic fields (Appendix~\ref{Mag_sim}).

Figure~\ref{fig:micro_magnet_neon}(b) shows the \( z \)-direction magnetic field gradient along the \( y \)-axis, \( \frac{\partial B_z}{\partial y} \), as a function of the distance between the \( z \)-center position of the Co magnet and the position of the electron, $\Delta z$. We found that the \( z \)-direction magnetic field gradient becomes maximum when $\Delta z \approx 146\,\text{nm}$. With this configuration, we obtain \( \frac{\partial B_z}{\partial y} = 0.36\,\text{mT/nm} \), resulting in a magnetic field gradient in the \( z \)-direction between the two potential minima, expressed in radian frequency as \( b_\perp = \frac{g \mu_B}{\hbar} \frac{\partial B_z}{\partial y} \cdot d = 2\pi \cdot 1\,\mathrm{GHz} \), where \( g \) is the free electron \( g \)-factor, \( \mu_B \) is the Bohr magneton and \( \hbar \) is the reduced Planck constant. This magnetic field gradient strength is sufficient to reach the strong coupling regime for spin-photon coupling and to achieve high-fidelity qubit gates.

Figure~\ref{fig:micro_magnet_neon}(c) shows the magnetic field offsets along the \( z \)- and \( y \)-axes introduced by the Co magnets at the position of the NbTiN resonators, which are at most \( B_z = 60\,\text{mT} \) and \( B_y = 130\,\text{mT} \), respectively. These values are low enough not to degrade the quality factor of the resonators for a NbTiN thickness of $20\,\text{nm}$~\cite{kroll2019magnetic,samkharadze2016high}. The \( B_y \) experienced by the electron at the positions of the potential minima is \( B_y = 90\,\text{mT}\). Thus, the total magnetic field that defines the Zeeman splitting in radian frequency is expressed as \( b_\parallel \approx \frac{g \mu_B}{\hbar} (B_y + B_\mathrm{ext}) \). Thus, to set the Zeeman splitting equal to the resonator resonance frequency of $4.8\,\mathrm{GHz}$, we need to apply \( B_\mathrm{ext} = 80\,\text{mT} \), which is sufficient to magnetize Co and perform qubit measurements~\cite{samkharadze2018strong,Harvey-Collard2022-jh,dijkema2025cavity}. In the presence of a magnetic field gradient, charge noise—which causes fluctuations in the charge detuning $\epsilon$—could lead to fluctuations in the spin splitting. Here, two Co magnets are placed symmetrically with a narrow gap relative to the electron positions. As a result, the Zeeman splitting at the two potential minima becomes equal, and the effect of charge noise at $\epsilon=0$ is largely suppressed.

\subsection{Spin-photon coupling and loss rate}
\label{sec:sp_couping_neon}

In this section, we estimate the spin-photon coupling strength, evaluate the loss rates of the spin state and the resonator, and discuss the feasibility of reaching the spin-photon strong coupling regime. For this purpose, we use the parameters of Resonator~1: an inductance of $L = 139\,\mathrm{nH}$ (Appendix~\ref{K_in}) and the measured resonance frequency $f_\mathrm{r}=\omega_\mathrm{r}/2\pi=4.81$~GHz. The characteristic impedance of the resonator is then calculated as $Z_0=2f_\mathrm{r}L=1337~\Omega$.

According to Ref.~\cite{samkharadze2016high}, the zero-point fluctuation (ZPF) of the rms voltage amplitude between the two ends of the resonator can be calculated as:
\begin{equation}
	V_0 = \frac{1}{\sqrt{2}} \frac{2L}{\pi} \sqrt{ \frac{ 2 \hbar \omega_r}{L}} \omega_r = 12 \,\mu\mathrm{V},
\end{equation}
and thus the charge-photon coupling strength can be calculated as
\begin{equation}
	\frac{g_c}{2\pi} = \frac{\alpha  e V_0 \cos \theta}{h}, \label{eq:gc_neon}
\end{equation}
where $\theta=\arctan(\epsilon/2t_c)$ and $e V_0 /h = 3\,\mathrm{GHz}$. Here, we use \(\alpha = 0.05\). This yields \(g_c / 2\pi = 150\,\mathrm{MHz}\) at \(\epsilon = 0\), a value comparable to those measured in semiconductor quantum dots using NbTiN resonators with similar designs~\cite{samkharadze2018strong,dijkema2025cavity}. Moreover, COMSOL simulations indicate that \(\alpha\) exceeds 0.1 for $d=100$~nm, when the neon layer thickness is below $300\,\text{nm}$, suggesting that the chosen value does not lead to an overestimation.

The high-frequency noise-induced loss rate is given by $\gamma =  \gamma_1 /2+\gamma_\phi$, where $\gamma_1=2\pi/T_1$, $\gamma_\phi=2\pi/T_\phi$, \(T_1\) is the energy relaxation time, and \(T_\phi\) is the pure dephasing time. The quasistatic noise-induced loss rate is given by
\(
\gamma^* =\frac{1}{T_2^*}
\),
where \(T_2^*\) is the inhomogeneous (quasistatic) dephasing time. We denote the loss rates for spin and charge by adding subscripts “\(\mathrm{s}\)” and “\(\mathrm{c}\)”, respectively, i.e., \(\gamma_{\mathrm{s}}\), \(\gamma_{\mathrm{s,1}}\), \(\gamma_{\mathrm{s,\phi}}\) and \(\gamma_{\mathrm{s}}^*\) for spin, and \(\gamma_{\mathrm{c}}\), \(\gamma_{\mathrm{c,1}}\), \(\gamma_{\mathrm{c,}\phi}\) and \(\gamma_{\mathrm{c}}^*\) for charge.

To assess the impact of different loss mechanisms, we consider three representative scenarios labeled as “thermal,” “$^{\mathrm{nat}}$Ne,” and “$^{22}$Ne,” corresponding to three distinct sets of assumptions regarding charge and spin loss rates. Charge loss rate values are taken from Refs.~\cite{zhou2022single,Li2025-em}. Spin loss rates are taken from Ref.~\cite{chen2022electron}.

We set \( \epsilon = 0 \) from here onwards for simplicity and to leverage the long coherence time at the charge sweet spot. To measure strong coupling between the electron spin state and a microwave photon in the resonator, we tune the spin and photon energies into resonance by setting the Zeeman splitting equal to the resonator resonance frequency, that is, the detuning of the spin state from the resonator \(\Delta_\mathrm{s} = b_\parallel - \omega_r = 0\). The spin-photon coupling is given by \(g_\mathrm{s} = \Lambda  g_\mathrm{c}\), where \(
\Lambda = \sin {\bar{\phi}},
\) with ${\bar{\phi}} = (\phi_+ + \phi_-)/2$ and $\phi_{\pm} = \arctan{[b_{\perp} /(2t_\mathrm{c} \pm b_{\parallel})]}$~\cite{Benito2019-pi}. Due to spin-charge coupling, the effective high-frequency noise-induced and quasi-static spin loss rate at the charge sweet spot become~\cite{Benito2019-pi}
\begin{equation}
	\gamma_\mathrm{s}' = \Lambda^2 \gamma_\mathrm{c} + (1-\Lambda^2) \gamma_\mathrm{s},
\end{equation}
and
\begin{equation}
	{\gamma^*_\mathrm{s}}'= \sqrt{  \left( \Lambda^2\frac{\gamma_\mathrm{c}^*}{b_\parallel}\right)^2+ \left( \frac{(\cos \phi_+ + \cos \phi_-) }{2}\gamma_\mathrm{s}^* \right)^2 },
\end{equation}
respectively.

The condition for spectroscopically observing strong spin--photon coupling is \(g_s \gg \gamma_\mathrm{s}', {\gamma^*_\mathrm{s}}', \kappa'\). As an example, for \(b_\parallel / 2\pi = \omega_r / 2\pi = 4.8\,\mathrm{GHz}\), \(\epsilon = 0\), \(2t_c / 2\pi = 8.0\,\mathrm{GHz}\), and \(b_\perp / 2\pi = 1\,\mathrm{GHz}\), we obtain \(\Lambda = 0.19\), yielding \(g_s / 2\pi = 28.5\,\mathrm{MHz}\). The total resonator decay rate, including Purcell decay induced by the coupling between the charge state and the resonator, is given by \(\kappa' = \kappa + \frac{g_c^2 \gamma_{\mathrm{c}}}{\Delta_\mathrm{c}^2}\), where \(\Delta_\mathrm{c} = \sqrt{\epsilon^2 + 4t_\mathrm{c}^2} - \omega_r\) is the detuning of the charge state from the resonator~\cite{benito2017input}. For these parameters, \(\kappa'/2\pi \approx \kappa/2\pi = 0.1\,\mathrm{MHz}\). The spin loss rate due to high-frequency noise is calculated to be \(\gamma_\mathrm{s}' / 2\pi \approx 7\,\mathrm{kHz}\). These parameters satisfy the condition for strong coupling. The corresponding cooperativity is estimated to be \(C = g_\mathrm{s}^2 / (\gamma_\mathrm{s}' \kappa') \gtrsim 10^6\), confirming the strong spin--photon coupling regime.

\subsection{Qubit gates}
\label{sec:Qubit_gate_neon}

To realize single-qubit gates, we can employ EDSR~\cite{nowack2007coherent,tokura2006coherent,pioro2008electrically,dijkema2025cavity} by applying MW signal to one of the Co/Ti/NbTiN electrodes and modulating the position of the electron along the $y$ axis. During EDSR, we detune the resonator from both the spin and charge states to suppress decoherence through the resonator (\(\Delta_\mathrm{c}, \Delta_\mathrm{s} \gg 0\)). The EDSR Rabi frequency is determined by $f^\mathrm{s}_\mathrm{R}=\Lambda f^\mathrm{c}_\mathrm{R}$, where $f^\mathrm{c}_\mathrm{R}$ is the charge Rabi frequency. Here, we use \( f^\mathrm{c}_\mathrm{R} = 10\,\mathrm{MHz} \), as reported in Ref.~\cite{zhou2022single}. The average gate fidelity of a $\pi$ gate can be calculated as~\cite{Benito2019-pi}
\begin{equation}
	F_1(\delta) = \frac{1}{6} \left[3 + e^{-2t_g \gamma_{s}'} + 2 e^{-t_g \gamma_{s}'} \cos(t_g \delta)\right],
\end{equation}
where \( t_g = 1 / (2 f_\mathrm{R}^s) \) and \(\delta\) is the detuning between the spin resonance frequency and the frequency of the MW applied to the electrode. When quasi-static noise is also taken into account, with $\delta$ following a Gaussian distribution with standard deviation $\sqrt{2} {\gamma^*_\mathrm{s}}'$, the average fidelity becomes
\begin{equation}
	\overline{F}_1 = \frac{1}{6} \left[3 + e^{-2t_g \gamma_{s}'} + 2 e^{-t_g \gamma_{s}'} e^{-\frac{(t_g {\gamma^*_\mathrm{s}}')^2}{2}}\right].
\end{equation}

Next, we consider a two-qubit gate. We focus on the iSWAP gate~\cite{Benito2019-al,Warren2019-mn}, which has been experimentally realized for semiconductor spin qubits~\cite{dijkema2025cavity}. In the dispersive regime, spins are coupled via virtual photons with \(\Delta_s = \beta g_s\) and \(\beta \gg 1\), suppressing photon-decay effects. The fidelity is~\cite{Benito2019-al}
\begin{equation}
	F_2=1-\frac{2\pi}{5g_s }\left( 2 \gamma_\mathrm{s}' \beta  +\frac{\kappa}{\beta }  \right).
\end{equation}

\begin{figure}[htbp]
	\centerfloat{
		\includegraphics[width=0.6\linewidth]{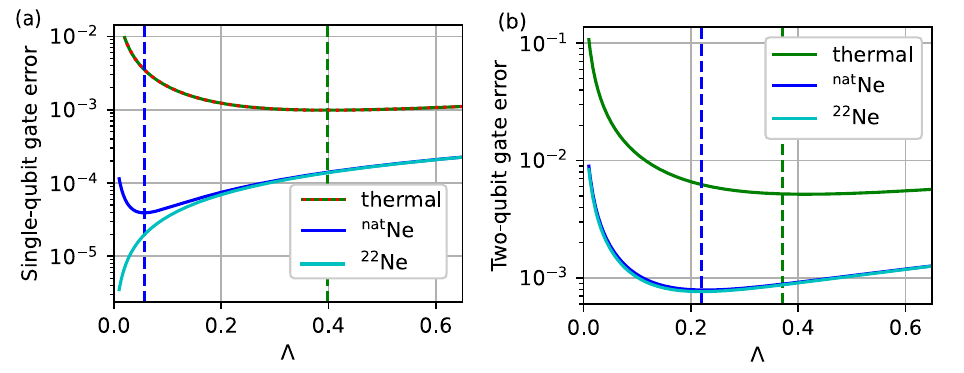}
	}
	\caption[Gate errors for single- and two-qubit operations]{%
		(a) Single-qubit gate error as a function of \(\Lambda\). (b) iSWAP gate error as a function of \(\Lambda\) for representative scenarios; \(\beta=10\) and \(\kappa/2\pi=0.1\,\mathrm{MHz}\).%
	}
	\label{fig:fidelity_neon}
\end{figure}

\subsection{Conclusion}
\label{sec:Conclusion_neon}

In conclusion, we experimentally demonstrated neon and electron depositions using NbTiN nanowire resonators, aiming for the realization of spin-photon coupling and high-fidelity spin qubit gates in the future. The presence of neon and electrons was confirmed by a decrease in the resonance frequency without compromising the resonator’s quality factor, supporting the suitability of NbTiN resonators for future qubit measurements. A closer examination of the resonator's response to electron deposition suggests that the Drude model fails to capture the system's behavior, likely due to electron localization induced by surface roughness in the neon. At present, the control over thickness and surface roughness in the neon deposition process is not well established and requires improvement. Additionally, we investigated the configuration of cobalt magnets needed for realizing spin-photon coupling and spin qubit gates. By combining the enhanced charge-photon coupling enabled by the high impedance of the NbTiN resonator with the calculated magnetic field gradient produced by the cobalt magnets, we estimated that spin-photon coupling can reach the strong coupling regime, allowing for the implementation of high-fidelity spin qubit gates.

Future work will involve the experimental implementation of integrated micromagnets and additional DC electrodes. To avoid degradation of the resonator’s quality factor in such implementations, mitigation strategies such as perforated ground planes~\cite{kroll2019magnetic} and DC filters~\cite{Mi2017-go,Harvey-Collard2020-dt} will be necessary. In this context, the high quality factor demonstrated here provides an important target and benchmark for evaluating the effectiveness of these approaches.

\section{Characterization of Tunnel Diode Oscillator for Qubit Readout Applications}\label{sec:tdo}

One of the key features required to realize a fault-tolerant scalable quantum computer is the integration of energy-efficient and reliable qubit control and readout electronics. Recently, qubit control electronics have been successfully integrated using cryogenic Complementary Metal-Oxide-Semiconductor (CMOS) technology \cite{Van_Dijk2019-na,bardin201929,pauka2021cryogenic,peng2022cryo,pellerano2022cryogenic} , and superconducting Josephson-junctions~\cite{howe2022digital}. Here, we focus on developing the readout electronics using tunnel-diode oscillator~\cite{chow1964principles,van1975tunnel} (TDO) circuits. TDO uses a tunnel diode as a negative resistance element~\cite{Reona1962-hv}, generating a microwave signal when connected to an LC circuit. Comparable to cryogenic CMOS devices (typical power consumption: 10~mW~\cite{bardin201929,pauka2021cryogenic,peng2022cryo,kang2022cryo,lee2022208,pellerano2022cryogenic}) and superconducting Josephson-junction circuits (100~$\mu$W~\cite{howe2022digital}), the TDO presented here consumes significantly less power, requiring only 1~$\mu$W.

\begin{figure}[htbp]
	\centerfloat{
		\includegraphics[width=\linewidth]{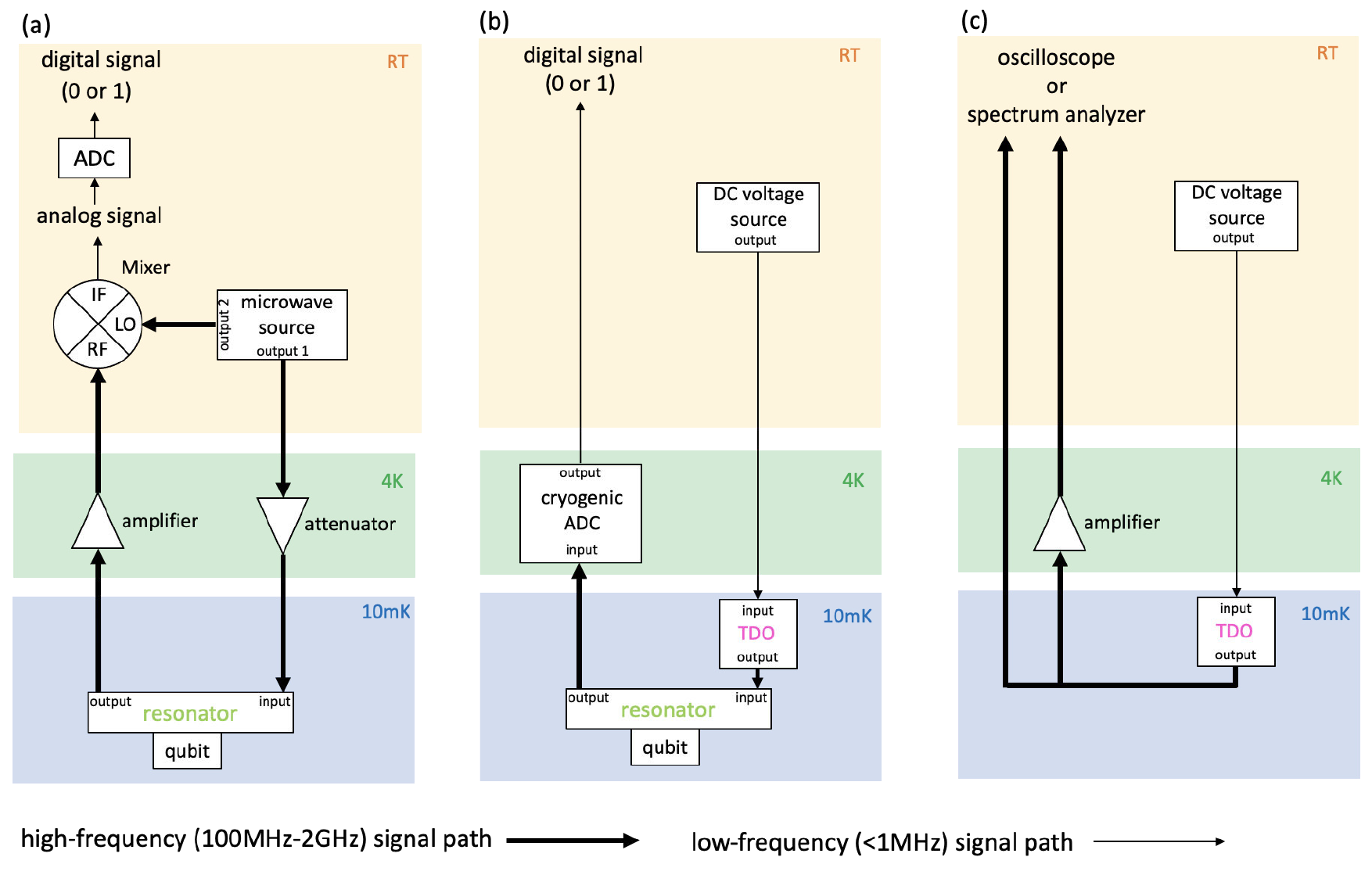}
	}
	\caption[Dispersive readout and TDO concept]{%
		(a, b) Schematic diagrams illustrating dispersive qubit readout using a resonator: the conventional method in (a) and the tunnel-diode oscillator (TDO) method in (b). The bandwidth of the twisted pair of DC lines is limited to 1 MHz but is sufficient for transmitting the final calculation results to RT and powering the TDO from RT. (c) Schematic of the setup used to characterize the TDO. The characterization includes data measured both with and without the amplifier at 4~K.%
	}
	\label{fig:motivation}
\end{figure}

This research aims to develop a technique that enables scalable qubit read-out schemes for quantum computers operating at cryogenic temperatures. The conventional method, illustrated in Fig.~\ref{fig:motivation}(a), is to generate a microwave signal with a room temperature (RT) microwave source and send it to a resonator dispersively coupled to the qubit at the 10~mK stage  (also called the mixing chamber, MC stage). The amplitude and phase of the reflected microwave signal depend on the qubit state, allowing us to determine the qubit state. This method is commonly used for qubits made from electrons in semiconductors~\cite{hanson2007spins,zwanenburg2013silicon,vigneau2023probing,urdampilleta2019gate,Burkard2023-ai} and a similar method has been proposed for electrons on helium~\cite{kawakami2023blueprint}.

In the conventional method, since the input signal is generated at RT, attenuators are placed at the 4~K stage to reduce thermal noise before it reaches the qubit. To prevent the reflected output signal from being overwhelmed by thermal noise at RT, it must be amplified at cryogenic temperatures. Cryogenic amplifiers typically have a lower noise figure, resulting in a higher signal-to-noise ratio. The resonant frequency of the resonator is typically in the range of 100~MHz to 2~GHz~\cite{schoelkopf1998radio,colless2013dispersive,vigneau2023probing}, and such high-frequency signals are transmitted through coaxial cables. A major scalability issue arises as the number of qubits increases because each qubit requires its own coaxial cable, which are relatively thick (feedthrough connector diameter: $\sim$1 cm, cable diameter: $\sim$1 mm) with attenuators and amplifiers also taking up the limited space inside a cryogenic refrigerator (diameter~$\sim$10-100~cm). Multiplexing can reduce the number of coaxial cables to some extent, but is limited by the resonator bandwidth~\cite{hornibrook2014frequency}.

To address this challenge, we propose placing the microwave source closer to the qubits. By integrating the microwave source and qubits on the same circuit board, we can replace bulky coaxial cables with compact on-board transmission lines. This integration significantly reduces the circuit's size, making it more suitable for scaling up the number of qubits. Since the qubits must be operated at cryogenic temperatures (typically 10~mK), the microwave source must function at the same cryogenic stage. In this regard, the TDO we developed operates at the MC stage as the qubit. CMOS technology offers the advantage of high output power, operates at higher frequencies ($\sim$GHz), and performs complex functions; however, it must be placed at the 4 K stage due to its high power consumption.~\cite{Van_Dijk2019-na, bardin201929,pauka2021cryogenic,peng2022cryo,pellerano2022cryogenic}. It should be noted that, while not as an oscillator, CMOS has been used at the 10~mK stage to implement other circuits for quantum device control, such as multiplexers and DC bias circuits. The TDO is powered by a DC voltage supplied through DC lines. For these scenarios, an analog-to-digital converter (ADC) at cryogenic temperatures~\cite{Braga2024-bx} (Fig.~\ref{fig:motivation}(b)) can be placed to remove the need for coaxial cables on the resonator output lines as well. This approach would enable faster feedback for quantum error correction~\cite{fowler2012surface} by eliminating the need to transmit signals to RT, requiring only the final computation results to be sent back. Additionally, placing the ADC at cryogenic temperatures removes the need for amplifiers, as thermal noise from RT would no longer be a concern.

In this work, we focus on the characterization of the TDO without utilizing an ADC or a resonator coupled to qubits (Fig.~\ref{fig:motivation}(c)). This initial step allows us to evaluate the performance and stability of the TDO as a standalone device under cryogenic conditions, providing a foundation for future integration with qubits. For qubit applications, microwaves are used for qubit state readout and manipulation. Amplitude stability is particularly important for readout~\cite{vigneau2023probing}, while phase noise also plays a crucial role in manipulation~\cite{Ball2016-zn}. Therefore, in this work, we characterized both amplitude stability and phase noise to evaluate the performance of the TDO in this context.

\subsection{Tunnel Diode Oscillator}\label{sec2}

We use the commercial BD-6 Ge-backward tunnel diode (American Microsemiconductor, Inc.) because it has the lowest power consumption among the commercially available options. TDOs with backward tunnel diodes demonstrate frequency stability of 0.001 ppm~\cite{van1975tunnel,van1981modeling}. Leveraging this precision, they have been employed in the past to investigate various physical properties, such as paramagnetic susceptibility in salts~\cite{clover1970magnetic}, penetration depth in superconductors~\cite{hashimoto2010evidence,fletcher2009evidence}, and the melting point of $^3$He~\cite{mikheev1989crystallization}. These results were obtained with long integration times using a frequency counter, whereas qubit readout demands precise measurements within shorter durations. This emphasizes the need to understand the frequency and amplitude stability achievable over shorter integration times. Furthermore, the oscillation frequencies, typically around 10 MHz~\cite{cole1969image,al2007features,hashimoto2010evidence,fletcher2009evidence,srikanth1999radio,elarabi2021cryogenic}, are relatively low for qubit readout applications. In the early stages of testing qubit readout, it is also desirable to have the capability to externally tune the frequency. To address this issue, we developed a TDO operating at around 140 MHz, utilizing a microfabricated superconducting spiral inductor. This frequency has reached the lower bound among experiments~\cite{verduijn2014radio} where the electron transition is measured through dispersive coupling to a resonator, making it applicable for qubit readout. By incorporating a varactor diode as a variable capacitor, we achieved a frequency tunability of 10 MHz without altering the amplitude. Additionally, this oscillator might also be used as a low-power, low-phase-noise clock source for cryo-ADCs.

\begin{figure}[htbp]
	\centerfloat{
		\includegraphics[width=0.7\linewidth]{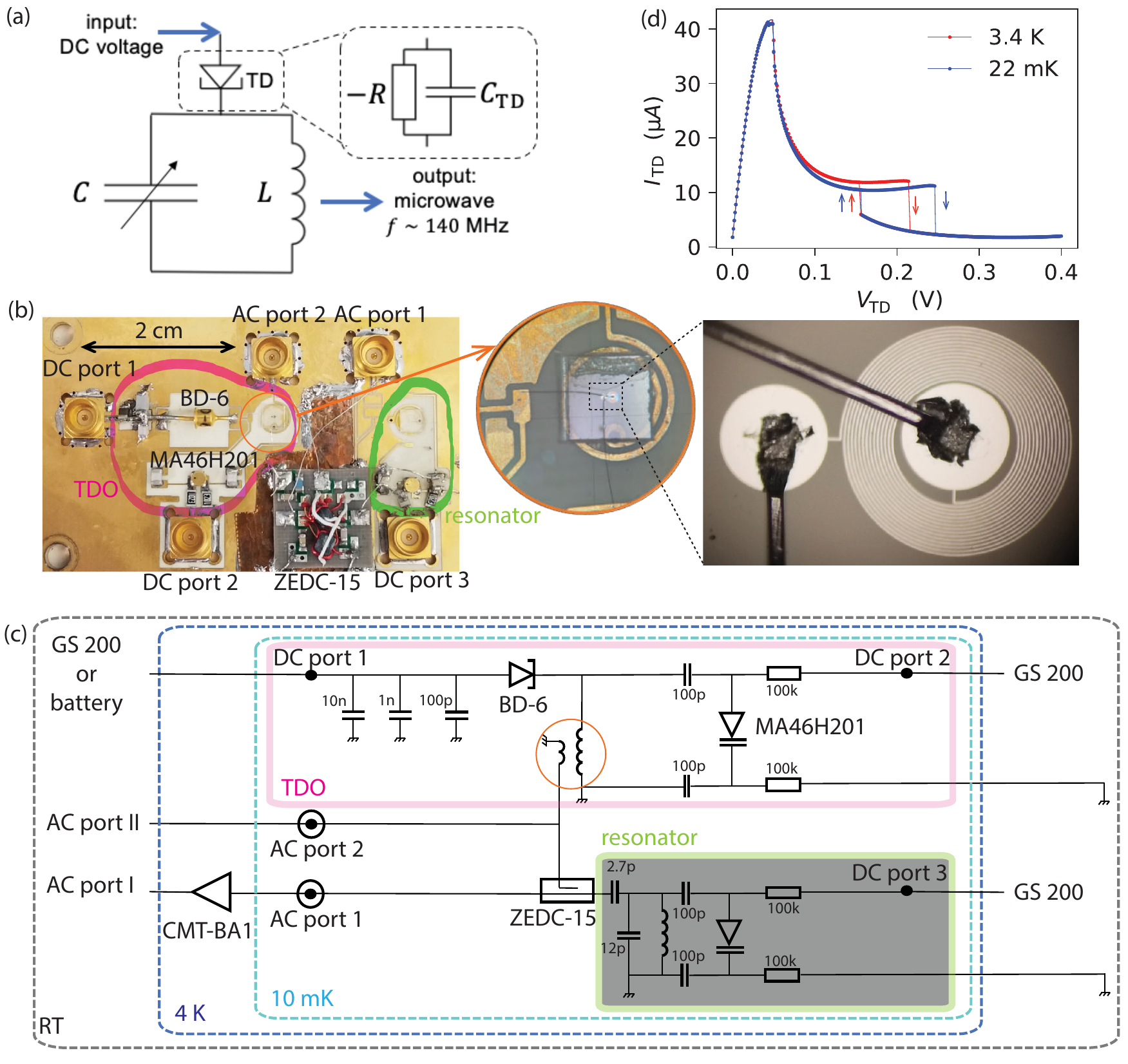}
	}
	\captionsetup{font=small}
	\caption[TDO circuit and device photograph]{%
		(a) Conceptual circuit illustration of the TDO. The tunnel diode (TD) can be modeled as a negative resistance \( -R \) and a capacitance \( C_\mathrm{TD} \).  The input voltage applied to the TD is converted into and output as a microwave signal with a frequency of  $\sim$140 MHz. (b) A photograph of the developed circuit. The pink box encloses the TDO, and the green box the resonator. The orange circle highlights a microfabricated 15-turn spiral inductor on a sapphire substrate, positioned on top of a 1.5-turn pickup coil, enabling a 10\% signal to be extracted from the TDO in terms of voltage. The pickup coil and the TDO are connected via a coupler (ZEDC-15). In this study, the resonator is detuned from its resonance, behaving as an open-ended element in the circuit. An enlarged photograph of the microfabricated 15-turn spiral inductor on a sapphire substrate is also shown. The diameter of its boding pad is 176 $\mu$m. (c) Circuit diagram of the device shown in (a). See the main text for details of the circuit components. The signal output from the TDO is measured at RT using a digital oscilloscope or a spectrum analyzer via AC port I or AC port II. For comparison, a signal from a commercial microwave source is sent to the cryogenic device through AC port II, and the reflected signal from the resonator is measured at RT using a digital oscilloscope or a spectrum analyzer via AC port I. The resonator is shaded in gray because its resonant frequency is set far outside the TDO's operating range, allowing it to function as a high-impedance element and remain inactive. (d) The I-V curve of the tunnel diode BD-6 at the mixing chamber stage temperature (MCT) of 22~mK and 3.4~K. Hysteresis is observed only during oscillation, depending on the sweep direction of the DC bias $V_\mathrm{TD}$ applied to the tunnel diode (indicated by arrows). When the bias is increased in the forward direction, oscillation occurs in the range of 0.04 V to 0.245 V, whereas in the reverse direction, the oscillation range is from 0.15 V to 0.04 V. %
	}
	\label{fig:circuit}
\end{figure}

Fig.~\ref{fig:circuit}(a) shows a conceptual circuit illustration explaining how a tunnel diode (TD) implements oscillations. When a TD is connected in an LC resonant circuit, the negative resistance (represented by $-R$ in Fig.~\ref{fig:circuit}(a)) cancels out the resistive losses, allowing sustained oscillations. The oscillation frequency is determined by the LC resonance condition:  
\begin{equation}
	f = \frac{1}{2\pi \sqrt{L(C + C_{\mathrm{TD}})}} \label{eq:freq}
\end{equation}
where \( C_{\mathrm{TD}} \) represents the capacitance of the TD. \( C \) is the sum of the parasitic capacitance of the circuit board and the capacitance of the varactor diode, and its capacitance value is tunable by changing the voltage applied to the varactor diode, \( V_{\mathrm{VD}} \). To achieve high-frequency oscillations, minimizing parasitic capacitance in both the circuit and its components is crucial. Additionally, a high-Q inductor is important, as the self-resonant frequency should be higher than the oscillation frequency.

It is important to carefully select suitable BD-6 tunnel diodes for two reasons: first, to ensure that they exhibit negative resistance at low temperatures, and second, to minimize parasitic capacitance. Although  the I-V characteristics measured at RT show little variation among different BD-6 tunnel diodes, some lose their negative resistance at low temperatures (see Appendix~\ref{secA:Indiv_diff}). Therefore, screening using low-temperature I-V measurements is necessary. BD-6 tunnel diodes also exhibit significant variation in parasitic capacitance. We measured 62 different BD-6 tunnel diodes and found that their parasitic capacitance at RT ranged from \( C_\mathrm{TD} = 1.3 \) pF to 29.4 pF. The tunnel diode used in this study has a parasitic capacitance of approximately $C_\mathrm{TD}=$7 pF (Table.~\ref{tab:capacitance}).

For high-frequency oscillation, the inductance should simultaneously achieve a high Q-factor and a resistance lower than the 5 k$\Omega$ negative resistance of the tunnel diode. Therefore, we  microfabricated 15-turn Nb spiral coil as an inductor, which was separately measured to be 95~nH at 4~K. Relatively high frequencies, around 100~MHz, have been achieved using a different approach based on a toroidal LC resonator~\cite{van1974sensitive}. However, due to its bulky nature, this approach is unsuitable for scalable qubit applications.

Fig.~\ref{fig:circuit}(b,c) shows the details of the developed circuit. This circuit board is thermally anchored to the 10~mK stage, which is referred to as the Mixing Chamber (MC) stage. The part enclosed by the pink line represents the TDO, which consists of the Nb spiral inductor and a variable capacitance formed by a varactor diode (MA46H201, MACOM). The varactor diode allows tuning of the LC resonator's capacitance, thereby varying the oscillation frequency of the TDO. The varactor diode is biased via DC port 2 using a DC voltage source (Yokogawa GS200). The oscillation signal is coupled to the pickup coil with a coupling ratio of 15:1.5.  In previous work~\cite{clover1970magnetic,al2007features,hashimoto2010evidence,fletcher2009evidence,srikanth1999radio,elarabi2021cryogenic}, the signal was extracted capacitively from the same line used to provide the DC bias to the tunnel diode. Instead, in this work, the signal is extracted inductively. This approach ensures that the DC bias line is used exclusively for providing the DC bias, thereby preventing the TDO oscillation from being affected by filters and other components further along the line. As a result, a more stable DC bias can be supplied to the tunnel diode. The tunnel diode is biased via DC port 1 using either the Yokogawa GS200 or a lead-acid battery. After the pickup coil, the signal is sent to AC port II and measured at RT. It is also routed through a directional coupler (ZEDC-15, Mini-Circuits). The coupling from the coupling port to the input/output port of the ZEDC-15 is 20~dB, as measured separately at the 4~K stage. The signal from the coupler is amplified at the 4~K stage using a cryogenic amplifier (CMT-BA1, Cosmic Microwave Technology) and measured at RT with a spectrum analyzer or a digital oscilloscope.

The part enclosed by the green line in Fig.~\ref{fig:circuit}(b, c) represents the resonator, which can be coupled to a qubit. In this work, the resonator's resonant frequency is set far outside the TDO's operating range, allowing it to act as a high-impedance element and remain unused. 

The operating point of the tunnel diode BD-6, where it exhibits negative resistance and oscillates, is approximately \( V \approx 0.1 \, \text{V} \) and \( I \approx 10 \, \mu\text{A} \) (Fig.~\ref{fig:circuit}(d)), resulting in a power consumption of about \( 1 \, \mu\text{W} \). When oscillations occur, the DC bias conditions are slightly modified due to the self-rectification of alternating currents by the TD. This explains the hysteresis observed in the I-V curve in Fig.~\ref{fig:circuit}(d) depending on the sweep direction of $V_\mathrm{TD}$ ~\cite{van1981modeling}.

\subsection{Frequency and power tunability}\label{sec:freq_power_tunability}

\begin{figure}[htbp]
	\centerfloat{
		\includegraphics[width=0.7\linewidth]{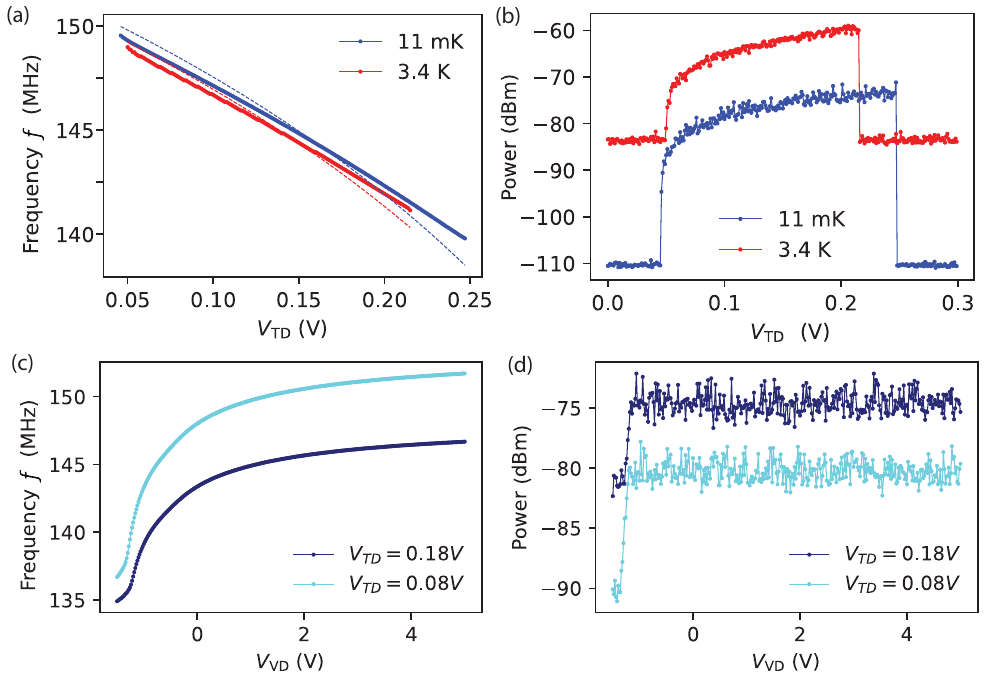}
	}
	\caption[Frequency and power tunability]{%
		The signal from the TDO obtained through AC port II was measured using a spectrum analyzer. (a,b) $V_\mathrm{VD}$ is set to 0. The sweep direction of $V_\mathrm{TD}$ is forward. 
		(a) Output frequency as a function of $V_\mathrm{TD}$ at MCT = 11~mK and 3.4~K. The dashed lines represent the fitted curves, which account for changes in depletion layer thickness induced by varying $V_\mathrm{TD}$. See the main text for details on the fitting procedure. 
		(b) Output power as a function of $V_\mathrm{TD}$ at MCT = 11~mK and 3.4~K. The difference in background levels is attributed to the use of different spectrum analyzers for measurements at MCT = 11~mK and 3.4~K. 
		(c) Output frequency as a function of $V_\mathrm{VD}$ at MCT = 11~mK.  
		(d) Output power as a function of $V_\mathrm{VD}$ at MCT = 11~mK. It remains constant for $V_\mathrm{VD} > -1.3$~V.%
	}
	\label{fig:freq_tunability}
\end{figure}

In this section, we demonstrate that the output power and frequency can be controlled by varying the voltage applied to the TD and that the frequency can be independently adjusted by varying the voltage applied to the varactor diode. The frequency variation due to $V_\mathrm{TD}$ is shown in Fig.~\ref{fig:freq_tunability}(a). The capacitance of the TD, \(C_{\mathrm{TD}}\), is voltage-dependent because the distance of the depletion layer in the p-n junction changes as a function of $V_\mathrm{TD}$. It is given by~\cite{chow1964principles} 
\begin{equation}
	C_{\mathrm{TD}} = C_0 \left( 1- V_{\mathrm{TD}}/V_{\mathrm{d}} \right)^{-1/2}, \label{eq:CTD}
\end{equation}
where $V_d=0.5$~V is the diffusion potential (estimated from a measurement) and $C_0$ is the capacitance of the TD when $V_\mathrm{TD}=0$. Eq.~\ref{eq:freq} with Eq.~\ref{eq:CTD} has been fitted, as shown by the dashed lines in Fig.~\ref{fig:freq_tunability}(a). The capacitance slightly decreases when the MCT is lowered from 3.4 K to 11 mK. The fitting results yield $C = 5.8$~pF and $C_0 = 5.7$~pF at MCT = 11~mK, and $C = 5.9$~pF and $C_0 = 5.8$~pF at MCT = 3.4~K. Fig.~\ref{fig:freq_tunability}(b) demonstrates that the output power can be tuned by approximately 10 dB by adjusting \( V_\mathrm{TD} \), as the operating point of the TD shifts along its I-V curve.  The microwave power that would be delivered to the resonator is around -90~dBm, which is comparable to the optimized power range for qubit readout of electrons used as qubits in semiconductors~\cite{gonzalez2015probing}.

Fig.~\ref{fig:freq_tunability}(c) shows the frequency tunability as a function of $V_\mathrm{VD}$. Considering the additional changes due to \( V_{\mathrm{TD}} \), we observe a total frequency tunability of 10~MHz. From this, we estimate that \( C \) varies from 7.9~pF to 5.3~pF as \( V_{\mathrm{VD}} \) is changed from -1.5~V to 5~V (Table. \ref{tab:capacitance}).

Fig.~\ref{fig:freq_tunability}(d) shows that for $V_\mathrm{VD} > -1.3$~V, the output power remains unaffected by $V_\mathrm{VD}$, indicating that the resistive component of the varactor remains unchanged (leakage current stays significantly small within this voltage range as seen in Fig.~\ref{fig:varactor}). Fig.~\ref{fig:freq_tunability}(d) shows that for \( V_\mathrm{VD} > -1.3 \)~V, the output power remains unaffected by \( V_\mathrm{VD} \), indicating that the resistive component of the varactor remains unchanged. This is consistent with the fact that the leakage current was measured to be significantly low within this voltage range, as shown in Fig.~\ref{fig:varactor}.
Please note that we have observed a change in output power when $V_\mathrm{VD}$ is varied within the same voltage range at RT when using a similar circuit with a normal coil inductor. We attribute this difference to the fact that the leakage current becomes significantly suppressed at low temperatures. 

\begin{table}[htbp]
	\caption[Capacitance values at 11 mK]{The table presents capacitance values under different voltage conditions at MCT = 11~mK. At 3.4~K, the capacitance values are higher by approximately 0.1~pF compared to those at 11~mK in all cases.}
	\centering
	\small
	\setlength{\tabcolsep}{4pt} 
	\renewcommand{\arraystretch}{1.2}
	\begin{tabular}{c|ccc|ccc}
		\hline
		& \multicolumn{3}{c|}{\( C_{\text{TD}} \)} & \multicolumn{3}{c}{\( C \)} \\
		\hline
		Voltage (V) &
		\( V_{\text{TD}}=0 \) &
		\( V_{\text{TD}}=0.08 \) &
		\( V_{\text{TD}}=0.18 \) &
		\( V_{\text{VD}}=-1.5 \) &
		\( V_{\text{VD}}=0 \) &
		\( V_{\text{VD}}=5 \) \\
		\hline
		Capacitance (pF) &
		5.7 & 6.3 & 7.2 & 7.9 & 5.8 & 5.3 \\
		\hline
	\end{tabular}
	\label{tab:capacitance}
\end{table}

\subsection{Phase noise}\label{sec:phase_noise}

\begin{figure}[htbp]
	\centerfloat{
		\includegraphics[width=0.7\linewidth]{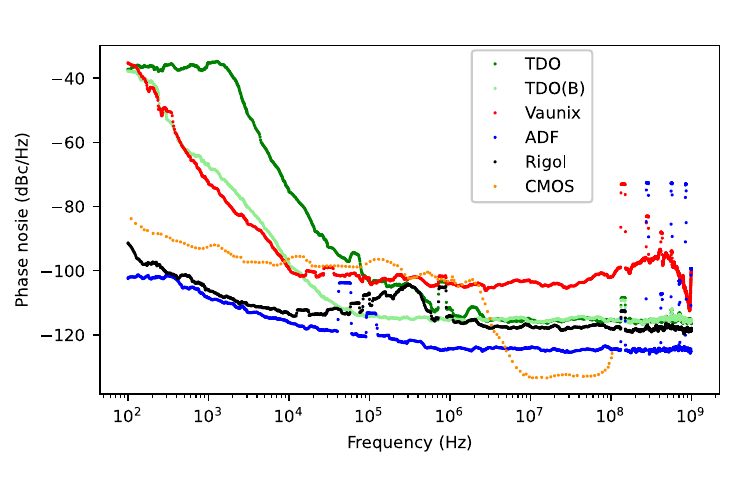}
	}
	\caption[Phase noise comparison]{%
		Measured phase noise relative to the carrier frequency (around 141.8 MHz) as a function of offset frequency with a bandwidth of 1~Hz. The phase noise of four different commercial microwave sources is shown: Rigol DSA815 (black), Analog Devices ADF4351 (blue), Vaunix Lab-Brick LMS-451D (red), along with the TDO powered by Yokogawa GS200 DC source (green) and by a lead-acid battery (light green). For comparison, the phase noise generated by a CMOS device at a carrier frequency of 600 MHz is also shown (orange)~\cite{Xue2023-cj}. The jumps observed below \(2 \times 10^5\) Hz are likely caused by noise from the pulse tube of the dilution refrigerator. The jumps around \(10^6\) Hz are attributed to interference from a nearby radio station transmitting at 810 kHz (Appendix~\ref{secA:DC_source}). The jumps above \(10^8\) Hz are due to harmonics of the carrier frequency.%
	}
	\label{fig:phase_noise}
\end{figure}

In this section, we report the phase noise of the signal from the TDO operating at MCT=11~mK through AC port I using a spectrum analyzer (R\&S FSV3030). Additionally, we sent the signal from the commercial microwave source through AC port II and measured the phase noise of the reflected signal through AC port I. The results are compared and shown in Fig.~\ref{fig:phase_noise}. Note that the phase noise of the commercial microwave source, measured directly on the spectrum analyzer without routing it to the device at low temperature, showed almost identical results. This indicates that the measured phase noise originates from the microwave sources and not from the device at low temperature.

We found that using a lead-acid battery to bias the TDO instead of the Yokogawa GS200 improves the phase noise as shown in Fig.~\ref{fig:phase_noise} (see also Appendix.~\ref{secA:DC_source} for details).  The measured phase noise of the TDO, powered by a battery, was $-115$~dBc/Hz at a 1~MHz offset, demonstrating comparable or superior performance to CMOS devices operating at the 4~K stage in a similar carrier frequency range~\cite{Xue2023-cj,Yang2019-bb,Lee2023-mx}. While CMOS devices achieve better phase noise in the lower- and higher-frequency ranges, the TDO provides better performance within the mid-frequency range. Moreover, with the battery as its power source, the TDO's phase noise is either comparable to or outperforms that of a commercial microwave source (Vaunix Lab-Brick LMS-451D) across the entire measured frequency range. 

The TDO has comparable or higher phase noise in the low-frequency range compared to commercial microwave sources. To improve this, enhancing the stability of the DC bias could help mitigate noise fluctuations. Additionally, the phase noise might be caused by two-level systems in the device, suggesting that improvements in the tunnel diode itself may be necessary. Another potential improvement is integrating a feedback mechanism to actively stabilize the frequency or implementing advanced filtering techniques to suppress unwanted noise components.

\subsection{Amplitude stability}\label{sec5}

\begin{figure}[htbp]
	\centerfloat{
		\includegraphics[width=0.7\linewidth]{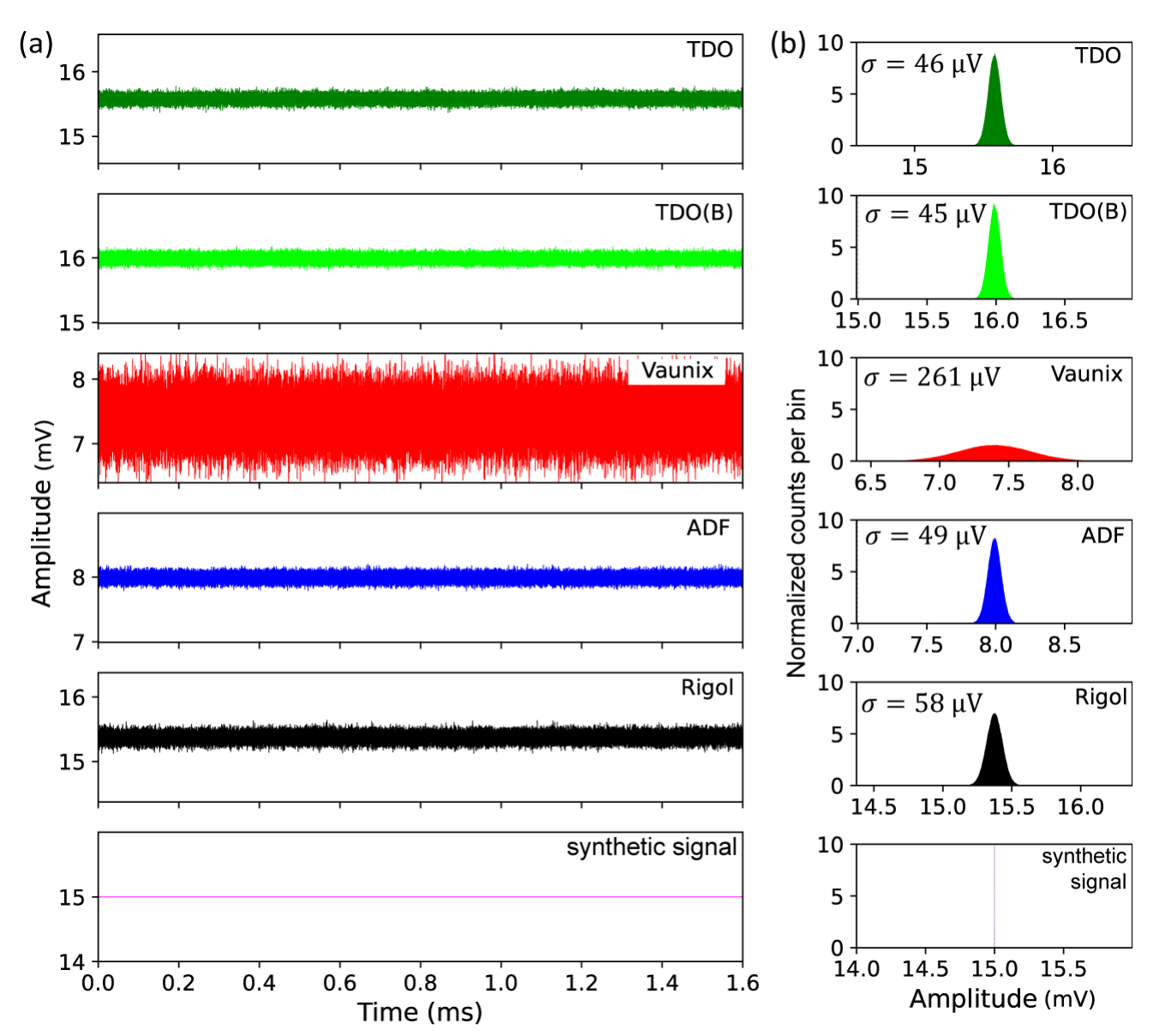}
	}
	\caption[Amplitude stability]{%
		The amplitude range is set to 2~mV for all the data for comparison. (a) The amplitude fluctuations over 1.6~ms are shown; see the main text for details. (b) Histograms of the amplitudes with 5000 bins and their standard deviations are also shown.%
	}
	\label{fig:amplitude-stability}
\end{figure}

In this section, we measured the amplitude of the signal from the TDO operating at MCT=11~mK through AC port I. The TDO generated a signal at 141.8~MHz with $V_\mathrm{TD}=0.18$~V and $V_\mathrm{VD}=-0.5$~V. The signal, recorded at a sampling rate of 20~GS/s for 1.6~ms, was passed through a digital bandpass filter centered at the carrier frequency with a 50~MHz bandwidth (see Appendix~\ref{secA:time-domain}). The amplitude was then extracted using a Hilbert transform and is plotted in Fig.~\ref{fig:amplitude-stability}(a), and its histogram is shown in Fig.~\ref{fig:amplitude-stability}(b). Similar to the phase noise measurement, a signal from a commercial microwave source was sent through AC port II, and the amplitude stability of the reflected signal, measured via AC port I, is shown in Fig.~\ref{fig:amplitude-stability} for comparison. For validation of our analytical method, a synthetic sine signal at 141.8~MHz generated is also plotted. Regardless of whether a lead-acid battery or the Yokogawa GS200 was used as the DC source, the amplitude stability of the TDO was superior to that of any commercial microwave source used in this study.

As with the phase noise measurements, the amplitude stability of the commercial microwave source, measured directly without routing it to the device at low temperature, showed identical results to those obtained when the signal was passed through the cryogenic device. This indicates that the observed amplitude stability is determined by the microwave sources themselves and not influenced by the device at low temperatures.

\subsection{Discussion}\label{sec:discuss}

We do not yet know the origin of the amplitude fluctuations, and identifying it will be the subject of future work. Since the oscilloscope used in this study has an 8-bit resolution, its resolution of \(1/2^8 = 0.39\%\) is comparable to the observed fluctuations of \(45 \, \mu\text{V}/15 \, \text{mV} = 0.3\%\). This suggests that the fluctuations might arise from the resolution limit of the oscilloscope rather than being intrinsic to the TDO. To investigate this further, improvements in the measurement setup, such as using higher-resolution ADCs, will be necessary. For qubit readout, particularly for electrons in semiconductors, where the qubit state is determined by observing the amplitude distribution~\cite{Burkard2023-ai,vigneau2023probing,urdampilleta2019gate}, it is unlikely that the TDO we developed would result in worse readout fidelity compared to the commercial microwave sources used here. However, looking forward, it will be necessary to evaluate the level of amplitude stability required for qubit readout using actual qubits.

Although the current study focuses on qubit readout, the TDO could potentially be utilized for qubit operation. However, its phase noise is currently not superior to that of commercial sources and could become a limiting factor if used for qubit operation~\cite{Ball2016-zn}. Further studies are needed to determine the phase noise requirements for qubit operation. Moreover, for qubit operation, the frequency of 140~MHz is too low. For operations involving electron spin states, GHz-range frequencies would be required. To achieve oscillation at higher frequencies, it is essential to reduce the parasitic capacitance of the TD and the circuit board. The parasitic capacitance of tunnel diodes exhibits significant variation, which is unlikely to be solely due to the pn-junction capacitance. Instead, this variation may originate from the diode packaging. Therefore, rather than relying on commercially available tunnel diodes, designing a tunnel diode specifically optimized for qubit readout or operation applications would be more effective. Minimizing the parasitic capacitance on the circuit board is also important. One possible approach is to fabricate the entire circuit board using microfabrication techniques. Another approach is to replace bulky varactor diodes with variable capacitors made from ferroelectric materials, such as strontium titanate (STO)~\cite{apostolidis2024quantum}. Ferroelectric materials like STO not only allow for variable capacitance but also eliminate leakage currents, reducing circuit losses and providing a significant advantage over traditional varactor diodes.

Considering that the cooling power of a typical cryogenic refrigerator's 4~K stage is approximately 200~mW, and its MC stage provides about $400 \, \mu\text{W}$ at 100~mK, it is feasible to place up to 20,000 microwave sources on the 4~K stage and 400 on the MC stage. These numbers are promising for realizing a scalable quantum computer; nevertheless, further improvements are necessary. In principle, lower power consumption of tunnel diodes can be fabricated by making the surface area of the tunnel diode smaller. While this would also lead to a reduction in output power, it can be compensated by adjusting the ratio of the number of turns in the spiral inductor to the number of turns in the pickup coil. However, the impact of back action on the oscillator must be carefully considered to ensure reliable operation. Furthermore, considering the need to apply a magnetic field for electron spin qubits, improvements such as using magnetic field-resilient superconducting materials like NbTiN instead of Nb and designing spiral inductors with these materials could be explored.

\subsection{Conclusion}\label{sec:concl}
In conclusion, we developed a 140~MHz tunnel diode oscillator with low power consumption of 1~$\mu$W and characterized its performance on the 10 mK stage. The achieved output power and frequency fall within the range typically used for qubit readout of electrons in semiconductors used as qubits. The amplitude stability is better than that of the commercial microwave sources measured in this study, which is encouraging for its application in qubit readout.


\chapter{Summary}

In this dissertation, we developed resonator-based readout techniques for FEB qubits on cryogenic substrates. Two experimental systems: an electron-on-helium system and an electron-on-neon system, were used. In addition, a cryogenic MW source was developed.

The main results of this dissertation are summarized as follows:

\begin{enumerate}
	
	\item { \textbf{(\RN{1}) Electron-on-helium:} \\
		\hspace*{2em}A high-Q LC tank circuit for electrons on helium was engineered and integrated at low temperature, using a Corbino-plate capacitor which yields a resonance near $\sim 121.9\ \mathrm{MHz}$ and a measured quality factor $Q \sim 311$. Using this circuit, quantum-capacitance detection of Rydberg transitions of FEs on liquid helium was demonstrated using microwave frequency modulation (FM-MW) combined with RF reflectometry. The observed sideband-based signal was explained by a model incorporating quantum capacitance and Landau–Zener transitions induced by frequency modulation. A capacitance sensitivity of $0.34\ \mathrm{aF}/\sqrt{\mathrm{Hz}}$ was achieved, which is sufficient for single-electron-scale detection in nanoscale devices and supports a scalable pathway toward qubit readout schemes based on FEs on helium. }
	
	\item {\textbf{(\RN{2}) Electron-on-neon:} \\
		\hspace*{2em}NbTiN superconducting nanowire resonators were demonstrated to maintain high performance after solid neon and electron deposition, validating suitability for electrons-on-neon devices. The resonators maintained quality factors on the order of $10^{5}$ after depositing solid neon and electrons. Representative measured parameters include a resonance frequency around $4.81\ \mathrm{GHz}$ and internal quality factor $Q_\mathrm{int}\approx 2.3\times 10^{5}$, establishing a robust starting point for the next device generation. Building on these experimental results, theoretical analysis suggests that, under optimal operating conditions such as operation at the charge sweet spot, single- and two-qubit gate fidelities approaching 99.99\% and 99.9\%, respectively, could be achievable even with natural neon, providing clear performance targets for future integrated devices.}

	\item {\textbf{(\RN{3}) Cryogenic MW source:} \\
		\hspace*{2em}A millikelvin-integrated TDO was developed and characterized as a compact cryogenic microwave source for scalable qubit readout. The design operates at $\sim 140\,\text{MHz}$ with tunable frequency and $\sim 1\,\mu\text{W}$ power consumption, making it suitable for continuous operation at the 10 mK stage for potentially large-scale qubit readout.}
	
\end{enumerate}

The methods developed in this dissertation establish an experimental and theoretical basis for manipulating and detecting quantum states of floating electrons on cryogenic substrates, paving the way toward the realization of qubits.

\printbibliography[title=References]

@article{cole1969image,
	title={Image-potential-induced surface bands in insulators},
	author={Cole, Milton W and Cohen, Morrel H},
	journal={Physical review letters},
	volume={23},
	number={21},
	pages={1238},
	year={1969},
	publisher={APS}
}

@article{kjaergaard2020superconducting,
	title={Superconducting qubits: Current state of play},
	author={Kjaergaard, Morten and Schwartz, Mollie E and Braum{\"u}ller, Jochen and Krantz, Philip and Wang, Joel I-J and Gustavsson, Simon and Oliver, William D},
	journal={Annual Review of Condensed Matter Physics},
	volume={11},
	number={1},
	pages={369--395},
	year={2020},
	publisher={Annual Reviews}
}

@article{kelly2015state,
	title={State preservation by repetitive error detection in a superconducting quantum circuit},
	author={Kelly, Julian and Barends, Rami and Fowler, Austin G and Megrant, Anthony and Jeffrey, Evan and White, Theodore C and Sank, Daniel and Mutus, Josh Y and Campbell, Brooks and Chen, Yu and others},
	journal={Nature},
	volume={519},
	number={7541},
	pages={66--69},
	year={2015},
	publisher={Nature Publishing Group UK London}
}

@article{arute2019quantum,
	title={Quantum supremacy using a programmable superconducting processor},
	author={Arute, Frank and Arya, Kunal and Babbush, Ryan and Bacon, Dave and Bardin, Joseph C and Barends, Rami and Biswas, Rupak and Boixo, Sergio and Brandao, Fernando GSL and Buell, David A and others},
	journal={Nature},
	volume={574},
	number={7779},
	pages={505--510},
	year={2019},
	publisher={Nature Publishing Group}
}

@article{bruzewicz2019trapped,
	title={Trapped-ion quantum computing: Progress and challenges},
	author={Bruzewicz, Colin D and Chiaverini, John and McConnell, Robert and Sage, Jeremy M},
	journal={Applied Physics Reviews},
	volume={6},
	number={2},
	year={2019},
	publisher={AIP Publishing}
}

@article{ruster2016long,
	title={A long-lived Zeeman trapped-ion qubit},
	author={Ruster, Thomas and Schmiegelow, Christian T and Kaufmann, Henning and Warschburger, Claudia and Schmidt-Kaler, Ferdinand and Poschinger, Ulrich G},
	journal={Applied Physics B},
	volume={122},
	number={10},
	pages={254},
	year={2016},
	publisher={Springer}
}

@article{wang2017single,
	title={Single-qubit quantum memory exceeding ten-minute coherence time},
	author={Wang, Ye and Um, Mark and Zhang, Junhua and An, Shuoming and Lyu, Ming and Zhang, Jing-Ning and Duan, L-M and Yum, Dahyun and Kim, Kihwan},
	journal={Nature Photonics},
	volume={11},
	number={10},
	pages={646--650},
	year={2017},
	publisher={Nature Publishing Group UK London}
}

@article{schafer2018fast,
	title={Fast quantum logic gates with trapped-ion qubits},
	author={Sch{\"a}fer, VM and Ballance, CJ and Thirumalai, K and Stephenson, LJ and Ballance, TG and Steane, AM and Lucas, DM},
	journal={Nature},
	volume={555},
	number={7694},
	pages={75--78},
	year={2018},
	publisher={Nature Publishing Group UK London}
}

@article{wintersperger2023neutral,
	title={Neutral atom quantum computing hardware: performance and end-user perspective},
	author={Wintersperger, Karen and Dommert, Florian and Ehmer, Thomas and Hoursanov, Andrey and Klepsch, Johannes and Mauerer, Wolfgang and Reuber, Georg and Strohm, Thomas and Yin, Ming and Luber, Sebastian},
	journal={EPJ Quantum Technology},
	volume={10},
	number={1},
	pages={32},
	year={2023},
	publisher={Springer Berlin Heidelberg}
}

@article{weiss2017quantum,
	title={Quantum computing with neutral atoms},
	author={Weiss, David S and Saffman, Mark},
	journal={Physics Today},
	volume={70},
	number={7},
	pages={44--50},
	year={2017},
	publisher={AIP Publishing}
}

@article{negretti2011quantum,
	title={Quantum computing implementations with neutral particles},
	author={Negretti, Antonio and Treutlein, Philipp and Calarco, Tommaso},
	journal={Quantum information processing},
	volume={10},
	pages={721--753},
	year={2011},
	publisher={Springer}
}

@article{evered2023high,
	title={High-fidelity parallel entangling gates on a neutral-atom quantum computer},
	author={Evered, Simon J and Bluvstein, Dolev and Kalinowski, Marcin and Ebadi, Sepehr and Manovitz, Tom and Zhou, Hengyun and Li, Sophie H and Geim, Alexandra A and Wang, Tout T and Maskara, Nishad and others},
	journal={Nature},
	volume={622},
	number={7982},
	pages={268--272},
	year={2023},
	publisher={Nature Publishing Group UK London}
}

@article{lyon2006spin,
	title={Spin-based quantum computing using electrons on liquid helium},
	author={Lyon, SA},
	journal={Physical Review A—Atomic, Molecular, and Optical Physics},
	volume={74},
	number={5},
	pages={052338},
	year={2006},
	publisher={APS}
}

@article{jennings2024quantum,
	title={Quantum computing using floating electrons on cryogenic substrates: Potential and challenges},
	author={Jennings, Ash and Zhou, Xianjing and Grytsenko, Ivan and Kawakami, Erika},
	journal={Applied Physics Letters},
	volume={124},
	number={12},
	year={2024},
	publisher={AIP Publishing}
}

@article{kajita1985wigner,
	title={Wigner crystallization of two dimensional electrons formed on the surface of solid neon},
	author={Kajita, Koji},
	journal={Journal of the Physical Society of Japan},
	volume={54},
	number={11},
	pages={4092--4095},
	year={1985},
	publisher={The Physical Society of Japan}
}

@article{platzman1999quantum,
	title={Quantum computing with electrons floating on liquid helium},
	author={Platzman, PM and Dykman, MI},
	journal={Science},
	volume={284},
	number={5422},
	pages={1967--1969},
	year={1999},
	publisher={American Association for the Advancement of Science}
}

@book{monarkha2004two,
	title={Two-dimensional interface electron systems},
	author={Monarkha, Yuriy and Kono, Kimitoshi and Monarkha, Yuriy and Kono, Kimitoshi},
	journal={Two-Dimensional Coulomb Liquids and Solids},
	pages={1--63},
	year={2004},
	publisher={Springer}
}

@article{grimes1980crystallization,
	title={Crystallization of electrons on the surface of liquid helium},
	author={Grimes, CC and Adams, G},
	journal={Surface Science},
	volume={98},
	number={1-3},
	pages={1--7},
	year={1980},
	publisher={Elsevier}
}

@article{grimes1979evidence,
	title={Evidence for a liquid-to-crystal phase transition in a classical, two-dimensional sheet of electrons},
	author={Grimes, CC and Adams, G},
	journal={Physical Review Letters},
	volume={42},
	number={12},
	pages={795},
	year={1979},
	publisher={APS}
}

@article{wigner1934interaction,
	title={On the interaction of electrons in metals},
	author={Wigner, Eugene},
	journal={Physical Review},
	volume={46},
	number={11},
	pages={1002},
	year={1934},
	publisher={APS}
}

@article{mehrotra1984density,
	title={Density-dependent mobility of a two-dimensional electron fluid},
	author={Mehrotra, R and Guo, CJ and Ruan, YZ and Mast, DB and Dahm, AJ},
	journal={Physical Review B},
	volume={29},
	number={9},
	pages={5239},
	year={1984},
	publisher={APS}
}

@article{cole1971electronic,
	title={Electronic surface states of a dielectric film on a metal substrate},
	author={Cole, Milton W},
	journal={Physical Review B},
	volume={3},
	number={12},
	pages={4418},
	year={1971},
	publisher={APS}
}

@article{jin2020quantum,
	title={Quantum electronics and optics at the interface of solid neon and superfluid helium},
	author={Jin, Dafei},
	journal={Quantum Science and Technology},
	volume={5},
	number={3},
	pages={035003},
	year={2020},
	publisher={IOP Publishing}
}

@article{grimes1976spectroscopy,
	title={Spectroscopy of electrons in image-potential-induced surface states outside liquid helium},
	author={Grimes, CC and Brown, TR and Burns, Michael L and Zipfel, CL},
	journal={Physical Review B},
	volume={13},
	number={1},
	pages={140},
	year={1976},
	publisher={APS}
}

@phdthesis{schuster2007circuit,
	title={Circuit quantum electrodynamics},
	author={Schuster, David Isaac},
	year={2007},
	publisher={Yale University}
}

@article{koolstra2019coupling,
	title={Coupling a single electron on superfluid helium to a superconducting resonator},
	author={Koolstra, Gerwin and Yang, Ge and Schuster, David I},
	journal={Nature communications},
	volume={10},
	number={1},
	pages={5323},
	year={2019},
	publisher={Nature Publishing Group UK London}
}

@article{burkard2020superconductor,
	title={Superconductor--semiconductor hybrid-circuit quantum electrodynamics},
	author={Burkard, Guido and Gullans, Michael J and Mi, Xiao and Petta, Jason R},
	journal={Nature Reviews Physics},
	volume={2},
	number={3},
	pages={129--140},
	year={2020},
	publisher={Nature Publishing Group UK London}
}

@article{wallraff2004strong,
	title={Strong coupling of a single photon to a superconducting qubit using circuit quantum electrodynamics},
	author={Wallraff, Andreas and Schuster, David I and Blais, Alexandre and Frunzio, Luigi and Huang, R-S and Majer, Johannes and Kumar, Sameer and Girvin, Steven M and Schoelkopf, Robert J},
	journal={Nature},
	volume={431},
	number={7005},
	pages={162--167},
	year={2004},
	publisher={Nature Publishing Group UK London}
}

@article{blais2021circuit,
	title={Circuit quantum electrodynamics},
	author={Blais, Alexandre and Grimsmo, Arne L and Girvin, Steven M and Wallraff, Andreas},
	journal={Reviews of Modern Physics},
	volume={93},
	number={2},
	pages={025005},
	year={2021},
	publisher={APS}
}

@phdthesis{koolstra2019trapping,
	title={Trapping a single electron on superfluid helium using a superconducting resonator},
	author={Koolstra, Gerwin},
	year={2019},
	school={The University of Chicago}
}

@article{ibberson2021large,
	title={Large dispersive interaction between a CMOS double quantum dot and microwave photons},
	author={Ibberson, David J and Lundberg, Theodor and Haigh, James A and Hutin, Louis and Bertrand, Benoit and Barraud, Sylvain and Lee, Chang-Min and Stelmashenko, Nadia A and Oakes, Giovanni A and Cochrane, Laurence and others},
	journal={PRX Quantum},
	volume={2},
	number={2},
	pages={020315},
	year={2021},
	publisher={APS}
}

@article{vigneau2023probing,
	title={Probing quantum devices with radio-frequency reflectometry},
	author={Vigneau, Florian and Fedele, Federico and Chatterjee, Anasua and Reilly, David and Kuemmeth, Ferdinand and Gonzalez-Zalba, M Fernando and Laird, Edward and Ares, Natalia},
	journal={Applied Physics Reviews},
	volume={10},
	number={2},
	year={2023},
	publisher={AIP Publishing}
}

@article{ahmed2018radio,
	title={Radio-frequency capacitive gate-based sensing},
	author={Ahmed, Imtiaz and Haigh, James A and Schaal, Simon and Barraud, Sylvain and Zhu, Yi and Lee, Chang-min and Amado, Mario and Robinson, Jason WA and Rossi, Alessandro and Morton, John JL and others},
	journal={Physical Review Applied},
	volume={10},
	number={1},
	pages={014018},
	year={2018},
	publisher={APS}
}

@article{apostolidis2020quantum,
	title={Quantum paraelectric varactors for radio-frequency measurements at mK temperatures},
	author={Apostolidis, P and Villis, BJ and Chittock-Wood, JF and Baumgartner, A and Vesterinen, V and Simbierowicz, S and Hassel, J and Buitelaar, MR},
	journal={arXiv preprint arXiv:2007.03588},
	year={2020}
}

@article{schuster2010proposal,
	title={Proposal for Manipulating and Detecting Spin and Orbital States of Trapped Electrons on Helium Using Cavity Quantum Electrodynamics},
	author={Schuster, DI and Fragner, A and Dykman, MI and Lyon, SA and Schoelkopf, RJ},
	journal={Physical review letters},
	volume={105},
	number={4},
	pages={040503},
	year={2010},
	publisher={APS}
}

@article{zhou2022single,
	title={Single electrons on solid neon as a solid-state qubit platform},
	author={Zhou, Xianjing and Koolstra, Gerwin and Zhang, Xufeng and Yang, Ge and Han, Xu and Dizdar, Brennan and Li, Xinhao and Divan, Ralu and Guo, Wei and Murch, Kater W and others},
	journal={Nature},
	volume={605},
	number={7908},
	pages={46--50},
	year={2022},
	publisher={Nature Publishing Group UK London}
}

@article{zhou2024electron,
	title={Electron charge qubit with 0.1 millisecond coherence time},
	author={Zhou, Xianjing and Li, Xinhao and Chen, Qianfan and Koolstra, Gerwin and Yang, Ge and Dizdar, Brennan and Huang, Yizhong and Wang, Christopher S and Han, Xu and Zhang, Xufeng and others},
	journal={Nature Physics},
	volume={20},
	number={1},
	pages={116--122},
	year={2024},
	publisher={Nature Publishing Group UK London}
}

@article{kawakami2023blueprint,
	title={Blueprint for quantum computing using electrons on helium},
	author={Kawakami, Erika and Chen, Jiabao and Benito, M{\'o}nica and Konstantinov, Denis},
	journal={Physical Review Applied},
	volume={20},
	number={5},
	pages={054022},
	year={2023},
	publisher={APS}
}

@article{zhang2012spin,
	title={Spin-orbit couplings between distant electrons trapped individually on liquid helium},
	author={Zhang, Miao and Wei, LF},
	journal={Physical Review B—Condensed Matter and Materials Physics},
	volume={86},
	number={20},
	pages={205408},
	year={2012},
	publisher={APS}
}

@article{dykman2023spin,
	title={Spin dynamics in quantum dots on liquid helium},
	author={Dykman, MI and Asban, Ofek and Chen, Qianfan and Jin, Dafei and Lyon, SA},
	journal={Physical Review B},
	volume={107},
	number={3},
	pages={035437},
	year={2023},
	publisher={APS}
}

@article{kawakami2013excitation,
	title={Excitation of a Si/SiGe quantum dot using an on-chip microwave antenna},
	author={Kawakami, E and Scarlino, Pasquale and Schreiber, LR and Prance, JR and Savage, DE and Lagally, MG and Eriksson, MA and Vandersypen, LMK},
	journal={Applied Physics Letters},
	volume={103},
	number={13},
	year={2013},
	publisher={AIP Publishing}
}

@article{pioro2008electrically,
	title={Electrically driven single-electron spin resonance in a slanting Zeeman field},
	author={Pioro-Ladriere, M and Obata, T and Tokura, Y and Shin, Y-S and Kubo, Toshihiro and Yoshida, K and Taniyama, T and Tarucha, S},
	journal={Nature Physics},
	volume={4},
	number={10},
	pages={776--779},
	year={2008},
	publisher={Nature Publishing Group UK London}
}

@article{pioro2008selective,
	title={Selective manipulation of electron spins with electric fields},
	author={Pioro-Ladriere, Michel and Obata, Toshiaki and Tokura, Yasuhiro and Shin, You-Sok and Kubo, Toshihiro and Yoshida, Katsuharu and Taniyama, Tomoyasu and Tarucha, Seigo},
	journal={Progress of Theoretical Physics Supplement},
	volume={176},
	pages={322--340},
	year={2008},
	publisher={Oxford University Press}
}

@article{mi2018coherent,
	title={A coherent spin--photon interface in silicon},
	author={Mi, Xiao and Benito, M{\'o}nica and Putz, Stefan and Zajac, David M and Taylor, Jacob M and Burkard, Guido and Petta, Jason R},
	journal={Nature},
	volume={555},
	number={7698},
	pages={599--603},
	year={2018},
	publisher={Nature Publishing Group UK London}
}

@article{benito2017input,
	title={Input-output theory for spin-photon coupling in Si double quantum dots},
	author={Benito, Monica and Mi, Xiao and Taylor, Jacob M and Petta, Jason R and Burkard, Guido},
	journal={Physical Review B},
	volume={96},
	number={23},
	pages={235434},
	year={2017},
	publisher={APS}
}

@article{samkharadze2018strong,
	title={Strong spin-photon coupling in silicon},
	author={Samkharadze, Nodar and Zheng, Guoji and Kalhor, Nima and Brousse, Delphine and Sammak, Amir and Mendes, UC and Blais, Alexandre and Scappucci, Giordano and Vandersypen, LMK},
	journal={Science},
	volume={359},
	number={6380},
	pages={1123--1127},
	year={2018},
	publisher={American Association for the Advancement of Science}
}

@article{reed2010fast,
	title={Fast reset and suppressing spontaneous emission of a superconducting qubit},
	author={Reed, Matthew D and Johnson, Blake R and Houck, Andrew A and DiCarlo, Leonardo and Chow, Jerry M and Schuster, David I and Frunzio, Luigi and Schoelkopf, Robert J},
	journal={Applied Physics Letters},
	volume={96},
	number={20},
	year={2010},
	publisher={AIP Publishing}
}

@article{jeffrey2014fast,
	title={Fast accurate state measurement with superconducting qubits},
	author={Jeffrey, Evan and Sank, Daniel and Mutus, JY and White, TC and Kelly, J and Barends, R and Chen, Y and Chen, Z and Chiaro, B and Dunsworth, A and others},
	journal={Physical review letters},
	volume={112},
	number={19},
	pages={190504},
	year={2014},
	publisher={APS}
}

@article{bronn2015broadband,
	title={Broadband filters for abatement of spontaneous emission in circuit quantum electrodynamics},
	author={Bronn, Nicholas T and Liu, Yanbing and Hertzberg, Jared B and C{\'o}rcoles, Antonio D and Houck, Andrew A and Gambetta, Jay M and Chow, Jerry M},
	journal={Applied Physics Letters},
	volume={107},
	number={17},
	year={2015},
	publisher={AIP Publishing}
}

@article{hu2012strong,
	title={Strong coupling of a spin qubit to a superconducting stripline cavity},
	author={Hu, Xuedong and Liu, Yu-xi and Nori, Franco},
	journal={Physical Review B—Condensed Matter and Materials Physics},
	volume={86},
	number={3},
	pages={035314},
	year={2012},
	publisher={APS}
}

@book{simons2004coplanar,
	title={Coplanar waveguide circuits, components, and systems},
	author={Simons, Rainee N},
	year={2004},
	publisher={John Wiley \& Sons}
}

@article{elarabi2021cryogenic,
	title={Cryogenic amplification of image-charge detection for readout of quantum states of electrons on liquid helium},
	author={Elarabi, Asem and Kawakami, Erika and Konstantinov, Denis},
	journal={Journal of Low Temperature Physics},
	volume={202},
	pages={456--465},
	year={2021},
	publisher={Springer}
}

@article{collin2002microwave,
	title={Microwave saturation of the Rydberg states of electrons on helium},
	author={Collin, E and Bailey, W and Fozooni, P and Frayne, PG and Glasson, P and Harrabi, K and Lea, MJ and Papageorgiou, G},
	journal={Physical review letters},
	volume={89},
	number={24},
	pages={245301},
	year={2002},
	publisher={APS}
}

@article{kawakami2019image,
	title={Image-charge detection of the Rydberg states of surface electrons on liquid helium},
	author={Kawakami, Erika and Elarabi, Asem and Konstantinov, Denis},
	journal={Physical Review Letters},
	volume={123},
	number={8},
	pages={086801},
	year={2019},
	publisher={APS}
}

@article{schoelkopf1998radio,
	title={The radio-frequency single-electron transistor (RF-SET): A fast and ultrasensitive electrometer},
	author={Schoelkopf, RJ and Wahlgren, P and Kozhevnikov, AA and Delsing, P and Prober, DE},
	journal={science},
	volume={280},
	number={5367},
	pages={1238--1242},
	year={1998},
	publisher={American Association for the Advancement of Science}
}

@book{chow1964principles,
	author    = {Chow, Woo Foung},
	title     = {Principles of Tunnel Diode Circuits},
	year      = {1964},
	publisher = {McGraw-Hill},
	address   = {New York},
	langid    = {english}
}

@article{van1975tunnel,
	title={Tunnel diode oscillator for 0.001 ppm measurements at low temperatures},
	author={Van Degrift, Craig T},
	journal={Review of Scientific Instruments},
	volume={46},
	number={5},
	pages={599--607},
	year={1975},
	publisher={American Institute of Physics}
}

@inproceedings{bardin201929,
	title={29.1 a 28nm bulk-cmos 4-to-8ghz!` 2mw cryogenic pulse modulator for scalable quantum computing},
	author={Bardin, Joseph C and Jeffrey, Evan and Lucero, Erik and Huang, Trent and Naaman, Ofer and Barends, Rami and White, Ted and Giustina, Marissa and Sank, Daniel and Roushan, Pedram and others},
	booktitle={2019 IEEE International Solid-State Circuits Conference-(ISSCC)},
	pages={456--458},
	year={2019},
	organization={IEEE}
}

@inproceedings{pellerano2022cryogenic,
	title={Cryogenic CMOS for qubit control and readout},
	author={Pellerano, Stefano and Subramanian, Sushil and Park, Jong-Seok and Patra, Bishnu and Mladenov, Todor and Xue, Xiao and Vandersypen, Lieven MK and Babaie, Masoud and Charbon, Edoardo and Sebastiano, Fabio},
	booktitle={2022 IEEE Custom Integrated Circuits Conference (CICC)},
	pages={01--08},
	year={2022},
	organization={IEEE}
}

@inproceedings{kang2022cryo,
	title={A cryo-CMOS controller IC with fully integrated frequency generators for superconducting qubits},
	author={Kang, Kiseo and Minn, Donggyu and Bae, Seunghun and Lee, Jaeho and Bae, Seongun and Jung, Gichang and Kang, Seokhyeong and Lee, Moonjoo and Song, Ho-Jin and Sim, Jae-Yoon},
	booktitle={2022 IEEE International Solid-State Circuits Conference (ISSCC)},
	volume={65},
	pages={362--364},
	year={2022},
	organization={IEEE}
}

@article{howe2022digital,
	title={Digital control of a superconducting qubit using a Josephson pulse generator at 3 K},
	author={Howe, Logan and Castellanos-Beltran, MA and Sirois, AJ and Olaya, David and Biesecker, John and Dresselhaus, PD and Benz, Samuel P and Hopkins, PF},
	journal={PRX quantum},
	volume={3},
	number={1},
	pages={010350},
	year={2022},
	publisher={APS}
}

@article{pauka2021cryogenic,
	title={A cryogenic CMOS chip for generating control signals for multiple qubits},
	author={Pauka, SJ and Das, K and Kalra, R and Moini, A and Yang, Y and Trainer, M and Bousquet, A and Cantaloube, C and Dick, N and Gardner, GC and others},
	journal={Nature Electronics},
	volume={4},
	number={1},
	pages={64--70},
	year={2021},
	publisher={Nature Publishing Group}
}

@article{peng2022cryo,
	title={A cryo-CMOS wideband quadrature receiver with frequency synthesizer for scalable multiplexed readout of silicon spin qubits},
	author={Peng, Yatao and Ruffino, Andrea and Yang, Tsung-Yeh and Michniewicz, John and Gonzalez-Zalba, Miguel Fernando and Charbon, Edoardo},
	journal={IEEE Journal of Solid-State Circuits},
	volume={57},
	number={8},
	pages={2374--2389},
	year={2022},
	publisher={IEEE}
}

@article{nakamura1999coherent,
	title={Coherent control of macroscopic quantum states in a single-Cooper-pair box},
	author={Nakamura, Yasunobu and Pashkin, Yu A and Tsai, JS},
	journal={nature},
	volume={398},
	number={6730},
	pages={786--788},
	year={1999},
	publisher={Nature Publishing Group UK London}
}

@article{nowack2007coherent,
	title={Coherent control of a single electron spin with electric fields},
	author={Nowack, Katja C and Koppens, FHL and Nazarov, Yu V and Vandersypen, LMK},
	journal={Science},
	volume={318},
	number={5855},
	pages={1430--1433},
	year={2007},
	publisher={American Association for the Advancement of Science}
}

@article{tokura2006coherent,
	title={Coherent single electron spin control in a slanting Zeeman field},
	author={Tokura, Yasuhiro and van der Wiel, Wilfred G and Obata, Toshiaki and Tarucha, Seigo},
	journal={Physical review letters},
	volume={96},
	number={4},
	pages={047202},
	year={2006},
	publisher={APS}
}

@article{huang2020superconducting,
	title={Superconducting quantum computing: a review},
	author={Huang, He-Liang and Wu, Dachao and Fan, Daojin and Zhu, Xiaobo},
	journal={Science China Information Sciences},
	volume={63},
	number={8},
	pages={180501},
	year={2020},
	publisher={Springer}
}

@article{goppl2008coplanar,
	title={Coplanar waveguide resonators for circuit quantum electrodynamics},
	author={G{\"o}ppl, Martin and Fragner, A and Baur, M and Bianchetti, Romeo and Filipp, Stefan and Fink, Johannes M and Leek, Peter J and Puebla, G and Steffen, Lars and Wallraff, Andreas},
	journal={Journal of Applied Physics},
	volume={104},
	number={11},
	year={2008},
	publisher={AIP Publishing}
}

@book{pozar2021microwave,
	title={Microwave engineering: theory and techniques},
	author={Pozar, David M},
	year={2021},
	publisher={John wiley \& sons}
}

@article{grytsenko2025characterization,
	title={Characterization of Tunnel Diode Oscillator for Qubit Readout Applications},
	author={Grytsenko, Ivan and van Haagen, Sander and Rybalko, Oleksiy and Jennings, Asher and Mohan, Rajesh and Tian, Yiran and Kawakami, Erika},
	journal={Journal of Low Temperature Physics},
	pages={1--19},
	year={2025},
	publisher={Springer}
}

@article{lim1986indium,
	title={Indium seals for low-temperature and moderate-pressure applications},
	author={Lim, CC},
	journal={Review of scientific instruments},
	volume={57},
	number={1},
	pages={108--114},
	year={1986},
	publisher={American Institute of Physics}
}

@article{samkharadze2016high,
	title={High-kinetic-inductance superconducting nanowire resonators for circuit QED in a magnetic field},
	author={Samkharadze, Nodar and Bruno, A and Scarlino, Pasquale and Zheng, G and DiVincenzo, DP and DiCarlo, L and Vandersypen, LMK},
	journal={Physical Review Applied},
	volume={5},
	number={4},
	pages={044004},
	year={2016},
	publisher={APS}
}

@article{dijkema2025cavity,
	title={Cavity-mediated iSWAP oscillations between distant spins},
	author={Dijkema, Jurgen and Xue, Xiao and Harvey-Collard, Patrick and Rimbach-Russ, Maximilian and de Snoo, Sander L and Zheng, Guoji and Sammak, Amir and Scappucci, Giordano and Vandersypen, Lieven MK},
	journal={Nature Physics},
	volume={21},
	number={1},
	pages={168--174},
	year={2025},
	publisher={Nature Publishing Group}
}

@article{kroll2019magnetic,
	title={Magnetic-field-resilient superconducting coplanar-waveguide resonators for hybrid circuit quantum electrodynamics experiments},
	author={Kroll, James G and Borsoi, F and Van Der Enden, KL and Uilhoorn, W and De Jong, D and Quintero-P{\'e}rez, M and Van Woerkom, DJ and Bruno, A and Plissard, SR and Car, D and others},
	journal={Physical Review Applied},
	volume={11},
	number={6},
	pages={064053},
	year={2019},
	publisher={APS}
}

@article{bretz2022high,
	title={High kinetic inductance NbTiN superconducting transmission line resonators in the very thin film limit},
	author={Bretz-Sullivan, Terence M and Lewis, Rupert M and Lima-Sharma, Ana L and Lidsky, David and Smyth, Christopher M and Harris, C Thomas and Venuti, Michael and Eley, Serena and Lu, Tzu-Ming},
	journal={Applied Physics Letters},
	volume={121},
	number={5},
	year={2022},
	publisher={AIP Publishing}
}

@article{probst2015efficient,
	title={Efficient and robust analysis of complex scattering data under noise in microwave resonators},
	author={Probst, Sebastian and Song, FB and Bushev, Pavel A and Ustinov, Alexey V and Weides, Martin},
	journal={Review of Scientific Instruments},
	volume={86},
	number={2},
	year={2015},
	publisher={AIP Publishing}
}

@misc{probst2024circlefit,
	author       = {S. Probst},
	title        = {circle fit},
	year         = {2024},
	note         = {Available online: \url{https://github.com/sebastianprobst/resonator_tools}},
	howpublished = {\url{https://github.com/sebastianprobst/resonator_tools}}
}

@article{konstantinov2007investigation,
	title={Investigation of microwave absorption of surface-state electrons on liquid 3He},
	author={Konstantinov, Denis and Isshiki, Hanako and Akimoto, Hikota and Shirahama, Keiya and Kono, Kimitoshi},
	journal={Journal of Low Temperature Physics},
	volume={148},
	number={3},
	pages={187--191},
	year={2007},
	publisher={Springer}
}

@article{jennings2025probing,
	title={Probing the Quantum Capacitance of Rydberg Transitions of Surface Electrons on Liquid Helium via Microwave Frequency Modulation},
	author={Jennings, Asher and Grytsenko, Ivan and Tian, Yiran and Rybalko, Oleksiy and Wang, Jun and Barabash, Itay Josef and Kawakami, Erika},
	journal={Physical Review Letters},
	volume={135},
	number={8},
	pages={087001},
	year={2025},
	publisher={APS}
}

@article{gonzalez2016gate,
	title={Gate-sensing coherent charge oscillations in a silicon field-effect transistor},
	author={Gonzalez-Zalba, M Fernando and Shevchenko, Sergey N and Barraud, Sylvain and Johansson, J Robert and Ferguson, Andrew J and Nori, Franco and Betz, Andreas C},
	journal={Nano letters},
	volume={16},
	number={3},
	pages={1614--1619},
	year={2016},
	publisher={ACS Publications}
}

@book{nielsen2010quantum,
	title={Quantum computation and quantum information},
	author={Nielsen, Michael A and Chuang, Isaac L},
	year={2010},
	publisher={Cambridge university press}
}

@article{gao2025establishing,
	title={Establishing a new benchmark in quantum computational advantage with 105-qubit zuchongzhi 3.0 processor},
	author={Gao, Dongxin and Fan, Daojin and Zha, Chen and Bei, Jiahao and Cai, Guoqing and Cai, Jianbin and Cao, Sirui and Chen, Fusheng and Chen, Jiang and Chen, Kefu and others},
	journal={Physical Review Letters},
	volume={134},
	number={9},
	pages={090601},
	year={2025},
	publisher={APS}
}

@article{chen2022electron,
	title={Electron spin coherence on a solid neon surface},
	author={Chen, Qianfan and Martin, Ivar and Jiang, Liang and Jin, Dafei},
	journal={Quantum Science and Technology},
	volume={7},
	number={4},
	pages={045016},
	year={2022},
	publisher={IOP Publishing}
}

@article{monarkha2010decay,
	title={Decay of excited surface electron states in liquid helium and related relaxation phenomena induced by short-wavelength ripplons},
	author={Monarkha, Yu P and Sokolov, SS and Smorodin, AV and Studart, Nelson},
	journal={Low Temperature Physics},
	volume={36},
	number={7},
	pages={565--575},
	year={2010},
	publisher={AIP Publishing}
}

@article{dykman2003qubits,
	title={Qubits with electrons on liquid helium},
	author={Dykman, MI and Platzman, PM and Seddighrad, P},
	journal={Physical Review B},
	volume={67},
	number={15},
	pages={155402},
	year={2003},
	publisher={APS}
}

@article{barenco1995elementary,
	title={Elementary gates for quantum computation},
	author={Barenco, Adriano and Bennett, Charles H and Cleve, Richard and DiVincenzo, David P and Margolus, Norman and Shor, Peter and Sleator, Tycho and Smolin, John A and Weinfurter, Harald},
	journal={Physical review A},
	volume={52},
	number={5},
	pages={3457},
	year={1995},
	publisher={APS}
}

@article{divincenzo1995two,
	title={Two-bit gates are universal for quantum computation},
	author={DiVincenzo, David P},
	journal={Physical Review A},
	volume={51},
	number={2},
	pages={1015},
	year={1995},
	publisher={APS}
}

@article{henriet2020quantum,
	title={Quantum computing with neutral atoms},
	author={Henriet, Lo{\"\i}c and Beguin, Lucas and Signoles, Adrien and Lahaye, Thierry and Browaeys, Antoine and Reymond, Georges-Olivier and Jurczak, Christophe},
	journal={Quantum},
	volume={4},
	pages={327},
	year={2020},
	publisher={Verein zur F{\"o}rderung des Open Access Publizierens in den Quantenwissenschaften}
}

@book{monarkha2013two,
	title={Two-dimensional Coulomb liquids and solids},
	author={Monarkha, Yuriy and Kono, Kimitoshi},
	volume={142},
	year={2013},
	publisher={Springer Science \& Business Media}
}

@article{leiderer2025surface,
	title={Surface Electrons on Solid Quantum Substrates: A Brief Review},
	author={Leiderer, Paul},
	journal={Journal of Low Temperature Physics},
	pages={1--20},
	year={2025},
	publisher={Springer}
}

@article{kajita1984new,
	title={A new two-dimensional electron system on the surface of solid neon},
	author={Kajita, K},
	journal={Surface Science},
	volume={142},
	number={1-3},
	pages={86--95},
	year={1984},
	publisher={Elsevier}
}

@article{nielsen2002simple,
	title={A simple formula for the average gate fidelity of a quantum dynamical operation},
	author={Nielsen, Michael A},
	journal={Physics Letters A},
	volume={303},
	number={4},
	pages={249--252},
	year={2002},
	publisher={Elsevier}
}

@incollection{cole1997surface,
	title={The Surface State Electron},
	author={Cole, Milton W},
	booktitle={Two-Dimensional Electron Systems: on Helium and other Cryogenic Substrates},
	pages={1--16},
	year={1997},
	publisher={Springer}
}

@article{kanai2023single,
	title={Single-electron qubits based on ring-shaped surface states on solid neon},
	author={Kanai, Toshiaki and Jin, Dafei and Guo, Wei},
	journal={arXiv e-prints},
	pages={arXiv--2311},
	year={2023}
}

@article{lee2024penetration,
	title={Penetration depth in dirty superconducting NbTiN thin films grown at room temperature},
	author={Lee, Yeonkyu and Yun, Jinyoung and Lee, Chanyoung and Sirena, M and Kim, Jeehoon and Haberkorn, N},
	journal={Applied Physics A},
	volume={130},
	number={7},
	pages={504},
	year={2024},
	publisher={Springer}
}

@article{chae2024elementary,
	title={An elementary review on basic principles and developments of qubits for quantum computing},
	author={Chae, Eunmi and Choi, Joonhee and Kim, Junki},
	journal={Nano Convergence},
	volume={11},
	number={1},
	pages={11},
	year={2024},
	publisher={Springer}
}

@article{vandersypen2004real,
	title={Real-time detection of single-electron tunneling using a quantum point contact},
	author={Vandersypen, LMK and Elzerman, JM and Schouten, RN and Willems van Beveren, LH and Hanson, R and Kouwenhoven, Leo P},
	journal={Applied Physics Letters},
	volume={85},
	number={19},
	pages={4394--4396},
	year={2004},
	publisher={AIP Publishing}
}

@article{schroer2012radio,
	title={Radio frequency charge parity meter},
	author={Schroer, MD and Jung, M and Petersson, KD and Petta, Jason R},
	journal={Physical review letters},
	volume={109},
	number={16},
	pages={166804},
	year={2012},
	publisher={APS}
}

@article{connors2020rapid,
	title={Rapid high-fidelity spin-state readout in Si/Si-Ge quantum dots via rf reflectometry},
	author={Connors, Elliot J and Nelson, JJ and Nichol, John M},
	journal={Physical Review Applied},
	volume={13},
	number={2},
	pages={024019},
	year={2020},
	publisher={APS}
}

@article{crippa2019gate,
	title={Gate-reflectometry dispersive readout and coherent control of a spin qubit in silicon},
	author={Crippa, Alessandro and Ezzouch, Rami and Apr{\'a}, A and Amisse, A and Lavieville, Romain and Hutin, L and Bertrand, Benoit and Vinet, M and Urdampilleta, Matias and Meunier, Tristan and others},
	journal={Nature communications},
	volume={10},
	number={1},
	pages={2776},
	year={2019},
	publisher={Nature Publishing Group UK London}
}

@article{jaynes2005comparison,
	title={Comparison of quantum and semiclassical radiation theories with application to the beam maser},
	author={Jaynes, Edwin T and Cummings, Frederick W},
	journal={Proceedings of the IEEE},
	volume={51},
	number={1},
	pages={89--109},
	year={2005},
	publisher={IEEE}
}

@book{chen2004electrical,
	title={The electrical engineering handbook},
	author={Chen, Wai Kai},
	year={2004},
	publisher={Elsevier}
}

@book{solymar2014electrical,
	title={Electrical properties of materials},
	author={Solymar, Laszlo and Walsh, Donald and Syms, Richard RA},
	year={2014},
	publisher={Oxford university press}
}

@book{chen2003vlsi,
	title={VLSI technology},
	author={Chen, Wai-Kai},
	year={2003},
	publisher={CRc Press}
}

@article{chang1975proximity,
	title={Proximity effect in electron-beam lithography},
	author={Chang, THP},
	journal={Journal of vacuum science and technology},
	volume={12},
	number={6},
	pages={1271--1275},
	year={1975},
	publisher={American Vacuum Society}
}

@article{owa2023review,
	title={Review of optical direct-write technology for semiconductor manufacturing},
	author={Owa, Soichi},
	journal={Journal of Micro/Nanopatterning, Materials, and Metrology},
	volume={22},
	number={4},
	pages={041402--041402},
	year={2023},
	publisher={Society of Photo-Optical Instrumentation Engineers}
}

@ARTICLE{Campbell2007-yx,
	title     = "Fabrication Engineering at the Micro and Nanoscale",
	author    = "Campbell, S",
	journal   = "(No Title)",
	publisher = "cir.nii.ac.jp",
	month     =  sep,
	year      =  2007
}

@article{london1951superfluids,
	title={Superfluids, Volume 1, Macroscopic Theory of Super Conductivity.},
	author={London, Fritz},
	year={1951}
}

@book{pobell2007matter,
	title={Matter and methods at low temperatures},
	author={Pobell, Frank},
	volume={2},
	year={2007},
	publisher={Springer}
}

@article{colless2013dispersive,
	title={Dispersive readout of a few-electron double quantum dot with fast rf gate sensors},
	author={Colless, JI and Mahoney, AC and Hornibrook, JM and Doherty, AC and Lu, H and Gossard, AC and Reilly, DJ},
	journal={Physical review letters},
	volume={110},
	number={4},
	pages={046805},
	year={2013},
	publisher={APS}
}

@article{padamsee2014superconducting,
	title={Superconducting radio-frequency cavities},
	author={Padamsee, Hasan S},
	journal={Annual review of nuclear and particle science},
	volume={64},
	number={1},
	pages={175--196},
	year={2014},
	publisher={Annual Reviews}
}

@article{krupka2006frequency,
	title={Frequency domain complex permittivity measurements at microwave frequencies},
	author={Krupka, Jerzy},
	journal={Measurement Science and Technology},
	volume={17},
	number={6},
	pages={R55},
	year={2006},
	publisher={IOP Publishing}
}

@article{berman1955thermal,
	title={Thermal conduction in artificial sapphire crystals at low temperatures I. Nearly perfect crystals},
	author={Berman, R and Foster, EL and Ziman, John Michael},
	journal={Proceedings of the Royal Society of London. Series A. Mathematical and Physical Sciences},
	volume={231},
	number={1184},
	pages={130--144},
	year={1955},
	publisher={The Royal Society London}
}

@article{apostolidis2024quantum,
	title={Quantum paraelectric varactors for radiofrequency measurements at millikelvin temperatures},
	author={Apostolidis, P and Villis, BJ and Chittock-Wood, JF and Powell, JM and Baumgartner, A and Vesterinen, V and Simbierowicz, S and Hassel, J and Buitelaar, MR},
	journal={Nature Electronics},
	volume={7},
	number={9},
	pages={760--767},
	year={2024},
	publisher={Nature Publishing Group UK London}
}

@article{wilen1988impedance,
	title={Impedance methods for surface state electrons},
	author={Wilen, L and Giannetta, R},
	journal={Journal of low temperature physics},
	volume={72},
	number={5},
	pages={353--369},
	year={1988},
	publisher={Springer}
}

@article{kawakami2021relaxation,
	title={Relaxation of the excited Rydberg states of surface electrons on liquid helium},
	author={Kawakami, Erika and Elarabi, Asem and Konstantinov, Denis},
	journal={Physical Review Letters},
	volume={126},
	number={10},
	pages={106802},
	year={2021},
	publisher={APS}
}

@article{shevchenko2010landau,
	title={Landau--zener--st{\"u}ckelberg interferometry},
	author={Shevchenko, Sergey N and Ashhab, Sahel and Nori, Franco},
	journal={Physics Reports},
	volume={492},
	number={1},
	pages={1--30},
	year={2010},
	publisher={Elsevier}
}

@article{Landau1932,
	title={Zur theorie der energieubertragung. II},
	author={Landau, Lev},
	journal={Physikalische Zeitschrift der Sowjetunion},
	volume={2},
	pages={46},
	year={1932}
}

@ARTICLE{Stuckelberg1932,
	author = "Zener, C.",
	journal = "R. Soc. London, Ser.",
	volume =  A137,
	number =  696,
	year =  1932
}

@article{Zener1932,
	title = {{Non-Adiabatic Crossing of Energy Levels}},
	year = {1932},
	journal = {Proceedings of the Royal Society A: Mathematical, Physical and Engineering Sciences},
	author = {Zener, C.},
	number = {833},
	month = {9},
	pages = {696--702},
	volume = {137},
	publisher = {The Royal Society},
	url = {http://rspa.royalsocietypublishing.org/cgi/doi/10.1098/rspa.1932.0165},
	doi = {10.1098/rspa.1932.0165},
	issn = {1364-5021}
}

@article{monarkha2006decay,
	title={Decay rate of the excited surface electron states on liquid helium},
	author={Monarkha, Yu P and Sokolov, SS},
	journal={Low Temperature Physics},
	volume={32},
	number={10},
	pages={970--972},
	year={2006},
	publisher={AIP Publishing}
}

@article{aassime2001radio,
	title={Radio-frequency single-electron transistor: Toward the shot-noise limit},
	author={Aassime, A and Gunnarsson, D and Bladh, K and Delsing, P and Schoelkopf, R},
	journal={Applied physics letters},
	volume={79},
	number={24},
	pages={4031--4033},
	year={2001},
	publisher={American Institute of Physics}
}

@article{gonzalez2015probing,
	title={Probing the limits of gate-based charge sensing},
	author={Gonzalez-Zalba, MF and Barraud, S and Ferguson, AJ and Betz, AC},
	journal={Nature communications},
	volume={6},
	number={1},
	pages={6084},
	year={2015},
	publisher={Nature Publishing Group UK London}
}

@article{oakes2023fast,
	title={Fast high-fidelity single-shot readout of spins in silicon using a single-electron box},
	author={Oakes, GA and Ciriano-Tejel, VN and Wise, DF and Fogarty, MA and Lundberg, T and Lain{\'e}, C and Schaal, S and Martins, F and Ibberson, DJ and Hutin, L and others},
	journal={Physical Review X},
	volume={13},
	number={1},
	pages={011023},
	year={2023},
	publisher={APS}
}

@article{iye1980mobility,
	title={Mobility of electrons in the surface state of liquid helium},
	author={Iye, Yasuhiro},
	journal={Journal of Low Temperature Physics},
	volume={40},
	number={5},
	pages={441--451},
	year={1980},
	publisher={Springer}
}

@article{mehrotra1987analysis,
	title={Analysis of the Sommer technique for measurement of the mobility for charges in two dimensions},
	author={Mehrotra, Ravi and Dahm, AJ},
	journal={Journal of low temperature physics},
	volume={67},
	number={1},
	pages={115--121},
	year={1987},
	publisher={Springer}
}

@article{smith1985direct,
	title={Direct measurement of the density of states of a two-dimensional electron gas},
	author={Smith, TP and Goldberg, BB and Stiles, PJ and Heiblum, M},
	journal={Physical Review B},
	volume={32},
	number={4},
	pages={2696},
	year={1985},
	publisher={APS}
}

@article{luryi1988quantum,
	title={Quantum capacitance devices},
	author={Luryi, Serge},
	journal={Applied Physics Letters},
	volume={52},
	number={6},
	pages={501--503},
	year={1988},
	publisher={American Institute of Physics}
}

@article{duty2005observation,
	title={Observation of quantum capacitance in the Cooper-pair transistor},
	author={Duty, Tim and Johansson, Goeran and Bladh, Kevin and Gunnarsson, David and Wilson, Christopher and Delsing, Per},
	journal={Physical review letters},
	volume={95},
	number={20},
	pages={206807},
	year={2005},
	publisher={APS}
}

@article{mizuta2017quantum,
	title={Quantum and tunneling capacitance in charge and spin qubits},
	author={Mizuta, Ryo and Otxoa, RM and Betz, AC and Gonzalez-Zalba, M Fernando},
	journal={Physical Review B},
	volume={95},
	number={4},
	pages={045414},
	year={2017},
	publisher={APS}
}

@article{lee2022208,
	title={A 208-mhz, 0.75-mw self-calibrated reference frequency quadrupler for a 2-ghz fractional-n ring-pll in 4-nm finfet cmos},
	author={Lee, Kyungmin and Jung, Jaehong and Kim, Seungjin and Oh, Seunghyun and Lee, Jongwoo and Park, Sung Min},
	journal={IEEE Transactions on Circuits and Systems II: Express Briefs},
	volume={70},
	number={8},
	pages={2719--2723},
	year={2022},
	publisher={IEEE}
}

@article{hanson2007spins,
	title={Spins in few-electron quantum dots},
	author={Hanson, Ronald and Kouwenhoven, Leo P and Petta, Jason R and Tarucha, Seigo and Vandersypen, Lieven MK},
	journal={Reviews of modern physics},
	volume={79},
	number={4},
	pages={1217--1265},
	year={2007},
	publisher={APS}
}

@article{zwanenburg2013silicon,
	title={Silicon quantum electronics},
	author={Zwanenburg, Floris A and Dzurak, Andrew S and Morello, Andrea and Simmons, Michelle Y and Hollenberg, Lloyd CL and Klimeck, Gerhard and Rogge, Sven and Coppersmith, Susan N and Eriksson, Mark A},
	journal={Reviews of modern physics},
	volume={85},
	number={3},
	pages={961--1019},
	year={2013},
	publisher={APS}
}

@article{urdampilleta2019gate,
	title={Gate-based high fidelity spin readout in a CMOS device},
	author={Urdampilleta, Matias and Niegemann, David J and Chanrion, Emmanuel and Jadot, Baptiste and Spence, Cameron and Mortemousque, Pierre-Andr{\'e} and B{\"a}uerle, Christopher and Hutin, Louis and Bertrand, Benoit and Barraud, Sylvain and others},
	journal={Nature nanotechnology},
	volume={14},
	number={8},
	pages={737--741},
	year={2019},
	publisher={Nature Publishing Group UK London}
}

@article{hornibrook2014frequency,
	title={Frequency multiplexing for readout of spin qubits},
	author={Hornibrook, JM and Colless, JI and Mahoney, AC and Croot, XG and Blanvillain, S and Lu, H and Gossard, AC and Reilly, DJ},
	journal={Applied Physics Letters},
	volume={104},
	number={10},
	year={2014},
	publisher={AIP Publishing}
}

@article{Braga2024-bx,
	title = {Design of a Low-Jitter 10 GHz PLL for a 12-bit 10-GSPS Cryogenic ADC for Quantum Readout in 22FDX},
	author = {Braga, Davide and Das, Kushal and Dick, Neil and Dyer, Ken and England, Troy and Fahim, Farah and Holm, Scott and Lu, Ping and Moini, Alireza and Nolet, Frederic and Rubinov, Paul and Sanderson, Len and Thompson, Barry and Wang, Xiaoran},
	abstractNote = {},
	doi = {10.2172/2305629},
	url = {https://www.osti.gov/biblio/2305629}, journal = {},
	number = {},
	volume = {},
	place = {United States},
	year = {2024},
	month = {2}
}

@article{fowler2012surface,
	title={Surface codes: Towards practical large-scale quantum computation},
	author={Fowler, Austin G and Mariantoni, Matteo and Martinis, John M and Cleland, Andrew N},
	journal={Physical Review A—Atomic, Molecular, and Optical Physics},
	volume={86},
	number={3},
	pages={032324},
	year={2012},
	publisher={APS}
}

@article{van1981modeling,
	title={Modeling of tunnel diode oscillators},
	author={Van Degrift, Craig T and Love, David P},
	journal={Review of Scientific Instruments},
	volume={52},
	number={5},
	pages={712--723},
	year={1981},
	publisher={American Institute of Physics}
}

@article{clover1970magnetic,
	title={Magnetic susceptibility measurements with a tunnel diode oscillator},
	author={Clover, RB and Wolf, WP},
	journal={Review of Scientific Instruments},
	volume={41},
	number={5},
	pages={617--621},
	year={1970},
	publisher={AIP Publishing}
}

@article{hashimoto2010evidence,
	title={Evidence for superconducting gap nodes in the zone-centered hole bands of KFe 2 As 2 from magnetic penetration-depth measurements},
	author={Hashimoto, K and Serafin, A and Tonegawa, S and Katsumata, R and Okazaki, R and Saito, T and Fukazawa, H and Kohori, Y and Kihou, K and Lee, CH and others},
	journal={Physical Review B—Condensed Matter and Materials Physics},
	volume={82},
	number={1},
	pages={014526},
	year={2010},
	publisher={APS}
}

@article{fletcher2009evidence,
	title={Evidence for a nodal-line superconducting state in LaFePO},
	author={Fletcher, JD and Serafin, A and Malone, L and Analytis, JG and Chu, J-H and Erickson, AS and Fisher, IR and Carrington, A},
	journal={Physical review letters},
	volume={102},
	number={14},
	pages={147001},
	year={2009},
	publisher={APS}
}

@article{mikheev1989crystallization,
	title={Crystallization thermometer with tunnel-diode oscillator for measurement of ultralow temperatures},
	author={Mikheev, VA and Movsesyan, GD and Babayan, KZ and Mina, RT and Maidanov, VA and Mikhin, NP and Chagovets, VK and Sheshin, GA},
	journal={Instruments and Experimental Techniques},
	volume={32},
	number={1 pt 2},
	pages={259--261},
	year={1989},
	publisher={Pleiades Publishing, Ltd.(Плеадес Паблишинг, Лтд)}
}

@article{al2007features,
	title={Features of a tunnel diode oscillator at different temperatures},
	author={Al-Harthi, S and Sellai, A},
	journal={Microelectronics journal},
	volume={38},
	number={8-9},
	pages={817--822},
	year={2007},
	publisher={Elsevier}
}

@article{srikanth1999radio,
	title={Radio-frequency impedance measurements using a tunnel-diode oscillator technique},
	author={Srikanth, H and Wiggins, J and Rees, H},
	journal={Review of scientific instruments},
	volume={70},
	number={7},
	pages={3097--3101},
	year={1999},
	publisher={American Institute of Physics}
}

@article{verduijn2014radio,
	title={Radio-frequency dispersive detection of donor atoms in a field-effect transistor},
	author={Verduijn, J and Vinet, M and Rogge, S},
	journal={Applied Physics Letters},
	volume={104},
	number={10},
	year={2014},
	publisher={AIP Publishing}
}

@article{van1974sensitive,
	title={A sensitive displacement transducer using an extremely reentrant 84 MHz cavity oscillator},
	author={Van Degrift, Craig T},
	journal={Review of Scientific Instruments},
	volume={45},
	number={9},
	pages={1171--1172},
	year={1974},
	publisher={American Institute of Physics}
}

@article{kawakami2016gate,
	title={Gate fidelity and coherence of an electron spin in an Si/SiGe quantum dot with micromagnet},
	author={Kawakami, Erika and Jullien, Thibaut and Scarlino, Pasquale and Ward, Daniel R and Savage, Donald E and Lagally, Max G and Dobrovitski, Viatcheslav V and Friesen, Mark and Coppersmith, Susan N and Eriksson, Mark A and others},
	journal={Proceedings of the National Academy of Sciences},
	volume={113},
	number={42},
	pages={11738--11743},
	year={2016},
	publisher={National Academy of Sciences}
}

@article{kha2015micromagnets,
	title={Do micromagnets expose spin qubits to charge and Johnson noise?},
	author={Kha, Allen and Joynt, Robert and Culcer, Dimitrie},
	journal={Applied Physics Letters},
	volume={107},
	number={17},
	year={2015},
	publisher={AIP Publishing}
}

@article{muller2022magnetic,
	title={Magnetic field robust high quality factor NbTiN superconducting microwave resonators},
	author={M{\"u}ller, Manuel and Luschmann, Thomas and Faltermeier, Andreas and Weichselbaumer, Stefan and Koch, Leon and Huber, Gerhard BP and Schumacher, Hans Werner and Ubbelohde, Niels and Reifert, David and Scheller, Thomas and others},
	journal={Materials for Quantum Technology},
	volume={2},
	number={1},
	pages={015002},
	year={2022},
	publisher={IOP Publishing}
}

@article{kim2025field,
	title={Field-resilient superconducting coplanar waveguide resonators made of Nb, NbTi, and NbTiN},
	author={Kim, Bongkeon and Yu, Chang Geun and Mal, Priyanath and Doh, Yong-Joo},
	journal={arXiv preprint arXiv:2509.20782},
	year={2025}
}

@article{Petersson2012,
    title = {{Circuit quantum electrodynamics with a spin qubit.}},
    year = {2012},
    journal = {Nature},
    author = {Petersson, K D and McFaul, L W and Schroer, M D and Jung, M and Taylor, J M and Houck, A A and Petta, J R},
    number = {7420},
    month = {10},
    pages = {380--3},
    volume = {490},
    publisher = {Nature Publishing Group},
    url = {http://www.nature.com/nature/journal/v490/n7420/full/nature11559.html?WT.ec_id=NATURE-20121018},
    doi = {10.1038/nature11559},
    issn = {1476-4687},
    pmid = {23075988},
    }

@article{Ono2002,
    title = {{Current rectification by Pauli exclusion in a weakly coupled double quantum dot system.}},
    year = {2002},
    journal = {Science},
    author = {Ono, K and Austing, D G and Tokura, Y and Tarucha, S},
    number = {5585},
    month = {8},
    pages = {1313--7},
    volume = {297},
    publisher = {American Association for the Advancement of Science},
    url = {http://science.sciencemag.org/content/297/5585/1313.abstract},
    doi = {10.1126/science.1070958},
    issn = {1095-9203},
    pmid = {12142438},
    }

@article{Goldman,
    title = {{Magnet designs for a crystal-lattice quantum computer}},
    year = {2000},
    journal = {Applied Physics A},
    author = {Goldman, J.R. and Ladd, T.D. and Yamaguchi, F. and Yamamoto, Y.},
    number = {1},
    pages = {11--17},
    volume = {71},
    publisher = {Springer-Verlag},
    url = {http://link.springer.com/article/10.1007/PL00021084},
    doi = {10.1007/PL00021084},
    issn = {1432-0630},
    }

@article{Sommer1971,
    title = {{Mobility of Electrons on the Surface of Liquid     He     4}},
    year = {1971},
    journal = {Phys. Rev. Lett.},
    author = {Sommer, W. T. and Tanner, David J.},
    number = {20},
    month = {11},
    pages = {1345--1349},
    volume = {27},
    publisher = {American Physical Society},
    url = {http://link.aps.org/doi/10.1103/PhysRevLett.27.1345},
    doi = {10.1103/PhysRevLett.27.1345},
    issn = {0031-9007}
}

@article{Lea2000,
    title = {{Could we Quantum Compute with Electrons on Helium?}},
    year = {2000},
    journal = {Fortschr. Phys.},
    author = {Lea, M.J. and Frayne, P.G. and Mukharsky, Yu.},
    number = {9-11},
    month = {9},
    pages = {1109--1124},
    volume = {48},
    publisher = {John Wiley {\&} Sons, Ltd},
    doi = {10.1002/1521-3978(200009)48:9/11<1109::AID-PROP1109>3.0.CO;2-I},
    issn = {00158208}
}

@phdthesis{Lambert1979ElectronsHelium,
    title = {{Electrons on the Surface of Liquid Helium}},
    year = {1979},
    booktitle = {University of California, Berkeley},
    author = {Lambert, David Kay},
    url = {https://books.google.co.jp/books/about/Electrons_on_the_Surface_of_Liquid_Heliu.html?id=HIBJAQAAMAAJ&redir_esc=y},
    school = {University of California, Berkeley}
}

@book{Andrei1997Two-DimensionalSubstrates,
    title = {Two-Dimensional Electron Systems : on Helium and other Cryogenic Substrates},
    year = {1997},
    author = {Andrei, Eva Y.},
    publisher = {Springer Netherlands},
    isbn = {9401512868}
}

@BOOK{Monarkha2004-un,
  title = "Two-dimensional coulomb liquids and solids",
  author = "Monarkha, Yuriy and Kono, Kimitoshi",
  publisher = "Springer",
  address = "Berlin, Germany",
  series = "Springer Series in Solid-State Sciences",
  month =  dec,
  year =  2004
}

@article{Monarkha2007-el,
title = "Decay rate of the excited states of surface electrons over liquid helium",
journal = "J. Low Temp. Phys.",
author = "Monarkha, Yu P and Sokolov, S S",
volume =  148,
pages = "157-161",
month =  aug,
year =  2007
}

@ARTICLE{Brenning2006-wc,
title = "An ultrasensitive radio-frequency single-electron transistor working up to 4.2 K",
author = "Brenning, Henrik and Kafanov, Sergey and Duty, Tim and Kubatkin, Sergey and Delsing, Per",
journal = "J. Appl. Phys.",
publisher = "American Institute of Physics",
volume =  100,
number =  11,
pages = "114321",
month =  dec,
year =  2006
}

@ARTICLE{Grimes1974-fg,
title = "Direct Spectroscopic Observation of Electrons in Image-Potential States Outside Liquid Helium",
author = "Grimes, C C and Brown, T R",
journal = "Phys. Rev. Lett.",
publisher = "American Physical Society",
volume =  32,
number =  6,
pages = "280-283",
month =  feb,
year =  1974
}

@ARTICLE{Harvey-Collard2022-jh,
title = "Coherent Spin-Spin Coupling Mediated by Virtual Microwave Photons",
author = "Harvey-Collard, Patrick and Dijkema, Jurgen and Zheng, Guoji and Sammak, Amir and Scappucci, Giordano and Vandersypen, Lieven M K",
journal = "Phys. Rev. X",
publisher = "American Physical Society",
volume =  12,
number =  2,
pages = "021026",
month =  may,
year =  2022
}

@article{Kanai2024-bo,
  title = {Single-Electron Qubits Based on Quantum Ring States on Solid Neon Surface},
  author = {Kanai, Toshiaki and Jin, Dafei and Guo, Wei},
  journal = {Phys. Rev. Lett.},
  volume = {132},
  issue = {25},
  pages = {250603},
  numpages = {6},
  year = {2024},
  month = {Jun},
  publisher = {American Physical Society},
  doi = {10.1103/PhysRevLett.132.250603},
  url = {https://link.aps.org/doi/10.1103/PhysRevLett.132.250603}
}

@article{Kamigaito2020,
  author        = {Kamigaito, Osamu},
  title         = {Circuit-model formulas for external-\(Q\) factor of resonant cavities with capacitive and inductive coupling},
  journal       = {arXiv preprint arXiv:2005.05843},
  year          = {2020},
  month         = may,
  archivePrefix = {arXiv},
  primaryClass  = {physics.acc-ph}
}

@ARTICLE{Xue2023-cj,
  title = "Cryo-CMOS modeling and a 600 MHz cryogenic clock generator for quantum computing applications",
  author = "Xue, Qiwen and Zhang, Yuanke and Wen, Mingjie and Zhai, Xiaohu and Chen, Yuefeng and Lu, Tengteng and Luo, Chao and Guo, Guoping",
  journal = "Chip",
  publisher = "Elsevier BV",
  volume =  2,
  number =  4,
  pages =  100065,
  month =  dec,
  year =  2023 ,
  doi = "10.1016/j.chip.2023.100065"
}

@INPROCEEDINGS{Yang2019-bb,
  author={Yang, Peilin and Guo, Yanshu and Jiang, Hanjun and Wang, Zhihua},
  booktitle={2019 IEEE Asian Solid-State Circuits Conference (A-SSCC)}, 
  title={A 360–456 MHz PLL frequency synthesizer with digitally controlled charge pump leakage calibration}, 
  year={2019},
  volume={},
  number={},
  pages={285-286},
  doi={10.1109/A-SSCC47793.2019.9056900}}

@ARTICLE{Lee2023-mx,
  title = "A 208-MHz, 0.75-mW self-calibrated reference frequency quadrupler for a 2-GHz fractional-N ring-PLL in 4-nm FinFET CMOS",
  author = "Lee, Kyungmin and Jung, Jaehong and Kim, Seungjin and Oh, Seunghyun and Lee, Jongwoo and Park, Sung Min",
  journal = "IEEE Trans. Circuits Syst. II Express Briefs",
  publisher = "Institute of Electrical and Electronics Engineers (IEEE)",
  volume =  70,
  number =  8,
  pages = "2719-2723",
  month =  aug,
  year =  2023,
  doi = "10.1109/TCSII.2022.3217756"
}

@ARTICLE{Burkard2023-ai,
  title = "Semiconductor spin qubits",
  author = "Burkard, Guido and Ladd, Thaddeus D and Pan, Andrew and Nichol, John M and Petta, Jason R",
  journal = "Rev. Mod. Phys.",
  publisher = "American Physical Society",
  volume =  95,
  number =  2,
  pages =  025003,
  month =  jun,
  year =  2023,
  doi = "10.1103/RevModPhys.95.025003"
}

@MISC{Reona1962-hv,
  title     = "Diode type semiconductor device",
  author    = "Reona Ezaki and Yuriko Kurose",
  howpublished = "{US Patent} 3,033,714",
  month     = may,
  year      = 1962,
  note      = "Filed: August 14, 1959. Granted: May 8, 1962"
}

@article{Ball2016-zn,
  author       = {Ball, H. and Oliver, W. and Biercuk, M.},
  title        = {The role of master clock stability in quantum information processing},
  journal      = {npj Quantum Information},
  year         = {2016},
  volume       = {2},
  pages        = {16033},
  doi          = {10.1038/npjqi.2016.33},
  url          = {https://doi.org/10.1038/npjqi.2016.33}
}

@PHDTHESIS{gao2008physics,
  title    = "The physics of superconducting microwave resonators",
  author   = "Gao, Jiansong",
  year     =  2008,
  school   = "California Institute of Technology",
  language = "en"
}

@book{Pozar1998,
	author = {Pozar, David M.},
	title = {Microwave engineering},
	publisher = {John Wiley \& Sons Inc.,},
	year = {1998},
	address = {New York},
	edition = {4th ed.}
}

@ARTICLE{Bottcher2022-dp,
  title = "Parametric longitudinal coupling between a high-impedance superconducting resonator and a semiconductor quantum dot singlet-triplet spin qubit",
  author = "Bøttcher, C G L and Harvey, S P and Fallahi, S and Gardner, G C and Manfra, M J and Vool, U and Bartlett, S D and Yacoby, A",
  journal = "Nat. Commun.",
  publisher = "Springer Science and Business Media LLC",
  volume =  13,
  number =  1,
  pages =  4773,
  month =  aug,
  year =  2022}

@ARTICLE{Warren2019-mn,
  title = "Long-distance entangling gates between quantum dot spins mediated by a superconducting resonator",
  author = "Warren, Ada and Barnes, Edwin and Economou, Sophia E",
  journal = "Phys. Rev. B.",
  publisher = "American Physical Society (APS)",
  volume =  100,
  number =  16,
  month =  oct,
  year =  2019
}

@ARTICLE{Borjans2020-do,
  title = "Resonant microwave-mediated interactions between distant electron spins",
  author = "Borjans, F and Croot, X G and Mi, X and Gullans, M J and Petta, J R",
  journal = "Nature",
  publisher = "Springer Science and Business Media LLC",
  volume =  577,
  number =  7789,
  pages = "195-198",
  month =  jan,
  year =  2020
}

@ARTICLE{Stockklauser2017-ao,
  title = "Strong Coupling Cavity QED with Gate-Defined Double Quantum Dots Enabled by a High Impedance Resonator",
  author = "Stockklauser, A and Scarlino, P and Koski, J V and Gasparinetti, S and Andersen, C K and Reichl, C and Wegscheider, W and Ihn, T and Ensslin, K and Wallraff, A",
  journal = "Phys. Rev. X",
  publisher = "American Physical Society",
  volume =  7,
  number =  1,
  pages =  011030,
  month =  "9~" # mar,
  year =  2017
}

@ARTICLE{Mi2017-sp,
  title = "Strong coupling of a single electron in silicon to a microwave photon",
  author = "Mi, X and Cady, J V and Zajac, D M and Deelman, P W and Petta, J R",
  journal = "Science",
  volume =  355,
  number =  6321,
  pages = "156-158",
  month =  "13~" # jan,
  year =  2017
}

@ARTICLE{Landig2018-wt,
  title = "Coherent spin-photon coupling using a resonant exchange qubit",
  author = "Landig, A J and Koski, J V and Scarlino, P and Mendes, U C and Blais, A and Reichl, C and Wegscheider, W and Wallraff, A and Ensslin, K and Ihn, T",
  journal = "Nature",
  publisher = "Nature Publishing Group",
  volume =  560,
  number =  7717,
  pages = "179-184",
  month =  "25~" # aug,
  year =  2018
}

@article{Benito2019-al,
  title = {Optimized cavity-mediated dispersive two-qubit gates between spin qubits},
  author = {Benito, M. and Petta, J. R. and Burkard, Guido},
  journal = {Phys. Rev. B},
  volume = {100},
  issue = {8},
  pages = {081412},
  numpages = {6},
  year = {2019},
  month = {Aug},
  publisher = {American Physical Society},
  doi = {10.1103/PhysRevB.100.081412},
  url = {https://link.aps.org/doi/10.1103/PhysRevB.100.081412}
}

@ARTICLE{Gao2008-ox,
  title = "Experimental evidence for a surface distribution of two-level systems in superconducting lithographed microwave resonators",
  author = "Gao, Jiansong and Daal, Miguel and Vayonakis, Anastasios and Kumar, Shwetank and Zmuidzinas, Jonas and Sadoulet, Bernard and Mazin, Benjamin A and Day, Peter K and Leduc, Henry G",
  journal = "Appl. Phys. Lett.",
  publisher = "AIP Publishing",
  volume =  92,
  number =  15,
  pages =  152505,
  month =  "14~" # apr,
  year =  2008
}

@article{Li2025-em,
  title       = {Noise-Resilient Host for Electron Qubit Operation up to 0.4 K},
  author      = {Li, Xinhao and Wang, Christopher S. and Dizdar, Brennan and Huang, Yizhong and Wen, Yutian and Guo, Wei and Zhang, Xufeng and Han, Xu and Zhou, Xianjing and Jin, Dafei},
  journal     = {arXiv preprint},
  year        = {2025},
  month       = feb,
  url         = {https://arxiv.org/abs/2502.01005},
  eprint      = {2502.01005},
  archivePrefix = {arXiv}
}

@ARTICLE{Benito2019-pi,
  title = "Electric-field control and noise protection of the flopping-mode spin qubit",
  author = "Benito, M and Croot, X and Adelsberger, C and Putz, S and Mi, X and Petta, J R and Burkard, Guido",
  journal = "Phys. Rev. B.",
  publisher = "American Physical Society (APS)",
  volume =  100,
  number =  12,
  pages =  125430,
  month =  "19~" # sep,
  year =  2019
}

@unpublished{jennings_inprep,
  author       = {A. Jennings and others},
  title        = {in preparation},
  note         = {in preparation},
  year         = {2025}
}

@ARTICLE{Gurioli1992-ms,
  title = "Thermal escape of carriers out of GaAs/AlxGa1-xAs quantum-well structures",
  author = "Gurioli, M and Martinez-Pastor, J and Colocci, M and Deparis, C and Chastaingt, B and Massies, J",
  journal = "Phys. Rev. B",
  publisher = "American Physical Society (APS)",
  volume =  46,
  number =  11,
  pages = "6922-6927",
  month =  "15~" # sep,
  year =  1992
}

@ARTICLE{Swank1963-ap,
  title = "Lifetime of the Excited F Center",
  author = "Swank, Robert K and Brown, Frederick C",
  journal = "Phys. Rev.",
  publisher = "American Physical Society (APS)",
  volume =  130,
  number =  1,
  pages = "34-41",
  month =  "1~" # apr,
  year =  1963
}

@ARTICLE{Lane2020-pc,
  title = "Integrating superfluids with superconducting qubit systems",
  author = "Lane, J R and Tan, D and Beysengulov, N R and Nasyedkin, K and Brook, E and Zhang, L and Stefanski, T and Byeon, H and Murch, K W and Pollanen, J",
  journal = "Phys. Rev. A",
  publisher = "American Physical Society",
  volume =  101,
  number =  1,
  pages =  012336,
  month =  "22~" # jan,
  year =  2020
}

@ARTICLE{Johansson2006-wv,
  title = "Readout methods and devices for Josephson-junction-based solid-state qubits",
  author = "Johansson, G and Tornberg, L and Shumeiko, V S and Wendin, G",
  journal = "J. Phys. Condens. Matter",
  publisher = "IOP Publishing",
  volume =  18,
  number =  21,
  pages = "S901",
  month =  "12~" # may,
  year =  2006
}

@ARTICLE{Chorley2012-aj,
  title = "Measuring the complex admittance of a carbon nanotube double quantum dot",
  author = "Chorley, S J and Wabnig, J and Penfold-Fitch, Z V and Petersson, K D and Frake, J and Smith, C G and Buitelaar, M R",
  journal = "Phys. Rev. Lett.",
  publisher = "American Physical Society (APS)",
  volume =  108,
  number =  3,
  pages =  036802,
  month =  "20~" # jan,
  year =  2012
}

@ARTICLE{Betz2015-zp,
  title = "Dispersively detected Pauli spin-blockade in a silicon nanowire field-effect transistor",
  author = "Betz, A C and Wacquez, R and Vinet, M and Jehl, X and Saraiva, A L and Sanquer, M and Ferguson, A J and Gonzalez-Zalba, M F",
  journal = "Nano Lett.",
  publisher = "American Chemical Society (ACS)",
  volume =  15,
  number =  7,
  pages = "4622-4627",
  month =  "8~" # jul,
  year =  2015
}

@article{bruno2015reducing,
  title={Reducing intrinsic loss in superconducting resonators by surface treatment and deep etching of silicon substrates},
  author={Bruno, A and De Lange, G and Asaad, S and Van Der Enden, KL and Langford, NK and DiCarlo, L},
  journal={Applied Physics Letters},
  volume={106},
  number={18},
  year={2015},
  publisher={AIP Publishing}
}

@article{barends2010minimal,
  title={Minimal resonator loss for circuit quantum electrodynamics},
  author={Barends, Rami and Vercruyssen, N and Endo, A and De Visser, PJ and Zijlstra, T and Klapwijk, TM and Diener, P and Yates, SJC and Baselmans, JJA},
  journal={Applied Physics Letters},
  volume={97},
  number={2},
  year={2010},
  publisher={AIP Publishing}
}

@article{li2025electron,
  title={Electron charge coherence on a solid neon surface},
  author={Li, Xinhao and Zou, Shan and Chen, Qianfan and Jin, Dafei},
  journal={arXiv preprint},
  year={2025},
  month = aug,
  url= {https://arxiv.org/abs/2507.20476},
  eprint = {2507.20476},
  archivePrefix = {arXiv}
}

@article{migone1986triple,
  title={Triple-point wetting of neon films},
  author={Migone, AD and Dash, JG and Schick, M and Vilches, OE},
  journal={Physical Review B},
  volume={34},
  number={9},
  pages={6322},
  year={1986},
  publisher={APS}
}

@article{Mi2017-go,
  title={Circuit quantum electrodynamics architecture for gate-defined quantum dots in silicon},
  author={Mi, X. and Cady, J. V. and Zajac, D. M. and Stehlik, J. and Edge, L. F. and Petta, J. R.},
  journal={Applied Physics Letters},
  volume={110},
  number={4},
  pages={043502},
  year={2017},
  publisher={AIP Publishing}
}

@article{Harvey-Collard2020-dt,
  title={On-Chip Microwave Filters for High-Impedance Resonators with Gate-Defined Quantum Dots},
  author={Harvey-Collard, Patrick and Zheng, Guoji and Dijkema, Jurgen and Samkharadze, Nodar and Sammak, Amir and Scappucci, Giordano and Vandersypen, Lieven M. K.},
  journal={Physical Review Applied},
  volume={14},
  number={3},
  pages={034025},
  year={2020},
  publisher={American Physical Society}
}

@ARTICLE{Van_Dijk2019-na,
	title = "Impact of Classical Control Electronics on Qubit Fidelity",
	author = "van Dijk, J P G and Kawakami, E and Schouten, R N and Veldhorst, M and Vandersypen, L M K and Babaie, M and Charbon, E and Sebastiano, F",
	journal = "Phys. Rev. Appl.",
	publisher = "American Physical Society",
	volume = 12,
	number = 4,
	pages = "044054",
	month = oct,
	year = 2019,
	doi = "10.1103/PhysRevApplied.12.044054"
}

\appendix
\chapter{Appendix for the helium experiment}

\section{Details of the circuits}\label{sec:details_circuit}

\begin{figure}[htbp]
	\centerfloat{
		\includegraphics[width=0.6\linewidth]{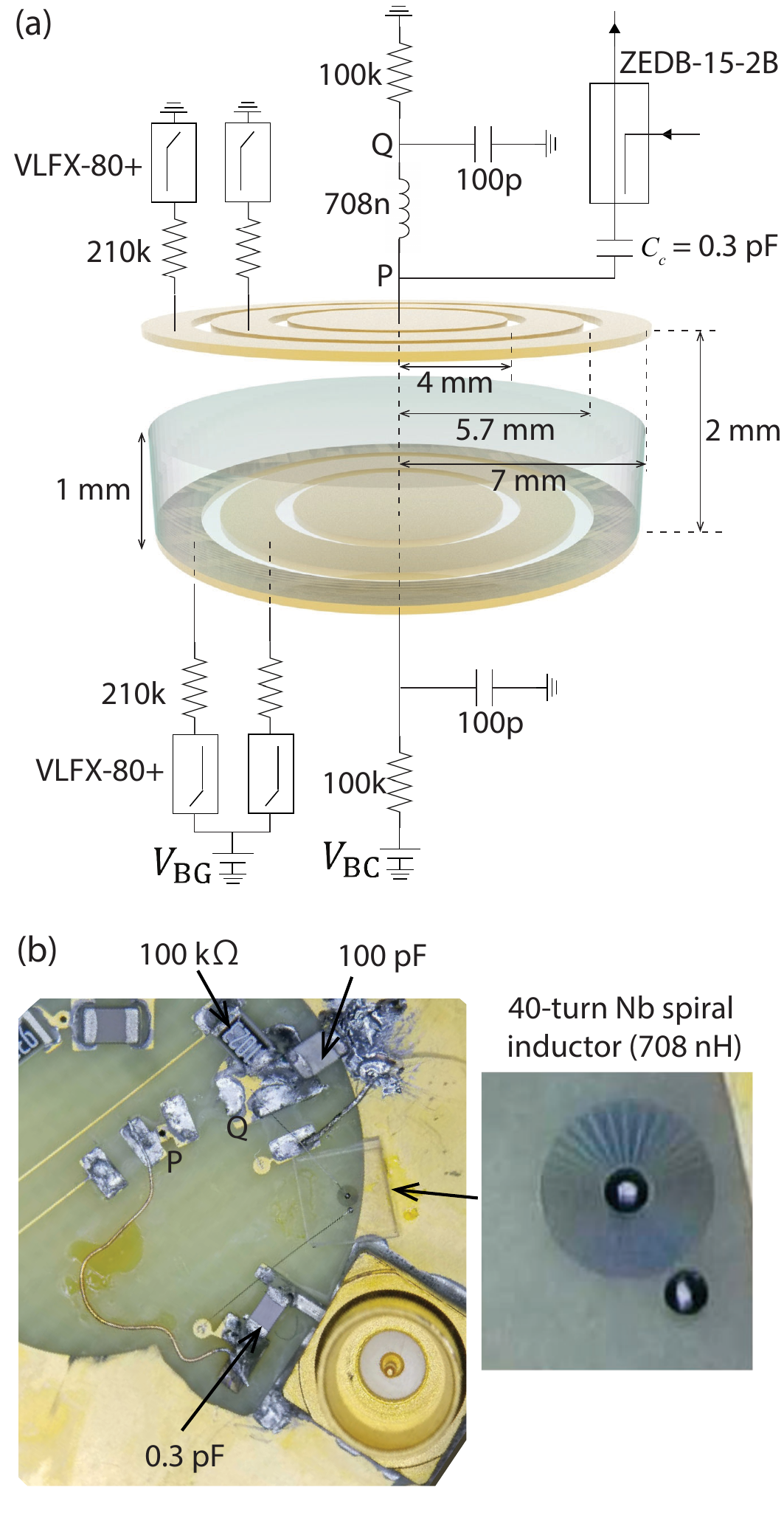}
	}
	\caption[Details of the helium experiment circuits]{%
		(a) The VLFX-80+ low-pass filters from Mini-Circuits attenuate signals above 80 MHz to reduce high-frequency noise and minimize losses in the LC resonant circuit. All electrical components shown are mounted on the top and bottom plates of the Corbino electrodes inside the experimental cell, except for the ZEDB-15-2B directional coupler, which is located outside the cell at the mixing chamber plate.
		(b) Photograph of the top-plate electrode. The inset shows the microfabricated 40-turn Nb spiral inductor on a sapphire substrate. Points $P$ and $Q$ in (a) correspond to the same points as $P$ and $Q$ in (b), respectively.%
	}
	\label{fig:circuit_details}
\end{figure}

The circuit details inside the leak-tight cell, including the RF coupler, are shown in Fig.~\ref{fig:circuit_details}. The incident signal is generated by a signal generator (E8267D, Keysight), passed through cryogenic attenuators totaling \(-20~\mathrm{dB}\) (\(-10~\mathrm{dB}\) at the \(1~\mathrm{K}\) stage and \(-10~\mathrm{dB}\) at the \(100~\mathrm{mK}\) stage), and applied to the coupled port of the RF coupler (Mini-Circuits, ZEDB-15-2B). The resulting signal \( V_\mathrm{RF} \cos(2\pi f_\mathrm{RF} t) \) is delivered to the LC circuit. The reflected signal is returned from the coupler's output port and amplified by a cryogenic amplifier (CITLF-3, Cosmic Microwaves) operating at 4~K and is measured as $V_\mathrm{o}$ with a spectrum analyzer (FSV3030, Rohde \& Schwarz). 

MW signals are generated by a signal generator (SMB100, Rohde \& Schwarz), and the frequency is multiplied by 12 using a frequency multiplier (WR6.5AMC-I, Virginia Diodes Inc.) at room temperature, from which the signal is transmitted to the cryogenic stage via a rectangular waveguide.

\section{Reflectance change}\label{sec:reflectance_change}

A parallel LC circuit weakly coupled to the external environment via a coupling capacitor $C_c$ as shown in Fig.~1(a) has a reflection coefficient given by~\cite{oakes2023fast,Kamigaito2020,blais2021circuit,schuster2007circuit}
\begin{equation}
	\Gamma_\mathrm{ref}=1 - \frac{2Q_\mathrm{tot}/Q_\mathrm{ext}}{1+j2Q_\mathrm{tot} \left(\frac{f}{f_0} -1  \right)},
	\label{eq:Gamma_quality_factor}
\end{equation}
where the internal quality factor is  $Q_\mathrm{int} = \omega_0 C_\mathrm{t} R$, the external quality factor is $Q_\mathrm{ext} = C_\mathrm{t} / (Z_0 \omega_0 C_c^2)$, the total quality factor is then expressed as 
\(
Q_\mathrm{tot} = \omega_0 C_\mathrm{t} \left( \frac{1}{R} + \frac{(\omega_r C_c Z_0)^2}{Z_0} \right)^{-1}.
\) Here, $R$ represents the circuit loss, which is obtained through fitting. The resonance frequency is $f_0=\omega_0/2\pi =1/2\pi \sqrt{LC_\mathrm{t}}$ with $C_\mathrm{t}=C+C_\mathrm{c}$.

When the total capacitance $C_\mathrm{t}$ changes to $C_\mathrm{t}+\Delta C$, the difference in the reflection coefficient at the probe frequency $f_0$ is given by
\begin{equation}
	\begin{aligned}
		\Gamma_\mathrm{ref}(C_\mathrm{t}+\Delta C) -  \Gamma_\mathrm{ref}(C_\mathrm{t}) 
		& \approx- j\frac{4Q_\mathrm{tot}^2 \left( \frac{f_0'}{f_0} -1   \right) / Q_\mathrm{ext}}{1 + \left(2Q_\mathrm{tot} \left( \frac{f_0'}{f_0} -1  \right)\right)^2} \\
		& \approx -j\frac{2Q_\mathrm{tot}^2}{Q_\mathrm{ext}} \frac{\Delta C }{C_\mathrm{t}},
	\end{aligned}
	\label{eq:GammaDiff}
\end{equation}
where the shifted resonance frequency is given by
$
f_0'  = \frac{1}{2 \pi  \sqrt{ L (C_\mathrm{t}+\Delta C)}}.
$

\section{Quantum capacitance}\label{sec:quantum_C}

Near zero-detuning \( \epsilon = h(f_\mathrm{Ry} - f_\mathrm{MW}) \sim 0 \), the eigenstates \( \ket{n_z = 1} \) and \( \ket{n_z = 2} \) transform into \( \ket{-} \) and \( \ket{+} \) (see Fig.~1(c) of the main text). The population of the state \( \ket{-} \) is given by  \( P_-=(1+\chi) /2 \) and that of \( \ket{+} \) \( P_+ = (1-\chi) /2 \), where the difference in population is defined as 
\begin{equation}
	\chi = P_- - P_+=\tanh \left( \frac{\Delta E}{2 k_B T} \right)
\end{equation}
for $\epsilon \sim 0$, where $T$ is the electron temperature and $\Delta E =E_+-E_-= \sqrt{\epsilon^2 + (2t_c)^2}$ is the energy difference between \( \ket{-} \) and \( \ket{+} \). \( 2 t_c \) is the Rydberg transition rate. The change in charge induced on the top central electrode by a single electron, relative to \( \ket{n_z = 1} \), is given by
\begin{equation}
	Q_1 = \Delta q \left(|\braket{n_z=2|-} |^2 P_-+ |\braket{n_z=2|+} |^2 P_+  \right),
\end{equation}
where $\Delta q$ is the induced charge difference between $\ket{n_z = 2}$ and $\ket{n_z = 1}$. Knowing that the probability of finding the states $\ket{\pm}$ in $\ket{n_z = 2}$ is given by  $|\braket{n_z=2|\pm} |^2=\frac{1}{2} \left( 1 \mp \frac{\epsilon}{\Delta E}   \right) $, $Q_1$ can be rewritten as
\begin{equation}
	Q_1(\epsilon)= \Delta q \left( \frac{1}{2} +\chi \frac{\epsilon}{2 \Delta E} \right) .
\end{equation}
Thus, the capacitance change by a single electron is given by
\begin{equation}
	C_1 (\epsilon) = \frac{dQ_1}{dV} = \Delta q \frac{dQ_1}{d\epsilon} = C_\mathrm{quantum} + C_\mathrm{tunnel},
\end{equation}
where we used $d\epsilon = \Delta q dV$. The quantum capacitance is defined as 
$C_\mathrm{quantum} = \chi \Delta q^2 \frac{\partial}{\partial \epsilon} \frac{\epsilon}{2 \Delta E}$, and the tunneling capacitance is given by $C_\mathrm{tunnel} = \Delta q^2 \frac{\epsilon}{2 \Delta E} \frac{\partial \chi}{\partial \epsilon}$. The tunneling capacitance becomes non-negligible when population redistribution processes, such as relaxation, occur at a rate comparable to or faster than the probing frequency \cite{gonzalez2016gate}. In our case, since the probing frequency 120~MHz is much higher than the relaxation rate 1~MHz, the tunneling capacitance can be ignored. Thus, the capacitance induced by a single electron becomes
\begin{equation}
	C_1 (\epsilon) \approx C_\mathrm{quantum} = \chi \Delta q^2\frac{(2t_c)^2}{2\Delta E ^3} .
\end{equation}
The charge induced by many electrons is given by $Q_N = \int Q_1(\epsilon^0 -\epsilon) n(\epsilon) d\epsilon$ with the electron number distribution $n(\epsilon)$. Therefore, the capacitance change by many electrons is given by 
\begin{equation}
	C_N(\epsilon^0) =  \Delta q \frac{dQ_N}{d\epsilon^0}= \int C_1(\epsilon^0 -\epsilon) n(\epsilon) d\epsilon.
\end{equation}
Now, we introduce MW frequency modulation. Due to this modulation, the detuning becomes time-dependent: $\epsilon^0 (t) = h (f_\mathrm{Ry}^0 - f_\mathrm{MW}^c - f_\mathrm{ma} \cos(2\pi f_\mathrm{mf}t))$. For sufficiently small \( f_\mathrm{ma} \) (i.e., below 470~MHz; see Fig.~4), as the quantum capacitance is linearly proportional to $f_\mathrm{ma}$ the following approximation holds:
\begin{align}
	C_N(t) & = C( h (f_\mathrm{Ry}^0 - f_\mathrm{MW}^c - f_\mathrm{ma} \cos(2\pi f_\mathrm{mf}t)) ) \\
	& \approx   C_0 + \delta C \cos(2\pi f_\mathrm{mf}t),
\end{align}
where \( C_0 = C_N( h (f_\mathrm{Ry}^0 - f_\mathrm{MW}^c )) \) and

\begin{equation}
	\delta C = -h f_\mathrm{ma} \frac{dC_N(\epsilon^0)}{d\epsilon} \Big|_{\epsilon^0 = h (f_\mathrm{Ry}^0 - f_\mathrm{MW}^c )}. \label{eq:delta_C}
\end{equation}
Using Eq.~\ref{eq:GammaDiff} and by inserting $\Delta C= C_N(t)$, the reflected signal \( V_\mathrm{o}= \Gamma_\mathrm{ref} (C_\mathrm{t}+C_N(t))V_\mathrm{RF} \cos ( 2\pi f_\mathrm{RF} t) \) can be rewritten as

\begin{equation}
	\begin{aligned}
		V_\mathrm{o} \approx &\ \Gamma_\mathrm{ref} (C_\mathrm{t} )V_\mathrm{RF} \cos ( 2\pi f_\mathrm{RF} t)  \\
		& -j\frac{2Q_\mathrm{tot}^2}{Q_\mathrm{ext}} \frac{C_0 }{C_\mathrm{t}} V_\mathrm{RF}\cos ( 2\pi f_\mathrm{RF} t) \\
		&  -j\frac{2Q_\mathrm{tot}^2}{Q_\mathrm{ext}} \frac{\delta C }{C_\mathrm{t}} V_\mathrm{RF} \cos(2\pi f_\mathrm{mf}t) \cos ( 2\pi f_\mathrm{RF} t).
	\end{aligned}    \label{eq:Vo}
\end{equation}

From Eq.~\ref{eq:delta_C} and Eq.~\ref{eq:Vo}, the signal containing the frequencies \( f_\mathrm{RF} \pm f_\mathrm{mf} \), which corresponds to the sideband peak, is given by

\begin{equation}
	\frac{Q_\mathrm{tot}^2}{Q_\mathrm{ext}} \frac{|\delta C| }{C_\mathrm{t}} V_\mathrm{RF} = \frac{Q_\mathrm{tot}^2}{Q_\mathrm{ext}} \frac{V_\mathrm{RF} }{C_\mathrm{t}}  h f_\mathrm{ma} \left| \frac{dC_N(\epsilon^0)}{d\epsilon} \Big|_{\epsilon^0 = h (f_\mathrm{Ry}^0 - f_\mathrm{MW}^c )} \right|.
	\label{eq:sidepeak}
\end{equation}
Note that the Rydberg transition can also be detected under continuous-wave (CW) microwave irradiation without FM~\cite{jennings_inprep}. In this case, Eq.~\ref{eq:Vo} reduces to a form without the last term. At \( f = f_0 \), the reflection signal is purely real (first term of Eq.~\ref{eq:Vo}), while the contribution from quantum capacitance appears as a small change in the imaginary part (second term). As a result, amplitude changes are subtle and difficult to resolve using amplitude-based measurements, necessitating phase-sensitive techniques such as lock-in detection. In contrast, FM produces sideband peaks at distinct frequencies, allowing changes in quantum capacitance to be detected via amplitude measurements. Furthermore, this approach enables systematic control of capacitance modulation, making it particularly well suited for evaluating sensitivity.

\section{Simulation of Electric Field Distribution and Rydberg transition frequency}\label{sec:Ez_fRy_distribution}

We computationally modeled two dimensional electrons system using a rectangular grid in cylindrical coordinates $(r, z, \theta)$. Owing to the cylindrical symmetry of the system, the $\theta$-dependence is omitted, and the simulation is carried out in the $r$–$z$ plane. 

The simulation domain spans 2~mm in the vertical direction, corresponding to the space between the bottom-plate and top-plate electrodes, and 7.5~mm in the radial direction, corresponding to the radius of the outer electrode of the Corbino electrodes. The vertical ($z$) direction is discretized into 200 segments and the radial ($r$) direction into 500 segments. This results in a grid resolution of $2~\mathrm{mm}/200$ in $z$ and $7~\mathrm{mm}/500$ in $r$.

The relaxation method was used to calculate the Green's function~\cite{wilen1988impedance} of an electron at any radial position at $z = 1$~mm (i.e., at grid index $N = 100$). The electric field generated by a surface electron (SE) located at $(r, z) = (r_M, 1~\mathrm{mm})$ is then calculated, where $r_M = 7.5~\mathrm{mm} \times M/500$.

\begin{figure}[htbp]
	\centerfloat{
		\includegraphics[width=0.6\linewidth]{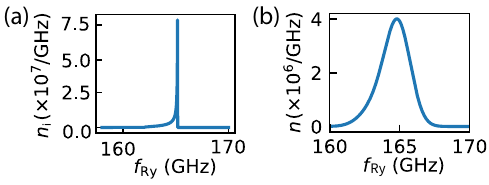}
	}
	\caption[Distribution of the Rydberg transition frequency]{%
		(a) Estimated distribution of the number of electrons $n_i(f_\mathrm{Ry})$ as a function of the Rydberg transition frequency $f_\mathrm{Ry}$, arising from spatial variations of the vertical electric field $E_z$ across the cell. (b) Distribution of the number of electrons $n$ as a function of the Rydberg transition frequency $f_\mathrm{Ry}$.%
	}
	\label{fig:Scewed_f_Ry_distribution}
\end{figure}

The saturated electron density $n_\mathrm{s}$ is determined by adding electrons to the disc until the vertical component of the electric field produced by the SEs exactly cancels the vertical component of the electric field from the electrodes, $E_z$, at a height of $z = 1$~mm. Figure~2(a) shows the numerically calculated $n_\mathrm{s}$ and $E_z$ as a function of $r$.
The Stark shift induced by $E_z$ is evaluated by numerically solving the Schr\"{o}dinger equation. The number of electrons in each radial disc segment is calculated over the entire grid, leading to an estimate of the broadening $n_\mathrm{i}(f_\mathrm{Ry})$ due to electric field variation across the cell (Fig.~\ref{fig:Scewed_f_Ry_distribution}(a)).
This is then convolved with a Gaussian distribution of 1~GHz width to account for inhomogeneous broadening from other sources (e.g., spatial variations in $E_z$ due to a non-horizontal liquid helium surface or distortions in the top and bottom electrodes), resulting in the total electron distribution $n(f_\mathrm{Ry})$, as shown in Figure~\ref{fig:Scewed_f_Ry_distribution}(b).
Finally, a convolution with $C_1$ is applied to obtain $C_N$ using Eq.~4 and Eq.~5, where $2t_c/h = 0.83$~MHz is Rydberg transition rate and $T=160$~mK is the temperature, resulting in Fig.~2(b).

\section{MW power dependence}\label{sec:power_dependence}

\begin{figure}[htbp]
	\centerfloat{
		\includegraphics[width=0.6\linewidth]{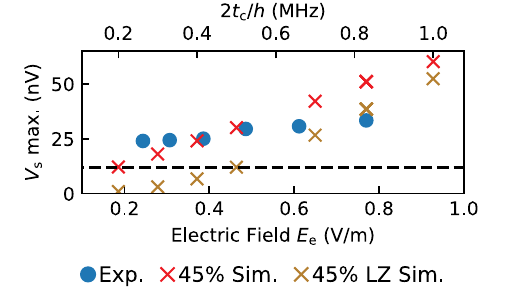}
	}
	\caption[Sideband amplitude versus MW electric field]{%
		Maximum sideband amplitude as a function of the Electric field irradiated to the electrons $E_\mathrm{e}$ (bottom axis) and the corresponding Rydberg transition rate \( 2t_c /h\) (top axis). Blue points (red crosses) represent the experimental (simulation) data. Bronze crosses represent the simulation results including the LZ effects and the dashed black line is the noise level $V_\mathrm{n}=12$~nV. Measurement conditions: \( f_\mathrm{mf} = 1 \)~kHz, \( f_\mathrm{ma} = 768 \)~MHz, and \( V_\mathrm{RF} = 9~\mu\mathrm{V} \).%
	}
	\label{fig:power_dependence}
\end{figure}

Figure~\ref{fig:power_dependence} shows the maximum sideband peak amplitude as a function of the corresponding electric field irradiated on the surface electrons (SEs), obtained by varying the microwave power sent from room temperature. The microwave power was controlled using a room-temperature variable attenuator. For all experiments presented in the main text, the attenuation was fixed at $-30~\mathrm{dB}$, corresponding to the highest power shown in Fig.~\ref{fig:power_dependence}. At lower power, nonadiabatic transitions become more prominent, leading to a reduced signal relative to simulations that do not include LZ effects. However, such a reduction was not observed in our experiments. The discrepancy may arise from imperfections in power delivery, such as microwave reflections reducing the attenuator’s effectiveness or the signal approaching the noise floor at high attenuation levels, making accurate power control difficult.

The applied MW intensity cannot be directly measured at low temperature but a calibration was performed with the attenuator at room temperature. The MW intensity was measured as a function of the MW frequency using a WR-6 waveguide detector on the opposite side of the experimental cell, a distance of $r_\mathrm{d}=31.26$~mm from the MW waveguide. Using the area of the WR-6 waveguide, the intensity at the detector for the highest MW power applied in Fig.\ref{fig:power_dependence} is $I_\mathrm{d} = 0.73$~mW/m$^2$. In the far-field regime, this gives an electric field of $E_\mathrm{d}=\sqrt{2 I_\mathrm{d} Z_\mathrm{vac}}$, where $Z_\mathrm{vac}=377~\Omega$ is the impedance of free space . The distance from the waveguide to the electrons at the center of the disc $r_\mathrm{e}=11.26$~mm, and the electric field at the electrons is then $E_\mathrm{e}=E_\mathrm{d}\times \frac{r_\mathrm{d}}{r_\mathrm{e}} = 2$~V/m. 

We can compare this to the $E_\mathrm{e}$ calculated from the Rydberg transition rate~\cite{platzman1999quantum} as $E_\mathrm{e} =  2t_c/e z_{12}=0.77$~V/m, where $z_{12}=4.6$~nm is the Rydberg transition moment\cite{platzman1999quantum}. The electric field calculated from the Rydberg transition rate being within an order of magnitude of the calibration measurement is reasonable as the calibration was performed at room temperature in air, the MW waveguide is in a closed cell between two sets of electrodes which have a distance less than the Fraunhofer distance, causing several standing wave resonances, and the electrons may not be at the exactly same solid angle as the detector with regard to the waveguide.

Using higher microwave power than that employed in this work is challenging with the present setup due to heating effects. Moreover, stronger microwave fields can induce excitations to higher Rydberg states~\cite{kawakami2021relaxation}, rendering the two-level approximation invalid.

\section{Comparison with Conventional Detection Techniques and Future improvements}\label{sec:appendix_compare}

Here, we compare the RF reflectometry method devel-
oped in this work with two previously demonstrated tech-
niques: detection as a current in Ref.~\cite{kawakami2019image} and as a voltage in Ref.~\cite{kawakami2021relaxation}. The current measurement described in Ref.~\cite{kawakami2021relaxation} operates at a signal frequency of around 250\,kHz. As the frequency increases, the impedance of the cable’s capacitance decreases, causing the signal to leak to ground and making it difficult to extract. Therefore, the measurement frequency must remain low. However, low-frequency measurements typically suffer from high noise levels, and it is generally difficult to find low-noise cryogenic amplifiers operating in this frequency range. Nevertheless, it remains the simplest and most direct approach. By contrast, the voltage measurement scheme in Ref.~\cite{kawakami2021relaxation} exhibits a broadband response that remains flat between 500\,kHz and 100\,MHz~\cite{elarabi2021cryogenic}. This makes the voltage-based approach more suitable for real-time signal detection. In contrast, RF reflectometry is not inherently broadband, but it is better suited for future applications targeting single-electron detection. In the following, we compare the signal amplitude obtained with the voltage-based approach to that of our present RF reflectometry experiments.

For simplicity, we consider the voltage signal induced by a single electron at $\epsilon=0$. Using the RF reflectometry, the quantum capacitance of a single electron is given by Eq.~4. At $\epsilon=0$ and $k_\mathrm{B} T \gg  t_\mathrm{c}$, this reduces to

\begin{equation}
	C_1 = \Delta q^2 \frac{1}{4 k_\mathrm{B} T}.
\end{equation}

Substituting into Eq.~9, the voltage signal for the RF reflectometry scheme becomes:
\begin{equation}
	V_\mathrm{s} = \left( G  Q_\mathrm{int}\frac{\Delta q V_\mathrm{RF}}{16k_B T} \right) \frac{ \Delta q}{C_\mathrm{t}} \label{eq:V_s_1}
\end{equation}
at critical coupling. In contrast, in Ref.~\cite{kawakami2021relaxation}, the signal voltage is simply:
\begin{equation}
	V_\mathrm{s} = \frac{ \Delta q}{C_\mathrm{t}}.
\end{equation}
Note that in Ref.~\cite{kawakami2021relaxation}, the impedance matching circuit includes attenuation and amplification stages that compensate each other, resulting in a net gain close to unity. In our present experimental conditions, the prefactor \(
G  \frac{Q_\mathrm{tot}^2}{Q_\mathrm{ext}} \frac{\Delta q V_\mathrm{RF}}{4k_B T}
\) is approximately 0.01. Therefore, the sensitivity of RF reflectometry is currently lower. 
However, in future implementations aimed at single-electron detection, a different device architecture would be used. Instead of a bulky parallel-plate capacitor, a nanofabricated structure would be employed, allowing the electron to be positioned much closer to a nanofabricated electrode~\cite{kawakami2023blueprint}. As a result, $\Delta q$ can be increased from the present value of $10^{-5}e$ to approximately $10^{-2}e$. Additionally, lowering the temperature from 160\,mK to 10\,mK enhances the prefactor to $\sim 100$, indicating that RF reflectometry will become the superior approach in that regime. Although the device geometry changes, the total capacitance—primarily set by large structures such as bonding pads—is expected to remain comparable. Thus, the LC circuit sensitivity estimated in this work should remain applicable. Note that in conventional semiconductor quantum dot experiments, the typical charge variation is 
\(\Delta q \approx 0.01\text{--}0.89e\)~\cite{Johansson2006-wv,colless2013dispersive,gonzalez2016gate,Betz2015-zp,ibberson2021large,ahmed2018radio} 
depending on the device geometry. The operating temperature is typically around 10~mK, and the relevant capacitance 
\(C_1\) is on the order of a few to hundreds of attofarads~\cite{colless2013dispersive,Chorley2012-aj,gonzalez2015probing,ahmed2018radio}. 
The above-discussed future implementations aimed at single-electron detection using electrons on helium are expected 
to approach the lower end of these \(C_1\) values.

The voltage signal in Eq.~\ref{eq:V_s_1} is proportional to both the resonance frequency \( f_0 \) and  \( R \) since \( Q_\mathrm{int} = \omega_0 C_\mathrm{t} R \) and thus by increasing these factors we can also enhance the voltage signal. In this work's setup, an 80\,MHz cutoff was chosen (Fig.~\ref{fig:circuit_details}) to enable Sommer--Tanner measurements~\cite{Sommer1971} between the middle and outer electrodes for confirming electron deposition. Lower cutoff filters are preferable in future implementations to suppress circuit loss.  In addition, replacing the Corbino electrodes—fabricated on an FR4 substrate—with a low-loss, RF-compatible substrate such as Rogers may further reduce circuit loss and enhance \( R \). Increasing the resonance frequency \(f_0\) requires reducing the inductance \(L\), as the total capacitance \(C_\mathrm{t}\)—which would be dominated by the bonding pads—is difficult to reduce further.  However, the achievable \(f_0\) is ultimately limited by the self-resonance of the inductor, which arises from its parasitic capacitance.  Moving to higher frequencies will eventually necessitate a transition from a lumped-element design to a distributed circuit. This shift prohibits multiplexing and eliminates one of the key advantages of the present method—its potential for compact integration~\cite{jennings2024quantum}.

\chapter{Appendix for the neon experiment}

\section{Simulation for Neon and Electron Deposition}
\label{sec:neon_electron_sim}

\begin{figure}[htbp]
	\centerfloat{
		\includegraphics[width=0.7\linewidth]{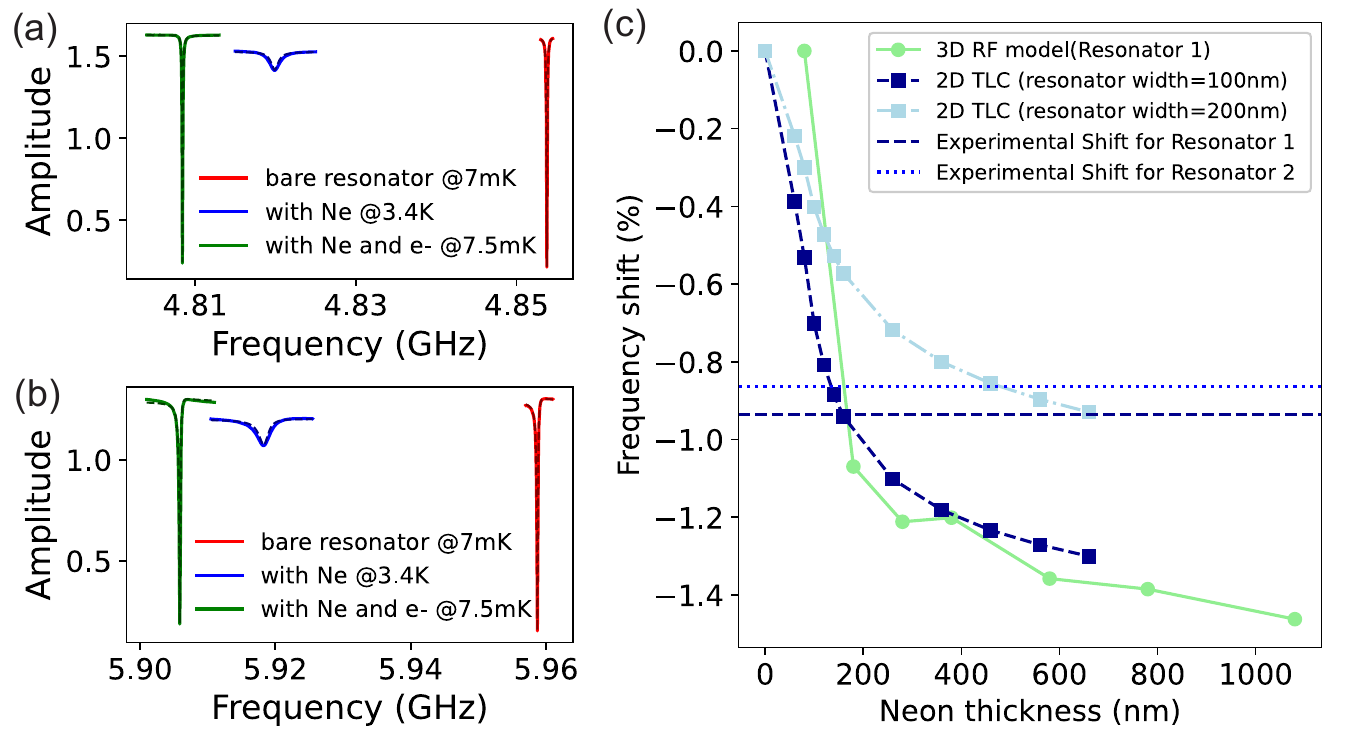}
	}
	\caption[Neon thickness and resonator response]{%
		(a) Resonance peaks measured for the bare Resonator~1 at $7\,\mathrm{mK}$ (red), with neon at $3.4\,\mathrm{K}$ (blue), and with neon and electrons at $7.5\,\mathrm{mK}$ (green). The corresponding internal quality factors ($Q_\mathrm{int}$: $2.34 \times 10^5$, $4.66 \times 10^3$, $2.22 \times 10^5$), external quality factors ($Q_\mathrm{ext}$: $3.19 \times 10^4$, $4.33 \times 10^3$, $3.36 \times 10^4$), and resonance frequencies ($f_\mathrm{r}$: $4.854\,\mathrm{GHz}$, $4.82\,\mathrm{GHz}$, $4.808\,\mathrm{GHz}$) are given for red, blue, and green, respectively. (b) Resonance peaks measured for the bare Resonator~2 at $7\,\mathrm{mK}$ (red), with neon at $3.4\,\mathrm{K}$ (blue), and with neon and electrons at $7.5\,\mathrm{mK}$ (green). The corresponding internal quality factors ($Q_\mathrm{int}$: $1.91 \times 10^5$, $3.9 \times 10^3$, $1.42 \times 10^5$), external quality factors ($Q_\mathrm{ext}$: $2.16 \times 10^4$, $3.49 \times 10^3$, $1.97 \times 10^4$), and resonance frequencies ($f_\mathrm{r}$: $5.959\,\mathrm{GHz}$, $5.918\,\mathrm{GHz}$, $5.906\,\mathrm{GHz}$) are given for red, blue, and green, respectively. (a, b) The dotted lines represent the fits. (c) Resonance frequency shift obtained from the COMSOL simulation as a function of the neon thickness. The vertical dotted and dashed lines represent the maximum measured resonance frequency shifts of Resonators~1 and 2 caused by the neon and electron deposition, $-0.94\%$ and $-0.86\%$, respectively. Resonance frequency shifts for Resonator~1 are simulated using 3D RF model.%
	}
	\label{fig:neon_deposition}
\end{figure}

Figure~\ref{fig:neon_deposition}(a,b) represents resonant peaks measured for Resonators~1 and 2 with and without neon and electrons at $3.1\,\text{K}$ and around $7\,\text{mK}$, and Fig.~\ref{fig:neon_deposition}(c) shows the resonance frequency shifts calculated as a function of neon thickness using COMSOL simulations. The neon thickness is defined from the bottom of the etched Si substrate. For the “3D RF model” described in Appendix~\ref{sec:3D_RF_COMSOL}, since the resonator is defined on a 2D plane located 80\,nm above the bottom of the etched Si substrate,
the graph starts from a neon thickness of $80\,\mathrm{nm}$.

The resonance frequency shift measured around $7\,\mathrm{mK}$ can be attributed to both the effects of neon and electron deposition. By comparing the experimental and simulation results, we can thus set an upper bound for the neon thickness. 
To obtain a reliable estimate of the thickness, we employed and compared two independent simulation methods: a 3D electromagnetic simulation using COMSOL (Appendix.~\ref{sec:3D_RF_COMSOL}) and a simplified 2D transmission line calculator (TLC, Appendix~\ref{sec:TLC}). The 3D model captures detailed geometric and field distribution effects, while the TLC approach provides fast, intuitive access to the dependence on dielectric loading. Consistent results from both methods increase our confidence in the estimated neon thickness.

Thus, the neon thickness deposited on Resonator~1 was estimated to be $160\,\text{nm}$ in both models. For Resonator~2, the resonance frequency is closer to the simulation result for a width of $150\,\text{nm}$ than to that for $200\,\text{nm}$ (Table~\ref{tab:reso_table_neon}). Therefore, using the simulation result from the TLC (Appendix~\ref{sec:TLC}) for a width of $150\,\text{nm}$, the neon thickness was estimated to be $270\,\text{nm}$.

\subsection{ 3D RF model}\label{sec:3D_RF_COMSOL}

The simulation was performed using COMSOL Multiphysics v.~6.2. We began the simulation by creating a 3D model with the “Radio Frequency” module loaded for the physics setup, within which the “Electromagnetic Waves, Frequency Domain (emw)” interface is used. The study type was set to Frequency Domain and MUMPS direct solver was used to compute the response of the system subjected to harmonic excitation over a range of frequencies.

We use a 3D model with dimensions of \( 2000\,\mathrm{\mu m} \times 2000\,\mathrm{\mu m} \times 800\,\mathrm{\mu m} \), where the center of the resonator is roughly positioned at the origin. For simulating the resonator response in the absence of neon, the region \( 301.3 < z < 800\,\mathrm{\mu m} \) is defined as vacuum, and the region \( 0 < z < 300\,\mathrm{\mu m} \) is defined as silicon (the relative permittivity of silicon is set to \( 11.4 \)). Thirteen layers with thickness of $100\,\text{nm}$ were added from \( 300 < z < 301.3\,\mathrm{\mu m} \) each of which could be set as neon ($\epsilon_r=1.244$) or vacuum to simulate a neon layer of various thicknesses.

We set a 2D plane at \( z=300\,\mathrm{\mu m} \), where the resonators, the feedline, and the grounds are defined. The transmission of the electromagnetic field is solved using lumped ports 1 and 2, which are defined between each end of the feedline and the ground plane, each with a characteristic impedance of 50\,\(\Omega\). The transmission parameter \( S_{21} \) is then plotted as a function of the frequency \( f \) of the microwave signal sent through the feedline.

We account for the kinetic inductance of NbTiN by modeling it as a complex conductivity within the “Transition Boundary Condition” applied to the resonator. According to Ref.~\cite{gao2008physics}, the current density \( \vec{J} \) can be expressed as
\[
\vec{J} =  \sigma \vec{E},
\]
where \( \vec{E} \) is the electric field, and the conductivity \( \sigma \) for a thin film is given by
\begin{equation}
	\sigma = \frac{1}{i \mu_0 \omega \lambda^2}, \label{eq:sigma}
\end{equation}
where \( \omega = 2\pi f \), \( \mu_0 \) is the magnetic permeability, and \( \lambda \) is the magnetic penetration depth. In the COMSOL simulation, the electrical conductivity for the “Transition Boundary Condition” of the resonator is set according to Eq.~\ref{eq:sigma}. When the sample thickness is comparable to or smaller than the skin depth, as is the case here, we deselect the “Electrically thick layer” option and specify a layer thickness of \( D = 20\,\mathrm{nm} \).

The feedline and grounds are modeled as “Perfect Electric Conductor,” even though they are also made of NbTiN, because their kinetic inductance is expected to have a negligible impact on the resonator response for this geometry. The quality factor of the resonator computed in this COMSOL simulation is determined solely by the external quality factor, as we treat silicon as a lossless dielectric in the simulation.

Table~\ref{tab:reso_table_neon} shows the comparison between the measured data and the COMSOL simulation with \( \lambda = 390\,\mathrm{nm} \). Compared to Resonator~1, Resonators~2 and~3 exhibit deviations from the simulation results based on the intended width and gap dimensions. This is believed to be due to the width of Resonators~2 and 3 being smaller than intended, resulting in lower frequencies compared to the simulation.

\subsection{``Transmission Line Calculator''  COMSOL model }\label{sec:TLC}

We modeled the nanowire resonator as a coplanar waveguide using the ``Transmission Line Calculator'' COMSOL model. Although the actual resonator is not a straight nanowire, in this model, it is approximated as a transmission line aligned with the $x'$-axis (designating this extending direction as the $x'$-axis). Electromagnetic fields, specifically transverse electromagnetic (TEM) waves, propagate along the $x'$-axis. Here, we also assume that $x'$ components of electric and magnetic fields are small and the propagating mode is deduced
from separate magnetic and electric analyses.

The width of the resonator is defined along the $y'$-axis as 150\,nm and the thickness along the $z$ axis as 20\,nm, and it extends indefinitely along the $x'$-axis. In this model, we define metal, silicon, neon, and electrons in the $y'$-\(z\) plane. 

\subsubsection{Simulation of neon thickness}\label{sec:simu_Ne_TLC}

To estimate the neon thickness, the neon thickness is considered in the \(z\)-direction, growing sequentially. We obtained the shunt capacitance per unit length, denoted as \( C_\mathrm{l} \). The frequency shift due to the presence of neon was then calculated using the formula:
\(
\Delta f = \sqrt{\frac{C_\mathrm{l}^\mathrm{w/o \, Ne}}{C_\mathrm{l}^\mathrm{w/ \, Ne}}} - 1
\), where \( C_\mathrm{l}^\mathrm{w/o \, Ne} \) and \( C_\mathrm{l}^\mathrm{w/ \, Ne} \) represent the shunt capacitance per unit length without and with neon, respectively.

\subsubsection{Simulation with the presence of electrons: Drude model }\label{sec:simu_with_e}

To incorporate the effect of electrons, we first use the Drude model~\cite{Andrei1997Two-DimensionalSubstrates} based on the same 2D framework described above. The neon thickness is \(270\,\mathrm{nm}\), and the surface electrons are assumed to be located \(2.5\,\mathrm{nm}\) above the neon surface, spreading along the \(x'\)-\(y'\) plane. The equation of motion for an electron is given by
\begin{equation}
	m_e \left(\frac{\partial \mathbf{v}}{\partial t} + \frac{\mathbf{v}}{\tau} \right) = -e \mathbf{E}, 
	\label{eq:eq_motion_0}
\end{equation}
where \(\mathbf{E} = \mathbf{E}_0 \exp(i \omega t)\) is the electric field acting on the electron~\footnote{Note that we use the convention \(\exp(i \omega t)\) for the time dependence of fields. While the physics community commonly adopts \(\exp(-i \omega t)\), the engineering convention \(\exp(i \omega t)\) is used here for consistency with COMSOL.}, and \(\mathbf{v}\) is the electron velocity. Here, \(\tau\) is the scattering time, and \(m_e\) is the electron mass. The solution to Eq.~\ref{eq:eq_motion_0} is
\begin{equation}
	\mathbf{v} = -\frac{e \tau}{m_e} \frac{1}{1 + i \omega \tau} \mathbf{E}_0 \exp(i \omega t).
\end{equation}
The corresponding 2D current density is \(\mathbf{j} = -e n_e \mathbf{v} = \sigma^{\mathrm{2D}} \mathbf{E}_0 \exp(i \omega t)\), leading to the well-known Drude model:
\begin{equation}
	\sigma^{\mathrm{2D}} = \frac{e^2 n_e \tau}{m_e} \frac{1}{1 + i \omega \tau}.
	\label{eq:Drude_conductivity}
\end{equation}
To integrate this into the COMSOL model, the electron conductivity is defined as
\begin{equation}
	\sigma_{ij} = \frac{1}{\delta z} \frac{e^2 n_e \tau}{m_e}  
	\frac{1}{1 + i \omega \tau}
\end{equation}
for \(i = j = x'\) or \(i = j = y'\), and \(\sigma_{ij} = 0\) otherwise, where $\delta z = 1\,\text{nm}$ is the arbitrarily defined thickness of the area considered as the electron layer. To check the validity of this choice, simulations were also conducted with $\delta z = 2.5\,\text{nm}$, and it was confirmed that the differences in the results were negligibly small. Please note that, as the $x'$ component of the electric and magnetic fields is small, $\sigma_{y'y'}$ mainly contributes to the resonance peak change. We obtained the shunt capacitance per unit length and calculated the frequency shift as 
\(
\Delta f = \sqrt{\frac{C_\mathrm{l}^\mathrm{w/o \, e^-}}{C_\mathrm{l}^\mathrm{w/ \, e^-}}} - 1
\), where \( C_\mathrm{l}^\mathrm{w/o \, e^-} \) and \( C_\mathrm{l}^\mathrm{w/ \, e^-} \) represent the shunt capacitance per unit length without and with electrons, respectively. The attenuation constant \(\alpha\) was extracted from the complex propagation constant \(\gamma = \alpha + i \beta\), computed in the simulation. The internal quality factor due to the electrons is then obtained via~\cite{Pozar1998}
\begin{equation}
	Q_e=\frac{\pi}{2 \alpha l}.
\end{equation}

\begin{figure}[htbp]
	\centerfloat{
		\includegraphics[width=0.7\linewidth]{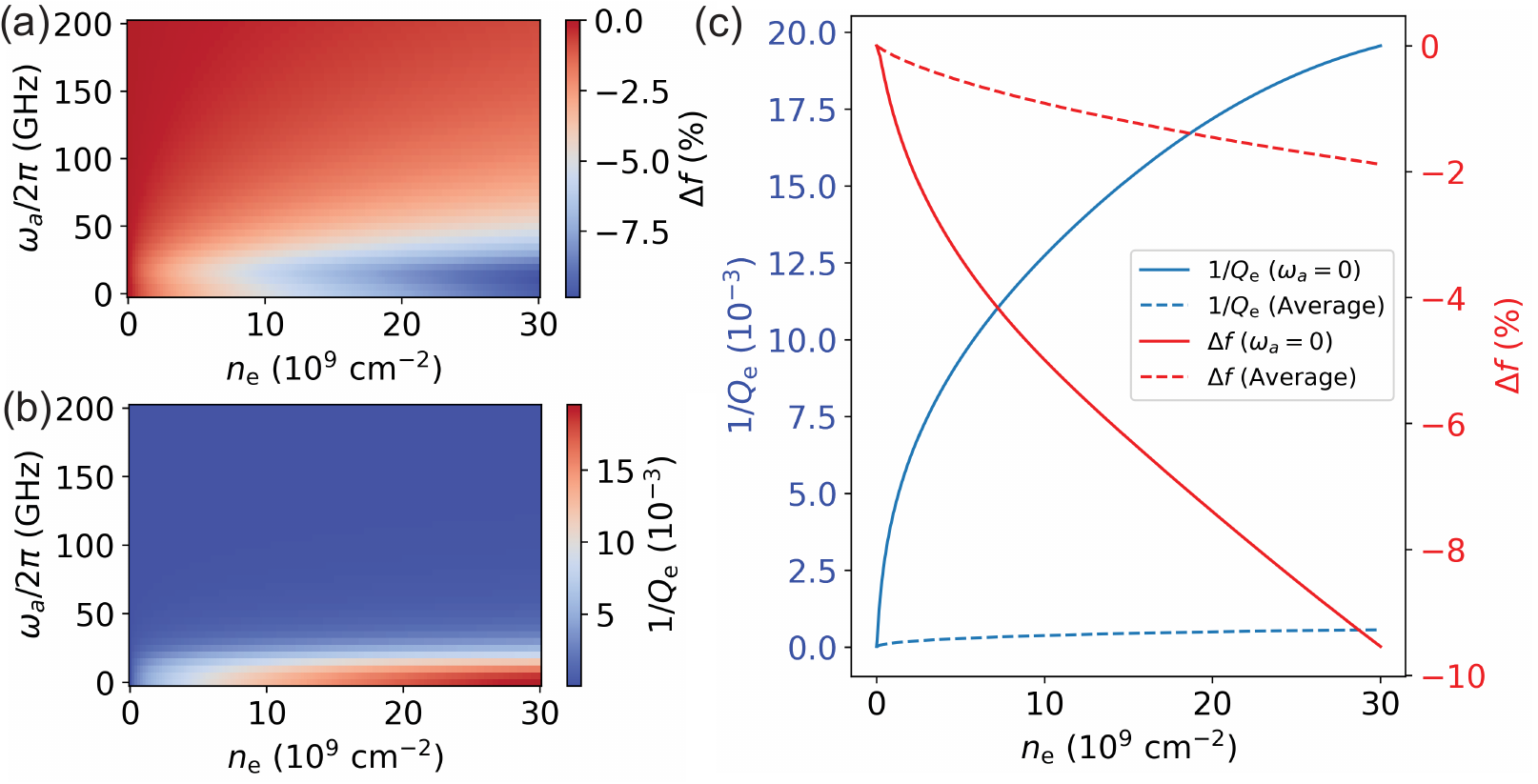}
	}
	\caption[Simulated $\Delta f$ and $1/Q_e$]{%
		(a,b) The resonance frequency shift \(\Delta f\) in (a) and the inverse quality factor \(1/Q_e\) in (b), due to the presence of electrons, are obtained from simulations using Eq.~\ref{eq:cond_COMSOL_omega_a} as a function of the electron density \(n_e\) and the confinement frequency \(\omega_a/2\pi\). 
		(c) Simulated \(1/Q_e\) (blue line) and \(\Delta f\) (red line) as functions of \(n_e\) for \(\omega_a/2\pi = 0\,\mathrm{GHz}\) (solid line) and average over \(\omega_a/2\pi \in [0, 200]\,\mathrm{GHz}\) with weights \(\propto\exp(
		\hbar\omega_a/k_BT)\) (dashed line). 
	}
	\label{fig:f_Q_Sim}
\end{figure}

Using this approach, we simulated the frequency shift \(\Delta f\) and electron-induced internal loss \(1/Q_e\) with \(\tau = 1.9\,\mathrm{ps}\), corresponding to the \(\omega_a = 0\) case in Fig.~\ref{fig:f_Q_Sim}. As discussed in the main text, the resulting \(1/Q_e\) from the simulation is nearly an order of magnitude larger than the experimentally measured value, which remains around \(4 \times 10^{-4}\) at \(3.4\,\mathrm{K}\) and shows no discernible dependence on electron density (Fig.~\ref{fig:e_deposition_neon}(c)).

\subsubsection{Simulation with the presence of electrons: Lorentz model}\label{sec:simu_with_e_Lorentz}

To explain the discrepancy between the measured and simulated internal quality factors, we consider electron localization induced by surface roughness. As one possible modeling approach, we assume that all electrons are confined in harmonic potential wells characterized by a frequency \(\omega_a\). The electron dynamics are then governed by the equation of motion:
\begin{equation}
	m_e \left(\frac{\partial \mathbf{v}}{\partial t} + \frac{\mathbf{v}}{\tau} \right) = -e \mathbf{E} - m_e \omega_a^2  \mathbf{x},
	\label{eq:eq_motion}
\end{equation}
where \(\mathbf{x}\) is the electron's position, taking the origin as the center of the harmonic potential it experiences, with a confinement frequency \(\omega_a\). Here, \(\tau\) is the scattering time, and \(m_e\) is the electron mass. The solution to Eq.~\ref{eq:eq_motion} is
\begin{equation}
	\mathbf{v} =i \omega \mathbf{x}= -\frac{e \tau}{m_e} \frac{1}{1 + i \left( \omega - \frac{\omega_a^2}{\omega} \right) \tau} \mathbf{E}_0 \exp(i \omega t).
\end{equation}
The resulting modified surface conductivity is given by:

\begin{equation}
	\sigma ^{\mathrm{2D}}= \frac{e^2 n_e \tau}{m_e} \frac{1}{1 + i \left( \omega - \frac{\omega_a^2}{\omega} \right) \tau}.
	\label{eq:2D_conductivity}
\end{equation}
To integrate this into the COMSOL model, the electron conductivity is now redefined as
\begin{equation}
	\sigma_{ij} = \frac{1}{\delta z} \frac{e^2 n_e \tau}{m_e}  
	\frac{1}{1 + i \left( \omega - \frac{\omega_a^2}{\omega} \right) \tau}. \label{eq:cond_COMSOL_omega_a}
\end{equation}
Using the same approach as described above, we simulated the frequency shift \(\Delta f\) and electron-induced internal loss \(1/Q_\mathrm{e}\) with \(\tau = 1.9\,\mathrm{ps}\), for trap frequencies \(\omega_a/2\pi\) ranging from \(0\) to \(200\,\mathrm{GHz}\) (Fig.~\ref{fig:f_Q_Sim}(a,b)). Introducing a finite \(\omega_a\) enhances the simulated resonance frequency shift around \(\omega_a \sim \sqrt{\omega/\tau}\), where the imaginary part of \(\sigma^{\mathrm{2D}}\) peaks. Meanwhile, the degradation of the simulated \(Q_{\mathrm{int}}\) is reduced as \(\omega_a\) increases, since the real part of \(\sigma^{\mathrm{2D}}\) decreases monotonically with increasing \(\omega_a\).

\begin{figure}[htbp]
	\centerfloat{
		\includegraphics[width=0.7\linewidth]{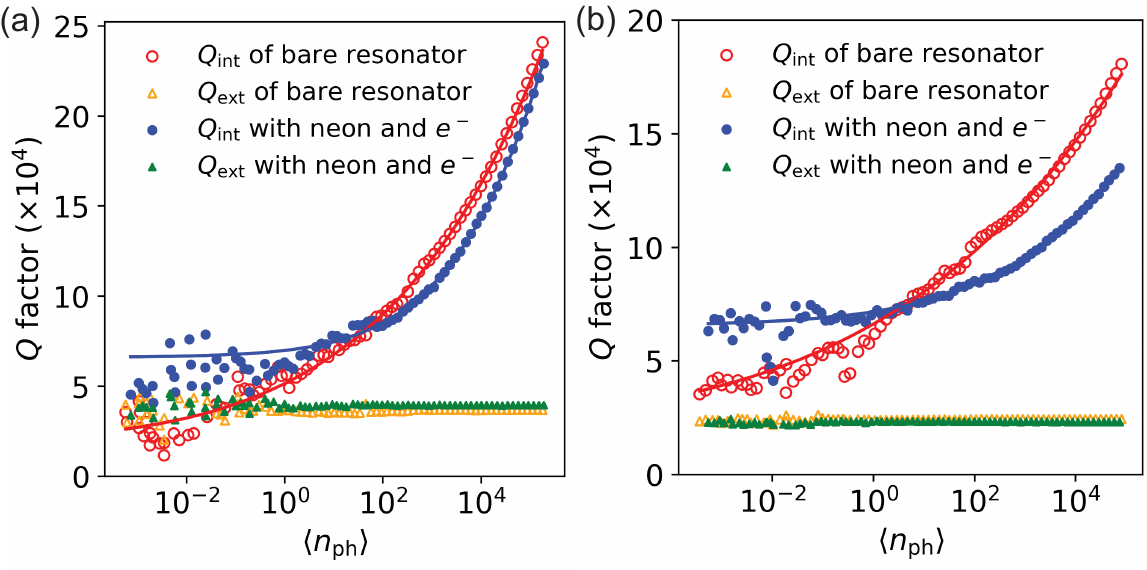}
	}
	\caption[Power dependence of $Q$ vs $\langle n_{\mathrm{ph}}\rangle$]{%
		Measured quality factor power dependence for Resonator~1 in (a) and Resonator~2 in (b) as a function of the number of photons in the resonators, \( \langle n_{\mathrm{ph}} \rangle \). The red open circles and blue-filled circles represent \( Q_\mathrm{int} \) of the bare resonator and the resonator with electrons and neon, respectively. The orange open triangles and green-filled triangles represent \( Q_\mathrm{ext} \) of the bare resonator and the resonator with electrons and neon, and we found that they are independent of \( \langle n_{\mathrm{ph}} \rangle \) with no observable changes before and after the deposition of electrons and neon. The red and blue solid lines are fits to \( Q_\mathrm{int} \) using Eq.~\ref{eq:TLS_loss}. (a) For Resonator~1, \( Q_{\mathrm{TLS, 0}}/F \) increases from \( (1.94\pm0.87)\times10^4 \) to \( (6.64\pm0.92)\times10^4 \), \( n_\mathrm{sat} \) increases from \( (1\pm5)\times10^{-3} \) to \( (3.0\pm1.2)\times10^{2} \), and \( \beta \) increases from \( 0.265\pm0.051 \) to \( 0.377\pm0.063 \) after the deposition of neon and electrons. Estimating \( Q_{\mathrm{other}} \) from the fit is difficult due to the lack of data at high photon numbers. However, we find that \( Q_{\mathrm{other}} \) changes from \( (2 \pm 49) \times 10^7 \) to \( (2 \pm 37) \times 10^7 \) after the deposition of neon and electrons. (b) For Resonator~2, \( Q_{\mathrm{TLS, 0}}/F \) increases from \( (2.2\pm2.2)\times10^4 \) to \( (6.8\pm1.6)\times10^4 \), \( n_\mathrm{sat} \) increases from \( (1\pm14)\times10^{-5} \) to \( (4.7\pm2.8)\times10^{2} \), and \( \beta \) increases from \( 0.190\pm0.060 \) to \( 0.256\pm0.059 \) after the deposition of neon and electrons. Additionally, \( Q_{\mathrm{other}} \) changes from \( (2\pm10)\times10^6 \) to \( (2\pm6)\times10^6 \) after the deposition of neon and electrons. For both resonators, the errors are estimated at the 95\% confidence level. The large uncertainties in \( n_{\mathrm{sat}} \) and \( Q_{\mathrm{other}} \) arise because we did not reach a high enough power to fully saturate the TLS. However, the values of \( n_{\mathrm{sat}} \) and \( Q_{\mathrm{other}} \) did not affect the fitting results of the other parameters.%
	}
	\label{fig:Qint_vs_nph}
\end{figure}

To explain the experimentally observed absence of degradation in \( Q_\mathrm{int} \), we take the following approach. We assume that the number of traps is significantly larger than the number of electrons and that the trap depths \(\omega_a\) are uniformly distributed. The thermal energy at \(T = 3.4\,\mathrm{K}\) is \(k_B T \approx 70\,\mathrm{GHz}\cdot h\). Electrons trapped in deeper potential wells (\(\hbar \omega_a \gg k_B T\)) remain trapped longer, whereas those in shallower traps escape more quickly. Following previous studies on thermal escape time from quantum wells~\cite{Gurioli1992-ms,Swank1963-ap}, we assume that the thermal escape time of an electron from a trap scales as \(\exp(\hbar \omega_a / k_B T)\), and similarly, the probability of an electron occupying a trap at a given \(\omega_a\) follows the same dependence. We calculate and plot \(\Delta f\) and \(Q_\mathrm{e}\) by integrating over trap depths ranging from \(\omega_a/2\pi = 0\) to \(200\,\mathrm{GHz}\). As shown in Fig.~\ref{fig:f_Q_Sim}(c), this model successfully reproduces the experimentally observed behavior where \(Q_\mathrm{int}\) does not degrade at \(\Delta f = -0.9\%\), yielding \(1/Q_\mathrm{e} = 3.7\times10^{-4}\). Note that the effective electron density \(n_e\) in this model refers only to electrons trapped within this frequency range, estimated to be approximately \(9.6\times10^{9}\)\,$\mathrm{cm}^{-2}$ for \(\Delta f = -0.9\%\). Electrons trapped deeper than \(200\,\mathrm{GHz}\) do not significantly contribute to \(\Delta f\) or \(Q_\mathrm{int}\) within this framework as seen in Fig.~\ref{fig:f_Q_Sim}(a,b).

\section{Power dependence of $Q_\mathrm{int}$}
\label{Power_Qint}

Figure~\ref{fig:Qint_vs_nph} shows \( Q_{\mathrm{int}} \) and \( Q_{\mathrm{ext}} \) of Resonators~1 and 2 as a function of the average photon number \( \langle n_{\mathrm{ph}} \rangle \) in the resonator, measured for the bare resonator, and with neon and electrons present. Since this measurement was performed at $10\,\mathrm{mK}$, we assume that the effect of electrons is minimal under our surface conditions. As shown in Fig.~\ref{fig:e_deposition_neon}, the quality factor does not degrade upon electron loading. Thus, the observed variations in \( Q_{\mathrm{int}} \) are primarily associated with the presence or absence of neon. The solid lines in Fig.~\ref{fig:Qint_vs_nph} represent fitting curves based on the following TLS loss model:

\begin{equation}
	\frac{1}{Q_\mathrm{int}} = \frac{F}{Q_\mathrm{TLS,0}\sqrt{1 + \left( \frac{\langle n_{\mathrm{ph}} \rangle}{n_{\mathrm{sat}}} \right)^{\beta}}} + \frac{1}{Q_{\mathrm{other}}},
	\label{eq:TLS_loss}
\end{equation}
where \( F \) is the participation ratio of the electric energy in the region where the TLS exists, \( 1/Q_{\mathrm{TLS},0} \) is the intrinsic TLS loss in the zero-photon and zero-temperature limit, \( n_{\mathrm{sat}} \) is the saturation photon number, and \( \beta \) is an empirical parameter that accounts for TLS dynamics. Since the data points with lower \( \langle n_{\mathrm{ph}} \rangle \) have a lower signal-to-noise ratio (SNR), we set the fitting weight \( w_i \) of each data point \( i \) proportional to its SNR. The fitting results demonstrate a good fit and suggest that \( Q_{\mathrm{int}} \) of the NbTiN resonators is limited by two-level system (TLS) loss ~\cite{Gao2008-ox,muller2022magnetic,bruno2015reducing,barends2010minimal}. With neon and electrons, \( Q_{\mathrm{TLS},0}/F \) increases by a factor of \(\sim3\) compared to the bare resonator. This is likely due to the gap between the resonator and the ground being filled with neon, which has a higher dielectric constant than vacuum~\cite{Lane2020-pc}. As a result, the participation ratio \( F \) decreases~\cite{Gao2008-ox}. Additionally, \( n_\mathrm{sat} \) also increases, indicating that more microwave power is required to saturate the TLS due to the presence of neon. However, the observed changes are larger than expected based on neon’s relative permittivity of 1.244. This suggests that some other mechanism might also be involved, and further study is needed.

\section{Two-tone measurement}
\label{2_tone}
\begin{figure}[htbp]
	\centerfloat{
		\includegraphics[width=0.7\linewidth]{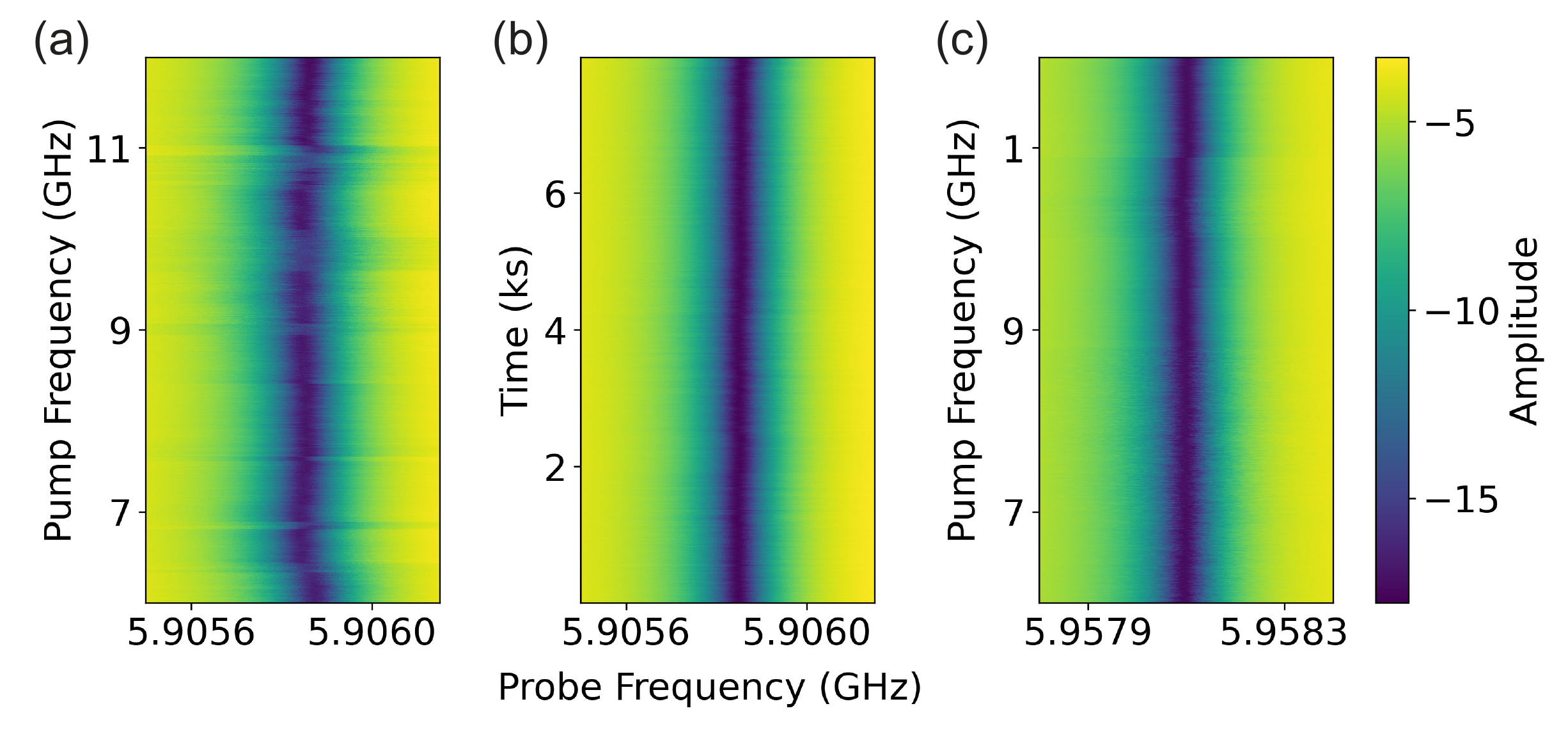}
	}
	\caption[Two-tone spectroscopy at $10\,\mathrm{mK}$]{%
		Resonance peaks of Resonator 2 measured at $10\,\mathrm{mK}$. We observed that the resonance frequency changed only when both neon and electrons were present. (a) Two-tone measurement with both neon and electrons. 
		(b) Stability measurement of the resonance peak over time. (c) Two-tone measurement following the removal of electrons and neon.%
	}
	\label{fig:two-tone}
\end{figure}

Figure~\ref{fig:two-tone} shows the two-tone measurement on Resonator~2. At room temperature, a probe signal with a frequency near the resonance frequency is combined with a pump signal in the $6-12\,\mathrm{GHz}$ range and injected into port~1. The pump signal excites from the ground orbital state to the orbital excited state when their energy difference corresponds to the pump signal frequency, causing a shift in the resonance frequency.

\section{Micromagnet Simulation}
\label{Mag_sim}

The numerical calculation of the local magnetic fields in different directions is done by modeling each Co micromagnet as a uniformly magnetized rectangular block. We treat the block as a sum of many small magnetic dipoles and integrate their fields over the block’s length, width, and thickness~\cite{Goldman}. 

In this work, we evaluate two Co blocks with magnetization $M=1.7\,\text{T}$, dimensions $1.5\,\mu\text{m}\times1.5\,\mu\text{m}\times t$, where the $t$ represents for different thickness to be simulated. The magnetic field along the axis of interest was calculated with the Magpylib Python package.

\begin{figure}[htbp]
	\centerfloat{
		\includegraphics[width=0.7\linewidth]{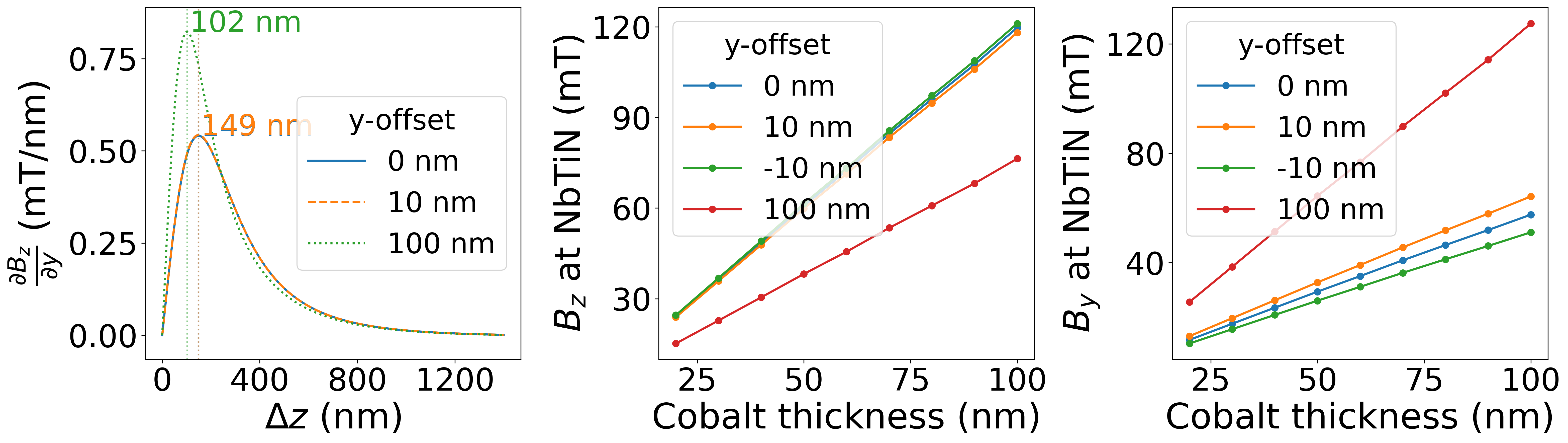}
	}
	\caption[Magnet misalignment simulation]{%
		(a) $\frac{\partial B_z}{\partial y}$ as a function of $\Delta z$ for a 100-nm-thick cobalt magnet with misalignment along the $y$-direction. (b) $B_z$ and (c) $B_y$ at the resonator position as functions of cobalt thickness, both with misalignment along the $y$-direction.%
	}
	\label{fig:mag_sim}
\end{figure}

As shown in Fig.~\ref{fig:micro_magnet_neon}(b), $d B_z/d y$ is maximized for $\Delta z \approx 146$~nm. We also simulate the displacement sensitivity. A 10 nm shift of the micromagnets along $y$ changes $dB_z/dy$ only slightly (orange dashed line in Fig.~\ref{fig:mag_sim}(a)), this indicates small misalignments have little effect. A larger 100 nm shift (green dotted line) increases $dB_z/dy$ at the same $\Delta z$. Thus, within normal fabrication tolerances, the gradient—and therefore the coupling strength—can be maintained or even improved as the electron moves closer to one of the magnets.

The absolute magnetic fields with misalignment along the $y$-direction were also simulated (Fig.~\ref{fig:mag_sim}(b,c)). The fields at the resonator position stay well below the NbTiN critical field. The field at the resonator position remains well below the critical field of NbTiN. Therefore, the micromagnets are not expected to degrade the resonator’s quality factor. In addition, we note that displacement along the $x$-direction has a negligible impact on both the magnetic field and its gradient.

\section{Kinetic inductance}\label{K_in}
From separate measurements at room temperature, we estimated the magnetic penetration depth at zero temperature for the NbTiN film with a thickness of \( D = 20\,\mathrm{nm} \), used in the experiments, to be \( \lambda = 390\,\mathrm{nm} \). The inductance per square \( L_{\square} \) is calculated as \(
L_\square = \frac{\mu_0 \lambda^2}{D} =9.6 \, \mathrm{pH}.
\)
The total inductance of the resonator is determined by multiplying \( L_\square \) by the number of squares that fit within the geometry. Thus, the kinetic inductance of Resonator~1 is calculated as \(
L = \frac{L_\square l}{w} = 9.6\, \mathrm{pH} \cdot \frac{1.45 \, \mathrm{mm}}{100 \, \mathrm{nm}} = 139 \, \mathrm{nH},
\) where $l$ and $w$ are the length and the width of the resonator, respectively. As shown in Table~\ref{tab:reso_table_neon}, the resonance frequencies obtained from both simulation and experiment are in good agreement for Resonator 1, thereby validating this inductance estimation.

\chapter{Appendix for the TDO experiment}

\section{Individual differences in BD-6 tunnel diodes}\label{secA:Indiv_diff}

\begin{figure}[htbp]
	\centerfloat{
		\includegraphics[width=0.9\linewidth]{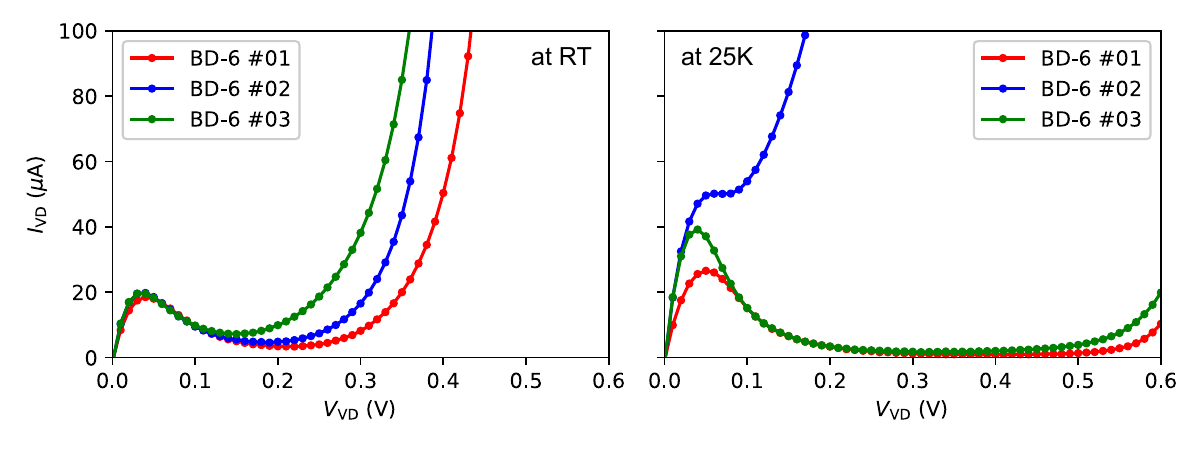}
	}
	\caption[I-V curves of BD-6 tunnel diodes]{%
		I-V curves of three different BD-6 tunnel diodes measured at RT and 25 K. The dots represent measurement points, and the lines connect these points to guide the eye. This measurement was performed using bare tunnel diodes, not those connected to an LC resonator for oscillation. As a result, no oscillation occurs, and thus no hysteresis is observed, unlike in Fig.~\ref{fig:circuit}(d)%
	}
	\label{fig:Indiv_diff}
\end{figure}

Fig.~\ref{fig:Indiv_diff} compares the I-V curves of different BD-6 tunnel diodes measured at 25~K and RT. As stated in the BD-6 data sheet, the peak current at RT is 20~$\mu$A, which is consistent across all BD-6 diodes. However, at 25~K, significant variations among individual devices become apparent. Notably, BD-6 \#02 does not exhibit negative resistance at low temperatures, meaning that it cannot be used for the cryogenic microwave source. The BD-6 diode used in this study exhibits a low-temperature I-V characteristic similar to that of BD-6 \#03 but is a different individual device. Typically, a temperature of 25~K is low enough to evaluate and select a suitable BD-6 tunnel diode.

\section{Temperature dependence}\label{secA1}

We conducted detailed measurements of the temperature dependence of the oscillation frequency using a similar circuit. This circuit utilized a BD-6 diode with a parasitic capacitance of 1.7~pF and a 50-turn Nb spiral inductor with an inductance of 1.28~$\mu$H, and did not include a varactor diode. The resulting oscillation frequency was 107~MHz. The circuit described in this work exhibited a comparable temperature dependence based on preliminary measurements.

This circuit is thermally anchored to the 10~mK stage, or the Mixing Chamber (MC) stage, and the MC stage temperature (MCT) was controlled while measuring a thermometer that is also thermally anchored to the MC stage. The sensitivity was found to be 0.3~kHz/mK in the MCT range of 60~mK to 120~mK, with no detectable dependence below 60~mK, aside from a slight hysteresis. This result is significantly better than the temperature dependence reported in previous studies above 1~K~\cite{van1974sensitive}.  

The observed temperature dependence between 60~mK and 120~mK is unlikely to be due to changes in the tunnel diode's I-V curve, as no significant variation in its characteristics was observed in this range. Therefore, it is improbable that frequency shifts result from a change in the operating point. In prior studies measuring above 1~K~\cite{van1974sensitive}, temperature dependence was attributed to the surface impedance, thermal expansion, and magnetic susceptibility induced by iron impurities in the copper used in the inductor. In contrast, this work uses a superconducting material, Nb, with a critical temperature $T_c$ of 6 K, in the inductor. Given that the measured temperature range is significantly below $T_c$, it is unlikely that the temperature dependence is primarily due to changes in the properties of Nb, though some contributions cannot be ruled out. Alternatively, the temperature dependence might arise from changes in the physical properties of the metals used in the PCB substrate.

Moreover, we note that we did not directly measure the temperature of the circuit itself, and its actual temperature may differ from MCT. The absence of detectable temperature dependence below 60~mK may be due to the fact that the diode’s temperature is likely saturated by the liberated heat and does not cool further.  Please also note that the TDO circuit described in the main text is similarly thermally anchored to the MC stage, and its temperature was measured using a thermometer separately thermally anchored to the MC stage. However, while the qubit must be cooled to 10~mK, the TDO circuit does not necessarily need to be at 10~mK. For the purposes discussed in this work, this is not a concern.

\section{Varactor diode IV curve}\label{secA:varactor}

We measured the IV curve of the varactor diode at 11~mK. The range in which the leak current is suppressed matches the range in which the output power remains constant (Fig.~\ref{fig:freq_tunability}).

\begin{figure}[htbp]
	\centerfloat{
		\includegraphics[width=0.6\linewidth]{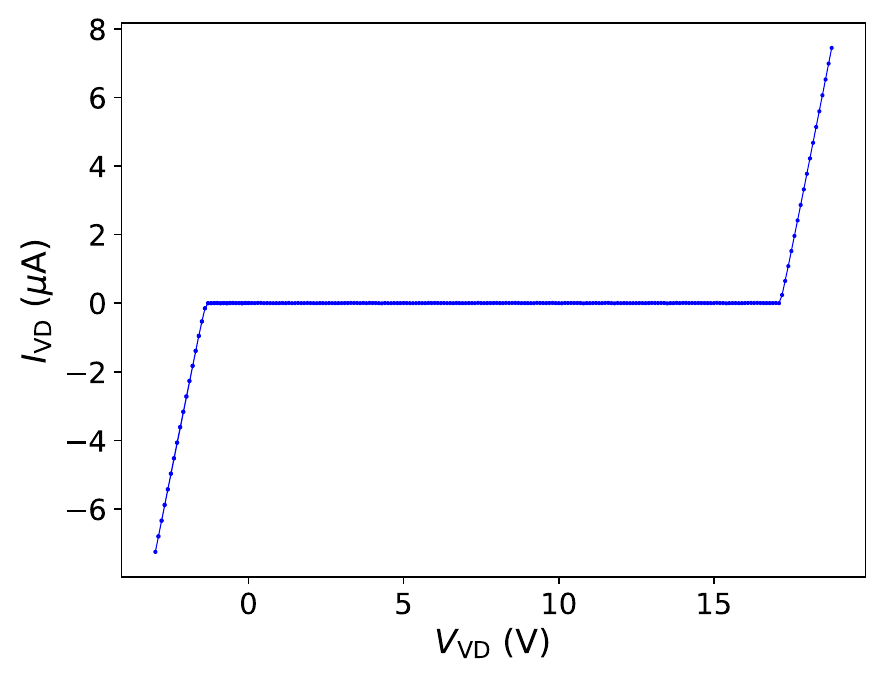}
	}
	\caption[IV curve of the varactor diode]{%
		IV curve of the varactor diode  (MA46H201, MACOM) measured at 11 mK. For $-1.3~V<V_\mathrm{VD}<17~V$, the leakage current is significantly small.%
	}
	\label{fig:varactor}
\end{figure}

\section{DC voltage source stability}\label{secA:DC_source}

\begin{figure}[htbp]
	\centerfloat{
		\includegraphics[width=0.9\linewidth]{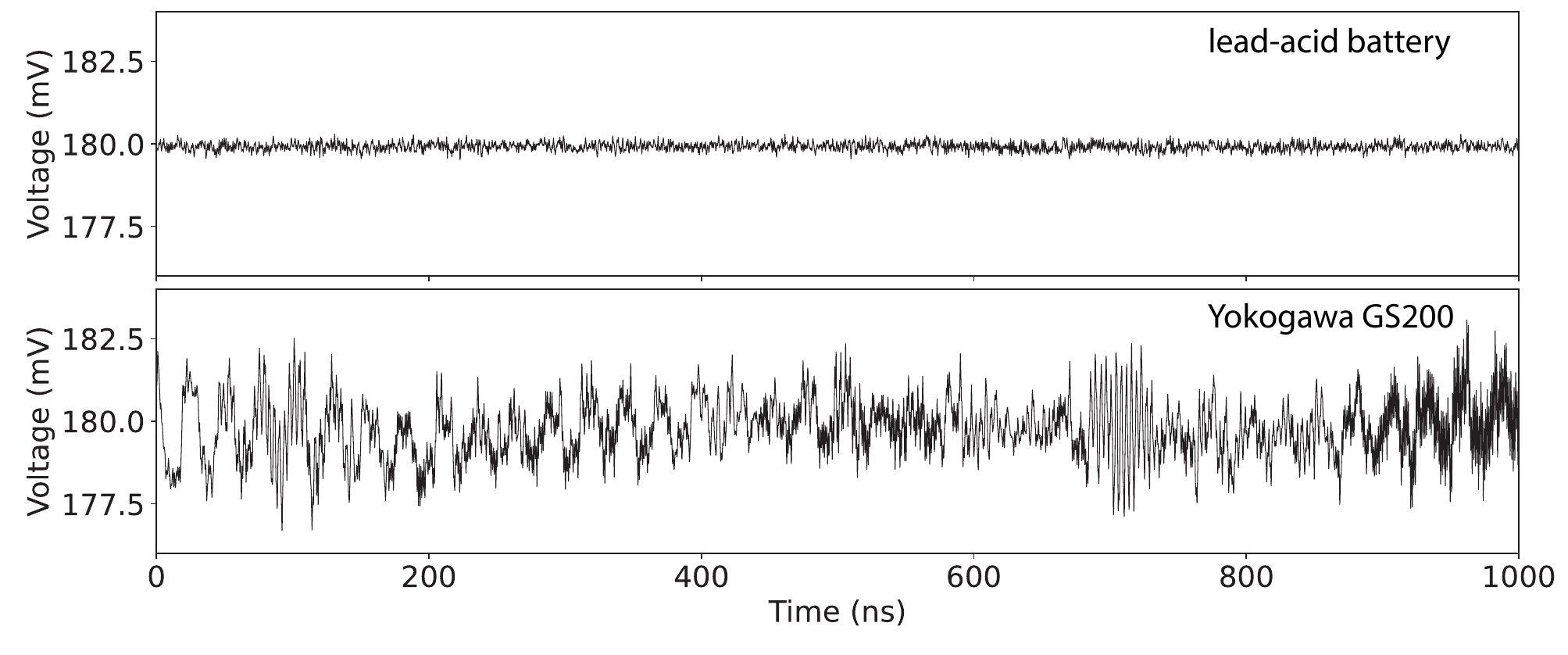}
	}
	\caption[DC source voltage stability comparison]{%
		Comparison of voltage stability between the Yokogawa GS200 and a lead-acid battery%
	}
	\label{fig:DC_comparasion}
\end{figure}

We discovered that the DC voltage source (Yokogawa GS200) was significantly affected by strong electromagnetic interference from a nearby radio station broadcasting at 810~kHz. The impact of the 810~kHz signal was directly observed in the output voltage of the Yokogawa GS200 (Fig.~\ref{fig:DC_comparasion}) and also in the frequency stability of the TDO.

Despite attempts to mitigate the interference using ferrite coils, shielding the lines used for delivering the DC voltage, and shielding the laboratory where the experiment was conducted, the 810~kHz radio signal in the lab was too strong to be removed, likely due to the radio station being only 200~m away. As a result, we decided to switch to using a lead-acid battery, which is not affected by the 810~kHz radio signal. Consequently, as shown in Fig.~\ref{fig:phase_noise}, the phase noise of the TDO was significantly improved.

\section{Time-domain data}\label{secA:time-domain}

\begin{figure}[htbp]
	\centerfloat{
		\includegraphics[width=0.9\linewidth]{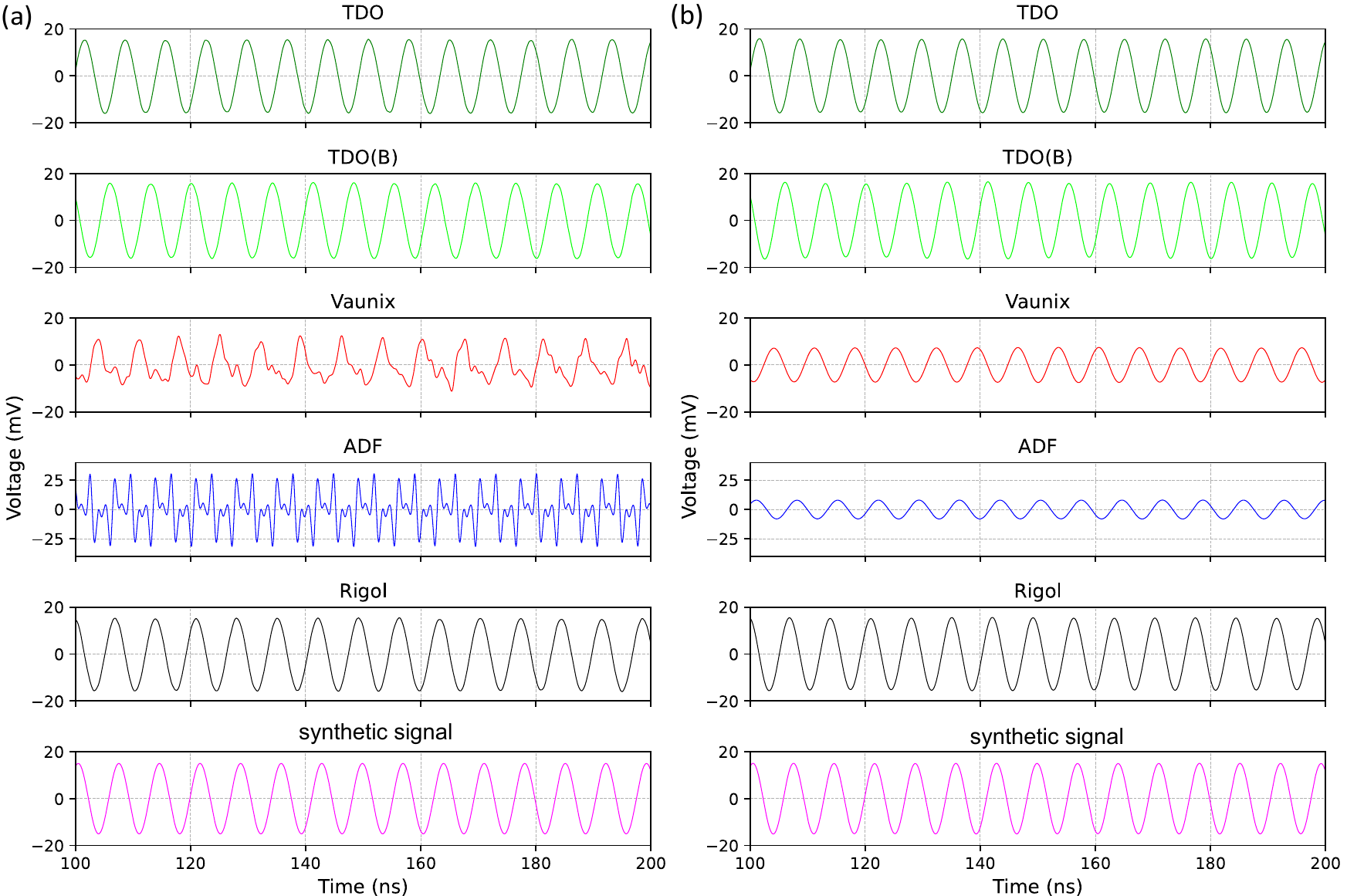}
	}
	\caption[Time-domain oscilloscope traces]{%
		Time-domain data measured via AC port II using a 20~GS/s sampling rate oscilloscope (LeCroy WaveRunner 9054). The data include signals from the TDO powered by Yokogawa GS200 DC source (green) and a lead-acid battery (light green), as well as four different commercial microwave sources: Rigol DSA815 (black), Analog Devices ADF4351 (blue), and Vaunix Lab-Brick LMS-451D (red). The respective carrier frequencies are $f_0=$141.800~MHz, 141.841~MHz, 141.802~MHz, 141.806~MHz, and 141.803~MHz. For the TDO, $V_\mathrm{TD}=0.18$~V, and $V_\mathrm{VD}=-0.5$~V. For a consistency check, a synthetic sine signal at 141.8~MHz with 15~mV amplitude is also plotted. The raw signal is shown in (a), while (b) presents the signal after applying a digital bandpass filter with a 50~MHz bandwidth centered on the carrier frequency.%
	}
	\label{fig:time-domain}
\end{figure}

The time-domain data were sampled at 20~GS/s using an 8-bit oscilloscope over a duration of 1.6~ms. 100~ns segments are shown in Fig.~\ref{fig:time-domain}. The signals from the ADF and Vaunix generators contain many harmonics, but their influence is eliminated after applying a digital filter (Fig.~\ref{fig:time-domain}). Such harmonics could become an issue when generating signals for multiplexing using these generators. On the other hand, the signal from the TDO shows no observable harmonics, giving it an advantage in this regard.

\section{Power spectra}\label{secA:power_spect}

\begin{figure}[htbp]
	\centerfloat{
		\includegraphics[width=0.9\linewidth]{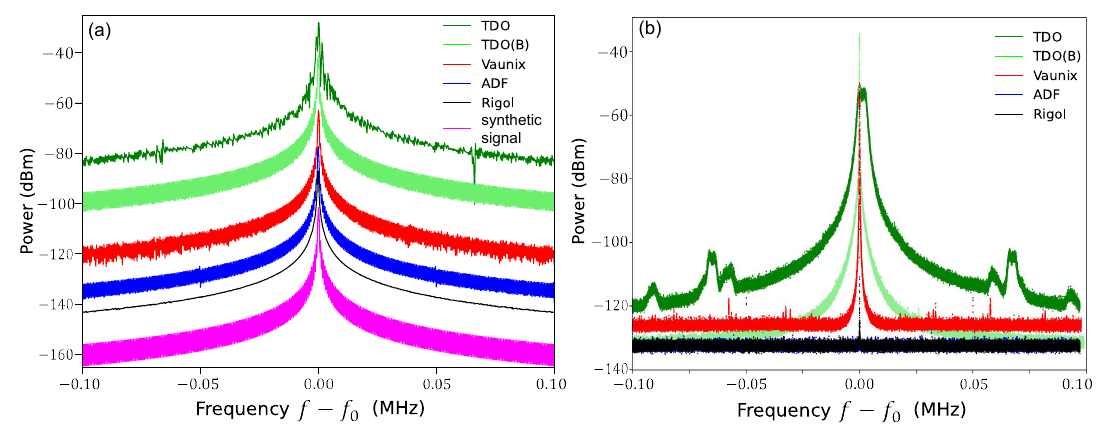}
	}
	\caption[Power spectra from time- and frequency-domain measurements]{%
		(a) Power spectra calculated from the time-domain signal measured at 20~GS/s over a duration of 1.6~ms (see Fig.~\ref{fig:time-domain}). The carrier frequency is the same as in Fig.~\ref{fig:time-domain}. For clarity, 15 dB offsets have been introduced, as the original data are at similar dBm levels. (b) Power spectra measured using a spectrum analyzer (R\&S FSV3030) with a resolution bandwidth of 5~Hz. The carrier frequencies are $f_0 = $ 141.80~MHz and 141.86~MHz for the TDO powered by Yokogawa GS200 DC source (green) and a lead-acid battery (light green), respectively, and $f_0 = $ 141.80~MHz, 141.81~MHz, and 141.80~MHz for the Rigol DSA815 (black), Analog Devices ADF4351 (blue), and Vaunix Lab-Brick LMS-451D (red), respectively. In both (a) and (b), the x-axes represent the frequency offset from the center carrier frequency $f_0$.%
	}
	\label{fig:power_spectra}
\end{figure}

Fig.~\ref{fig:power_spectra}(a) shows the power spectra obtained from the time-domain signal in Fig.~\ref{fig:time-domain} with an integration time of 1.6~ms, while Fig.~\ref{fig:power_spectra}(b) shows the power spectra measured using a spectrum analyzer with an RBW of 5~Hz. In Fig.~\ref{fig:power_spectra}(a), little difference is observed between the TDO, commercial microwave sources, and even the synthetic signal. This is because the high-frequency phase noise (see Fig.~\ref{fig:phase_noise}) is low and relatively similar for all cases. The differences observed in Fig.~\ref{fig:power_spectra}(b) must originate from variations in low-frequency phase noise.

To obtain the power spectra shown in Fig.~\ref{fig:power_spectra}(a), the discrete Fourier transform with zero-padding was computed using the following formula:

\begin{equation}
	X_k = \sum_{n=0}^{N-1} x_n \exp\left( -j 2 \pi \frac{nk}{N} \right),
\end{equation}
where the time-domain data $x_n$ is shown in Fig.~\ref{fig:time-domain}(b) and 
$X_k$ represents the Fourier coefficients. The root-mean-square amplitude in volts was calculated as:

\begin{equation}
	\bar{X_k} = \frac{1}{\sqrt{2}} \frac{|X_k|}{N/2},
\end{equation}
where $|X_k|$ is the magnitude of the Fourier coefficient and $N$ is the total number of samples. The power spectrum in dBm, assuming a 50~$\Omega$ system, was computed using:

\begin{equation}
	P = 10 \log_{10} \left( \frac{\bar{X_k}^2}{0.001 \cdot 50} \right),
\end{equation}
where $\bar{X_k}^2/50$ represents the power in watts.

\end{document}